\documentclass[final,1p,times,twocolumn]{elsarticle}

\usepackage{epsfig}
\usepackage{bm}
\usepackage{epstopdf}

\usepackage{amssymb}
\usepackage{amsmath}

 \usepackage{lineno}

\journal{Progress of Particle and Nuclear Physics}

\makeatletter
\@addtoreset{equation}{section}

\makeatother

\newcommand{\LL}{\Lambda\Lambda}
\newcommand{\sLL}{{\scriptscriptstyle \LL}}
\newcommand{\Kmp}{K^-p}
\newcommand{\Kbn}{\bar{K}^0n}
\newcommand{\sKmp}{{\scriptscriptstyle \Kmp}}
\newcommand{\sKbn}{{\scriptscriptstyle \Kbn}}
\newcommand{\calS}{\mathcal{S}}
\newcommand{\DBLL}{\Delta B_{\Lambda\Lambda}}
\newcommand{\reff}{r_\mathrm{eff}}
\newcommand{\fm}{\mathrm{fm}}
\newcommand{\MeV}{\mathrm{MeV}}

\usepackage[normalem]{ulem}  
\usepackage{color} 

\renewcommand\sout{\bgroup \color{red} \ULdepth=-.5ex \ULset}
\newcommand{\Comment}[1]{}

\usepackage{longtable}
  \makeatletter
    
    \@addtoreset{figure}{section}
  \makeatother
 \makeatletter
    
   \@addtoreset{table}{section}
  \makeatother

\begin{document}

\begin{frontmatter}



\title{Exotic Hadrons from Heavy Ion Collisions\tnoteref{Report}}
\tnotetext[Report]{Report No.:YITP-16-120}

\author[Kangwon]{Sungtae Cho}
\ead{sungtae.cho@kangwon.ac.kr}
\author[YITP]{Tetsuo Hyodo}
\ead{hyodo@yukawa.kyoto-u.ac.jp}
\author[Metropolitan]{Daisuke Jido}
\ead{jido@tmu.ac.jp}
\author[Texas]{Che Ming Ko}
\ead{ko@comp.tamu.edu}
\author[Yonsei]{Su Houng Lee}
\ead{suhoung@yonsei.ac.kr}
\author[Tokyotech]{Saori Maeda}
\ead{s-maeda@th.phys.titech.ac.jp}
\author[Kyoto]{Kenta Miyahara}
\ead{miyahara@ruby.scphys.kyoto-u.ac.jp}
\author[YITP]{Kenji Morita}
\ead{kmorita@yukawa.kyoto-u.ac.jp}
\author[SaoPaulo]{Marina Nielsen}
\ead{mnielsen@if.usp.br}
\author[YITP]{Akira Ohnishi}
\ead{ohnishi@yukawa.kyoto-u.ac.jp}
\author[JAEA]{Takayasu Sekihara}
\ead{sekihara@post.j-parc.jp}
\author[FIAS]{Taesoo~Song}
\ead{song@fias.uni-frankfurt.de}
\author[Tokyotech]{Shigehiro Yasui}
\ead{yasuis@th.phys.titech.ac.jp}
\author[RIKEN]{Koichi Yazaki}
\ead{koichiyzk@yahoo.co.jp}
\author[]{\\(ExHIC Collaboration)}

\address[Kangwon]{Division of Science Education, Kangwon National University, Chuncheon 200-701, Korea}
\address[YITP]{Yukawa Institute for Theoretical Physics, Kyoto University, Kyoto, 606-8317, Japan}
\address[Metropolitan]{Department of Physics, Tokyo Metropolitan University, Hachioji 192-0397, Japan}
\address[Texas]{Cyclotron Institute and Department of Physics and Astronomy, \\ Texas A\&M University, College Station, Texas 77843, USA}
\address[Yonsei]{Department of Physics and Institute of Physics and Applied Physics, Yonsei University, Seoul 03722, Korea}
\address[Tokyotech]{Department of Physics, Tokyo Institute of Technology, Tokyo 152-8551, Japan}
\address[Kyoto]{Department of Physics, Graduate School of Science, Kyoto University, Kyoto 606-8502, Japan}
\address[SaoPaulo]{Instituto de F\'isca, Universidade de S\~ao Paulo, C.P. 66318, 05389-970 S\~ao Paolo, SP, Brazil}
\address[JAEA]{Advanced Science Research Center, Japan Atomic Energy Agency, Tokai, Ibaraki 319-1195, Japan}
\address[FIAS]{Frankfurt Institute for Advanced Studies and Institute for Theoretical Physics, Johann Wolfgang Goethe Universtit\"at, Frankfurt am Main, Germany}
\address[RIKEN]{RIKEN Nishina Center, Hirosawa 2-1, Wako, Saitama 351-0198, Japan}



\begin{abstract}
High energy heavy ion collisions are  excellent ways 
for producing heavy hadrons and composite particles, including the light (anti)nuclei.
With upgraded detectors at the Relativistic Heavy Ion Collider (RHIC) and the Large Hadron Collider (LHC), it has become possible to measure hadrons beyond their ground states. Therefore, heavy ion collisions provide a new method for studying exotic hadrons that are either molecular states made of various hadrons or compact system consisting of multiquarks. Because their structures are related to the fundamental properties of Quantum Chromodynamics (QCD), studying exotic hadrons is currently one of the most active areas of research in hadron physics.  
Experiments  carried out at various  accelerator facilities have  indicated that some exotic hadrons may have already been produced.   
The present review is a summary of the current understanding of a selected set of exotic particle candidates that can be potentially measured in heavy ion collisions.  It also includes discussions on the production of resonances, exotics and hadronic molecular states in these collisions based on the coalescence model and the statistical model.  A more detailed discussion is given on the results from these models, leading to the conclusion that the yield of a hadron that is a compact multiquark state is typically an order of magnitude smaller than if it is an excited hadronic state with normal quark numbers or a loosely bound hadronic molecule.  Attention is also given to some of the proposed heavy exotic hadrons that could be produced with sufficient abundance in heavy ion collisions  because of the significant numbers of charm and bottom quarks that are produced at RHIC and even larger numbers  at LHC,  making it possible to study them in these experiments.  Further included in the discussion are the general formalism for the coalescence model that involves resonance particles and 
its implication on the present estimated yield for resonance production. Finally, a review is given on recent studies to constrain the hadron-hadron interaction through correlation measurements in heavy ion collisions and their implications on the interpretation and the possible existence of exotic states in hadronic interactions.

\end{abstract}


\end{frontmatter}


\tableofcontents


\section{Introduction} 
\label{sec:introduction}

High-energy heavy-ion collisions provide a unique opportunity to study the properties of the high energy density QCD matter formed at the initial stage of the  collisions. It is now well established from the study of jet quenching and anisotropic flow~\cite{Braun-Munzinger:2014pya} that the produced high energy density matter is strongly interacting with a very small specific viscosity.  
These collisions at the same time provide the suitable conditions for producing weakly bound hadronic states such as the light nuclei, hypernuclei~\citep{Abelev:2010rv} and anti-nuclei~\citep{Agakishiev:2011ib}.    An interesting  experimental finding is that the yields of these  nuclei follow the statistical model predictions with temperature and chemical potentials that are fitted to the yields of the ground state particles~\cite{Adam:2015vda}.    These results thus  suggest that the final abundance of  ground state hadrons and  light nuclei are already determined near the energy density at which quarks and gluons hadronize.   However, when analyzing particles beyond the ground states, one finds that their yields sometimes deviate from the statistical model predictions, which, on the other hand,  may  reveal valuable information about the evolution of the hadronic system after the hadronization of quarks and gluons and/or the structure of these particles.  For example, resonances with large width that are reconstructed from daughter particles are found to be less produced than the statistical model prediction~\citep{Adam:2016bpr}, suggesting that hadronic interactions and their freeze-out  conditions play important roles in determining the final yields. Moreover, depending on their  quantum numbers, the yields of excited states can be either suppressed~\citep{KanadaEn'yo:2006zk} or enhanced~\citep{Cho:2014xha} relative to those predicted by  the statistical model.  These results suggest the importance of the structure of a  hadron on its yield in heavy ion collisions.  
 
\begin{figure}[tbp]
\begin{center}
\includegraphics[bb=0 0 306 503,angle=90,height=5cm]{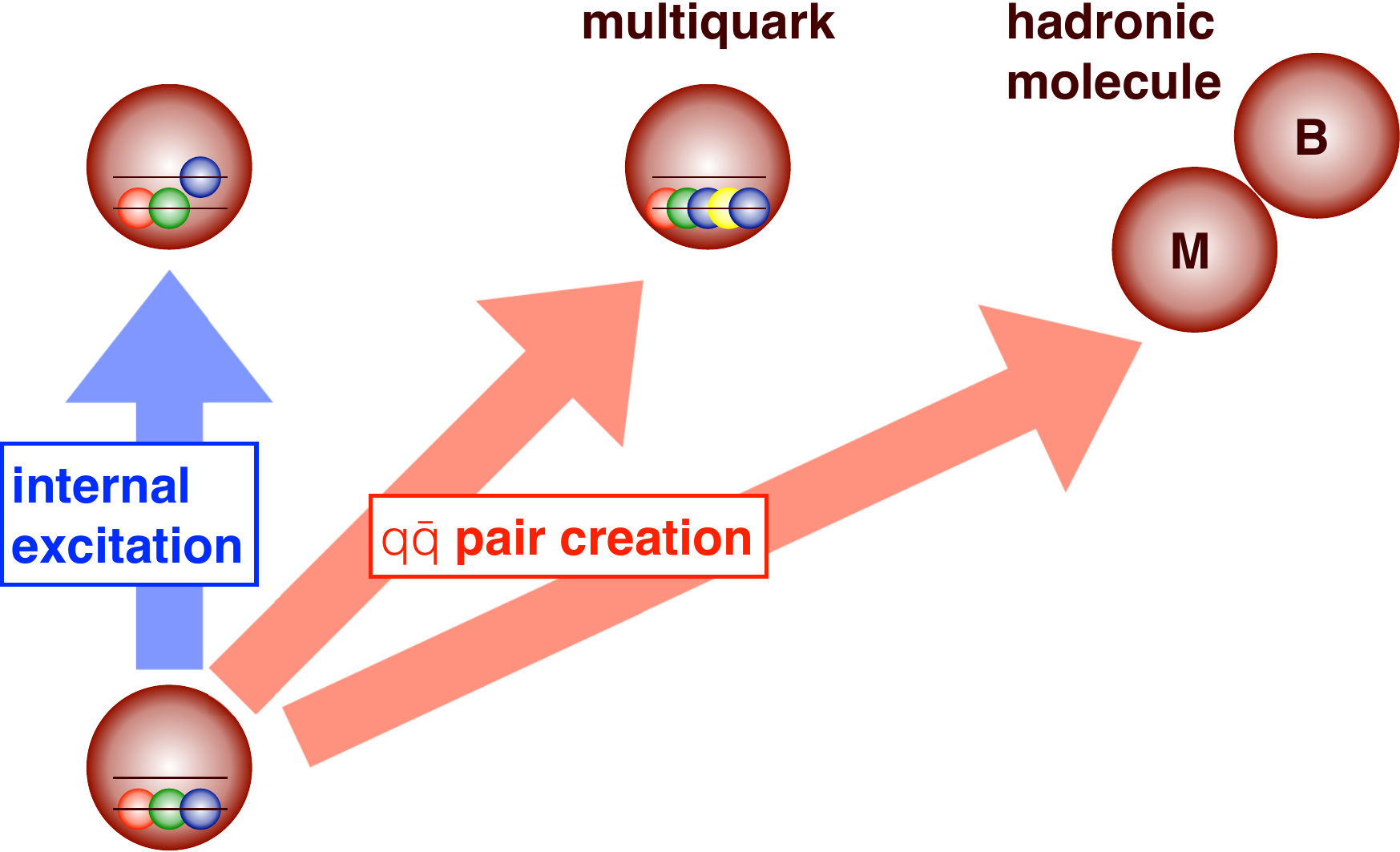}
\caption{Schematic figure for a cryptoexotic state. }
\label{fig:schematic}
\end{center}
\end{figure}

During the past decade, there has also been a revival of hadron physics research due to the observation of many exotic state candidates.  These findings started from the observation of  $D_{sJ}$(2317)~\cite{Aubert:2003fg} and X(3872)~\citep{Choi:2003ue} and continued on to the recently observed pentaquark states  at the LHC~\cite{Aaij:2015tga}.  Hadrons of such exotic structures have been proposed since the early days of bag models as the color confinement does not rule out the existence of a multiquark configuration  in a color singlet state~\citep{Jaffe:1976ig,Jaffe:1976ih}.  A multiquark configuration can be either a flavour exotic state or a cryptoexotic state.  For the latter, one cannot distinguish its structure among an internally excited state, compact multiquark state or a molecular configuration, as shown in Fig.~\ref{fig:schematic} based just on its quantum numbers. Even for flavor exotic states, a multiquark configuration can always be decomposed into meson and baryon states, making it impossible to discriminate a compact multiquark state from a hadronic molecular state just from its quantum numbers.
Moreover, some observed structures from an elementary process that have been interpreted as exotic states could be caused by kinematical effects.
It is worth to note that the existence of stable tetraquark configurations can also influence the properties of QCD  at finite temperature and density as the tetraquark condensation may lead to a second chiral phase transition~\cite{Pisarski:2016ukx}.

In Refs.~\citep{Cho:2010db,Cho:2011ew} published by the authors of present review, it is   
proposed that heavy-ion collisions at ultrarelativistic energies provide a unique opportunity to study exotic particle candidates.  Due to  the abundant number of heavy 
quarks and antiquarks produced in these collisions, various exotic hadrons could be formed, and 
their yields have been estimated by using 
the quark coalescence model.  When the parameters in the model are fit to 
the yields of ground state hadrons as predicted from the statistical model, it has been found 
that the yield of a hadron is typically an order of magnitude smaller when it is a compact multiquark state than that of an excited hadronic state with normal quark numbers and/or a molecular configuration. Combined with the fact that only resonances with large natural width are affected by the subsequent hadronic evolution, it is thus possible to determine if an exotic hadron produced in  
relativistic heavy ion collisions is a compact multiquark state or has a hadronic molecular configuration. Also, owing to the significant numbers of charm and bottom quarks produced at RHIC and even larger numbers at LHC, some of the proposed and recently measured heavy exotic hadrons could be produced with sufficient abundance for experimental detection, making it possible to study these new exotic hadrons in heavy ion collisions. 
Moreover, the structures in the invariant mass spectrum of a specific decay channel of, for example, a B meson, which are generated by kinematic effects, cannot be produced statistically and thus will not appear in heavy ion collisions. Therefore, heavy ion collisions also make it 
possible to discriminate such effects from real resonances.   

Another important recent development is the measurement of two-particle momentum correlations in relativistic heavy ion collisions. In recent experiments, correlations have been measured for particle pairs such as $p\bar{p}$~\cite{Adamczyk:2015hza}, $\bar{p}\bar{p}$~\cite{Adamczyk:2015hza}, 
$p\Lambda$~\cite{NA49_protonlambda,STAR_p-lambda2006,HADES_p-lambda2016},
$\bar{p}\Lambda (p\bar{\Lambda})$~\cite{STAR_p-lambda2006}, and $\Lambda\Lambda$~\cite{Ahn:1998fj,Yoon:2007aq,Adamczyk:2014vca,Adam:2015nca,Kim:2013vym} in addition to $\pi\pi$, $KK$ and $pp$. The two-particle correlation is generated by quantum statistics and final state interactions, and it also depends on the size and lifetime of the emission source~\cite{Koonin:1977fh,Lednicky:1981su,Bauer:1993wq,Lednicky:2005tb,Lisa:2005dd,Pratt:2008qv}. Therefore, one can use, on the one hand, the two-particle correlation to determine the source size if the interaction between the two particles is known. On the other hand, it is possible to extract information on the interactions
between two particles by using the experimental data on their correlation functions~\cite{Greiner:1989ig,Ohnishi:1998at,Kisiel:2014mma,Shapoval:2014yha,Ohnishi:2016elb,Morita:2014kza,Morita:2016auo}, if the property of their emission source is known. The latter provides, in particular, a unique opportunity to explore interactions between short-lived hadrons, such as the meson-meson, hyperon-hyperon and antibaryon-antibaryon interactions, which can serve as crucial inputs for understanding possible exotic hadronic states, such as the hadronic molecules and dibaryons. While the $\Lambda\Lambda$ interaction is accessible via the double $\Lambda$ hypernuclei~\cite{Takahashi:2001nm}, there is no other ways to access other short-lived hadron-hadron interactions experimentally. In heavy ion collisions, the freeze-out conditions are well studied and the chaotic source assumption is known to work reasonably. It is thus  possible to constrain various hadron-hadron interactions from the experimental data on two-particle correlations.

This review is organized as follows. It begins with a survey of the current status on our understanding of exotic hadrons. This is followed by a discussion on the general issues related to particle production in heavy-ion collisions. An updated account is then given on the yields of some potentially exotic hadrons that could be measured in experiments on heavy-ion collisions. 
Next, we will present the new development in the coalescence model for resonances.
Also, reviews are given on the basic theoretical framework and the current status on the study of  hadron-hadron interactions using two-particle correlations. 

\section{Current status of exotic hadrons}
\label{sec:exotic_hadrons}
\subsection{Light hadrons}
\label{sec:light_hadrons}

\subsubsection{Scalar mesons} 
\label{sec:scalar_meson}

The scalar mesons $f_{0} (980)$ and $a_{0} (980)$, together with
$f_{0} (500)~ (\sigma)$ and $K_{0}^{\ast} (800)~ (\kappa)$, have been
thought to have some exotic structures, since they exhibit an inverted
mass spectrum compared to what is expected if they have simple  $q \bar{q}$
configurations. Model studies have suggested that $f_{0} (980)$ and
$a_{0} (980)$ could be compact $q q \bar{q} \bar{q}$
systems~\cite{Jaffe:1976ig, Jaffe:1976ih}, $K \bar{K}$ molecules in
$s$ wave~\cite{Weinstein:1982gc, Weinstein:1983gd}, and dynamically
generated states in the $K \bar{K}$ and $\pi \pi$/$\pi \eta$
coupled-channel chiral dynamics~\cite{Oller:1997ng, Oller:1998hw,
  Oller:1998zr}.  However, the RHIC data seem to rule out a dominant tetraquark configuration for $f_{0}(980)$ as shown in previous studies by the authors~\cite{Cho:2010db,Cho:2011ew}.  Therefore, the structure of these scalar mesons  is still highly controversial.

Recently the so-called compositeness has been introduced to hadron
physics for investigating the internal structure of hadrons in
terms of the hadronic molecular configuration~\cite{Hyodo:2013nka,
  Sekihara:2014kya}.  The compositeness is defined as the two-body
composite part of the normalization of the total wave function, and
with this quantity it has been found that the $f_{0} (980)$ resonance
in the coupled-channel chiral dynamics is indeed dominated by the $K
\bar{K}$ composite state~\cite{Sekihara:2014kya}.
The $K \bar{K}$ compositeness of $f_{0} (980)$ and $a_{0} (980)$
has been evaluated in Ref.~\cite{Sekihara:2014qxa}
from experimental observations, 
and is found to have a large absolute value for $f_0(980)$
and a small but nonnegligible absolute value for $a_0(980)$.
Moreover, the spatial
structure, which reflects the hadronic spatial expanse inside
the system, has been theoretically studied for $f_{0} (980)$ by using
the finite volume method~\cite{Sekihara:2012xp}, and the distance
between $K$ and $\bar{K}$ inside $f_{0} (980)$ has been estimated as
$2.6$--$3.0 \text{ fm}$.

\subsubsection{$\Lambda(1405)$}  
\label{sec:Lambda(1405)}

The $\Lambda(1405)$ resonance is a negative parity excited state of $\Lambda$ baryon in the energy region between the $\pi\Sigma$ and $\bar{K}N$ thresholds~\cite{Agashe:2014kda}. While it should be described as a $p$-wave excited baryon in constituent quark models, the observed mass is too small in comparison with other negative parity baryons~\cite{Isgur:1978xj}. Rather, the $\Lambda(1405)$ is considered to be a meson-baryon molecule, as initiated in a study using the phenomenological model with vector meson exchange~\cite{Dalitz:1959dn,Dalitz:1960du}. In recent studies, the meson-baryon molecule picture is understood on the basis of chiral SU(3) dynamics~\cite{Kaiser:1995eg,Oset:1998it,Oller:2000fj,Lutz:2001yb,Hyodo:2011ur,Kamiya:2016jqc}. The $\bar{K}N$ molecular nature of the $\Lambda(1405)$ is also supported by the lattice QCD simulation~\cite{Hall:2014uca}, as well as the analysis with the weak-binding relation~\cite{Kamiya:2015aea}. A remarkable feature of this resonance is the two-pole nature~\cite{Minireview}, which stems from the attraction in the singlet and octet channels in the SU(3) basis~\cite{Jido:2003cb} and the $\bar{K}N$ and $\pi\Sigma$ channels in the isospin basis~\cite{Hyodo:2007jq}. A recent striking experimental achievement is the precise measurement of the kaonic hydrogen by SIDDHARTA~\cite{Bazzi:2011zj,Bazzi:2012eq}, which significantly reduces the uncertainty of the subthreshold extrapolation of the $\bar{K}N$ amplitude in the $\Lambda(1405)$ region~\cite{Ikeda:2011pi,Ikeda:2012au}.

The spatial structure of the $\Lambda(1405)$ has been studied by evaluating its form factor in the chiral unitary approach~\cite{Sekihara:2008qk,Sekihara:2010uz}. Here the form factor is evaluated at the higher energy pole, which gives a dominant contribution to the $\Lambda(1405)$. By switching off the decay into the $\pi\Sigma$ channel, the mean distance between $\bar{K}$ and $N$ in the $\Lambda(1405)$ is obtained as $\langle r^{2}\rangle\sim 2.8$ fm$^{2}$, which is in fair agreement with the estimation by the finite volume method~\cite{Sekihara:2012xp}. By using the effective single-channel $\bar{K}N$ potential constrained by the SIDDHARTA data, the mean distance is evaluated as $\langle r^{2}\rangle\sim 2.1$ fm$^{2}$~\cite{Miyahara:2015bya}.

From the experimental viewpoint, the $\Lambda(1405)$ has been observed in the low energy exclusive reactions. Traditionally, the kaon and pion beams have been used~\cite{Thomas:1973uh,Hemingway:1984pz}, and several high-statistics data are recently available in photoproductions by the LEPS collaboration~\cite{Niiyama:2008rt} and the CLAS collaboration~\cite{Moriya:2013eb,Moriya:2013hwg}, and in the proton-proton collisions by the HADES collaboration~\cite{Agakishiev:2012xk}. It is remarkable that the spin-parity $1/2^{-}$ is experimentally determined by the CLAS collaboration~\cite{Moriya:2014kpv}.

So far, the $\Lambda(1405)$ is not observed in high-energy inclusive processes, such as heavy ion collisions. This is mainly because its decay is dominated by the $\pi\Sigma=\{\pi^{+}\Sigma^{-},\pi^{0}\Sigma^{0},\pi^{-}\Sigma^{+}\}$ mode, except for a tiny fraction of the radiative decays into $\gamma \Lambda$ and $\gamma\Sigma^{0}$. In order to measure the $\pi\Sigma$ modes, it is necessary to detect at least one neutral particle ($\pi^{0}$ or neutron) in the weak decay of $\Sigma$, which is in general difficult. If the detection of neutrons is possible in heavy ion collisions, the $\Lambda(1405)$ can be reconstructed from the $\pi^{+}\Sigma^{-}$ mode where $\Sigma^{-}$ decays into $n \pi^{-}$ by almost 100\%. Alternatively, the $\pi^{-}\Sigma^{+}$ mode is also feasible,
where the $\Sigma^+$ has 50\% probability of decaying into $n\pi^+$.

\subsubsection{Dibaryons} 
\label{sec:dibaryons}

The $S=-2$ H dibaryon  was first predicted in Ref.~\cite{Jaffe:1976yi} as the color spin interaction is the most attractive and the Pauli principle between quarks does not operate in the $J=0,I=0$ channel in flavour SU(3).
Extensive experimental searches ruled out the possibility of a deeply bound H dibaryon~\cite{Yamamoto:2000wf,Takahashi:2001nm}. The two-$\Lambda$ binding energy in the double $\Lambda$ hypernucleus observed in the Nagara event is determined to be $B_{\Lambda\Lambda}({}_{\Lambda\Lambda}^6\mathrm{He})=6.91\pm0.16$ MeV~\citep{Takahashi:2001nm,Ahn:2013poa}, which sets the H mass range as $M_H > 2M_\Lambda - B_{\Lambda\Lambda}({}_{\Lambda\Lambda}^6\mathrm{He})$.
There is still a possibility that the H particle exists 
as a shallow bound state or as a resonance.
Experimentally, a bump structure was observed
between the $\Lambda\Lambda$ and the $N\Xi$  threshold  
($10-15$ MeV above the $\Lambda\Lambda$ threshold)
in the $\Lambda\Lambda$ invariant mass spectrum from the $(K^-,K^+)$ reaction~\citep{Yoon:2007aq},
while no clear signal was found so far
in the $\Lambda p \pi^-$ invariant mass analysis from heavy ion collisions by ALICE~\citep{Adam:2015nca}
as well as in the $\Lambda\Lambda$ invariant mass spectrum from $\Upsilon(1S)$ and $\Upsilon(2S)$ decays by Belle~\citep{Kim:2013vym}.
Lattice calculations show that
the H dibaryon becomes bound in the massive pion cases~\citep{Beane:2010hg,Inoue:2010es},
and it may evolve to a resonance near the $\Xi N$ threshold at a smaller pion mass and with the SU(3)$_f$ breaking effects, by combining physical hadron masses with the SU(3) symmetric potentials at unphysical quark masses~\cite{Inoue:2011ai}.
The latter result is consistent with the old prediction of the quark cluster model calculation~\cite{Takeuchi:1990qj} that takes account of the instanton-induced interaction effects~\cite{Kobayashi:1970ji,'tHooft:1976fv}. Chiral extrapolations of the lattice QCD results to the physical point indicate an unbound H dibaryon with respect to the $\Lambda\Lambda$ threshold~\cite{Shanahan:2011su,Haidenbauer:2011za,Yamaguchi:2016kxa}.
A constituent quark model calculation also shows that a compact multiquark configuration would be highly unlikely~\cite{Park:2016cmg}.
The $\Lambda\Lambda$ correlation in heavy-ion collisions
and its implication to the $\Lambda\Lambda$ interaction are discussed in 
Sec.~\ref{Sec:LLint}.

As in the H particle case,
stable or resonance dibaryon states may appear in those channels
in which the color spin interaction is attractive~\cite{Oka:1988yq,Gal:2015rev}
and the Pauli principle does not operate,
since the repulsive core of the baryon-baryon interaction is
mainly due to the Pauli principle between quarks in the presence of the color spin interaction
in the quark-quark force~\cite{Oka:1981rj}.
The SU(3)$_f$ breaking and channel coupling effects are also important,
since a bound or resonance state is sensitive to the threshold.
The channels with the most attractive color spin interactions
with $S=0, -1, -2, -3$ are
$\Delta\Delta (I=0,J^\pi=3^+)$,
$N\Sigma^*-\Delta\Sigma (I=1/2, J^\pi=2^+)$,
$\Lambda\Lambda-N\Xi-\Sigma\Sigma (I=0,J^\pi=0^+)$,
and $N\Omega-\Lambda\Xi^*-\Sigma^*\Xi-\Sigma\Xi^* (I=1/2, J^\pi=2^+)$, respectively~\cite{Gal:2015rev}.
Among these channels,
the third ($S=-2$) is the H particle channel,
and recent lattice and experimental dibaryon studies
suggest that $N\Omega$ and $\Delta\Delta$ may have bound states.

The $N \Omega$ channel was also found to be attractive in the quark model mainly due to the changes in the spatial wave function as the two baryons merge and the absence of any repulsion from the color spin interaction~\citep{Goldman:1987ma}.
In fact, recent lattice calculations find an attractive potential in the S-wave spin 2 channel that allows for a bound state with binding energy in the order of 18.9 MeV~\cite{Etminan:2014tya}, although the  pion mass in the simulation is still larger than the physical one.  
The $N\Omega$ correlation function can tell 
if the $N\Omega$ bound state exists~\cite{Morita:2016auo}
as discussed in Sec.~\ref{Sec:pOme}.


The nature of the low-mass enhancement in the $\pi \pi$-invariant mass spectrum from the double-pion fusion reaction~\cite{Abashian:1960zz,Booth:1961zz}, called the Abashian-Booth-Crowe (ABC) effect, is recently found to be  related to a possible resonance structure called the  $d^*(2380)$~\cite{Bashkanov:2008ih,Adlarson:2011bh,Adlarson:2012fe,Adlarson:2014pxj,Adlarson:2014ozl,Kukulin:2008sx}.
The resonance structure has a width of only 70 MeV and about 80 MeV below the $2 \Delta$ threshold~\cite{Adlarson:2014tcn}. There are works that claim the state is a dibaryon state~\cite{Chen:2014vha}
but from a constituent quark model analysis, it is problematic if a compact multiquark configuration can be stable in that channel~\cite{Park:2015nha}.

\subsection{Heavy hadrons}
\label{sec:heavy_hadrons}

\subsubsection{$D_{s0}^{\ast} (2317)$} 
\label{sec:Ds}

Recent progress in experimental facilities, particularly the B
factories~\cite{Bevan:2014iga}, has provided an opportunity to
investigate the hadron spectroscopy involving heavy charm and bottom quarks.  One of the most important tasks in the
heavy-hadron spectroscopy is to understand how the properties of the
exotic hadron candidates change with the heavy quark masses because one expects the existence of heavy exotic states that are analogous  to
the corresponding light exotic hadrons, such as the light scalar mesons discussed in
Sec.~\ref{sec:scalar_meson}.

In this respect, the charmed and strange scalar meson $D_{s0}^{\ast}
(2317)$, with quantum numbers $( I, J^{P} ) = ( 0, 0^{+} )$, is of
special interest.  The $D_{s0}^{\ast} (2317)$ was first observed by BaBar through its isospin violating $\pi ^{0} D_{s}^{+}$ decay mode~\cite{Aubert:2003fg}, which was followed by its confirmation by
CLEO~\cite{Besson:2003cp}, Belle~\cite{Krokovny:2003zq}, and
FOCUS~\cite{Vaandering:2004ix}.  An important feature of $D_{s0}^{\ast} (2317)$ is that its  mass $M_{D_{s0}^{\ast} (2317)} = 2317.7 \pm
0.6 \text{ MeV}$~\cite{Agashe:2014kda} is about 160~MeV below
the prediction of a quark model for the charmed
meson~\cite{Godfrey:1985xj, Godfrey:1986wj}, implying that it has some exotic
configuration besides an ordinary $q \bar{q}$ configuration.  There have thus been extensive discussions on the $D_{s0}^{\ast}
(2317)$, such as a $c \bar{s}$
state~\cite{Dai:2003yg, Bali:2003jv, Dougall:2003hv,
Hayashigaki:2004gq, Narison:2003td}, two-meson molecular
state~\cite{Barnes:2003dj, Szczepaniak:2003vy, Kolomeitsev:2003ac, 
Guo:2006fu, Faessler:2007gv, Gamermann:2006nm}, $D$-$K$
mixing~\cite{vanBeveren:2003kd}, tetraquark state~\cite{Cheng:2003kg,
Terasaki:2003qa, Maiani:2004vq, Bracco:2005kt}, or a mixture of
two-meson and tetraquark states~\cite{Browder:2003fk}.  Also, in terms of the compositeness, the $K D$ molecular component of
the $D_{s0}^{\ast} (2317)$ has been studied experimentally via some 
observables~\cite{Albaladejo:2016hae}, and a lattice
simulation~\cite{Torres:2014vna} has indicated the existence of a 
dominant $K D$
molecular component of about $70 \% $ in the $D_{s0}^{\ast} (2317)$ wave function.

\subsubsection{Charmonium-like states} 
\label{sec:charmoniumlike}

In recent years, there has been a remarkable progress in the heavy meson spectroscopy~\cite{Swanson:2006st,Nielsen:2009uh,Brambilla:2010cs,Hosaka:2016pey,Agashe:2014kda}. Plenty of new states, called $XYZ$, are observed above the open charm/bottom thresholds. Because the properties of these states are not well described in the conventional constituent quark model, the $XYZ$ states are expected to have an exotic structure. Among many interesting states, here we summarize the current status of $X(3872)$ and the charged charmonium-like states, $Z_{c}^{\pm}$.

One of the most intensively studied states is the $X(3872)$. It is firstly observed by the Belle collaboration in the $B$ decay~\cite{Choi:2003ue}. Subsequently, the $X(3872)$ is confirmed by the CDF collaboration~\cite{Acosta:2003zx}, the D0 collaboration~\cite{Abazov:2004kp}, the BaBar collaboration~\cite{Aubert:2004ns}, the LHCb collaboration~\cite{Aaij:2011sn}, and the CMS collaboration~\cite{Chatrchyan:2013cld}.
The mass $M_{X(3872)}$ and width $\Gamma_{X(3872)}$ are given by~\cite{Agashe:2014kda}
\begin{align}
    M_{X(3872)}
    &= 
    3871.69\pm 0.17\text{ MeV} ,\quad
    \Gamma_{X(3872)}
    <1.2\text{ MeV}
\end{align}
It is worth noting that the mass is very close to the threshold of the $D^{0}\bar{D}^{*0}$ state
\begin{align}
    M_{D^{0}}+M_{\bar{D}^{*0}}
    &= 
    3871.8\pm 0.12\text{ MeV} .
\end{align}
The $C$-parity $C=-1$ is determined by measuring the $J/\psi \gamma$ decay~\cite{Aubert:2008ae,Bhardwaj:2011dj}. The spin-parity can be studied from the angular distribution of the final states. Eventually, the LHCb collaboration determines $J^{PC}=1^{++}$ based on the angular correlations~\cite{Aaij:2013zoa}. No charged partners are observed in the $J/\psi \pi^{\pm}\pi^{0}$ mode.

While the quantum numbers of the $X(3872)$ can in principle be given by the $\bar{c}c$ configuration, the proximity of the $D^{0}\bar{D}^{*0}$ state urges us to consider the hadronic molecule interpretation. In fact, the phenomenological Lagrangian analysis of the radiative decay~\cite{Dong:2008gb}
indicates the dominance of the $D^{0}\bar{D}^{*0}$ molecular component.  On the other hand, an analysis of the prompt production cross section~\cite{Bignamini:2009sk} shows that the $\bar{c}c$ component is required in addition to the molecule component. In addition, lattice study in Ref.~\cite{Padmanath:2015era} indicates that a candidate for the $X(3872)$ is found only when both the $\bar{c}c$ and $D\bar{D}^{*}$ operators are included in the analysis.

The charged charmonium-like states $Z_{c}^{\pm}$ are of particular interest, because a light quark-antiquark pair is required in addition to $\bar{c}c$ as the valence component. At present, altogether eight charged charmonium-like states have been reported, although not all the states are  firmly established. Below we briefly overview the experimental status of these charged charmonium-like states. The basic properties are summarized in Table~\ref{tbl:Zc}. 

The first candidate is called $Z_{c}(4430)$, observed in the $\pi^{\pm}\psi(2S)$ spectrum of the $B$ decay into $K \pi^{\pm}\psi(2S)$ by the Belle collaboration~\cite{Choi:2007wga}. Although the BaBar collaboration did not confirm the $Z_{c}(4430)$~\cite{Aubert:2008aa}, the Dalitz plot analysis of the $B$ decay by Belle reconfirmed the original findings~\cite{Mizuk:2009da}. The spin-parity $J^{P}=1^{+}$ is favored, based on the full amplitude analysis by Belle~\cite{Chilikin:2013tch}. The $Z_{c}(4430)$ is finally confirmed by LHCb~\cite{Aaij:2014jqa} with more than ten times higher statistics than the original observation. The analysis of LHCb~\cite{Aaij:2014jqa} also suggests the possibility of additional state $Z_{c}(4240)$ in the $\pi^{-}\psi(2S)$ spectrum. Two other charged states $Z_{c}(4050)$ and $Z_{c}(4250)$ were found in the $\pi^{+}\chi_{c1}$ mass distribution in $\bar{B}^{0}\to K^{-}\pi^{+}\chi_{c1}$ by Belle~\cite{Mizuk:2008me}. These states were not confirmed in the analysis by BaBar~\cite{Lees:2011ik}. 

The $Z_{c}(3900)$ state is observed in the $\pi^{\pm}J/\psi$ spectrum of the process $e^{+}e^{-}\to \pi^{+}\pi^{-}J/\psi$ by the BESIII collaboration~\cite{Ablikim:2013mio}. The state was confirmed in the same decay mode by Belle~\cite{Liu:2013dau} and by the analysis of CLEO-c data~\cite{Xiao:2013iha}. BESIII also looked for other decay modes. The $Z_{c}(3900)$ signal was not found in $h_{c}\pi^{\pm}$ spectrum~\cite{Ablikim:2013wzq}, while the structure is observed in the $(D\bar{D}^{*})^{\pm}$ mode~\cite{Ablikim:2013xfr}. Angular analysis by BESIII favors the spin-parity assignment of $J^{P}=1^{+}$~\cite{Ablikim:2013xfr}. A recent lattice QCD study finds no convincing signal for the $Z_{c}(3900)$ candidate~\cite{Prelovsek:2013xba,Prelovsek:2014swa}. The result by the HAL QCD collaboration indicates the interpretation of $Z_{c}(3900)$ as a threshold cusp effect~\cite{Ikeda:2016zwx}.

A narrow state was observed by Belle in the $h_{c}\pi^{\pm}$ spectrum, which is called $Z_{c}(4020)$~\cite{Ablikim:2013wzq}. The state is also found in the $(D^{*}\bar{D}^{*})^{\pm}$ mode~\cite{Ablikim:2013emm}. In the study of the $J/\psi \pi^{+}$ spectrum of the $\bar{B}^{0}\to K^{-}J/\psi \pi^{+}$ decay by Belle, the $Z_{c}(4200)$ was observed~\cite{Chilikin:2014bkk}. Favored spin-parity assignment turns out to be $J^{P}=1^{+}$. Belle also found a signal of $Z_{c}(4055)$ in the $J/\psi \pi^{+}$ channel~\cite{Wang:2014hta}.

Among many charged states, we will concentrate on the structure of the most established states, $Z_{c}(3900)$ and $Z_{c}(4430)$. In the molecular interpretation, these are considered to be $\bar{D}D^{*}$ and $D_{1}\bar{D}$ molecules, respectively.
A novel idea based on a strong diquark-antidiquark correlation is also considered for the structure of  X and Z states~\cite{Brodsky:2014xia}.

\begin{table}[tbp]
\caption{Summary of the charged charmonium-like states. The mass and width are taken from PDG~\cite{Agashe:2014kda}. $Z_{c}(3900)$, $Z_{c}(4020)$ and $Z_{c}(4430)$ are listed in the PDG summary Table.}
\begin{center}
\begin{tabular}{c|l|l|c|c|c}
\hline
State & Mass [MeV] & Width [MeV] & Decay mode & $J^{P}$ & Reference \\ 
\hline
$Z_{c}(3900)$ & $3886.6\pm 2.4$ & $28.1\pm 2.6$ 
& $J/\psi \pi^{\pm}$, $(D\bar{D}^{*})^{\pm}$ & $1^{+}$ & \cite{Ablikim:2013mio,Liu:2013dau,Xiao:2013iha,Ablikim:2013xfr} \\
$Z_{c}(4020)$ & $4024.1\pm 1.9$ & $13\pm 5$ 
& $h_{c}\pi^{\pm},(D^{*}\bar{D}^{*})^{\pm}$ & - & \cite{Ablikim:2013wzq,Ablikim:2013emm} \\
$Z_{c}(4050)$ & $4051\pm 14^{+20}_{-41}$ & $82^{+21}_{-17}{}^{+47}_{-22}$ 
& $\chi_{c1}\pi^{\pm}$ & - & \cite{Mizuk:2008me} \\
$Z_{c}(4055)$ & $4054\pm 3\pm 1$ & $45\pm 11\pm 6$ 
& $\psi(2S)\pi^{\pm}$ & - & \cite{Wang:2014hta} \\
$Z_{c}(4200)$ & $4196^{+31}_{-29}{}^{+17}_{-13}$ & $370\pm 70{}^{+70}_{-132}$
& $J/\psi \pi^{\pm}$ & $1^{+}$ & \cite{Chilikin:2014bkk} \\
$Z_{c}(4240)$ & $4239\pm 18{}^{+45}_{-10}$ & $220\pm 47{}^{+108}_{-74}$
& $\psi(2S) \pi^{\pm}$ & $0^{-}$ & \cite{Aaij:2014jqa} \\
$Z_{c}(4250)$ & $4248^{+44}_{-29}{}^{+180}_{-35}$ & $177^{+54}_{-39}{}^{+316}_{-61}$ 
& $\chi_{c1}\pi^{\pm}$ & - & \cite{Mizuk:2008me} \\
$Z_{c}(4430)$ & $4478^{+15}_{-18}$ & $181\pm 31$ 
& $\psi(2S)\pi^{\pm}$ & $1^{+}$ & \cite{Choi:2007wga,Mizuk:2009da,Chilikin:2013tch,Aaij:2014jqa} \\
\hline
\end{tabular}
\end{center}
\label{tbl:Zc}
\end{table}%

\subsubsection{Charged bottomonium-like states} 
\label{sec:bottomoniumlike}


As bottom analogue of charged charmonium-like state, $Z_{c}^{+}$,
charged bottomonium-like state called $Z_{b}$ was reported in Belle~\cite{Belle:2011aa}.
They are $Z_{b}(10610)^{+}$ and $Z_{b}(10650)^{+}$ with spin-parity $J^{P}=1^{+}$.
The neutral state, $Z_{b}(10610)^{0}$, was also reported~\cite{Krokovny:2013mgx}.
Because the masses of the two $Z_{b}$'s are very close to $B\bar{B}^{\ast}$ ($B^{\ast}\bar{B}$) and $B^{\ast}\bar{B}^{\ast}$ thresholds, respectively,
it might be natural to regard those states as a $B\bar{B}^{\ast}$ ($B^{\ast}\bar{B}$) hadronic molecule and a $B^{\ast}\bar{B}^{\ast}$ hadronic molecule, respectively~\cite{Bondar:2011ev,Voloshin:2011qa,Ohkoda:2011vj}.
However, $Z_{b}(10610)^{+}$ and $Z_{b}(10650)^{+}$ decay to $B\bar{B}^{\ast}$ ($B^{\ast}\bar{B}$) and $B^{\ast}\bar{B}^{\ast}$, and hence they cannot be 
simple $B\bar{B}^{\ast}$ ($B^{\ast}\bar{B}$) and $B^{\ast}\bar{B}^{\ast}$ bound states.
The hadronic molecule structure can be studied in the decay properties of $Z_{b}^{+} \rightarrow \Upsilon \pi^{+}$ in view of the heavy quark symmetry~\cite{Bondar:2011ev,Voloshin:2011qa,Ohkoda:2012rj,Ohkoda:2013cea} and within a phenomenological Lagrangian approach~\cite{Dong:2012hc}.
However, the situation may be much different, because the observed experimental peaks of $Z_{b}$'s could be explained by the cusp effect~\cite{Swanson:2014tra}.

In relativistic heavy ion collisions, enhancement of the scattering amplitude of $\Upsilon+\pi$ can be seen via
\begin{eqnarray}
 \Upsilon + \pi_{\mathrm{thermal}} \rightarrow Z_{b} \rightarrow \Upsilon + \pi, \hspace{0.5em} B+\bar{B}^{\ast},
\end{eqnarray}
with the intermediate $Z_{b}$ states, where $\pi_{\mathrm{thermal}}$ is the thermal pion in the hadron phase.


\begin{table}[tbp]
\caption{Summary of the charged bottomonium-like states.}
\begin{center}
\begin{tabular}{c|l|l|c|c|c}
\hline
State & Mass [MeV] & Width [MeV] & Decay mode & $J^{P}$ & Reference \\ 
\hline
$Z_{b}(10610)^{+}$ & $10607.2 \pm 2.0$ & $18.4 \pm 2.4$ 
& $\pi^{\pm} \Upsilon(ns)$, $\pi^{\pm} h_b$ & $1^{+}$ & \cite{Belle:2011aa} \\
$Z_{b}(10610)^{0}$ & $10609 \pm 4 \pm 4$ & - 
& $\Upsilon(2,3s) \pi^0$  & $1^{+}$ & \cite{Krokovny:2013mgx} \\
$Z_{b}(10650)^{+}$ & $10652.2 \pm 1.5$ & $11.5\pm 2.2$ 
& $\pi^{\pm} \Upsilon(ns)$, $\pi^{\pm} h_b$ & $1^+$ & \cite{Belle:2011aa} \\
\hline
\end{tabular}
\end{center}
\label{tbl:Zb}
\end{table}%

\begin{table}[tbp] 
\caption{Summary of exotic hadrons with light flavors. Shown are
the mass ($m$), isospin ($I$), spin and parity
($J^P$), the quark structure ($2q/3/q/6q$ and $4q/5q/8q$),
molecular configuration (Mol.) and corresponding oscillator
frequency ($\omega_{\rm Mol.}$). For
the $\omega_\mathrm{Mol.}$, it is fixed by the binding energies B
of hadrons ($\omega \simeq 6 \times \mathrm{B}$, marked (B)) or
their mean square distances $\langle r^2\rangle$ ($\omega \simeq
3/2\mu \langle r^2 \rangle$, marked (R)). In the case of three-body molecular
configurations for exotic dibaryons, $\omega_\mathrm{Mol.}$ is that for the subsystem.
}\label{tbl:summary_light}
\begin{center}
\begin{tabular}{ccccccc}
\hline
Particle & $m$ [MeV] & $(I,J^{P})$ & $q\bar{q}$/$qqq$ ($L$) & multiquark & Mol. ($L$) & $\omega_{\mathrm{Mol}}$ [MeV] \\ 
\hline
$f_{0}(980)$ & 980 & $(0, 0^{+})$ & $q\bar{q}$ ($P$) ($s\bar{s}$ ($P$)) & $qs\bar{q}\bar{s}$ & $\bar{K}K$ ($S$) & 67.8(B) \\ 
$a_{0}(980)$ & 980 & $(1, 0^{+})$ & $q\bar{q}$ ($P$) & $qs\bar{q}\bar{s}$ & $\bar{K}K$ ($S$) & 67.8(B) \\ 
$K(1460)$ & 1460 & $(1/2, 0^{-})$ & --- & $qq\bar{q}\bar{s}$ ($P$) & $\bar{K}KK$ ($P$) & 69.0(R) \\ 
$\Lambda(1405)$ & 1405 & $(0, 1/2^{-})$ & $uds$ ($P$) & $udsq\bar{q}$ & $\bar{K}N$ ($S$) & 20.5(R) \\ 
$\Delta \Delta$ & 2380 & $(0, 3^{+})$ & --- & $q^{6}$ & --- & --- \\ 
$\Lambda\Lambda$-$N\Xi$ ($H$) & 2245& $(0, 0^{+})$ & --- & $uuddss$ & $N\Xi$ ($S$)& 73.2(B) \\ 
$N\Omega$ & 2592 & $(1/2, 2^{+})$  & --- & $uudsss$ & --- & --- \\ 
\hline
\end{tabular}
\end{center}
\end{table}%

\begin{table}[tbp] 
\caption{Summary of exotic hadrons with heavy flavors. The notations are the same as those in Table \ref{tbl:summary_light}.}
\begin{center}
\begin{tabular}{ccccccc}
\hline
Particle & $m$ [MeV] & $(I ,J^{P})$ & $q\bar{q}$/$qqq$ $(L)$ & multiquark & Mol. ($L$) & $\omega_{\mathrm{Mol}}$ [MeV] \\ 
\hline
$D_{s}(2317)$ & 2317 & $(0, 0^{+})$ & $c\bar{s}$ ($P$) & $c\bar{s}q\bar{q}$ & $DK$ ($S$) & 273(B) \\ 
$X(3872)$ & 3872 & $(0, 1^{+})$ & $c\bar{c}$ ($P$) & $c\bar{c}q\bar{q}$ & $D\bar{D}^{\ast}$ ($S$) & 3.6(B) \\ 
$Z_{c}(3900)$ & 3900 & $(1, 1^{+})$ & --- & $c\bar{c}u\bar{d}$ & --- & --- \\  
$Z_{c}(4430)$ & 4430 & $(1, 1^{+})$ & --- & $c\bar{c}u\bar{d}$ & $D_{1}\bar{D}^{\ast}$ ($S$) & 13.5(B) \\ 
$Z_{b}(10610)$ & 10610 & $(1, 1^{+})$ & --- & $b\bar{b}u\bar{d}$ & --- & --- \\ 
$Z_{b}(10650)$ & 10650 & $(1, 1^{+})$ & --- & $b\bar{b}u\bar{d}$ & --- & --- \\ 
$X(5568)$ & 5568 & $(1, 0^{+})$ & --- & $s\bar{b}u\bar{d}$ & --- & --- \\ 
$P_{c}(4380)$ & 4380 & $(1/2, 3/2^{-})^{\,b}$ & --- & $c\bar{c}uud$ ($S$) & $\bar{D}\Sigma_{c}^{\ast}$ ($S$) & 60(B) \\ 
$P_{c}(4450)$ & 4450 & $(1/2, 5/2^{+})^{\,b}$ & --- & $c\bar{c}uud$ ($P$) & --- & --- \\ 
\hline
\end{tabular}
\end{center}
\label{tbl:summary_heavy}
\end{table}%

\begin{table}[tbp] 
\caption{Summary of other exotic hadrons.  The notations are the same as those in Table \ref{tbl:summary_light}. More information for each particle is available in Refs.~\cite{Cho:2011ew,Maeda:2015hxa}.}
\begin{center}
\begin{tabular}{cccccccc}
\hline
Particle & $m$ [MeV] & $(I,J^{P})$ & $q\bar{q}$/$qqq$ & multiquark ($L$) & Mol. ($L$) & $\omega_{\mathrm{Mol}}$ [MeV] & Ref. \\ 
\hline
$\Theta(1530)$ & 1530 & $(0,1/2^{+})$ & --- & $qqqq\bar{s}$ ($P$) & --- & --- & \cite{Diakonov:1997mm} \\
$\bar{K}KN$ & 1920 & $(1/2,1/2^{+})$ & --- & $qqqs\bar{s}$ ($P$) & $\bar{K}KN$ & 42(R) & \cite{Jido:2008kp} \\
$\bar{K}NN$ & 2352 & $(1/2,0^{-})$ & $q^{5}s$ ($P$) & $q^{6}s\bar{q}$ ($S$) & $\bar{K}NN$ & 20.5(T) & \cite{Akaishi:2002bg}  \\
$\Omega\Omega$ & 3228 & $(0, 0^{+})$ & --- & $s^{6}$ & --- & --- & \cite{Zhang:2000sv}  \\
\hline
$T_{cc}^{1}$ & 3797 & $(0,1^{+})$ & --- & $ud\bar{c}\bar{c}$ & --- & --- & \cite{Zouzou:1986qh} \\
$\bar{D}N$ & 2790 & $(0,1/2^{-})$ & --- & $qqqq\bar{c}$ & $\bar{D}N$ & 6.48(R) & \cite{Yasui:2009bz} \\
$\bar{D}^{\ast}N$ & 2919 & $(0,3/2^{-})$ & --- & $qqqq\bar{c}(D)$ & $\bar{D}^{\ast}N$ & 6.48(R) & \cite{Yamaguchi:2011xb} \\
$\Theta_{cs}$ & 2980 & $(1/2,1/2^{+})$ & --- & $qqqs\bar{c}$ ($P$) & --- & --- & \cite{Lipkin:1987sk,Gignoux:1987cn} \\
$H_{c}^{++}$ & 3377 & $(1,0^{+})$ & --- & $qqqqsc$ & --- & ---- & \cite{Lee:2009rt} \\
$\bar{D}NN$ & 3734 & $(1/2,0^{-})$ & --- & $q^{7}\bar{c}$ & $\bar{D}NN$ & 6.48(T) & \cite{Yamaguchi:2013hsa} \\
$\Lambda_{c}N$ & 3225 & $(1/2,1^{+})$ & --- & $cuduud$ & $\Lambda_{c}N$ & 4.24(R) & \cite{Maeda:2015hxa} \\
$\Lambda_{c}NN$ & 4164 & $(0, 3/2^{+})$ & --- & $cuduududd$ & $\Lambda_{c}NN$ & 33.16(R) & \cite{Maeda:2015hxa} \\
$T_{cb}^{0}$ & 7123 & $(0,0^{+})$ & --- & $ud\bar{c}\bar{b}$ & --- & --- & \cite{Lee:2009rt} \\
\hline
\end{tabular}
\end{center}
\label{tbl:summary_others}
\end{table}%

\subsubsection{New states: $P_c$ and $X(5568)$} 
\label{sec:new_states}

The existence of hidden-charm pentaquark states has been predicted
in many theoretical works~\cite{Wu:2010jy,Yuan:2012wz,Yang:2011wz,Xiao:2013yca,Uchino:2015uha}. 
Therefore, it was with great excitement that the theoretical community
heard about the LHCb observation of two hidden-charm pentaquark-like structures
$P^+_c(4380)$ and $P^+_c(4450)$ in the $J/\psi p$ invariant mass distribution in
the decay $\Lambda^0_b \to J/\psi p K^-$~\cite{Aaij:2015tga}. 
They used an amplitude analysis of the three-body final-state, and the masses and 
widths obtained are $M_{P_c(4380)} = (4380 \pm 8 \pm
29)$~MeV, $\Gamma_{P_c(4380)} = (205 \pm 18 \pm 86)$~MeV, $M_{P_c(4450)} =
(4449.8 \pm 1.7 \pm 2.5)$~MeV, and $\Gamma_{P_c(4450)} = (39 \pm 5 \pm 19)$ MeV. The 
significance of the lower mass and higher mass states is 9 $\sigma$ and 12 $\sigma$, 
respectively. The preferred  spin-parity assignments are $J^P= 3/2^\pm$ or 
$5/2^\mp$. 

Before the LHCb observation, the predicted masses for genuine pentaquark states, 
with both negative and positive parities~\cite{Yuan:2012wz}, or meson-baryon
bound states with  $J^P=3/2^-$~\cite{Wu:2010jy,Yang:2011wz,Xiao:2013yca}, cover the 
observed masses of the two $P^+_c$ structures.
After the observation of the two $P^+_c$ structures, many theoretical works 
appeared proposing various explanations for these structures 
~\cite{Chen:2015loa,Roca:2015dva,He:2015cea,Huang:2015uda,Meissner:2015mza,Xiao:2015fia,Chen:2016heh,Shimizu:2016rrd,Maiani:2015vwa,Anisovich:2015cia,Ghosh:2015ksa,Wang:2015epa,Wang:2015ixb,Chen:2015moa,Lebed:2015tna,Zhu:2015bba,Chen:2016otp,Scoccola:2015nia,Mironov:2015ica}.
It was even suggested that the observed structures could be due to kinematical 
triangle singularities~\cite{Guo:2015umn,Liu:2015fea,Mikhasenko:2015vca}. Of course 
this possibility needs to be examined by future experiments.

Since the masses of these  two $P^+_c$ structures are very close to the mass 
thresholds of the $\bar{D}\Sigma_c^*$ and $\bar{D}^* \Sigma_c$, a very natural
explanation is that the observed structures could be meson-baryon molecules~\cite{Chen:2015loa,Roca:2015dva,He:2015cea,Huang:2015uda,Meissner:2015mza,Xiao:2015fia,Chen:2016heh,Shimizu:2016rrd}. Other possible explanations are: diquark-diquark-antiquark 
pentaquarks~\cite{Maiani:2015vwa,Anisovich:2015cia,Ghosh:2015ksa,Wang:2015epa,Wang:2015ixb}, compact diquark-triquark pentaquarks~\cite{Chen:2015moa,Lebed:2015tna,Zhu:2015bba,Chen:2016otp},  topological soliton model~\cite{Scoccola:2015nia}, genuine multiquark states other than molecules~\cite{Mironov:2015ica}, etc.
For more comprehensive discussions, see Refs.~\cite{Chen:2016qju,Burns:2015dwa,Oset:2016lyh}.

The  D0  Collaboration   has  recently  announced the observation  of  
a narrow enhancement of the experimental data in the $B_{s}^0\pi^\pm$ mass 
spectrum in  the energy around 5.6~GeV. The enhancement was interpreted
as a  new state: $X^\pm(5568)$~\cite{D0:2016mwd}.  The mass and  width for
this    state    have    been     found    to    be    $m=5567.8\pm2.9
(\mbox{sta})^{+0.9}_{-1.9}(\mbox{syst}$   MeV/c$^2$         and
$\Gamma=21.9\pm6.4  (\mbox{sta})^{+5.0}_{-2.5}(\mbox{syst})$ MeV/c$^2$,
respectively~\cite{D0:2016mwd}.  The isospin  of $X(5568)$  is clearly
one. Its  spin-parity is not  yet known  although a scalar  four quark
interpretation has been suggested in Ref.~\cite{D0:2016mwd}.
The $X(5568)$ would be a very interesting addition to 
the list of undoubtedly exotic mesons, since its wave function consists of 
four different flavors: $u$, $b$, $d$ and $s$ quarks. However,  the LHCb 
Collaboration has not confirmed the observation of the $X(5568)$.
In their analysis~\cite{Aaij:2016iev} no structure is found
in the $B_{s}^0\pi^\pm$ mass spectrum from the  $B_{s}^0\pi^+$ threshold up to 
$M_{B_s^0\pi^+}\leq 5700$ GeV. 

The announcement of the exotic state $X(5568)$  stimulated the theoretical 
interest and  several  theoretical works have been done to investigate the 
properties of such state. There are studies based on QCD sum rules~\cite{Agaev:2016mjb,Zanetti:2016wjn,Wang:2016mee,Chen:2016mqt,Agaev:2016ijz,Dias:2016dme,Wang:2016wkj,Tang:2016pcf,Agaev:2016urs,Albuquerque:2016nlw},
quark models~\cite{Wang:2016tsi,Liu:2016ogz,Xiao:2016mho,Stancu:2016sfd,Lu:2016zhe,Chen:2016npt,Ali:2016gdg,He:2016yhd}, 
rescattering effects~\cite{Liu:2016xly}, coupled-channel analysis~\cite{Albaladejo:2016eps} and more general arguments~\cite{Burns:2016gvy,Guo:2016nhb}.
In Ref.~\cite{Burns:2016gvy} various interpretations for the $X(5568)$ signal were 
considered and the authors concluded that threshold, cusp, tetraquark and molecular
models were all unfavored. In Ref.~\cite{Guo:2016nhb} additional arguments, based on 
general properties of QCD, were provided to question the existence of the $X(5568)$.
In Refs.~\cite{Zanetti:2016wjn,Wang:2016tsi,Chen:2016npt,Ali:2016gdg} although it 
was possible to find a tetraquark state with $J^P=0^+$ and the same quark content as 
the $X(5568)$, the obtained masses were around 200 MeV higher than the announced
 $X(5568)$ mass~\cite{Wang:2016tsi,Ali:2016gdg}, or too high to be candidate of 
$X(5568)$~\cite{Zanetti:2016wjn,Lu:2016zhe,Chen:2016npt}. However, the width of the 
state can be of the same order as the width reported by D0~\cite{Dias:2016dme}.  
In Ref.~\cite{Chen:2016npt} no molecular structure could be obtained to explain the 
$X(5568)$ state, while in Ref.~\cite{Albuquerque:2016nlw} molecular and tetraquark 
states could be obtained but with masses around 5200 MeV.

In all other calculations it was possible to explain the properties of the $X(5568)$.
In particular, in Refs.~\cite{Agaev:2016mjb,Wang:2016mee,Chen:2016mqt,Agaev:2016ijz,Wang:2016wkj,Tang:2016pcf,Agaev:2016urs,He:2016yhd}
 the results of the calculations for  the properties  of $X(5568)$  are in excellent 
agreement  with the experimental  value.  In Refs.~\cite{Agaev:2016mjb,Wang:2016mee,Agaev:2016ijz,Wang:2016wkj,Tang:2016pcf,Agaev:2016urs,He:2016yhd} 
$J^{P} =  0^{+}$ was assumed while in Ref.~\cite{Chen:2016mqt} scalar as
well as  axial tetraquark currents were considered.   In
Ref.~\cite{Wang:2016tsi}  a model  using multiquark  interactions has  been
used and  a 150 MeV higher  mass is found for  $X(5568)$, although the
systematic   errors  still   allow  their   state  to   be  related   to
$X(5568)$. Another multiquark  model calculations using the color-magnetic
interaction has been presented in Refs.~\cite{Liu:2016ogz,Stancu:2016sfd} with very 
good agreement with the experimental  value. The possibility of
explaining the enhancement in the  data as near threshold rescattering
effects has been  studied in Ref.~\cite{Liu:2016xly}. The $B \bar  K$ and $B^*
\bar   K$   molecular   interpretations   have   been   suggested   in
Ref.~\cite{Xiao:2016mho}. In Ref.~\cite{Albaladejo:2016eps}, using a 
coupled-channel analysis, it is possible to find a pole that can be associated with 
the $X(5568)$ state, although the cutoff used is much larger than the normal one.

Clearly, more  analysis are required to clarify this situation from the experimental 
side as well as from the theoretical side.

\section{Yields of particles} 
\label{sec:yields}

In relativistic heavy ion collisions,  a variety of hadrons and their resonances are produced.  To describe the yields of these particles, a number of approaches have been used. These include the microscopic transport model~\cite{Cassing:2009vt,Lin:2004en} and the macroscopic hydrodynamic model~\cite{Hirano:2005wx,Song:2007ux}. In the transport model, hadron production from the produced quark-gluon plasma is treated either kinetically via parton to hadron reactions~\cite{Cassing:2009vt} or via the quark coalescence~\cite{Lin:2004en}. In the hydrodynamic approach, hadrons are produced through the statistical hadronization model~\cite{Andronic:2005yp}.   

On the other hand, the schematic model, which is based on isentropic boost invariant longitudinal and accelerated transverse expansions, has been adopted in Ref.~\citep{Cho:2015exb} to understand the production of exotic hadrons in heavy ion collisions and to quantitatively evaluate their yields from the dynamically expanding quark-gluon plasma.
In this approach, both the initial QGP and final hadronic matter are treated as 
noninteracting free gases.  For the crossover  
transition between these two phases of matter, it starts at the critical temperature $T_C$ and ends at the hadronization temperature $T_H$.  During 
this phase transition, the system expands from the critical volume $V_C$ at $T_C$ to the
hadronization volume $V_H$ at $T_H$ while maintains a constant entropy.

\begin{figure}[!h]
\begin{center}
\includegraphics[bb=0 0 840 643,width=0.53\textwidth]{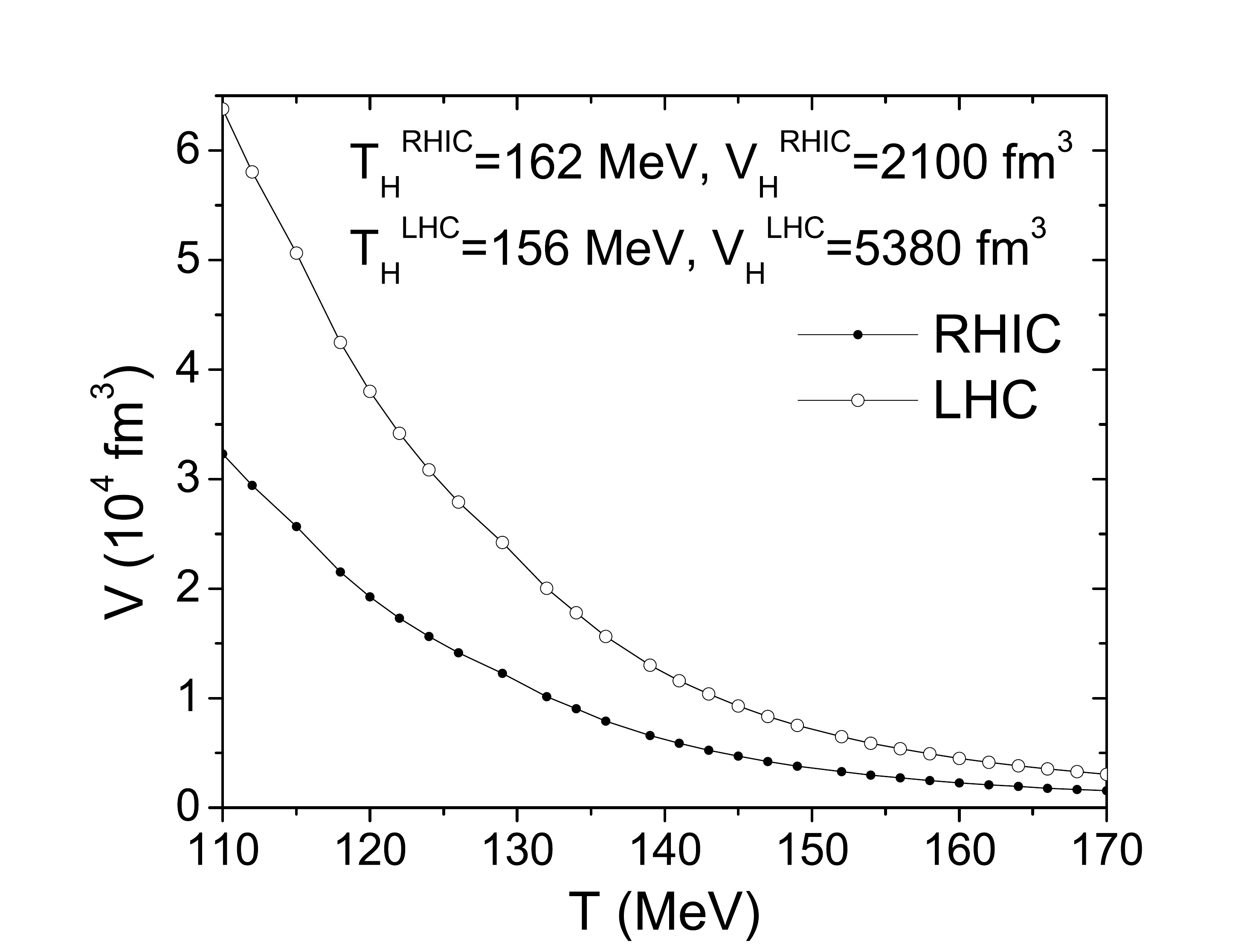}
\end{center}
\caption{Temperature dependence of the volume of the matter produced at RHIC (LHC) during the entropy conserving expansion with the hadronization temperature
$T_H=162~(156)$ MeV and the volume $V_H=2100~(5380)$ fm$^3$.} \label{Volevol}
\end{figure}

In the present review on the yields of hadrons in relativistic heavy ion collisions, we update the results reported in Refs.~\cite{Cho:2010db,Cho:2011ew}
by using the schematic model of Ref.~\citep{Cho:2015exb}. Following this reference, 
we take the same hadronization temperature $T_H=162~(156)$
MeV and  volume $V_H=2100~(5380)$ fm$^3$ for RHIC~(LHC)
as used in the statistical hadronization model analysis of the experimental data~\cite{Andronic:2012dm, Stachel:2013zma}. 
The critical temperature $T_C$ and volume $V_C$ are then determined by tracing back in time and using the entropy conservation condition $s_HV_H=s_CV_C$,
where $s_C$ and $s_H$ are the entropy density of the system at
$T_C$ and $T_H$, respectively. The information on the entropy
density of the system $s=(\epsilon+p)/T$ at different temperatures
is available from the lattice results $p(T)/T^4$ and
$(\epsilon-3p)(T)/T^4$ for the pion mass 135 MeV~\cite{Borsanyi:2010cj}.  Figure \ref{Volevol} shows the temperature dependence of the volume of the matter produced at RHIC (LHC) during its isentropic expansion.

To determine the critical temperature $T_C$ and volume
$V_C$, we consider here two possibilities. The first scenario is to
take the  critical temperature to be the  same as the hadronization
temperature, $T_C=T_H$, and also for the critical volume
$V_C=V_H$. Since hadrons are continuously produced during the
crossover phase transition, we require that the number of hadrons
produced via the statistical hadronization is equal to that produced
via coalescence of constituent quarks at the end of the phase transition.

In the second scenario, we assume that the critical temperature $T_c$ and the oscillator frequency, used in the harmonic oscillator model for the wave functions of composite systems, are the same at RHIC and LHC, and determine the respective volumes 
by requiring that the numbers of various hadrons produced via the quark coalescence 
at the beginning of phase transition ($T_C$) are equal to those produced at the end of
transition ($T_H$) via  the statistical hadronization.

Although hadron yields in the statistical hadronization model only depends on the hadronization temperature $T_H$ and volume $V_H$, those in the quark coalescence model also depend on the the number of  quarks in the system and the oscillator frequency in the wave function of the produced hadron.   
The critical temperature $T_C$ and volume $V_C$ 
can thus be determined by choosing an appropriate value for the oscillator frequency. 
Using the rho and omega mesons as examples, we  obtain the critical
temperature $T_C=166$ MeV at RHIC and LHC.  Hence, in the second scenario, we find that the critical temperature at LHC drops from 166 MeV to 156 MeV, while that at
RHIC changes slightly from 166 MeV to 162 MeV during the crossover transition.

\begin{table}[htbp] 
\caption{Statistical and coalescence model parameters for Scenario 1 and 2 at
RHIC (200 GeV), LHC (2.76 TeV) and LHC (5.02 TeV),
and those given in Refs.~\cite{Cho:2010db,Cho:2011ew}.
Quark masses are taken to be 
$m_{q}=350~\MeV$,
$m_{s}=500~\MeV$,
$m_{c}=1500~\MeV$
and 
$m_{b}=4700~\MeV$.
In Refs.~\cite{Cho:2010db,Cho:2011ew}, light quark masses were taken to be $m_q=300~\MeV$.
}\label{Tab:pars}
\begin{tabular}{l|cc|cccc|cc}
\hline
\hline
	& \multicolumn{2}{c|}{RHIC} & \multicolumn{2}{c}{LHC (2.76 TeV)} & \multicolumn{2}{c|}{LHC (5.02 TeV)}
	& RHIC & LHC (5 TeV)\\
	& Sc. 1 & Sc. 2
	& Sc. 1 & Sc. 2 
	& Sc. 1 & Sc. 2
	& \multicolumn{2}{c}{Refs~\cite{Cho:2010db,Cho:2011ew}}\\
\hline
$T_H~(\MeV)$   &\multicolumn{2}{c|}{162} &\multicolumn{4}{c|}{156} & \multicolumn{2}{c}{175} \\
$V_H~(\fm^3)$  &\multicolumn{2}{c|}{2100}&\multicolumn{4}{c|}{5380} & 1908 & 5152 \\ 
$\mu_B~(\MeV)$ &\multicolumn{2}{c|}{24}  &\multicolumn{4}{c|}{0} & 20 & 0 \\ 
$\mu_s~(\MeV)$ &\multicolumn{2}{c|}{10}  &\multicolumn{4}{c|}{0} & 10 & 0 \\ 
$\gamma_c$     &\multicolumn{2}{c|}{22}  &\multicolumn{2}{c}{39} & \multicolumn{2}{c|}{50} & 6.40 & 15.8 \\ 
$\gamma_b$     &\multicolumn{2}{c|}{$4.0\times10^7$} 
& \multicolumn{2}{c}{$8.6\times10^8$} 
& \multicolumn{2}{c|}{$1.4\times10^9$} 
& $2.2\times10^6$ & $3.3\times10^7$ \\ 
\hline
$T_C~(\MeV)$		& 162  & 166 & 156  & 166 & 156  & 166 & \multicolumn{2}{c}{175} \\
$V_C~(\fm^3)$		& 2100 & 1791 & 5380 & 3533 & 5380 & 3533 & 1000 & 2700 \\
$\omega (\MeV)$		& 590 & 608 & 564 & 609 & 564 & 609 & \multicolumn{2}{c}{550} \\
$\omega_s (\MeV)$	& 431 & 462 & 426 & 502 & 426 & 502 & \multicolumn{2}{c}{519} \\
$\omega_c (\MeV)$	& 222 & 244 & 219 & 278 & 220 & 279 & \multicolumn{2}{c}{385} \\
$\omega_b (\MeV)$	& 183 & 202 & 181 & 232 & 182 & 234 & \multicolumn{2}{c}{338} \\
$N_u=N_d$		& 320 & 302 & 700 & 593 & 700 & 593 & 245 & 662\\
$N_s=N_{\bar{s}}$	& 183 & 176 & 386 & 347 & 386 & 347 & 150 & 405\\
$N_c=N_{\bar{c}}$ & \multicolumn{2}{c|}{4.1} & \multicolumn{2}{c}{11} & \multicolumn{2}{c|}{14} & 3 & 20 \\ 
$N_b=N_{\bar{b}}$ & \multicolumn{2}{c|}{0.03} & \multicolumn{2}{c}{0.44} & \multicolumn{2}{c|}{0.71} & 0.02 & 0.8 \\ 
\hline
$T_F~(\MeV)$		&\multicolumn{2}{c|}{119}   &\multicolumn{4}{c|}{115}  & \multicolumn{2}{c}{125}\\
$V_F~(\fm^3)$		&\multicolumn{2}{c|}{20355} &\multicolumn{4}{c|}{50646}& 11322& 30569 \\
$N_K$			&\multicolumn{2}{c|}{67.5}  &\multicolumn{4}{c|}{134}  & 142$^\dagger$& 363$^\dagger$\\
$N_{\bar{K}}$		&\multicolumn{2}{c|}{59.6}  &\multicolumn{4}{c|}{134}  & 127$^\dagger$& 363$^\dagger$\\
$N_N$			&\multicolumn{2}{c|}{20}    &\multicolumn{4}{c|}{32}   &62$^\dagger$&150$^\dagger$\\
$N_\Delta$		&\multicolumn{2}{c|}{18}    &\multicolumn{4}{c|}{28}   & --- & --- \\
$N_\Lambda$		&\multicolumn{2}{c|}{3.8}   &\multicolumn{4}{c|}{6.5}  & --- & --- \\
$N_\Xi$			&\multicolumn{2}{c|}{2.6}   &\multicolumn{4}{c|}{4.4}  & 4.7 & 13 \\
$N_\Omega$		&\multicolumn{2}{c|}{0.37}  &\multicolumn{4}{c|}{0.62} &0.81 & 2.3 \\

$N_D=N_{\bar{D}}$	&\multicolumn{2}{c|}{1.5}   & \multicolumn{2}{c}{4.0} & \multicolumn{2}{c|}{5.2} & 1.0 & 6.9\\
$N_{D^*}=N_{\bar{D}^*}$	&\multicolumn{2}{c|}{2.0}   & \multicolumn{2}{c}{5.4} & \multicolumn{2}{c|}{6.9} & 1.5 & 10\\
$N_{D_1}=N_{\bar{D}_1}$	&\multicolumn{2}{c|}{0.20}  & \multicolumn{2}{c}{0.49} & \multicolumn{2}{c|}{0.63} & 0.19 & 1.3\\
$N_{B}=N_{\bar{B}}$	&\multicolumn{2}{c|}{$8.1\times10^{-3}$}&\multicolumn{2}{c}{0.12}&\multicolumn{2}{c|}{0.20}&$5.3\times10^{-3}$ & 0.21\\
$N_{B^*}=N_{\bar{B}^*}$	&\multicolumn{2}{c|}{$1.9\times10^{-2}$}& \multicolumn{2}{c}{0.27} & \multicolumn{2}{c|}{0.45} & $1.2\times10^{-2}$  & 0.49\\
$N_{\Lambda_c}$		&\multicolumn{2}{c|}{0.17}&\multicolumn{2}{c}{0.36} & \multicolumn{2}{c|}{0.46}& --- & --- \\
$N_{\Sigma_c}$		&\multicolumn{2}{c|}{0.2}&\multicolumn{2}{c}{0.41} & \multicolumn{2}{c|}{0.52}& --- & --- \\
$N_{\Sigma_c^*}$	&\multicolumn{2}{c|}{0.28}&\multicolumn{2}{c}{0.56} & \multicolumn{2}{c|}{0.71}& --- & --- \\
$N_{\Xi_c}$		&\multicolumn{2}{c|}{0.11}&\multicolumn{2}{c}{0.25} & \multicolumn{2}{c|}{0.32}&0.10 & 0.65 \\
\hline
\end{tabular}
\\
$^\dagger$ Values contain feed down contributions.
\end{table}


In Table~\ref{Tab:pars}, we tabulate the critical temperature $T_C$ and
volume $V_C$ at the beginning of the crossover transition between
the quark-gluon plasma and hadronic matter, and the hadronization
temperature $T_H$ and volume $V_H$ at the end of the mixed
phase or hadronization for the two scenarios. Also given in
Table~\ref{Tab:pars} are the baryon and strange chemical
potentials evaluated in the statistical hadronization model~\cite{Andronic:2012dm}. 
The small strange chemical potential at RHIC, where the produced QGP has small net baryons or baryon chemical potential $\mu_B$, indicates that there is an approximate chemical equilibrium of strangeness at RHIC, and this is due to the short equilibration time and net zero strangeness in the QGP.
Because of the higher energy, longer lifetime, and almost zero net baryons of QGP at LHC, a complete chemical equilibrium is reached for strangeness, resulting in a zero strange chemical potential.

\subsection{Statistical model}

Both the statistical hadronization model and the coalescence model
have been used  to evaluate the yields of exotic hadrons produced from
heavy ion collisions. The statistical model has been very
successful in explaining the relative yields of normal hadrons in
relativistic heavy ion collisions. In the statistical
hadronization model, the number of produced hadrons of a given
type $h$ is given by~\cite{Andronic:2005yp}
\begin{equation}
\label{Eq:Stat}
N_h^{stat}=V_H\frac{g_h}{2\pi^2}\int_0^\infty\frac{p^2 dp}
{\gamma_h^{-1} e^{E_h/T_H} \pm 1}\approx \frac{\gamma_hg_hV_H}
{2\pi^2}m_h^2T_H K_2\Big(\frac{m_h}{T_H}\Big)\approx
\gamma_hg_hV_H\left( \frac{m_hT_H}{2\pi}\right)^{3/2}e^{-m_i/T_H},
\end{equation}
with $g_h$ being the degeneracy of the hadron, $\gamma_h$ the
fugacity, $K_2$ the modified Bessel function of the second kind,
and $V_H$ and $T_H$, respectively, the volume and temperature of
the source for the statistical production of hadrons. The fugacity $\gamma_h$ of hadron species $h$,  is generally 
expressed as
\begin{equation}
\gamma_h= \gamma_c^{n_c+n_{\bar{c}}} \gamma_b^{n_b+n_{\bar{b}}}
e^{(\mu_B B + \mu_s S)/T_H}\ ,
\end{equation}
where $B$, $S$, $n_c (n_{\bar{c}})$ and $n_b (n_{\bar{b}})$ are the
baryon number, strangeness, (anti-)charm quark number, and
(anti-)bottom quark number of the hadron, respectively.
Values of the strangeness chemical potentials for heavy  ion collision at RHIC and LHC are listed in
Table \ref{Tab:pars}. Since charm and bottom quarks are mostly
produced from initial hard scattering, their numbers are much
larger than those expected from a chemically equilibrated QGP.
Therefore, we consider the fugacity $\gamma_h>1$ for both charmed
and bottom hadrons. The fugacity of $n$-multiple charm quark
hadrons is the product of $n$ charm quark fugacity $\gamma_c$, or
$\gamma_c^n$, and also same for bottomed hadrons.

As shown  in Ref.~\cite{Cho:2011ew}, the charm and bottom fugacities $\gamma_c$ and
$\gamma_b$ can be determined by requiring that the total yield of charm or bottom
hadrons in the statistical hadronization model to be the same as
the total charm $N_c$ or bottom $N_b$ quark number from initial
hard nucleon-nucleon scattering. With the values $N_c=4.1$ and
$N_b=0.03$ for heavy ion collisions at RHIC at 200 GeV, which will be
explained in detail in the next session, we obtain $\gamma_c=22$
and $\gamma_b=4.0\times 10^6$. For examples, when $N_c=4.1$ and
$N_b=0.03$,
\begin{eqnarray}
N_c &= & N_{D}+N_{D^*}+\frac12\left(N_{D_s}+N_{\bar{D}_s}\right)
    +\frac12\left(N_{\Lambda_c}+N_{\bar{\Lambda}_c}\right)
    \nonumber \\
&=&1.48+2.05+\frac{0.45+0.40}{2}+\frac{0.17+0.12}{2}=4.1,\nonumber \\
N_b &= &N_{\bar{B}}+N_{\bar{B}^*}+\frac12\left(N_{\bar{B}_s}+
N_{B_s}\right)+\frac12\left(N_{\Lambda_b}+N_{\bar{\Lambda}_b}\right)
\nonumber \\
&=&8.05\times10^{-3}+1.85\times10^{-2}+\frac{2.56+2.26}{2}\times10^{-3}
+\frac{1.25+0.93}{2}\times10^{-3}=0.03.
\end{eqnarray}
In the above evaluation, the average yield of heavy anti-strange
and strange mesons as well as that of heavy anti-baryons and
baryons have been considered in order to average out the effect of
strangeness and baryon chemical potentials. A similar analysis for
LHC at 2.76 (5.02) TeV based on the charm and bottom quark numbers
$N_c=11~(14)$ and $N_b=0.44~(0.71)$ gives the charm and bottom
fugacities $\gamma_c=39~(50)$ and $\gamma_b=8.6\times
10^8~(1.4\times10^9)$, respectively.
In Table~\ref{Tab:pars}, we give the fugacities
needed for the evaluation of exotic hadron yields at RHIC and LHC
in heavy ion collisions.


Also shown in Table~\ref{Tab:pars} are the yields of various kinds
of hadrons obtained in the statistical hadronization model,
Eq.~\eqref{Eq:Stat}. We have evaluated here only the yield of directly
produced hadrons, and have not taken into account the feed-down
contributions. For examples, the yield of nucleons $N_{N}$
includes only directly produced $N$ at chemical freeze-out, and
those from the strong decay of $\Delta$ have not been included.

\subsection{Coalescence model}
\label{subsec:2C}

The coalescence model describes the production of hadrons through
the recombination or coalescence of constituents. It is based on the sudden approximation by considering the overlap between the Wigner function of the produced
particle and the density matrix of the constituents.  The model has been
successful in explaining the enhanced production of baryons
compared to that of mesons at mid-rapidity in the intermediate
transverse momentum~\cite{Greco:2003mm} and  the quark number scaling of
elliptic flows \cite{Molnar:2003ff} in heavy ion collisions. It has also been
extensively adopted to investigate hadron production from the
quark-gluon plasma in relativistic heavy ion
collisions~\cite{Hwa:2003bn, Greco:2003xt, Greco:2003mm,
Fries:2003vb, Fries:2003kq}. In this model, the number of hadron of type $h$
produced from the coalescence of $n$ constituents is given by~\cite{Greco:2003xt}
\begin{equation}
N_h^\mathrm{coal}=g_h \int \left[\prod_{i=1}^n\frac{1}{g_i}\frac{
p_i\cdot d\sigma_i}{(2\pi)^3}\frac{\mathrm{d}^3{\bf p}_i}{E_i}
f(x_i,p_i)\right]f^W(x_1,\cdots, x_n:p_1,\cdots,p_n),
\end{equation}
where $g_h$ is the degeneracy of the hadron, $g_i$ is that of its
$i$th constituent, and d$\sigma_i$ is an element of a space-like
hypersurface. The covariant phase-space distribution function of
the constituents, $f(x_i,p_i)$ are normalized to their numbers,
i.e.,
\begin{equation}
\int p_i\cdot d\sigma_i\frac{\mathrm{d}^3{\bf p}_i}{(2\pi)^3E_i}
f(x_i,p_i)=N_i,
\end{equation}
and the Wigner function of the produced hadron, $f^W(x_1
...x_n:p_1 ...p_n)$ is defined by
\begin{equation}
f^W(x_1,\cdots,x_n:p_1,\cdots,p_n)=\int\prod_{i=1}^n dy_i
e^{ip_iy_i}\psi^*\Big(x_1+\frac{y_1}{2},\cdots,x_n+\frac{y_n}{2}\Big)
\psi\Big(x_1-\frac{y_1}{2},\cdots,x_n-\frac{y_n}{2}\Big),
\end{equation}
in terms of its wave function $\psi(x_1,\cdots,x_n)$.

In the non-relativistic limit and using a spherically
symmetric harmonic oscillator function for the hadron
wave function, the above equation becomes~\cite{Chen:2003tn,
Chen:2007zp},
\begin{equation}\label{ncoal}
N_h^{\rm coal}=g_h \prod_{j=1}^n \frac{N_j}{g_j} \prod_{i=1}^{n-1}
{\int d^3 y_i d^3k_i f_i(k_i) f^W(y_i,k_i) \over \int d^3 y_i d^3
k_i f_i(k_i)},
\end{equation}
where $f^W(y_i,k_i)$ is the Wigner function associated with the
internal (relative) wave function with its internal (relative)
spatial and momentum coordinates $y_i$ and $k_i$. For heavy ion collisions at the RHIC and LHC energies, one can assume that the momentum distributions $f_j(p_j)$ are Boltzmann  at temperature $T$
for the transverse momentum $p_{j,T}$ and the strong Bjorken correlation of equal
momentum-energy rapidities, $Y_j =
\log[(E_j+p_{j,z})/(E_j-p_{j,z})]/2$ and space-time rapidities,
$\eta_j = \log[(t_j+z_j)/(t_j-z_j)]/2$ for the longitudinal momentum, i.e.,
\begin{equation}
\label{Eq:therm} f_j(p_j) \propto \delta(Y_j-\eta_j) \exp
\left(-\frac{p_{j,T} ^2}{2m_j T}\right).
\end{equation}
Using the relation
\begin{equation}
\prod_{j=1}^n \exp \left(-\frac{p_{j,T} ^2}{2m_j T}\right)
 = \exp\left (-\frac{P_T ^2}{2MT}\right) \prod_{i=1}^{n-1} \tilde{f}_i(k_i),
\end{equation}
with $P_T$ and $M$ being the total transverse momentum and 
mass, respectively, the 2-dimensional momentum
distribution functions of the constituents become 
$\tilde{f}_i(k_i)\propto e^{-k_{i}^2/(2\mu_i T)}$ in the Jacobi coordinates. In the above,  the reduced
constituent masses $\mu_i$ are defined by
\begin{equation}
\frac{1}{\mu _{i}}=\frac{1}{m_{i+1}}+\frac{1}{\sum_{j=1}^{i}{m_{j}}},
\label{reduced-mass1}
\end{equation}
or, explicitly
\begin{equation}
\mu_1=\frac{m_1 m_2}{m_1+m_2}, ~~\mu_2=\frac{m_3(m_1+
m_2)}{m_1+m_2+m_3}, ~~\mu_3=\frac{m_4(m_1+m_2+
m_3)}{m_1+m_2+m_3+m_4},~~{\rm and~so~on.} \label{reduced-mass2}
\end{equation}
The rapidity variables can be simplified at midrapidities, or
$Y=\eta\sim0$ as $\eta_j \simeq z_j/t_j$ and $Y_j \simeq
p_{j,z}/m_j$ in the non-relativistic limit. Therefore, as long as
the time $t_j$ when the coalescence occurs after the collision is
large compared with the internal time scale of the hadron,
$1/\omega$, or $t_j \gg 1/\omega$, one can omit in the Wigner
function $f^W$ the contribution from the longitudinal momentum. In
this case, the 3-dimensional momentum integrations
in Eq.~\eqref{ncoal} reduces to 2-dimensional ones over the Wigner
functions $f^W(y_i,k_i)$ and $\tilde{f}_i(k_i)$ in transverse
momentum $k_i$.  

For a uniform distribution of particles in the emission source, the Wigner functions for the $s$-wave, $p$-wave, and
$d$-wave are given explicitly as
\begin{eqnarray}
 f^W_s(y_i,k_i) & = & 8 \exp\left(
-\frac{y_i^2}{\sigma_i^2}-k_i^2\sigma_i^2\right), \nonumber \\
 f^W_p(y_i,k_i) & = & \bigg( \frac{16}{3} \frac{y_i^2}{\sigma_i^2}
-8+\frac{16}{3} \sigma_i^2 k_i^2 \bigg)\exp\left(
-\frac{y_i^2}{\sigma_i^2}-k_i^2 \sigma_i^2\right) \nonumber\\
f^W_d(y_i,k_i) & = & \frac{16}{30}\bigg[4\frac{y_i^4}{\sigma_i^4}
-20\frac{y_i^2}{\sigma_i^2}+15-20\sigma_i^2k_i^2+4\sigma_i^4k_i^4
+16y_i^2k_i^2-8(\vec y_i\cdot\vec k_i)^2\bigg] \nonumber\\
&&  \times \exp{\Big(
-\frac{y_i^2}{\sigma_i^2}-k_i^2\sigma_i^2\Big)}, \label{Eq:spdWig}
\end{eqnarray}
with the parameters $\sigma_i=1/\sqrt{\mu_i \omega}$ related to
the reduced constituent masses $\mu_i$ and the oscillator
frequency $\omega$.

After carrying out the phase-space integrals in Eq.~\eqref{ncoal}, one gets the coalescence factor for each relative coordinate,
\begin{equation}
F(\sigma_i, \mu_i, l_i, T)\equiv \frac{\int d^3y_i d^2k_i
\tilde{f}_i(k_i)f^W (y_i,k_i)}{\int d^2 k_i \tilde{f}_i(k_i)}=
\frac{(4\pi \sigma_i ^2)^{3/2}}{1+2\mu_i T \sigma_i ^2} \frac{(2
l_i) !!}{(2l_i +1)!!} \left[\frac{2\mu_i T \sigma_i ^2}{1+2\mu_i T
\sigma_i ^2}\right]^{l_i}, \label{cf}
\end{equation}
with $l_i$ being the angular momentum of the wave function
associated with the relative coordinate $y_i$. The final expression for the yield of hadrons in the
coalescence model is then 
\begin{eqnarray}
N_h^{\rm coal}&\simeq& gV \prod_{j=1}^n \frac{N_j}{g_j V}
\prod_{i=1}^{n-1} F(\sigma_i, \mu_i, l_i, T) \nonumber\\
&\simeq& gV \prod_{j=1}^n \frac{N_j}{g_j V} \prod_{i=1}^{n-1}
\frac{(4\pi\sigma_i^2)^{3/2}}{1+2\mu_iT\sigma_i^2}\frac{(2l_i)!!}
{(2l_i+1)!!} \left[ \frac{(2\mu_i T\sigma_i^2)}{(1+2\mu_{i}
T\sigma_{i}^{2})} \right]^{l_i} \nonumber \\
&\simeq& \frac{gV(M\omega)^{3/2}}{(4\pi)^{3/2}}
\frac{(2T/\omega)^L}{(1+2T/\omega)^{n+L-1}} \prod_{j=1}^n
\frac{N_j(4\pi)^{3/2}}{g_j V (m_j\omega)^{3/2}}\prod_{i=1}^{n-1}
\frac{(2l_i)!!}{(2l_i+1)!!}, \label{nquark}
\end{eqnarray}
where $l_i$ is $0$, $1$, and $2$ for an $s$-wave, a $p$-wave
and a $d$-wave constituent, respectively, and $L=\sum_{i=1}^{n-1}
l_i$, and $M=\sum_{i=1}^n m_i$. In the above, the relation
$\mu_i\sigma_i^2=1/\omega$ has been used to convert the main dependence on $l_i$
into the form of the orbital angular momentum sum $L$. When
$L\geq2$, the factor in Eq.~\eqref{nquark} depends on the way $L$
is decomposed into $l_i$, e.g., when $L=2$ and $n=3$ a factor
$4/9$ has to be considered for the combination $(l_1, l_2)=(1,1)$,
while a factor $8/15$ for $(l_1, l_2)=(2,0)$.

\subsubsection{Quark coalescence}


To evaluate the yields of hadrons produced from the QGP at
the critical temperature $T_c$ when the volume is $V_C$ in the
coalescence model, one needs to find the appropriate oscillator
frequency $\omega$. One can choose the oscillator frequencies for light,
strange, charmed, and bottom hadrons, denoted by $\omega_q$, $\omega_s$,
$\omega_c$ and $\omega_b$, respectively,  in the quark coalescence to 
reproduce the yields of certain normal hadrons in the
statistical model. These oscillator frequencies can then be used to
predict the yields of exotic hadrons.

For hadrons composed of only up and down quarks, one fixes the oscillator frequency $\omega_q$ to obtain in the coalescence model a similar yield of omega mesons as in the
statistical model. This is shown in
Table~\ref{Tab:pars} together with the numbers of light quarks, $N_u=N_d$, and
that of strange quarks, $N_s$, in the quark-gluon plasma, which are obtained from the statistical
model based on the hadronization temperature $T_H$ and volume
$V_H$ given in the table.

The parameter $\omega_s$ needed to evaluate the yield of hadrons
composed of light and strange quarks in the coalescence model is
determined by fitting the statistical model prediction for
$\Lambda(1115)$ including the contribution from resonance decays
\cite{Cho:2011ew}. Considering states that decay dominantly to
$\Lambda(1115)$, one obtains the following result for heavy ion
collisions for scenario 1 at RHIC:
\begin{eqnarray}
N_{\Lambda(1115)}^\mathrm{stat,total} & =&
N_{\Lambda(1115)}^\mathrm{stat}+\frac{1}{3}
N_{\Sigma(1192)}^\mathrm{stat}+N_{\Xi(1318)}^\mathrm{stat}+
\left(0.87+\frac{0.11}{3}\right)N_{\Sigma(1385)}^\mathrm{stat}+
N_{\Xi(1530)}^\mathrm{stat}+N_{\Omega^{-}(1672)}^\mathrm{stat} \nonumber \\
&=&3.83+\frac{1}{3}\times7.77+2.59+\left(0.87+\frac{0.11}{3}\right)
\times5.79 +1.66+0.37=16.3, \label{stat-feed1}
\end{eqnarray}
where 0.87 and $0.11/3$ in the parentheses represent the branching
ratios of $\Sigma(1385)\rightarrow \Lambda +\pi$ and $\Sigma(1385)
\rightarrow \Sigma^0 + \pi $ in the $\Sigma(1385)$ decay,
respectively. All numbers are calculated based on values 
in Table \ref{Tab:pars} for $T_H$ and $V_H$ with $\mu_s=10$ MeV and $\mu_B=24$
MeV at RHIC. With the constituent quark
masses $m_{u,d}=350$ MeV and $m_s=500$ MeV for light and strange
quarks, the above result can be reproduced by the coalescence model with $\omega_s=431$ MeV
after taking into account the contributions from the decay of same resonances 
as in Eq.~\eqref{stat-feed1}. Specifically, one has in the
coalescence model
\begin{equation}
N_{\Lambda(1115)}^\mathrm{coal,total}
=1.60+\frac{1}{3}\times4.81+1.28+\left(0.87+\frac{0.11}{3}\right)\times9.62
+2.57+0.51= 16.3. \label{coal-feed1}
\end{equation}
The same method is used to determine the oscillator frequencies $\omega_s$
for the two scenarios at both RHIC and LHC, and the values are given in Table \ref{Tab:pars}.

The oscillator frequency for charmed hadrons, $\omega_c$, is
determined from reproducing the yield of $\Lambda_c(2286)$ including
the feed-down contributions in the statistical model~\cite{Oh:2009zj} but without
considering the effect of diquarks~\cite{Lee:2007wr}. Resonances included in the feed-down contributions include $\Sigma_c(2455)$,
$\Sigma_c(2520)$, and $\Lambda_c(2625)$  as states of higher
masses are negligible. For scenario 1 at RHIC, one then obtains 
\begin{eqnarray}
N_{\Lambda_c(2286)}^\mathrm{stat,total} &=& N_{\Lambda_c(2286)
}^\mathrm{stat}+N_{\Sigma_c(2455)}^\mathrm{stat}+N_{
\Sigma_c(2520)}^\mathrm{stat}+0.67\times N_{\Lambda_c(2625)}^\mathrm{
stat} \nonumber \\
&=& 0.169+0.198+0.277+0.67 \times0.048=0.676. \label{stat-feed2}
\end{eqnarray}
from the statistical model and 
\begin{equation}
N_{\Lambda_c(2286)}^\mathrm{coal,total}=0.064+0.193+0.385+0.67
\times0.051=0.676 \label{coal-feed2}
\end{equation}
from the coalescence model, if  the charm quark mass $m_{c}=1500$
MeV and the value $\omega_c=222$ MeV are used. The 
oscillator frequencies $\omega_c$ for other scenarios at both RHIC
and LHC are summarized in Table \ref{Tab:pars}, where it is seen that the different total charm quark
number 11 at LHC 2.76 TeV energy and 14 at 5.02 TeV energy yield
similar oscillator frequencies in both scenarios.

The oscillator frequency for bottom hadrons is fixed by considering the yield of $\Lambda_b(5620)$ and contributions from
$\Sigma_b(5810)$ and $\Sigma_b^*(5830)$ decays in the statistical
model  and fitting  the results to that in the coalescence model. That
is,
\begin{eqnarray}
N_{\Lambda_b(5620)}^\mathrm{stat,total}&=&N_{\Lambda_b(5620)}^\mathrm{
stat}+N_{\Sigma_b(5810)}^\mathrm{stat}+N_{\Sigma_b(5830)}^\mathrm{
stat} \nonumber\\
&=&1.25\times10^{-3}+1.20\times10^{-3}+2.13\times10^{-3}=4.58\times10^{-3},
\label{stat-feed3} \\
N_{\Lambda_b(5620)}^\mathrm{coal,total}&=&
4.6\times10^{-4}+1.37\times10^{-3}+2.75\times10^{-3}=4.58\times10^{-3},
\label{coal-feed3}
\end{eqnarray}
yielding $\omega_b=183$ MeV for RHIC in scenario 1 using the
bottom quark mass $m_{b}=4700$ MeV. The oscillator frequencies for
bottom hadrons for the other scenario at both RHIC and
LHC are obtained using the same method, and they are summarized in Table \ref{Tab:pars}.  Again, the different total bottom quark number 0.44
at LHC 2.76 TeV energy and 0.71 at 5.02 TeV energy lead to
almost same oscillator frequencies in both scenarios.

Using the values of $\omega_q$ for normal hadrons, it has been confirmed  from Eq.~\eqref{nquark} that the addition of an $s$-wave,
 a $p$-wave, or $d$-wave quark yields, respectively, a
coalescence or suppression factor for the yields~\cite{Cho:2011ew}
\begin{eqnarray}
\frac{1}{g_i}\frac{N_i}{V} \frac{(4 \pi \sigma_i^2)^{3/2} }{(1+2
\mu_i T \sigma_i^2)}  \sim & 0.168
\nonumber \\
\frac{1}{g_i}\frac{N_i}{V} \frac{2}{3} \frac{(4 \pi
\sigma_i^2)^{3/2} 2 \mu_i T\sigma_i^2 }{(1+2 \mu_i T
\sigma_i^2)^2} \sim & 0.040
\nonumber \\
\frac{1}{g_i}\frac{N_i}{V} \frac{8}{15} \frac{(4 \pi
\sigma_i^2)^{3/2} (2 \mu_i T\sigma_i^2)^2 }{(1+2 \mu_i T
\sigma_i^2)^3} \sim & 0.011, \label{coal-factors}
\end{eqnarray}
for scenario 1 at RHIC. The coalescence factors for the other scenario
are similar to those in Eq.~\eqref{coal-factors}.  They show that the
$d$-wave coalescence is more suppressed than the $p$-wave
coalescence, which is further suppressed relative to the $s-$wave
coalescence~\cite{KanadaEn'yo:2006zk}. Since the production of
multiquark hadrons involves more $s$-, $p$-, and $d$-wave
coalescence factors, their yields are  therefore generally suppressed.

\subsubsection{Hadron coalescence}

Since weakly bound hadronic molecules are expected to be 
continuously produced from the constituent hadrons and dissociated
by interactions with other hadrons during the hadronic stage in
heavy ion collisions, their yields in the coalescence model are determined at the end of the hadronic evolution at the kinetic
freeze-out temperature $T_{F}$ and volume $V_F$. 
For the oscillator frequency in the wave function of a hadronic molecule, it can be determined from its relation to the 
mean square distance $\langle r^2\rangle$ between the two
constituent hadrons. For a hadronic molecule in
the relative $s$-wave state, the oscillator frequency is given by
$\omega=3/(2\mu_R\langle{r^2}\rangle)$ with the reduced mass
$\mu_R=m_1 m_2/(m_1+m_2)$. The mean square distance of the
hadronic molecule is also related to its binding energy $B$ via
the scattering length $a_0$ of the two interacting constituent
hadrons, that is, $B \simeq \hbar^2/(2\mu_R a_0^2)$ and
$\langle{r^2}\rangle \simeq a_0^2/2$. These relations are valid
when the binding energy is small and the scattering length is
large compared to the range of the hadronic interaction. For a
weakly bound two-body state, this leads to the simple relation
$\omega=6B$. One notes that $\langle{r^2}\rangle$ is not the
mean squared radius from the center of mass but is rather the mean
square distance in the relative coordinate between the two hadrons.  
The oscillator frequency for $f_0(980)$, for instance, obtained from its binding energy of 
$B^{f_0}=M_{K^+}+M_{\bar{K}_0}-M_{f_0(980)}
=493.7+497.6-980=11.3$ MeV, is $\omega_{f_0(980)}=6\times
B^{f_0}=67.8$ MeV. 

The oscillator frequencies for all hadronic molecules 
evaluated in the above described  method are summarized in the tables for exotic
hadrons. The number of constituent hadrons used in calculating the yield of hadronic molecules in the hadron
coalescence are determined from the statistical model at the
hadronization temperature and volume,
and are given in Table \ref{Tab:pars}.

Also shown in Table \ref{Tab:pars} are the temperature $T_F$ and volume $V_F$ of the hadronic matter 
at kinetic freeze-out, which are determined from requiring  the yield of
well-known hadronic molecules, such as the deuteron, from the hadron
coalescence at kinetic freeze-out temperature and volume to be
equal to that from the statistical hadronization model 
at hadronization temperature $T_H$ and volume  $V_H$. Since the
statistical hadronization model explains very well the yield of
deuterons~\cite{Stachel:2013zma}, it is necessary for the hadron
coalescence model to explain also the deuteron production in order
for it to be able to predict the production of exotic hadron
molecules.

Because the temperature and volume are related in the isentropic expansion of the system in heavy ion collisions
as already shown in Fig. \ref{Volevol}, the kinetic
freeze-out temperature and volume can thus be simultaneously determined by fitting the  yield of deuteron from the hadron coalescence model to that in the
statistical hadronization model.  This is achieved by taking the oscillator frequency
in the deuteron wave function to be $\omega_d=6\times
B^d=6\times2.2=13.2$ MeV  and using the number of nucleons in Table
\ref{Tab:pars}.  As shown in the table, the kinetic freeze-out at RHIC
is found to take place at a higher temperature but a smaller system
volume than those at LHC.

\subsubsection{Heavy quark pair production}

Charm quark pairs are produced through nucleon-nucleon binary collisions in relativistic heavy-ion collisions.
Their numbers can be determined by using the Monte Carlo Glauber model to simulate nucleon-nucleon binary collisions~\cite{Song:2015sfa,Song:2015ykw}.
In a given binary collision, the probability to produce a charm quark pair is given by the ratio of the charm production cross section to its total inelastic scattering cross section.
For the charm production cross section, it can be obtained from fitting to the experimental data at various collision energies~\cite{Song:2015sfa}.
The energy-momentum of each charm quark pair produced from a nucleon-nucleon collision can be obtained using the PYTHIA event generator~\cite{Sjostrand:2006za}.
Although heavy quark production in PYTHIA is based on the leading-order calculations in pQCD, albeit taking into account the effects due to initial and final parton showers, it can be tuned to reproduce the transverse momentum spectrum and rapidity distribution of
charm quarks from Fixed-Order Next-to-Leading Logarithm (FONLL) calculations~\cite{Cacciari:1998it,Cacciari:2001td}.
As for the bottom quark,
it can be simply determined from the ratio
of the  cross section for bottom production to that for charm production in the FONLL calculations.

An effect that needed to be considered for heavy quark production in heavy ion collisions is the modification of the parton distribution functions in heavy nucleus.
For example, the parton distribution functions are known to decrease at small momentum fraction $x$ compared to those in a single nucleon. This so-called  shadowing effect suppresses the production of heavy quark pair.
Since the small-$x$ region in the parton distribution function can also contribute to heavy quark production as the collision energy increases, it is important to take into account the shadowing effect on heavy quark production in heavy ion collisions, and this can be included by using  the EPS09 package~\cite{Eskola:2009uj}.  It has been shown that recent experimental data on charm production at the LHC are much better described after including the shadowing effect~\cite{Song:2015ykw,Cao:2015hia}.

Using the above described method, the number of charm pairs produced at midrapidity in 0-10 \% central collision  at RHIC and LHC are given  as follows.  Without the shadowing effect, there are 4.5, 17, 23 pairs, respectively at 200 GeV, 2.76 TeV, 5.02 TeV.  With shadowing, the numbers are 4.1, 11, 14 pairs, respectively, which are the ones used in Table \ref{tbl:heavy_quark_number}.  For the bottom quarks, they are estimated by using the ratio of the bottom production cross section  to that of charm given by FONLL calculations, which are  0.75 \%, 4 \%, 5.1\%  at collision energies of 200 GeV, 2.76 TeV, 5.02 TeV, respectively.  

\begin{table}[tbp]
\caption{Estimates of heavy quark pairs $dN/dy$ at midrapidity in 0-10\% central collision at RHIC and LHC.}
\begin{center}
\begin{tabular}{cccc}
\hline
 & RHIC & LHC @2.76 TeV & LHC @5.02 TeV  \\
\hline
Without shadowing & & & \\
 $N_c=N_{\bar{c}}$  & 4.5 & 17 & 23 \\
 $N_b=N_{\bar{b}}$  & 0.034 & 0.68 & 1.2 \\
 With shadowing & &  &  \\
 $N_c=N_{\bar{c}}$  & 4.1 & 11 & 14 \\
 $N_b=N_{\bar{b}}$  & 0.031 & 0.44 & 0.71 \\ 
\hline
\end{tabular}
\end{center}
\label{tbl:heavy_quark_number}
\end{table}%

\subsection{Freeze-out conditions for molecular states} 
\label{sec:freezeout}

In this subsection, we will closely follow the discussion given in Ref.~\cite{Cho:2015exb} on  the freeze-out condition of a hadron from an expanding system without further elastic collisions.

Kinetic freeze-out of a particle of species $i$ from a matter
occurs when its scattering time $\tau^i_{scatt}$ becomes larger
than the expansion time of the system $\tau_{exp}$~\cite{Bondorf:1978kz}. 
The scattering time scale depends on the elastic scattering cross section with other particles as follows:
\begin{equation}
\tau_{scatt}^i=\frac{1}
{\sum_j\langle\sigma_{ij} v_{ij}\rangle n_j}, \label{tau_scatt}
\end{equation}
with $\langle\sigma_{ij} v_{ij}\rangle$ being the thermal
average of the product of the  cross section times the relative velocity between
particle species $i$ and $j$, and $n_j$ the density of particle $j$.
The expansion time is defined as
\begin{equation}
\tau_{exp}=\frac{1}{\partial\cdot u}, \label{tau_exp}
\end{equation}
with $u$ being the expansion velocity of the system and  can be approximated by the ratio
of the fireball volume $V$ to its change in time, $V/(dV/dt)$.

For a spherically symmetric expanding fireball with radius $R$,
the expanding time scale is  reduced to $R/(3dR/dt)$~\cite{Becattini:2014hla}.
If the radius expands with a constant velocity, $R=vt$, one finds that $\tau_{exp}=\frac{1}{3}t$.  
Assuming for simplicity that the system is composed
of only one species and that the cross section is 
independent of velocity, the freeze-out condition then becomes
\begin{equation}
\tau_{exp}=\tau_{scatt} \rightarrow \frac{R}{3dR/dt}=\frac{1}{n\sigma\langle v\rangle}. \label{criteria1}
\end{equation}
Although there is no general relation between $dR/dt$ and $\langle v \rangle$,  particularly in the presence of a collective flow, the condition for the kinetic freeze out becomes simple if the rate of change in the radius is close to the
average velocity of the particles in the system. In this case, it is simply given by~\cite{Becattini:2014hla}
\begin{equation}
\frac{N}{R_{fo}^2}=\frac{4\pi}{\sigma_{fo}}, \label{sigmatoR2}
\end{equation}
where the subscript "$fo$" stands for physical quantities at
kinetic freeze-out and $N$ is the total number of particles. It is seen that the two dimensional density determines the condition for
freeze-out, and this is  because the transverse total cross section determines
whether a particle still interacts when it escapes
from the medium.

On the other hand, the three dimensional density at the freeze-out goes as 
\begin{equation}
\frac{N}{R_{fo}^3}= \bigg(\frac{4 \pi}{\sigma_{fo}}
\bigg)^{3/2}\frac{1}{N^{1/2}} \label{three-density}
\end{equation}
and it decreases with the square root of the total number of particles.  This suggests that for higher collision energies and/or larger 
initial temperature and/or number of particles, the
three dimensional density at which freeze-out takes places becomes
smaller~\cite{Cho:2015exb}. 

The above result is a general one not restricted to spherically symmetric expansion.  
For a system that follows the boost invariant Bjorken picture $R_L=c \tau$ with transverse expansion $R_T=v \tau$ of constant velocity $v$, one again finds that at large time, $\tau_{exp}=\frac{1}{3}\tau$ with $\tau$ being the invariant time, which has been explicitly confirmed in a hydrodynamical calculation~\citep{Cho:2015exb}.  In this case, using $V=\pi R_T^2 R_L$ leads to  the freeze-out condition
\begin{equation}
\frac{1}{3} \tau =\frac{1}{n\sigma\langle v\rangle}. \label{criteria2}
\end{equation}
This then leads to the following density at kinetic freeze-out:
\begin{equation}
\frac{N}{(R_T^2 R_L)_{fo}}  = 
 \bigg(\frac{3 \pi }{\sigma_{fo}}
\bigg)^{3/2} \frac{1}{N^{1/2}}, \label{three-density2}
\end{equation}
where $\langle v \rangle=(v^2c)^{1/3}$, with $c$ being the velocity of light, has been assumed for simplicity.  
As can be seen from Eq.~\eqref{three-density} and Eq.~\eqref{three-density2}, the 
relation between the freeze-out density and the cross section and/or the total number of particles seems to have a universal behaviour in three dimensions.  
The freeze-out conditions of the constituents in the hadronic matter for light nuclei or
hadronic molecules, which are bound, plays an important role in determining their yields in the coalescence model.

For resonances with large decay width compared to the inverse lifetime of the hadronic phase, the freeze-out condition of its daughter particles will determine the decrease of its yield 
from the statistical model prediction. This can be understood by considering, for example, a simple rate equation for the $K^*$ meson during the hadronic stage,
\begin{eqnarray}
&&\frac{dN_{K^*}(\tau)}{d\tau}=\frac{1}{\tau^K_{scatt}}
N_{K}(\tau)-\frac{1}{\tau^{K^*}_{scatt}}N_{K^*}(\tau),  \label{NKvKsrateSimple}
\end{eqnarray}
with $1/\tau^{K^*}_{scatt}=\sum_{i}\langle\sigma_{K^* i}v_{K^*
i}\rangle n_i+ \langle \Gamma_{K^*} \rangle $, and $1/\tau^K_{scatt}=\sum_j\langle\sigma_{K
j}v_{K j}\rangle n_j$. Here $i$ and $j$ stand 
for mostly the light mesons such as the  pion and $\rho$ meson, i.e.,
$1/\tau^{K^*}_{scatt}=\langle\sigma_{K^*\rho\to K\pi}v_{K^*\rho}
\rangle n_{\rho}+\langle\sigma_{K^*\pi\to K\rho}v_{K^*\pi} \rangle
n_{\pi}+\langle\Gamma_{K^*}\rangle$
and 
$1/\tau^K_{scatt}=\langle\sigma_{K\pi\to K^*\rho}v_{K\pi} \rangle
n_{\pi}+\langle\sigma_{K\rho\to K^*\pi}v_{K\rho} \rangle
n_{\rho}+\langle\sigma_{K\pi\to K^*}v_{K\pi}\rangle n_{\pi}$ with
$\langle\Gamma_{K^*}\rangle$ being the thermally averaged decay
width of the $K^*$ meson~\cite{Cho:2015qca}. In the above, non-linear terms originated from the interaction between
$K^*$ mesons or kaons, like $K\bar{K}\to\rho\pi$, are not considered.   

Consider a simple picture where the total number of light mesons and $K$ mesons are 
fixed  as the system expands.  
The equilibrium number of $K^*$ mesons is given by the 
asymptotic value obtained by taking the right hand side of Eq.~\eqref{NKvKsrateSimple} to be zero, given as 
\begin{eqnarray}
N_{K^*}^{asym}(\tau)  =\frac{\sum_j\langle\sigma_{Kj}v_{Kj} \rangle
N_j}{\sum_j\langle\sigma_{K^*j}v_{K^*j} \rangle
N_{j}+V(\tau)\langle \Gamma_{K^*} \rangle} N_K. \label{ksasym}
\end{eqnarray}
At chemical freeze-out, this value should correspond to that given by the statistical model.  As the system expands, while the total number of light hadrons and $K$ meson remain fixed, the $K^*$ number decreases due to decay  as the freeze-out volume $V(\tau)$ increases, 
leading to a suppression factor that depends on the freeze-out
volume, a result borne out in the measured $K^*$ number in heavy ion collision~\cite{Adam:2016bpr}.  This mechanism becomes
relevant only for particles that have natural decay width, which leads to terms in the rate equation that are proportional to their numbers and thus scale with the volume of the system.   
For bound states composed of hadrons, they do not have natural decay widths and are thus not affected by this  suppression mechanism.    Although the kinetic freeze-out condition for hadrons depend on their elastic scattering cross sections,  a universal kinetic freeze-out temperature is used in Ref.~\cite{Cho:2015exb}, and it is  determined by requiring the deuteron yield from the coalescence model at this temperature to reproduce the experimental value,  which has been found to follow the statistical model prediction at the chemical freeze-out point.

\subsection{Yields of hadrons} 
\label{subsec:yields}

This section summarizes the expected yields of exotic hadrons in
central Au+Au collisions at $\sqrt{s_{NN}}=200$ GeV at RHIC,
central Pb+Pb collisions at $\sqrt{s_{NN}}=2.76$ TeV at LHC,
and central Pb+Pb collisions at $\sqrt{s_{NN}}=5.02$ TeV at LHC.
We show the results
for normal quark ($\bar{q}q$, $3q$),
multiquark,
and hadronic molecule configurations,
calculated from the coalescence model in addition to those estimated
from the statistical model.
These results are shown in Tables
~\ref{tbl:yields_light},
~\ref{tbl:yields_heavy},
~\ref{tbl:yields_other_I}
and
~\ref{tbl:yields_other_II}.
We also give some discussions on the obtained results.

\begin{table}[tbp]
\caption{Summary of particle yields for light hadrons (cf.~Table~\ref{tbl:summary_light}).}
\label{tbl:yields_light}
\begin{center}
\begin{tabular}{c|c|c|c|c|c|c}
\hline
\hline
 \multicolumn{7}{c}{RHIC} \\
\hline
 Particle & $q\bar{q}/qqq$ & multiquark & $q\bar{q}/qqq$ & multiquark & Mol. & Stat.  \\
\cline{2-5}
 & \multicolumn{2}{|c}{{\small scenario 1}} & \multicolumn{2}{|c}{{\small scenario 2}} & \multicolumn{1}{|c}{} & \multicolumn{1}{|c}{}  \\
\hline
$f_{0}(980)$ & 2.1 (0.7) & $3.9 \times 10^{-2}$ & 2.1 (0.7) & $4.0 \times 10^{-2}$ & 1.7 & 3.5 \\
$a_{0}(980)$ & 6.4 & $1.2\times10^{-1}$& 6.4 & $1.2\times10^{-1}$ & 5.2 & 10 \\
$K(1460)$ & --- & $5.8\times10^{-2}$ & --- & $5.7\times10^{-2}$ & $1.3\times10^{-1}$ & $6.3\times10^{-1}$ \\
$\Lambda(1405)$ & $4.7\times10^{-1}$ & $2.3\times10^{-2}$ & $4.5\times10^{-1}$ & $2.4\times10^{-2}$ & $7.3 \times 10^{-1}$ & $8.6\times10^{-1}$  \\
$\Delta \Delta$ & --- & $4.2\times10^{-3}$ & --- & $5.3\times10^{-3}$ & --- & $1.8\times10^{-2}$ \\
$\Lambda\Lambda$-$N\Xi$ ($H$) & --- & $4.7\times10^{-4}$ & --- & $5.0\times10^{-4}$ & $1.6\times10^{-3}$ & $4.9\times10^{-3}$  \\
$N\Omega$ & --- & $1.7\times10^{-3}$ & --- & $1.9\times10^{-3}$ & $1.4\times10^{-3}$ & $6.7\times10^{-3}$ \\
\hline
\hline
 \multicolumn{7}{c}{LHC (2.76 TeV)} \\
\hline
 Particle & $q\bar{q}/qqq$ & multiquark & $q\bar{q}/qqq$ & multiquark & Mol. & Stat.  \\
\cline{2-5}
 & \multicolumn{2}{|c}{{\small scenario 1}} & \multicolumn{2}{|c}{{\small scenario 2}} & \multicolumn{1}{|c}{} & \multicolumn{1}{|c}{}  \\
\hline
$f_{0}(980)$ & 4.3 (1.2) & $5.4\times10^{-2}$ & 4.1 (1.2) & $6.0\times10^{-2}$ & 3.2 & 6.6  \\
$a_{0}(980)$ & 13 & $1.6\times10^{-1}$ & 12 & $1.8 \times 10^{-1}$ & 9.5 & $20$\\
$K(1460)$ & --- & $8.2 \times 10^{-2}$ & --- & $8.0\times10^{-2}$ & $1.9\times10^{-1}$ & 1.0 \\
$\Lambda(1405)$ & $7.5\times10^{-1}$ & $2.9\times10^{-2}$ & $7.0\times10^{-1}$ & $3.2 \times 10^{-2}$ & 1.1 & 1.4  \\
$\Delta \Delta$ & -- & $5.8 \times 10^{-3}$ & -- & $1.0 \times 10^{-2}$ & --- & $1.9 \times 10^{-2}$ \\
$\Lambda\Lambda$-$N\Xi$ ($H$) & --- & $5.0 \times 10^{-4}$ & --- & $6.1 \times 10^{-4}$ & $1.8 \times 10^{-3}$ & $5.9 \times 10^{-3}$  \\
$N\Omega$ & --- & $1.8 \times 10^{-3}$ & --- & $2.3 \times 10^{-3}$ & $1.6\times10^{-3}$ & $7.8 \times 10^{-3}$  \\
\hline
\hline
 \multicolumn{7}{c}{LHC (5.02 TeV)} \\
\hline
 Particle & $q\bar{q}/qqq$ & multiquark & $q\bar{q}/qqq$ & multiquark & Mol. & Stat.  \\
\cline{2-5}
 & \multicolumn{2}{|c}{{\small scenario 1}} & \multicolumn{2}{|c}{{\small scenario 2}} & \multicolumn{1}{|c}{} & \multicolumn{1}{|c}{}  \\
\hline
$f_{0}(980)$ & 4.3 (1.2) & $5.4 \times 10^{-2}$ & 4.1 (1.2) & $6.0 \times 10^{-2}$ & 3.2 & 6.6 \\
$a_{0}(980)$ & 13 & $1.6\times10^{-1}$ & 12 & $1.8 \times 10^{-1}$ & 9.5 & 20 \\
$K(1460)$ & --- & $8.2 \times 10^{-2}$ & --- & $8.0 \times 10^{-2}$ & $1.9 \times 10^{-1}$ & 1.0  \\
$\Lambda(1405)$ & $7.5\times10^{-1}$ & $2.9 \times 10^{-2}$ & $7.0\times10^{-1}$ & $3.2 \times 10^{-2}$ & 1.1 & 1.4 \\
$\Delta \Delta$ & --- & $5.8 \times 10^{-3}$ & --- & $1.0 \times 10^{-2}$ & --- & $1.9 \times 10^{-2}$  \\
$\Lambda\Lambda$-$N\Xi$ ($H$) & --- & $5.0 \times 10^{-4}$ & --- & $6.1 \times 10^{-4}$ & $1.8 \times 10^{-3}$ & $5.9 \times 10^{-3}$ \\
$N\Omega$ & --- & $1.8 \times 10^{-3}$ & --- & $2.3 \times 10^{-3}$ & $1.6\times10^{-3}$ & $7.8 \times 10^{-3}$ \\
\hline
\end{tabular}
\end{center}
\end{table}%

\begin{table}[tbp]
\caption{Summary of particle yields for heavy hadrons (cf.~Table~\ref{tbl:summary_heavy}).}
\label{tbl:yields_heavy}
\begin{center}
\begin{tabular}{c|c|c|c|c|c|c}
\hline
\hline
 \multicolumn{7}{c}{RHIC} \\
\hline
 Particle & $q\bar{q}/qqq$ & multiquark & $q\bar{q}/qqq$ & multiquark & Mol. & Stat.  \\
\cline{2-5}
 & \multicolumn{2}{|c}{{\small scenario 1}} & \multicolumn{2}{|c}{{\small scenario 2}} & \multicolumn{1}{|c}{} & \multicolumn{1}{|c}{}  \\
\hline
$D_{s}(2317)$ & $2.3 \times 10^{-2}$ & $2.4 \times 10^{-3}$ & $2.3 \times 10^{-2}$ & $2.5 \times 10^{-3}$ & $6.5 \times 10^{-3}$ & $6.6 \times 10^{-2}$ \\
$X(3872)$ & $5.4\times10^{-4}$ & $5.0 \times 10^{-5}$ & $5.6 \times 10^{-4}$ & $5.3 \times 10^{-5}$ & $9.1 \times 10^{-4}$ & $5.7 \times 10^{-4}$  \\
$Z_{c}(3900)$ & --- & $1.5 \times 10^{-4}$ & --- & $1.6 \times 10^{-4}$ & --- & $1.5 \times 10^{-3}$ \\
$Z_{c}(4430)$ & --- & $1.5 \times 10^{-4}$ & --- & $1.6 \times 10^{-5}$ & $5.0 \times 10^{-5}$ & $6.5 \times 10^{-5}$  \\
$Z_{b}(10610)$ & --- & $2.0 \times 10^{-9}$ & --- & $2.1 \times 10^{-9}$ & --- & $2.1 \times 10^{-8}$ \\
$Z_{b}(10650)$ & --- & $2.0 \times 10^{-9}$ & --- & $2.1 \times 10^{-9}$ & --- & $1.6 \times 10^{-8}$ \\
$X(5568)$ & --- & $5.1 \times 10^{-5}$ & --- & $5.2 \times 10^{-5}$ & --- & $2.3 \times 10^{-3}$  \\
$P_{c}(4380)$ & --- & $2.5 \times 10^{-5}$ & --- & $2.6 \times 10^{-5}$ & $2.9 \times 10^{-5}$ & $9.2 \times 10^{-5}$ \\
$P_{c}(4450)$ & --- & $1.5 \times 10^{-5}$ & --- & $1.5 \times 10^{-5}$ & --- & $9.1 \times 10^{-5}$ \\
\hline
\hline
 \multicolumn{7}{c}{LHC (2.76 TeV)} \\
\hline
 Particle & $q\bar{q}/qqq$ & multiquark & $q\bar{q}/qqq$ & multiquark & Mol. & Stat.  \\
\cline{2-5}
 & \multicolumn{2}{|c}{{\small scenario 1}} & \multicolumn{2}{|c}{{\small scenario 2}} & \multicolumn{1}{|c}{} & \multicolumn{1}{|c}{}  \\
\hline
$D_{s}(2317)$ & $5.2 \times 10^{-2}$ & $4.3 \times 10^{-3}$ & $5.0 \times 10^{-2}$ & $4.5 \times 10^{-3}$ & $1.4 \times 10^{-2}$ & $1.5 \times 10^{-1}$ \\
$X(3872)$ & $1.6\times10^{-3}$ & $1.1 \times 10^{-4}$ & $1.7 \times 10^{-3}$ & $1.3 \times 10^{-4}$ & $2.7 \times 10^{-3}$ & $1.7 \times 10^{-3}$ \\
$Z_{c}(3900)$ & --- & $3.4 \times 10^{-4}$ & --- & $4.0 \times 10^{-4}$ & --- & $4.3 \times 10^{-3}$ \\
$Z_{c}(4430)$ & --- & $3.4 \times 10^{-4}$ & --- & $4.0 \times 10^{-4}$ & $1.4 \times 10^{-4}$ & $1.7 \times 10^{-4}$  \\
$Z_{b}(10610)$ & --- & $1.3 \times 10^{-7}$ & --- & $1.5 \times 10^{-7}$ & --- & $1.9 \times 10^{-6}$ \\
$Z_{b}(10650)$ & --- & $1.3 \times 10^{-7}$ & --- & $1.5 \times 10^{-7}$ & --- & $1.5 \times 10^{-6}$ \\
$X(5568)$ & --- & $5.0 \times 10^{-4}$ & --- & $5.2 \times 10^{-4}$ & --- & $3.1 \times 10^{-2}$ \\
$P_{c}(4380)$ & --- & $5.0 \times 10^{-5}$ & --- & $5.8 \times 10^{-5}$ & $6.4 \times 10^{-5}$ & $2.1 \times 10^{-4}$ \\
$P_{c}(4450)$ & --- & $2.9 \times 10^{-5}$ & --- & $3.2 \times 10^{-5}$ & --- & $2.0 \times 10^{-4}$ \\
\hline
\hline
 \multicolumn{7}{c}{LHC (5.02 TeV)} \\
\hline
 Particle & $q\bar{q}/qqq$ & multiquark & $q\bar{q}/qqq$ & multiquark & Mol. & Stat.  \\
\cline{2-5}
 & \multicolumn{2}{|c}{{\small scenario 1}} & \multicolumn{2}{|c}{{\small scenario 2}} & \multicolumn{1}{|c}{} & \multicolumn{1}{|c}{}  \\
\hline
$D_{s}(2317)$ & $6.5 \times 10^{-2}$ & $5.4 \times 10^{-3}$ & $6.4 \times 10^{-2}$ & $5.7 \times 10^{-3}$ & $1.8 \times 10^{-2}$ & $1.9 \times 10^{-1}$ \\
$X(3872)$ & $2.5 \times 10^{-3}$ & $1.8 \times 10^{-4}$ & $2.7 \times 10^{-3}$ & $2.1 \times 10^{-4}$ & $4.5 \times 10^{-3}$ & $2.8 \times 10^{-3}$ \\
$Z_{c}(3900)$ & --- & $5.4 \times 10^{-4}$ & --- & $6.4 \times 10^{-4}$ & --- & $7.1 \times 10^{-3}$ \\
$Z_{c}(4430)$ & --- & $5.4 \times 10^{-4}$ & --- & $6.4 \times 10^{-4}$ & $2.3 \times 10^{-4}$ & $2.8 \times 10^{-4}$ \\
$Z_{b}(10610)$ & --- & $3.4 \times 10^{-7}$ & --- & $3.9 \times 10^{-7}$ & --- & $5.0 \times 10^{-6}$ \\
$Z_{b}(10650)$ & --- & $3.4 \times 10^{-7}$ & --- & $3.9 \times 10^{-7}$ & --- & $3.9 \times 10^{-6}$  \\
$X(5568)$ & --- & $7.9 \times 10^{-4}$ & --- & $8.2 \times 10^{-4}$ & --- & $5.0 \times 10^{-2}$ \\
$P_{c}(4380)$ & --- & $7.9 \times 10^{-5}$ & --- & $9.3 \times 10^{-5}$ & $1.0 \times 10^{-4}$ & $3.4 \times 10^{-4}$ \\
$P_{c}(4450)$ & --- & $4.7 \times 10^{-5}$ & --- & $5.0 \times 10^{-5}$ & --- & $3.4 \times 10^{-4}$ \\
\hline
\end{tabular}
\end{center}
\end{table}%

\begin{table}[tbp]
\caption{Summary of particle yields for other hadrons (I) (cf.~Table~\ref{tbl:summary_others}).}
\label{tbl:yields_other_I}
\begin{center}
\begin{tabular}{c|c|c|c|c|c|c}
\hline
\hline
\multicolumn{7}{c}{RHIC} \\
\hline
Particle & $q\bar{q}/qqq$ & multiquark & $q\bar{q}/qqq$ & multiquark & Mol. & Stat.  \\
\cline{2-5}
& \multicolumn{2}{|c}{{\small scenario 1}} & \multicolumn{2}{|c}{{\small scenario 2}} & \multicolumn{1}{|c}{} & \multicolumn{1}{|c}{}  \\
\hline
$\Theta(1530)$  & --- & $6.7 \times 10^{-3}$ & --- & $6.7 \times 10^{-3}$ & --- & $5.0 \times 10^{-1}$ \\
$\bar{K}KN$  & --- & $5.0 \times 10^{-3}$ & --- & $5.1 \times 10^{-3}$ & $4.2 \times 10^{-2}$ & $1.2 \times 10^{-1}$  \\
$\bar{K}NN$  & $7.3 \times 10^{-4}$ & $2.7 \times 10^{-5}$ & $7.4 \times 10^{-4}$ & $2.9 \times 10^{-5}$ & $3.9\times 10^{-3}$ & $5.8\times 10^{-3}$  \\
$\Omega\Omega$  & --- & $8.2 \times 10^{-6}$ & --- & $9.4 \times 10^{-6}$ & --- & $1.5 \times 10^{-5}$ \\
\hline
\hline
\multicolumn{7}{c}{LHC (2.76 TeV)} \\
\hline
Particle & $q\bar{q}/qqq$ & multiquark & $q\bar{q}/qqq$ & multiquark & Mol. & Stat.  \\
\cline{2-5}
& \multicolumn{2}{|c}{{\small scenario 1}} & \multicolumn{2}{|c}{{\small scenario 2}} & \multicolumn{1}{|c}{} & \multicolumn{1}{|c}{}  \\
\hline
$\Theta(1530)$  & --- & $8.2 \times 10^{-3}$ & --- & $8.5 \times 10^{-3}$  & --- & $6.8 \times 10^{-1}$ \\
$\bar{K}KN$  & --- & $6.0 \times 10^{-3}$ & --- & $6.6 \times 10^{-3}$ & $5.1 \times 10^{-2}$ & $1.5 \times 10^{-1}$ \\
$\bar{K}NN$  & $7.9 \times 10^{-4}$ & $2.3 \times 10^{-5}$ & $8.6 \times 10^{-4}$ & $3.0 \times 10^{-5}$ & $3.9\times 10^{-3}$ & $6.3 \times 10^{-3}$ \\
$\Omega\Omega$  & --- & $7.6 \times 10^{-6}$ & --- & $1.2 \times 10^{-5}$ & --- & $1.8 \times 10^{-5}$ \\
\hline
\hline
\multicolumn{7}{c}{LHC (5.02 TeV)} \\
\hline
Particle & $q\bar{q}/qqq$ & multiquark & $q\bar{q}/qqq$ & multiquark & Mol. & Stat.  \\
\cline{2-5}
& \multicolumn{2}{|c}{{\small scenario 1}} & \multicolumn{2}{|c}{{\small scenario 2}} & \multicolumn{1}{|c}{} & \multicolumn{1}{|c}{}  \\
\hline
$\Theta(1530)$  & --- & $8.2 \times 10^{-3}$ & ---& $8.5 \times 10^{-3}$ & --- & $6.8 \times 10^{-1}$ \\
$\bar{K}KN$  & --- & $6.0 \times 10^{-3}$ & --- & $6.6 \times 10^{-3}$ & $5.2 \times 10^{-2}$ & $1.5 \times 10^{-1}$ \\
$\bar{K}NN$  & $7.9 \times 10^{-4}$ & $2.3 \times 10^{-5}$ & $8.6 \times 10^{-4}$ & $3.0 \times 10^{-5}$ & $3.9 \times 10^{-3}$ & $6.3 \times 10^{-3}$ \\
$\Omega\Omega$  & --- & $7.6 \times 10^{-6}$ & --- & $1.2 \times 10^{-5}$ & --- & $1.8 \times 10^{-5}$ \\
\hline
\end{tabular}
\end{center}
\end{table}%

\begin{table}[tbp]
\caption{Summary of particle yields for other hadrons (II) (cf.~Table~\ref{tbl:summary_others}).}
\label{tbl:yields_other_II}
\begin{center}
\begin{tabular}{c|c|c|c|c|c|c}
\hline
\hline
\multicolumn{7}{c}{RHIC} \\
\hline
Particle & $q\bar{q}/qqq$ & multiquark & $q\bar{q}/qqq$ & multiquark & Mol. & Stat.  \\
\cline{2-5}
& \multicolumn{2}{|c}{{\small scenario 1}} & \multicolumn{2}{|c}{{\small scenario 2}} & \multicolumn{1}{|c}{} & \multicolumn{1}{|c}{}  \\
\hline
$T_{cc}^{1}$  & --- & $5.0 \times 10^{-5}$ & --- & $5.3 \times 10^{-5}$ & --- & $8.9 \times 10^{-4}$ \\
$\bar{D}N$  & --- & $2.6 \times 10^{-3}$ & --- & $2.6 \times 10^{-3}$ & $1.3 \times 10^{-2}$ & $1.0 \times 10^{-2}$ \\
$\bar{D}^{\ast}N$ & --- & $9.8 \times 10^{-4}$ & --- & $9.3 \times 10^{-4}$ & $1.1 \times 10^{-2}$ & $9.6 \times 10^{-3}$ \\
$\Theta_{cs}$  & --- & $7.4 \times 10^{-4}$ & --- & $7.4 \times 10^{-4}$ & --- & $6.4 \times 10^{-3}$ \\
$H_{c}$  & --- & $2.7 \times 10^{-4}$ & --- & $2.8 \times 10^{-4}$ & --- & $5.7 \times 10^{-4}$ \\
$\bar{D}NN$  & --- & $1.8 \times 10^{-5}$ & --- & $1.8 \times 10^{-5}$ & $9.4 \times 10^{-5}$ & $5.1 \times 10^{-5}$ \\
$\Lambda_{c}N$ & --- & $1.5\times10^{-3}$ & --- & $1.5\times10^{-3}$ & $5.0\times10^{-3}$ & $2.9\times10^{-3}$ \\
$\Lambda_{c}NN$ & --- & $6.7\times10^{-6}$ & --- & $6.7\times10^{-6}$ & $2.9\times10^{-6}$ & $9.8\times10^{-6}$ \\
$T_{cb}^{0}$  & --- & $9.3 \times 10^{-8}$ & --- & $9.9 \times 10^{-8}$ & --- & $1.6 \times 10^{-6}$ \\
\hline
\hline
\multicolumn{7}{c}{LHC (2.76 TeV)} \\
\hline
Particle & $q\bar{q}/qqq$ & multiquark & $q\bar{q}/qqq$ & multiquark & Mol. & Stat.  \\
\cline{2-5}
& \multicolumn{2}{|c}{{\small scenario 1}} & \multicolumn{2}{|c}{{\small scenario 2}} & \multicolumn{1}{|c}{} & \multicolumn{1}{|c}{}  \\
\hline
$T_{cc}^{1}$  & --- & $1.1 \times 10^{-4}$ & --- & $1.3 \times 10^{-4}$ & --- & $2.7 \times 10^{-3}$ \\
$\bar{D}N$  & --- & $4.3 \times 10^{-3}$ & --- & $4.2 \times 10^{-3}$ & $2.3 \times 10^{-2}$ & $1.9 \times 10^{-2}$ \\
$\bar{D}^{\ast}N$  & --- & $1.6 \times 10^{-3}$ & --- & $1.3 \times 10^{-3}$ & $2.0 \times 10^{-2}$ & $1.8 \times 10^{-2}$ \\
$\Theta_{cs}$  & --- & $1.2 \times 10^{-3}$ & --- & $1.2 \times 10^{-3}$ & --- & $1.2 \times 10^{-2}$ \\
$H_{c}$  & --- & $3.8 \times 10^{-4}$ & --- & $4.0 \times 10^{-4}$ & --- & $8.6 \times 10^{-4}$ \\
$\bar{D}NN$  & --- & $2.0 \times 10^{-5}$ & --- & $2.0 \times 10^{-5}$ & $1.1 \times 10^{-4}$ & $6.7 \times 10^{-5}$ \\
$\Lambda_{c}N$ & --- & $2.2\times10^{-3}$ & --- & $2.2\times10^{-3}$ & $7.0\times10^{-3}$ & $4.3\times10^{-3}$ \\
$\Lambda_{c}NN$ & --- & $6.7\times10^{-6}$ & --- & $6.5\times10^{-6}$ & $2.7\times10^{-6}$ & $9.9\times10^{-6}$ \\
$T_{cb}^{0}$  & --- & $1.1 \times 10^{-6}$ & --- & $1.3 \times 10^{-6}$ & --- & $2.7 \times 10^{-5}$ \\
\hline
\hline
\multicolumn{7}{c}{LHC (5.02 TeV)} \\
\hline
Particle & $q\bar{q}/qqq$ & multiquark & $q\bar{q}/qqq$ & multiquark & Mol. & Stat.  \\
\cline{2-5}
& \multicolumn{2}{|c}{{\small scenario 1}} & \multicolumn{2}{|c}{{\small scenario 2}} & \multicolumn{1}{|c}{} & \multicolumn{1}{|c}{}  \\
\hline
$T_{cc}^{1}$  & --- & $1.8 \times 10^{-4}$ & --- & $2.1 \times 10^{-4}$ & --- & $4.4 \times 10^{-3}$ \\
$\bar{D}N$  & --- & $5.3 \times 10^{-3}$ & --- & $5.3 \times 10^{-3}$ & $3.0 \times 10^{-2}$ & $2.4 \times 10^{-2}$ \\
$\bar{D}^{\ast}N$  & --- & $2.0 \times 10^{-3}$ & --- & $1.7 \times 10^{-3}$ & $2.6 \times 10^{-2}$ & $2.3 \times 10^{-2}$ \\
$\Theta_{cs}$  & --- & $1.5 \times 10^{-3}$ & --- & $1.4 \times 10^{-3}$ & --- & $1.6 \times 10^{-2}$ \\
$H_{c}$  & --- & $4.7 \times 10^{-4}$ & --- & $4.9 \times 10^{-4}$ & --- & $1.1 \times 10^{-3}$ \\
$\bar{D}NN$  & --- & $2.5 \times 10^{-5}$ & --- & $2.5 \times 10^{-5}$ & $1.5 \times 10^{-4}$ & $8.6 \times 10^{-5}$ \\
$\Lambda_{c}N$ & --- & $2.7\times10^{-3}$ & --- & $2.7\times10^{-3}$ & $9.1\times10^{-3}$ & $5.5\times10^{-3}$ \\
$\Lambda_{c}NN$ & --- & $8.2\times10^{-6}$ & --- & $8.0\times10^{-6}$ & $3.5\times10^{-6}$ & $1.3\times10^{-5}$ \\
$T_{cb}^{0}$  & --- & $2.3 \times 10^{-6}$ & --- & $2.7 \times 10^{-6}$ & --- & $5.6 \times 10^{-5}$ \\
\hline
\end{tabular}
\end{center}
\end{table}%

\begin{figure}[tbh]
\begin{center}
\includegraphics[bb=0 0 467 378,width=8cm]{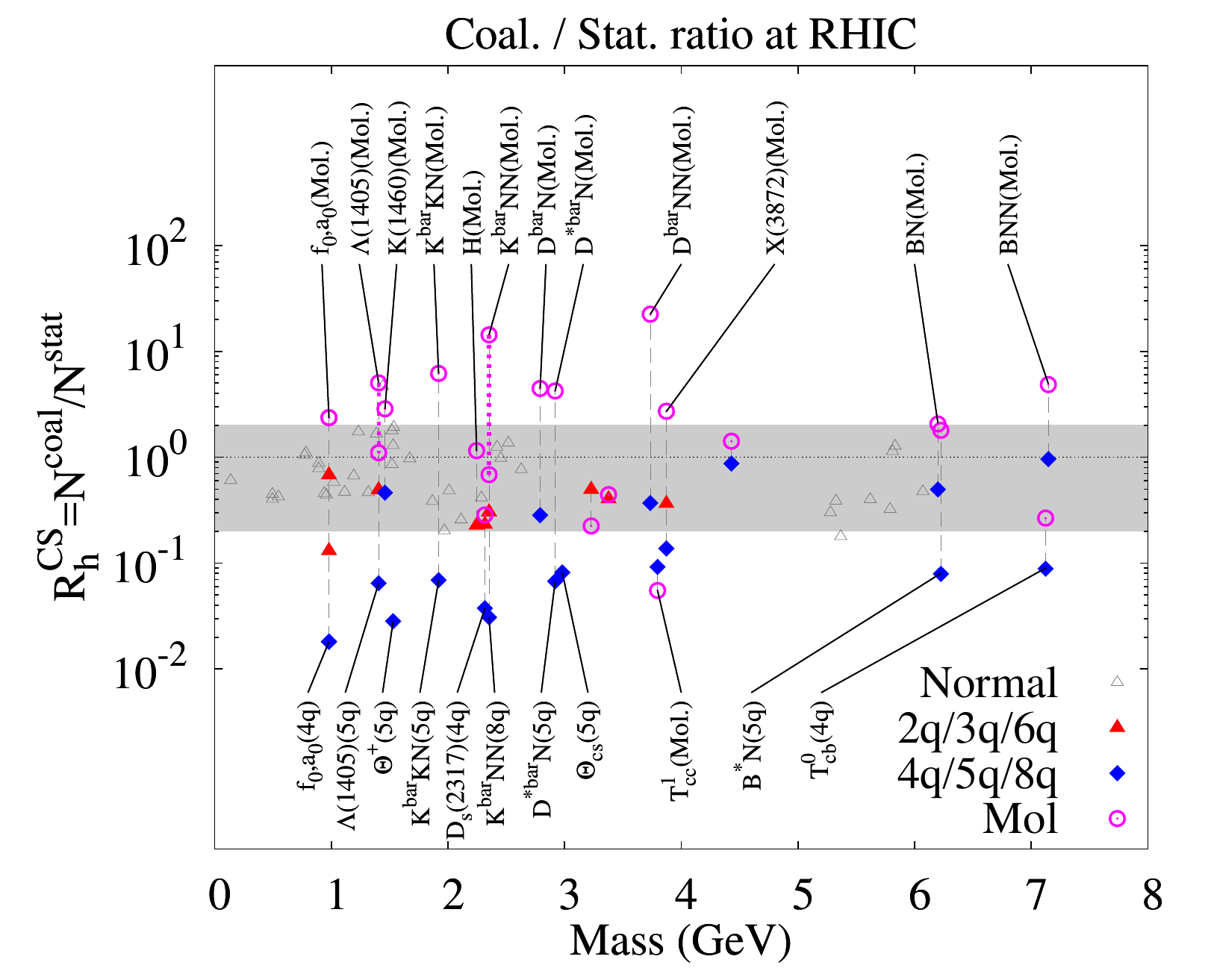}\\
\includegraphics[bb=0 0 360 252,width=8cm]{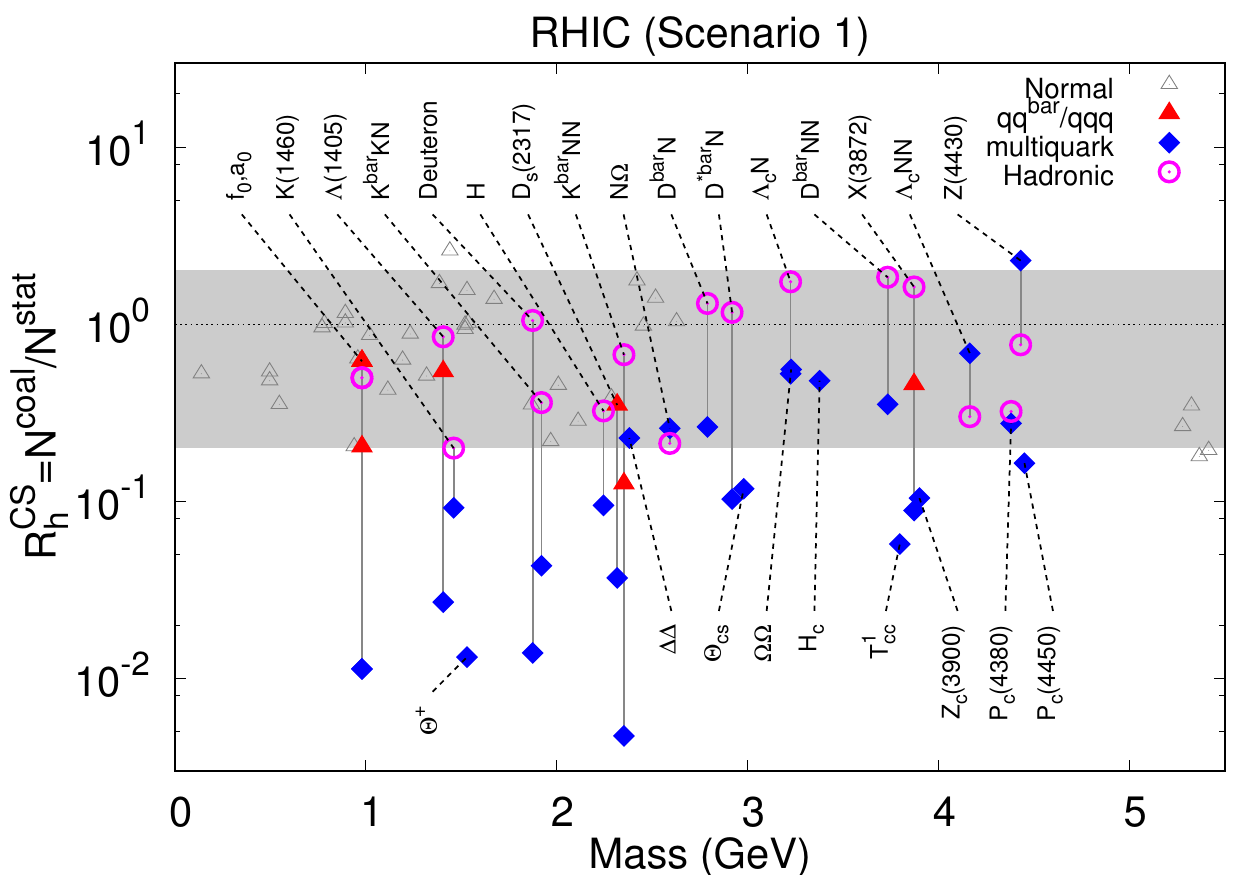}%
\end{center}
\caption{Coalescence-statistical yield ratio
from central Au+Au collisions at $\sqrt{s_{NN}}=200$ GeV at RHIC.
Upper panel is taken from Ref.~\cite{Cho:2011ew},
and lower panel shows the updated results.}
\label{Fig:RCS}
\end{figure}

In  Refs.~\cite{Cho:2010db,Cho:2011ew}, it was found that for most of the hadronic states, the yield from the
coalescence model for the compact multiquark state is smaller than
that for the usual quark configuration as a result of the
suppression due to the coalescence of additional quarks.
For the same state, the yield from the coalescence model
for a molecular configuration is larger than that from the statistical model prediction.
This is in contrast to high energy pp collisions, where molecular configurations with small
binding energies are difficult to produce at high transverse momentum $p_T$~\cite{Bignamini:2009sk}.
The upper panel of Fig.~\ref{Fig:RCS}, shows the coalescence-statistical yield ratio,
$R_{cs}=N_\mathrm{coal}/N_\mathrm{stat}$, given in Refs.~\cite{Cho:2010db,Cho:2011ew} using parameters  given in Table~\ref{Tab:pars} and assuming that the hadron coalescence takes place at $T_F=125~\MeV$ as well as including the resonance decay contributions to the  $K(\bar{K})$ and $N$ yields.
In this treatment, however, the coalescence model overestimates the deuteron yield,
which is known to follow the statistical model prediction at the chemical freeze-out temperature.

\begin{figure}[tbh]
\begin{center}
\includegraphics[bb=0 0 360 252,width=8cm]{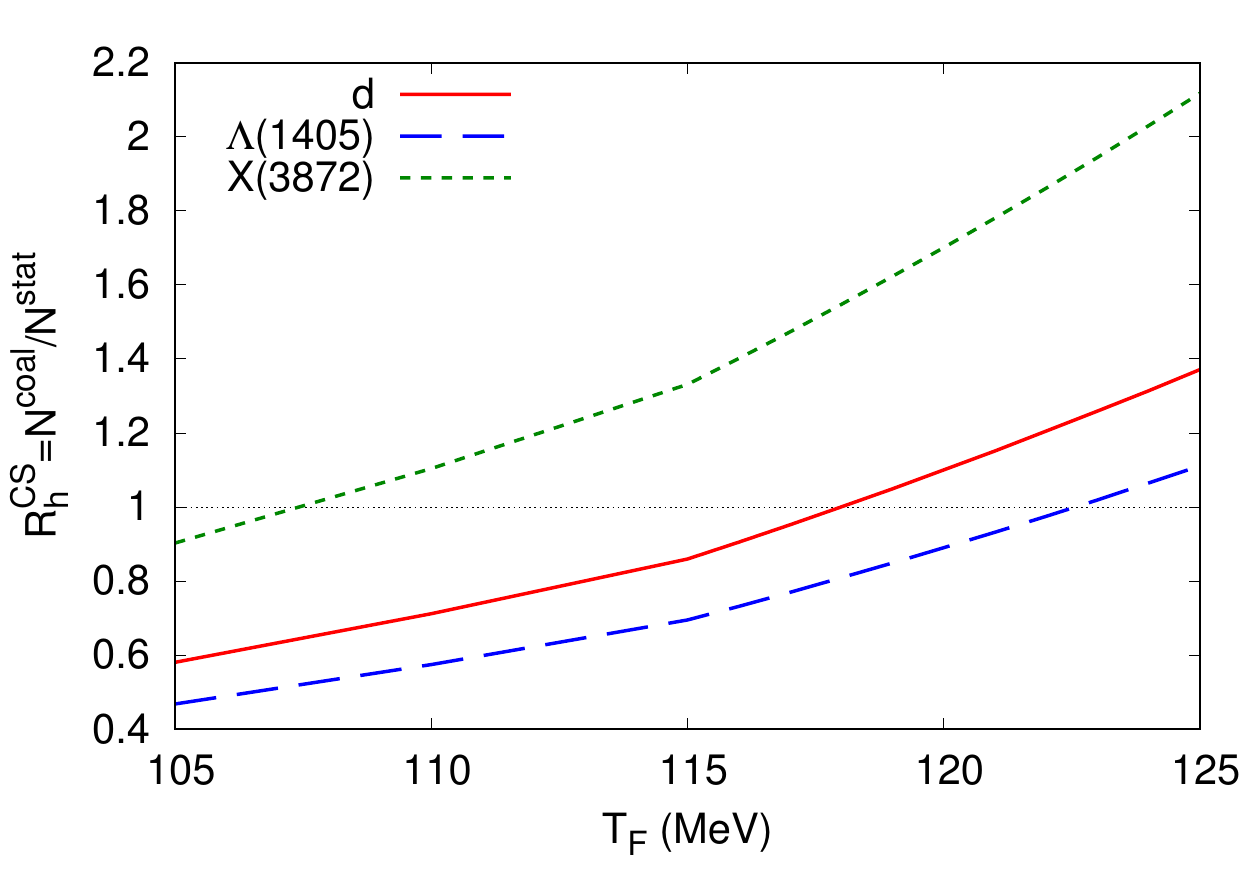}%
\end{center}
\caption{Freeze-out temperature dependence 
of the coalescence-statistical yield ratio
for deuteron, $\Lambda(1405)$ and $X(3872)$ at RHIC.}
\label{Fig:RCSd}
\end{figure}

In Fig.~\ref{Fig:RCSd}, we show the freeze-out temperature dependence 
of the coalescence-statistical yield ratio $R_h^{CS}$ for deuteron and $\Lambda(1405)$ at RHIC.
Requiring   $R_h^{CS}=1$ for the deuteron leads to a freeze-out temperature of deuteron $T_F=119~\MeV$.
According to  Eq.~\eqref{three-density2}, 
the density at which a particle freezes out is inversely proportional to its scattering cross section with other particles in the medium.  Since 
the  elastic cross section of a particle is related to its size, one would expect that the freeze-out density for a particle decreases as its  size increases.  This result suggests that the freeze-out temperature will be smaller for particles of larger size.
From the relation between the radius and the oscillator frequency $\omega$ in the wave function of a hadron, the root mean square radii for $\Lambda(1405), d, X(3872)$ are found to be 1.71, 1.77 and 2.36 fm, respectively.  Because of their different radii and thus sizes, these particles are expected to freeze-out at different temperatures.  Such subtleties are, however, neglected in previous studies, and using a common freeze-out temperature seems to still give results for hadronic molecules that are roughly consistent with the statistical model
results, i.e., $0.2 < R_h^{CS} < 2$. 

In the lower panel of Fig.~\ref{Fig:RCS}, we show the updated results in the present treatment described in this review.
Similar to the old results, the yield from the coalescence model for a compact multiquark state is  generally suppressed due to the coalescence of additional quarks,
as discussed above and in Refs.~\cite{Cho:2010db,Cho:2011ew}.
The yield from the coalescence model for a molecular configuration
strongly depends on its size.
Loosely bound hadronic molecules are more easily produced,
and tightly bound molecules have smaller size and their production is suppressed.
Thus, the yield of a hadronic molecule in heavy ion collisions  can be used as a measure of its spatial size.

\newpage

\section{Coalescence model for resonances}  
\label{sec:coalescence_resonant}

\subsection{Coalescence model}

One of the commonly used tools for calculating production probabilities of composite particles in high energy collisions is the coalescence model~\cite{Cho:2010db, Cho:2011ew, Sato:1981ez}. It is based on a sudden approximation and gives the production probability of 
a composite particle by the probability of finding it in the particle source formed by the collision just before the freeze-out stage. For a composite particle, $\gamma$, which is a bound state, $|\gamma \rangle\!\rangle $, of two particles, 
$a$ and $b$, the probability is given by
\begin{equation}
P_{\gamma} = \langle\!\langle \gamma |\hat{\rho}|\gamma \rangle\!\rangle,  \label{eq:coales}
\end{equation}
where $\hat{\rho}$ is the density matrix for the two particles in the source.

 Most of the composite particles produced in high energy collisions are, however, not stable bound states but resonances with non-negligible widths. When the particle $\gamma$ is a resonance due to the interaction between $a$ and $b$, a straightforward extension of Eq.~\eqref{eq:coales} would be to replace the bound state wave function by the resonance wave function 
 with the appropriate modification of the complex conjugation~\cite{Sekihara:2014kya}. The calculated probability, $P_{\gamma}$, would be complex, however, and its meaning becomes unclear in the case of broad resonances as we will see later in the numerical examples. Experimentally, the resonance is observed as a peak 
 in the invariant mass spectrum for the $a$ and $b$ scattering system. 
 
 In this section, the formulation for the resonance particle production in the way it is experimentally observed is discussed. 
 
  \subsection{Model for S-wave resonance}
 To make the discussion clear, we consider here a simple (Lee type) 
 model of a S-wave resonance~\cite{Lee:1954iq, Hyodo:2013nka}, where a particle $c$ is coupled to two particles, $a$ and $b$, giving rise to a resonance in the two particle scattering system. In the center-of-mass system, the Hamiltonian is given by
 \begin{equation}
 H = H_{0} + V,  \label{eq:H}
 \end{equation}
 where $H_{0}$ describes the system without the coupling which is given by $V$. The free two particle state with relative momentum ${\bf{k}}$ is denoted by $|{\bf{k}} \rangle$, while the one particle state is denoted by $|c \rangle$, and they are eigenstates of $H_{0}$ with eigenvalues $E_{k}$ and $E_{c}$, respectively, i.e. 
 \begin{equation}
 H_{0} |{\bf{k}} \rangle = E_{k} |{\bf{k}} \rangle, \quad H_{0} |c \rangle = E_{c} |c \rangle. \label{eq:eigen0}
 \end{equation}   
 The non-zero matrix elements of $V$ are expressed as 
 \begin{equation}
 \langle {\bf{k}} | V | c \rangle = \langle{\bf{k}}|V|c \rangle =  g v (k), \label{eq:V}
 \end{equation}
 where $g$ and $v(k)$ are assumed to be real. The one-particle state, $| c \rangle$, together with  the two-particle states, $| {\bf{k}} \rangle$, constitutes a complete set of the model space.
\begin{equation}
|c\rangle \langle c| + \int \frac{d{\bf{k}}}{(2\pi)^{3}} |{\bf{k}}\rangle \langle {\bf{k}}| = 1. \label{eq:cmp2}
\end{equation} 

The scattering state with the asymptotic relative momentum, ${\bf{p}}$, can be expressed as
 \begin{equation}
 |{\bf{p}} \rangle\!\rangle^{\pm} = \Bigl(1 + \frac{1}{E_{p} ^{\pm} - H_{0}} T(E_{p}^{\pm})\Bigr) |{\bf{p}} \rangle, \label{eq:swf}
 \end{equation} 
 where $E_{p} ^{\pm} = E_{p} \pm i\eta$ specifies the asymptotic boundary condition and the T-matrix, $T(E)$, satisfies the Lippmann Schwinger equation,
 \begin{equation}
 T(E) = V + V \frac{1}{E - H_{0}} T(E) = V + V \frac{1}{E - H} V, \label{eq:T}
 \end{equation} 
 for generally complex $E$. If the system has no bound state, the scattering states 
$| {\bf{p}} \rangle\!\rangle^{\pm}$ give the complete set and thus
\begin{equation}
\int \frac{d{\bf{p}}}{(2\pi)^{3}} |{\bf{p}}\rangle\!\rangle^{\pm} {^{\pm}}\langle\!\langle {\bf{p}}| = 1
= |c\rangle \langle c| + \int \frac{d{\bf{k}}}{(2\pi)^{3}} |{\bf{k}}\rangle \langle {\bf{k}}|. \label{eq:cmp1}
\end{equation}

The Lippmann-Schwinger equation for $T$, Eq.~\eqref{eq:T}, can be easily solved and the relevant matrix elements are given by
\begin{eqnarray} 
&&\langle{\bf{k}}|T(E)|c\rangle = \langle c|T(E)|{\bf{k}} \rangle = \frac{g v(k) (E-E_{c})}{E-E_{c} - \Sigma(E)}, \label{eq:Tck}  \\
&&\langle {\bf{k}}|T(E)|{\bf{k}}' \rangle = \frac{g^{2} v(k) v(k')}{E-E_{c} - \Sigma(E)},  \label{eq:Tkk}
\end{eqnarray} 
where $\Sigma(E)$ is the self-energy of the particle $c$,
\begin{equation}
\Sigma(E) = g^{2}\int \frac{d{\bf{k}}}{(2\pi)^{3}} \frac{v(k)^{2}}{E - E_{k}}.
\end{equation}
Eq.~\eqref{eq:swf} then gives the scattering state, $|{\bf{p}}\rangle\!\rangle ^{\pm}$, as
\begin{eqnarray}
&&\langle c |{\bf{p}} \rangle\!\rangle ^{\pm}  = \frac{g v(p)}{E_{p}^{\pm} - E_{c} - \Sigma (E_{p}^{\pm})}, \label{eq:sol1}
 \hspace{2cm} \\
&&\langle {\bf{k}} |{\bf{p}} \rangle\!\rangle ^{\pm}  = (2\pi)^{3} \delta ({\bf{k}} - {\bf{p}} ) 
+ \frac{g v(k) \langle c| {\bf{p}} \rangle\!\rangle^{\pm}}{E_{p}^{\pm} - E_{k}},  \label{eq:sol2} \hspace{0.5cm} 
\end{eqnarray}
A resonance state, $| r \rangle\!\rangle$, is an eigenstate of $H$ with a complex eigenvalue, $\mathcal{E}_{r}$, and satisfies
\begin{eqnarray}
(\mathcal{E}_{r} - H_{0}) |r \rangle\!\rangle = V |r \rangle\!\rangle, \label{eq:res}  \\
(\mathcal{E}_{r} - E_{k}) \langle {\bf{k}}|r \rangle\!\rangle = gv(k) \langle c|r \rangle\!\rangle, \label{eq:res1}  \\
(\mathcal{E}_{r} - E_{c}) \langle c|r \rangle\!\rangle = g \int \frac{d{\bf{k}}}{(2\pi)^{3}} v(k) \langle {\bf{k}}|r \rangle\!\rangle,  \label{eq: res2}
\end{eqnarray}
leading to the eigenvalue equation
\begin{eqnarray}
\mathcal{E}_{r} - E_{c} = \Sigma (\mathcal{E}_{r}). \label{eq: eigen}
\end{eqnarray}
which confirms that $\mathcal{E}_{r}$ is a pole of T-matrix in the complex $E$ plane.
In this model, the conjugate state, $|\tilde{r} \rangle\!\rangle$, which is an eigenstate of $H$ with the eigenvalue 
$\mathcal{E}_{r} ^{*}$, is related to $|r \rangle\!\rangle$ by complex conjugation,
i.e.
\begin{equation}
\langle {\bf{k}} | \tilde{r} \rangle\!\rangle = \langle {\bf{k}} | r \rangle\!\rangle ^{*}, \quad 
\langle c | \tilde{r} \rangle\!\rangle = \langle c | r \rangle\!\rangle ^{*}.  \label{eq:conj} 
\end{equation}
The normalization condition becomes
\begin{equation}
\langle c| r \rangle\!\rangle ^{2} + \int \frac{d{\bf{k}}}{(2\pi)^{3}} \langle {\bf{k}}| r \rangle\!\rangle^{2} = 1, \nonumber
\end{equation}
and determines $\langle c | r \rangle\!\rangle ^{2}$ as
\begin{equation}
\langle c| r \rangle\!\rangle ^{2} = (1 -  \Sigma' (\mathcal{E}_{r}))^{-1}. \label{eq:norm}
\end{equation}
The T-matrix can now be decomposed into the resonance and non-resonance terms as
 \begin{equation}
 T(E) = T^{r}(E) + T^{nr} (E), \quad T^{r}(E) =  \frac{V| r \rangle\!\rangle \langle\!\langle \tilde{r} |V}{E - \mathcal{E}_{r}}. 
 \label{eq:Tres}
 \end{equation}
 Depending on the boundary condition specified by $E_{p} ^{\pm}$, one can choose the resonance wave function $| r \rangle\!\rangle$ satisfying the corresponding boundary condition with ${\rm{Im}}\mathcal{E}_{r} = \mp \Gamma_{r} /2$ (half width of the resonance).
The resonance term of the T-matrix introduced in Eq.~\eqref{eq:Tres} has the matrix elements
\begin{eqnarray}
\langle {\bf{k}}|T^{r}(E)|c \rangle = \frac{g v(k) (\mathcal{E}_{r}-E_{c})}{(E-\mathcal{E}_{r}) 
(1-\Sigma'(\mathcal{E}_{r}))}  \label{eq:Tresck} \\
\langle {\bf{k}}|T^{r}(E)|{\bf{k}}' \rangle = \frac{g^{2} v(k) v(k')}{(E-\mathcal{E}_{r})(1-\Sigma'(\mathcal{E}_{r}))} 
\label{eq:Treskk}
\end{eqnarray}

 \subsection{Coalescence model for scattering states}
 A straightforward extension of Eq.~\eqref{eq:coales} to scattering states gives the probability, $P({\bf{p}})$, of finding the two particles, $a$ and $b$, with the relative momentum ${\bf{p}}$ as
 \begin{equation}
 P({\bf{p}}) = {^{-}}\langle\!\langle {\bf{p}}|\hat{\rho}|{\bf{p}}\rangle\!\rangle^{-}.  \label{eq:P}
 \end{equation}
 Using Eq.~\eqref{eq:swf}, one can decompose it into the free (background) term, $P^{(0)}$, the interaction term, $P^{(1)}$, and the interference term, $P^{(2)}$, as
\begin{align}
 P({\bf{p}}) =& P^{(0)}({\bf{p}}) + P^{(1)}({\bf{p}}) + P^{(2)}({\bf{p}}),  \label{eq:P012} \\
 P^{(0)}({\bf{p}}) =& \langle{\bf{p}}| \hat{\rho} | {\bf{p}}\rangle, \label{eq:P0}  \\
 P^{(1)}({\bf{p}}) =& \langle{\bf{p}}|T^{\dagger}(E_{p}^{-}) \frac{1}{E_{p}^{+} - H_{0}} \hat{\rho} \frac{1}{E_{p}^{-} - H_{0}}
 T(E_{p}^{-})|{\bf{p}}\rangle, \label{eq:P1} \\ 
 P^{(2)}({\bf{p}}) =& 2 {\rm{Re}} \langle{\bf{p}}|\hat{\rho} \frac{1}{E_{p}^{-} - H_{0}} T(E_{p}^{-}) |{\bf{p}}\rangle. \label{eq:P2a}\ .
\end{align}
 Since $P({\bf{p}})$ gives the invariant mass spectrum for the $(a,b)$ pairs, one  expects  the resonance structures to appear in $P^{(1)}({\bf{p}})$ and $P^{(2)}({\bf{p}})$. In order to get information on the resonance state from the production processes, one decomposes these quantities further into resonance and non-resonance parts, using the decomposition of the T-matrix, Eq.~\eqref{eq:Tres}.
\begin{align}
P^{(n)} ({\bf{p}}) =& P^{(n),r} ({\bf{p}}) + P^{(n),nr} ({\bf{p}}),  \quad n = 1,2 \label{eq:Pir}  \\
P^{(1),r}({\bf{p}}) 
 =& |\langle {\bf{p}} |\tilde{r} \rangle\!\rangle|^{2} 
 \langle\!\langle r| \frac{\mathcal{E}_{r}^{*} - H_{0}}{E_{p}^{+} - H_{0}} \hat{\rho} 
 \frac{\mathcal{E}_{r} - H_{0}}{E_{p}^{-} - H_{0}} 
 | r \rangle\!\rangle  \nonumber  \\
&+ 2{\rm{Re}}\left( \langle {\bf{p}}|{T^{nr}}^{\dagger} (E_{p}^{-}) \frac{1}{E_{p}^{+} - H_{0}}\hat{\rho} 
 \frac{\mathcal{E}_{r} - H_{0}}{E_{p}^{-} - H_{0}} |r \rangle\!\rangle \langle\!\langle \tilde{r} | {\bf{p}} \rangle \right),\label{eq:P1rr} \\
 P^{(2),r}({\bf{p}}) 
  =&  2 {\rm{Re}} \left( \langle {\bf{p}}|\hat{\rho} \frac{\mathcal{E}_{r} - H_{0}}{E_{p}^{-} - H_{0}} | r \rangle\!\rangle 
  \langle\!\langle \tilde{r} |{\bf{p}} \rangle \right),  \label{eq:P2rr}  \\
  P^{(1),nr}({\bf{p}}) =&
  \langle {\bf{p}}|{T^{nr}}^{\dagger}(E_{p}^{-}) \frac{1}{E_{p}^{+} - H_{0}} \hat{\rho} \frac{1}{E_{p}^{-} - H_{0}}
T^{nr}(E_{p}^{-})|{\bf{p}} \rangle, \label{eq:Pnr1} \\ 
 P^{(2),nr}({\bf{p}}) =&
 2 {\rm{Re}} \langle {\bf{p}}|\hat{\rho} \frac{1}{E_{p}^{-} - H_{0}} T^{nr}(E_{p}^{-}) |{\bf{p}} \rangle, \label{eq:P2nr}
\end{align}
 where Eq.~\eqref{eq:res} is used to get Eqs.~\eqref{eq:P1rr} and \eqref{eq:P2rr}. 
It is seen that the resonance parts, $P^{(1),r}({\bf{p}})$ and $P^{(2),r}({\bf{p}})$, carry the information on the properties of the resonance through its energy, $\mathcal{E}_{r}$, and wave functions, $ |r \rangle\!\rangle$ and $|\tilde{r} \rangle\!\rangle$. The main problem is how to extract these quantities 
 from the experimentally observed invariant mass spectrum given by $P({\bf{p}})$. As to be discussed later with numerical examples, it might be possible to obtain $P^{(0)}({\bf{p}})$, $P^{(1)}({\bf{p}})$ and $P^{(2)}({\bf{p}})$ separately through their angular distributions depending on the nature of the density matrix $\hat{\rho}$ describing the source. Once the interaction term, $P^{(1)}({\bf{p}})$, and the interference term, $P^{(2)}({\bf{p}})$, are separately obtained, the further decomposition of them into the resonance and non-resonance parts can in principle be done through their 
 $p = |{\bf{p}}|$ dependences, since the non-resonance parts have smooth energy dependence and are also small in magnitude compared with the resonance parts in the relevant region.
  
As for the density matrix $\hat{\rho}$ describing the source, one can assume that it has no matrix elements between 
$|c \rangle$ and $|{\bf{k}} \rangle$ and denote its non-zero matrix elements as
\begin{equation}
\langle {\bf{k}} | \hat{\rho} | {\bf{k}}' \rangle = \rho ( {\bf{k}}, {\bf{k}}' ), \quad \langle c | \hat{\rho} | c \rangle = \rho_{c}. \label{eq:density}
\end{equation}
$\rho_{c}$ then contributes only to $P^{(1)}$ in the decomposition \eqref{eq:P012} of $P$ and one has
\begin{align}
P^{(1)}({\bf{p}}) =& P^{(1)}_{c}({\bf{p}}) + P^{(1)}_{ab}({\bf{p}}),  \hspace{3.3cm}  \label{eq:P1cab} \\
P^{(1)}_{c}({\bf{p}}) =& \rho_{c} |\langle c| {\bf{p}} \rangle\!\rangle ^{-} |^{2} 
= \frac{\rho_{c} (g v(p))^{2}}{|E_{p}^{-} - E_{c} - \Sigma(E_{p}^{-})|^{2}}, \label{eq:P1c} \\
P^{(1)}_{ab}({\bf{p}})  
=&\frac{g^{4} v(p)^{2}  F^{(1)}(E_{p}^{-})}{|E_{p}^{-} - E_{c} - \Sigma(E_{p}^{-})|^{2}}, \label{eq:P1ab} \\
{\rm{with}} \quad & F^{(1)}(E) \equiv \int \frac{d {\bf{k}} d{\bf{k}}' }{(2\pi)^{6}} 
\frac{\rho({\bf{k}}, {\bf{k}}') v(k) v(k')}{(E^{*} - E_{k})(E - E_{k'})}. \hspace{0.5cm} \label{eq:F1} 
\end{align}
and 
\begin{align}
P^{(0)}({\bf{p}}) =& \rho ({\bf{p}}, {\bf{p}}), \label{eq:P0ab} \\
P^{(2)}({\bf{p}}) 
=& 2{\rm{Re}} \Bigl( \frac{g^{2} v(p) F^{(2)} ({\bf{p}}, E_{p}^{-})}{E_{p}^{-} - E_{c} - \Sigma(E_{p}^{-})}  \Bigr),  \label{eq:P2b} \\
{\rm{with}} \quad 
&F^{(2)}({\bf{p}}, E) \equiv \int \frac{d{\bf{k}}}{(2\pi)^{3}} \frac{\rho({\bf{p}}, {\bf{k}}) v(k)}{E-E_{k}}. 
\label{eq:F2}
\end{align}

Applying Eq.~\eqref{eq:P1rr} to calculate the resonance part, one notes that the contribution of $\rho_{c}$ contains divergences at $E_{p} = E_{c}$, which are of course cancelled by the corresponding divergences in the non-resonance part. In order to avoid this problem and get a more reasonable resonance part, one modifies Eq.~\eqref{eq:Tresck} by replacing $\mathcal{E}_{r}$ with $E$ in the numerator, i.e. 
\begin{equation}
\langle {\bf{k}}|T^{r}(E)|c \rangle = \frac{g v(k) ( E-E_{c})}{(E-\mathcal{E}_{r}) 
(1-\Sigma'(\mathcal{E}_{r}))},  \label{eq:Tresckm} 
\end{equation}
so as to eliminate the divergences without changing the residue at the pole.
One then obtains for the resonance parts
\begin{align}
P_{c}^{(1),r}({\bf{p}})  
=& \frac{g^{2}v(p)^{2}\rho_{c}}{|1-\Sigma'(\mathcal{E}_{r})|^{2} |E_{p} - \mathcal{E}_{r}|^{2}} 
 \left[2{\rm{Re}} \left( \frac{(1-\Sigma'(\mathcal{E}_{r}))(E_{p} - \mathcal{E}_{r})}{E_{p}^{-} - E_{c} - \Sigma (E_{p}^{-})} \right) - 1 \right],  \label{eq:P1crm} \\
P_{ab}^{(1),r}({\bf{p}})  
=& \frac{g^{4}v(p)^{2} F^{(1)} (E_{p}^{-})}{|1-\Sigma'(\mathcal{E}_{r})|^{2}|E_{p} - \mathcal{E}_{r}|^{2}} 
  \left[2{\rm{Re}} \left( \frac{(1-\Sigma'(\mathcal{E}_{r}))(E_{p} - \mathcal{E}_{r})}{E_{p}^{-} - E_{c} - \Sigma (E_{p}^{-})} \right) - 1\right],  \label{eq:P1abrm} \\
P^{(2),r} ({\bf{p}}) 
=& -2{\rm{Re}} \left(\frac{g^{2} v(p) F^{(2)}({\bf{p}},E_{p}^{-})}{(E_{p} - \mathcal{E}_{r}) (1 - \Sigma' (\mathcal{E}_{r}))} \right), 
\label{eq:P2rm}
\end{align}
where $F^{(1)}$ and $F^{(2)}$ are defined by Eqs.~\eqref{eq:F1} and \eqref{eq:F2}, respectively.

It is noticed here that the completeness of $|{\bf{p}}\rangle\!\rangle ^{-}$, Eq.~\eqref{eq:cmp2}, leads to the following sum rules for the integrated probabilities denoted by $\Pi$s.
\begin{align}
\Pi_{ab}^{(1)} + \Pi^{(2)} =&
\int \frac{d{\bf{p}}}{(2\pi)^{3}} \Bigl(P_{ab}^{(1)} ({\bf{p}}) + P^{(2)} ({\bf{p}}) \Bigr) = 0, \label{eq:sum1} \\
\Pi_{c}^{(1)} =& \int \frac{d{\bf{p}}}{(2\pi)^{3}} P_{c}^{(1)} ({\bf{p}}) = \rho_{c}, \label{eq:sum2}
\end{align}
where  
\begin{align}
\Pi =& \int \frac{d{\bf{p}}}{(2\pi)^{3}} P({\bf{p}}) = {\rm{Tr}} \hat{\rho}  
= \int \frac{d{\bf{p}}}{(2\pi)^{3}} \rho({\bf{p}},{\bf{p}}) + \rho_{c}  \nonumber \\
=& \int \frac{d{\bf{p}}}{(2\pi)^{3}} P^{(0)}({\bf{p}}) +\rho_{c} = \Pi^{(0)} + \Pi_{c}^{(1)}
\nonumber
\end{align}
obtained from Eqs.~\eqref{eq:P} and \eqref{eq:P0} have been used. The sum rules imply that the number of $(a,b)$ pairs is not affected by their mutual interactions but is increased by the decay of $c$.

\subsection{Numerical examples}

To see how the resonance appears in the probability, $P({\bf{p}})$, which gives the invariant mass spectrum of the $(a,b)$ pairs, calculated by the above formalism, some numerical examples are presented below. 

With the non-relativistic kinetic energy and a monopole form factor, i.e.
\begin{equation}
E_{k} = \frac{k^{2}}{2m}, \quad v(k) = \frac{1}{k^{2} + \mu^{2}}. \label{eq:keff}
\end{equation}
$\Sigma(E)$ is then given by 
\begin{eqnarray}
2m\Sigma (E) 
= - \frac{\lambda}{(\mu - i p_{E})^{2}}, \label{eq: sigma1}
\end{eqnarray}
where $\lambda = \frac{m^{2}g^{2}}{2\pi \mu}$ and $p_{E} = \pm \sqrt{2mE}$ with the appropriately chosen sign. 
For example, $p_{E_{p} ^{\pm}} = \pm \sqrt{2mE_{p}} = \pm p$ for real positive $E_{p}$ and $p$.  
The full off-shell T-matrix becomes
\begin{align}
2m\langle {\bf{k}}|T(E)|c \rangle = 2m \langle c|T(E)|{\bf{k}} \rangle  
=& \frac{\sqrt{8\pi \lambda \mu} (p_{E}^{2} - p_{c}^{2})}{k^{2}+\mu^{2}} 
 \left( p_{E} ^{2} - p_{c}^{2} + \frac{\lambda}{(\mu - i p_{E} )^{2}}\right)^{-1},  \\
2m\langle {\bf{k}}|T (E)|{\bf{k}}' \rangle  
=& \frac{8\pi \lambda \mu}{(k^{2} + \mu^{2})({k'}^{2} + \mu^{2})} 
 \left( p_{E} ^{2} - p_{c}^{2} + \frac{\lambda}{(\mu - i p_{E} )^{2}}\right)^{-1},  \label{eq: T1}
\end{align}
where $p_{c} ^{2} = 2m E_{c}$.
The poles of $T (E)$ are given by the solutions of a 4th order equation for the variable 
$p_{E} $, and one chooses one of them with positive real part and negative imaginary part for the resonance pole which will be denoted as $p_{r}$, i.e. $p_{r}^{2} = 2m \mathcal{E}_{r}$. In this notation, one has 
\begin{eqnarray}
\Sigma'(\mathcal{E}_{r}) = \frac{-i \lambda}{p_{r}(\mu - ip_{r})^{3}}, \quad  \nonumber
\end{eqnarray}
and $\langle c| r \rangle\!\rangle ^{2}$ given by Eq.~\eqref{eq:norm} becomes
\begin{equation}
 \langle c| r \rangle\!\rangle ^{2} = \biggl( 1 + \frac{i\lambda}{p_{r} (\mu - ip_{r})^{3}} \biggr)^{-1}. \label{eq:norm1}
 \end{equation} 
The density matrix given by Eq.~\eqref{Eq:therm} and used in the ExHIC papers~\cite{Cho:2010db, Cho:2011ew} is in the present notation
  \begin{equation}
  \rho({\bf{k}}, {\bf{k}}') = N (2\pi)^{4} \delta({\bf{k}} - {\bf{k}}') \delta(k_{z})  \label{eq:density0}
  \exp\left(-\beta \frac{k_{T} ^{2}}{2m}\right),
  \end{equation}
  where $k_{z}$ and ${\bf{k}}_{T}$ are the longitudinal and transverse components of ${\bf{k}}$, respectively, and the normalization $N$ is determined by the condition, 
 \begin{equation}
 \int \frac{d{\bf{k}}}{(2\pi)^{3}} \rho({\bf{k}},{\bf{k}}) = n_{ab}, 
 \quad {\rm{giving}} \quad N = \frac{2\pi \beta n_{ab}}{Vm},   \label{eq:density1n}
 \end{equation}
 where $(2\pi)^{3} \delta ({\bf{0}})$ has been replaced by the volume of the emission source $V$ and $n_{ab}$ is the number of pairs of the particles $a$ and $b$.  Inserting Eq.~\eqref{eq:density0} into Eqs.~\eqref{eq:F1} and 
\eqref{eq:F2}, however, one sees that $O(1/\eta)$ terms appear in both equations. The first delta function in Eq.~\eqref{eq:density0}, which gives rise to this problem, is based on the assumption that the size of the emission source is much larger than the sizes of hadrons. Although the assumption is justified for the bound state formation, it is not the case for scattering states. One  therefore chooses a Gaussian distribution in the transverse direction, i.e. 
\begin{eqnarray}
\rho({\bf{k}}, {\bf{k}}') = \tilde{N} (2\pi)^{2} \delta(k_{z} - k_{z}') \delta(k_{z})
 \exp(-\alpha ({\bf{k}_{T}} - {\bf{k}}'_{T})^{2}) \nonumber \\
\times \exp(-\beta \frac{({\bf{k}}_{T} + {\bf{k}}'_{T})^{2}}{8m}),  \label{eq:density1} \hspace{4cm} 
\end{eqnarray}
and the normalization condition \eqref{eq:density1n} determines $\tilde{N}$ as
\begin{equation}
 \tilde{N} = \frac{2\pi \beta n_{ab}}{ L m}, \label{eq: density2n} \hspace{5cm}
 \end{equation}
 where  $2\pi \delta(0)$ is replaced by the longitudinal length $L$ of the source. Since for large $\alpha$, $\exp(-\alpha({\bf{k}}_{T} - {\bf{k}}'_{T})^{2}) \rightarrow \frac{1}{4\pi \alpha} (2\pi)^{2} \delta ({\bf{k}}_{T} - {\bf{k}}'_{T} )$,  $4\pi \alpha L$ corresponds to the volume $V$ of the source in Eq.~\eqref{eq:density1n}.
 Substituting Eq.~\eqref{eq:density1} into Eqs.~\eqref{eq:F1} and \eqref{eq:F2}, and carrying out the angular parts of the integrals, 
one obtains 
\begin{align}
P_{ab} ^{(1)} ({\bf{p}}) =& \frac{(2\lambda \mu)^{2} F^{(1)}(E_{p}^{-})}{m^{2}(p^{2} + \mu^{2})^{2} |p^{2} - p_{c} ^{2} 
+ \frac{\lambda}{(\mu + ip)^{2}}|^{2}},  \nonumber \\
F^{(1)}(E_{p}^{-})  
=& \tilde{N} \int_{0}^{\infty} \int_{0} ^{\infty} \frac{dk dk' k k' e^{-\alpha_{+} (k^{2} + {k'}^{2})}
I_{0} (2\alpha_{-} k k')}{(k^{2} + \mu^{2})({k'}^{2} + \mu^{2})(E_{p}^{+} - E_{k})
(E_{p}^{-} - E_{k})},  \nonumber \\
P^{(2)} ({\bf{p}}) =&  {\rm{Re}} \Bigl(
\frac{4 \lambda \mu F^{(2)}({\bf{p}},E_{p}^{-}) }{m (p^{2} + \mu^{2}) (p^{2} - p_{c} ^{2} + \frac{\lambda}{(\mu + ip)^{2}})}  \Bigr), \nonumber \\
F^{(2)}({\bf{p}}, E_{p}^{-}) 
=& 2\pi \tilde{N} \delta (p_{z}) \int_{0} ^{\infty} \frac{dk k e^{-\alpha_{+} (p^{2} + k^{2})} I_{0} (2\alpha_{-} pk)}
{(k^{2} + \mu^{2}) (E_{p}^{-} - E_{k})} ,  \nonumber
\end{align}
where $\alpha_{\pm} = \alpha \pm \frac{\beta}{8m}$ and the variable $k_{T}$ has been changed to $k$ since they are the same for $k_{z} = 0$. 
It is  seen that the two terms have very different angular dependences. $P^{(1)}_{ab}$ is 
isotropic reflecting the S-wave nature of the interaction, while the $P^{(2)}$
contributes only on the $p_{z} = 0$ plane due to the strong anisotropy of the density matrix 
(Eq.~\eqref{eq:density1}). $P^{(1)}_{c} ({\bf{p}}) $ given by Eq.~\eqref{eq:P1c} is 
\begin{equation}
P^{(1)}_{c}({\bf{p}}) = \frac{8\pi \mu \lambda \rho_{c}}{(p^{2} + \mu^{2})^{2} 
 |p^{2} -p_{c} ^{2} + \frac{\lambda}{(\mu + ip)^{2}}|^{2}},  
\end{equation}
and has the isotropic angular dependence as $P_{ab} ^{(1)}$. Thus, in the direction of $p_{z}=0$, the interference term, $P^{(2)}$ is dominant while in the other direction, the probability is given by the interaction term, $P^{(1)} =P_{ab} ^{(1)} + P^{(1)}_{c}$. The resonance parts can be similarly calculated by Eqs.~\eqref{eq:P1crm}, \eqref{eq:P1abrm} and \eqref{eq:P2rm}.

For numerical examples, three sets of interaction parameters are considered with  the set A generating a resonance similar to $\Lambda (1405)$ while the sets B and C  giving typical examples of broad and narrow resonances, respectively. Thus, $(a,b)$ is $(\pi, \Sigma)$ and $c$ is the $\bar{K} N$ bound state with its coupling to $\pi \Sigma$ switched off.  The interaction parameters are given in Table \ref{Table4.1} and the resulting properties of the resonances (pole positions in the complex $E$ and $p$ planes, $\mathcal{E}_{r}$ and 
$p_{r}$, and the square of the overlap between the resonance state and the one-particle state, 
$\langle c| r \rangle\!\rangle ^{2}$ (Eq. \eqref{eq:norm1})) are given in Table \ref{Table4.2}. The square of the overlap for a bound state would represents the probability of the one particle state to remain in the bound state and therefore would be a real positive number less than or equal to 1. In the case of a resonance state, however, the probability interpretation is not applicable and the magnitude of deviation from 1 gives a measure of the contribution from $(a,b)$ scattering states in forming the resonance.

\begin{table}
\caption{Interaction parameters}\label{Table4.1}
\begin{center}
 \begin{tabular} {ccccc}
\hline
\hline
  set & $m ({\rm{GeV}})$ & $\mu ({\rm{GeV}})$ & $\lambda ({\rm{GeV}} ^{4})$ & $E_{c} ({\rm{GeV}})$ \\
\hline
 A  &  $0.125$ & $0.5$ & $3.0 \times 10^{-3}$ &  $0.12$ \\
 B  & $0.125$ & $0.5$  & $6.25 \times 10^{-3}$ & $0.15$ \\
 C &  $0.125$ &$0.5$ & $6.25 \times 10^{-4}$ & $0.10$ \\
\hline
\hline
\end{tabular}
\end{center} 
\end{table}
 
\begin{table}
\caption{Resonance properties}\label{Table4.2}
\begin{center}
\begin{tabular} {cccc}
\hline
\hline
set & $\mathcal{E}_{r} ({\rm{GeV}})$ & $p_{r} ({\rm{GeV}})$ & $ \langle c| r \rangle\!\rangle ^{2}$ \\ \hline
 A  &  $0.080 - i 0.026$ & $0.143 - i 0.023 $ & $1.157 - i 0.116$ \\
 B  & $0.047 - i 0.065$ & $0.126 - i 0.064 $ & $1.758 - i 0.383$ \\
 C &  $0.092 - i 0.0052$ & $0.152 - i 0.0043 $ & $1.024 - i 0.019$ \\
\hline
\hline
\end{tabular}
\end{center} 
\end{table}

 The density matrix parameters are chosen to be similar to those used
in \ref{subsec:yields} with the parameters given in Table \ref{Tab:pars}
for RHIC~\cite{Cho:2010db, Cho:2011ew}
and are given in Table \ref{Table4.3}. Thus the temperature $1/\beta$ and the volume $V = 4\pi \alpha L$ are 
the freeze-out temperature $T_{F} (0.119 {\rm{GeV}})$ and volume $V_{F} (20355 {\rm{fm}}^{3})$, respectively, and $n_{\pi \Sigma}$ is estimated by the statistical model at the hadronization temperature, $T_{H} (0.162 {\rm{GeV}})$, and volume, $V_{H} (2100 {\rm{fm}}^{3})$. $\rho_{c}$ is calculated by the coalescence model with the $\bar{K} N$ density matrix 
 chosen analogously to that for $\pi \Sigma$ and the bound state wave function taken to be that 
 of the harmonic oscillator ground state with the oscillator frequency $\omega = 20.5 {\rm{MeV}}$. 
To see the relative importance of the three terms, one takes here 
 $n_{\pi \Sigma} = 1$ and show only the ratio $n_{\bar{K} N}/n_{\pi \Sigma}$ and the resulting 
 $\rho_{c}$.

\begin{table}
\caption{Density matrix parameters}\label{Table4.3}
 \begin{center}
 \begin{tabular} {ccccc}
\hline
\hline
 $\beta ({\rm{GeV}}^{-1})$ & $ L ({\rm{GeV}}^{-1}) $ & $\alpha ({\rm{GeV}}^{-2})$ & 
$n_{\bar{K} N}/n_{\pi \Sigma}$ & $\rho_{c}$ \\ \hline 
 $ 8.4 $ & $ 1055 $ & $ 200 $ & $ 1.00 $ & $0.0025$ \\ \hline
\hline
\end{tabular}
\end{center}
\end{table}
 
\begin{figure}
\begin{center}
\includegraphics[bb=0 0 260 168,width=6cm]{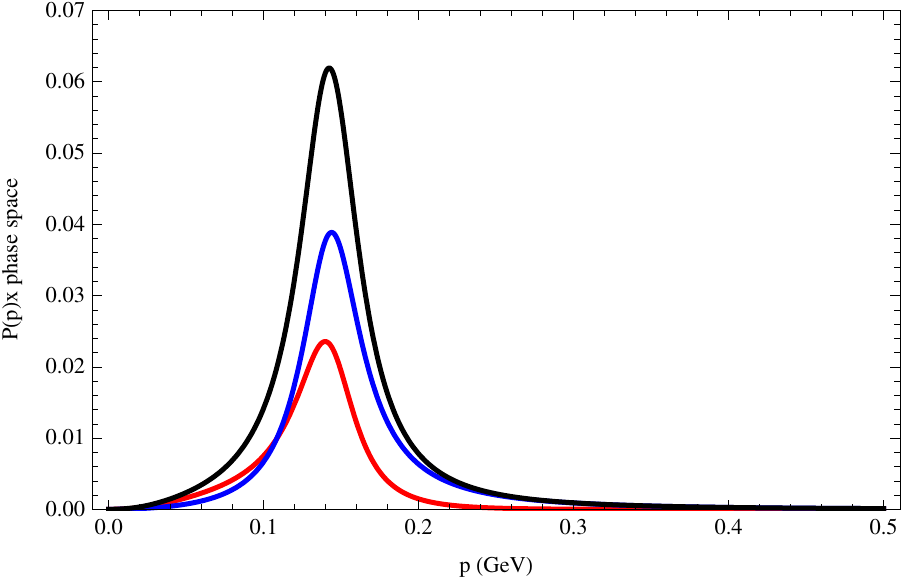}~~%
\includegraphics[bb=0 0 260 163,width=6cm]{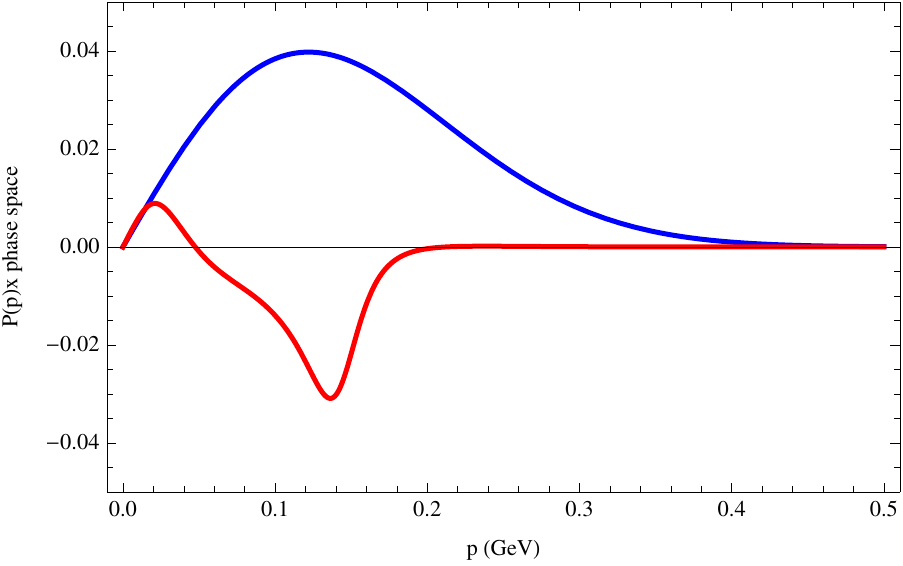}\\
\includegraphics[bb=0 0 260 168,width=6cm]{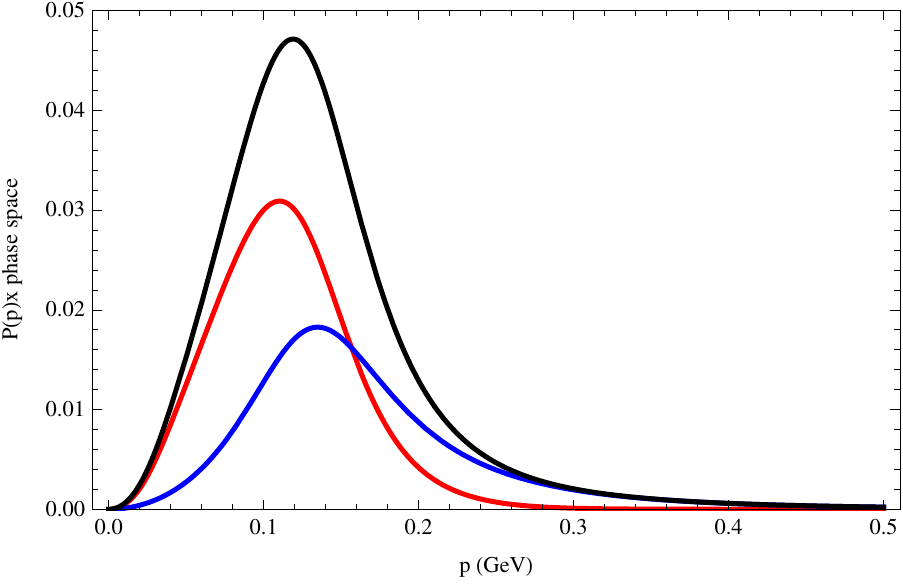}~~%
\includegraphics[bb=0 0 260 163,width=6cm]{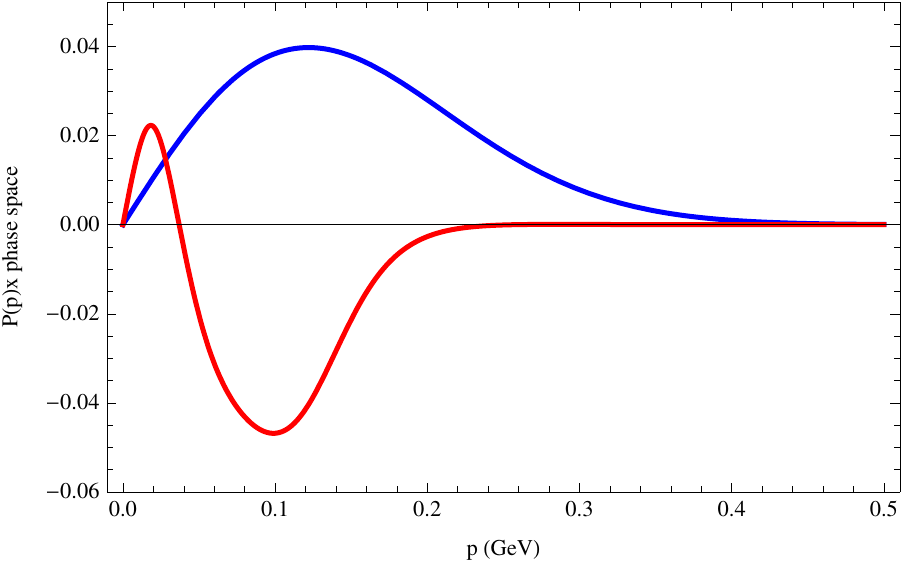}\\
\includegraphics[bb=0 0 260 168,width=6cm]{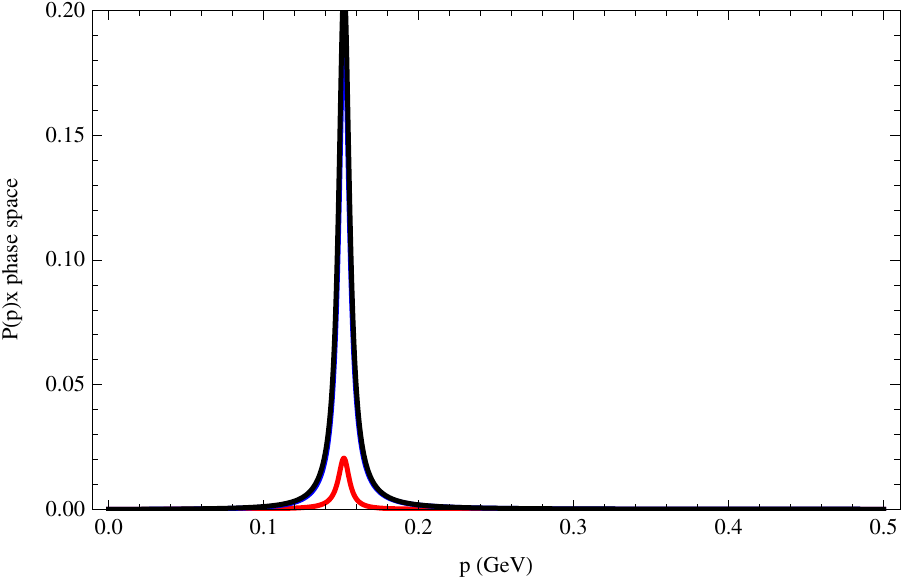}~~%
\includegraphics[bb=0 0 260 163,width=6cm]{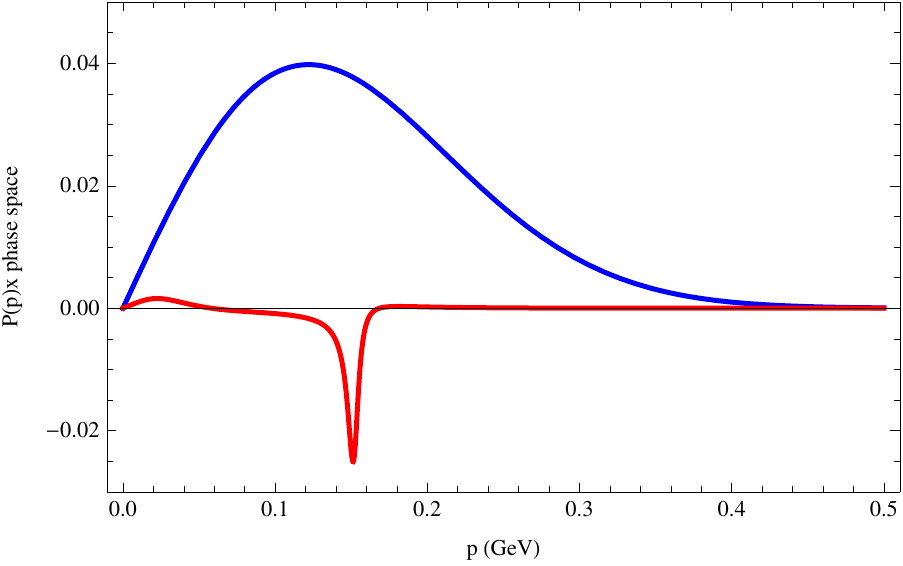}\\
\end{center}
\caption{Left: $P_{ab} ^{(1)}$ (red), $P^{(1)}_{c}$ (blue) 
and the sum $P^{(1)} = P_{ab} ^{(1)} + P_{c}^{(1)}$ (black) 
multiplied by the phase space factor, $p^{2}/2\pi^{2}$,
as functions of $p$
for the sets A (top), B (middle), C (bottom).
Right: $P^{(0)}/50$ (blue) and $P^{(2)}$ (red) multiplied by the phase space factor, $p/2\pi$, as a function of $p$
for the sets A (top), B (middle), C (bottom).
}\label{Fig.4.1}
\end{figure}

 
Shown in the left panels of Fig.\ref{Fig.4.1} are the two terms, $P_{ab} ^{(1)}$ and $P_{c}^{(1)}$, which have the same 
 isotropic angular dependence, and the sum  $P^{(1)} = P_{ab} ^{(1)} + P_{c}^{(1)}$, multiplied by the phase space factor $p^{2}/2\pi^{2}$ as functions of  $p$, for the interaction parameter sets A , B and C.
  
One sees that $P_{ab}^{(1)}$ 
 and $P_{c}^{(1)}$ give similar but slightly different spectra, the peak position of the former being visibly shifted downward  in the cases of the sets A ($\Lambda (1405)$ ) and B (broad resonance). The integrated values of the former, $\Pi_{ab}^{(1)}$, are $0.0017$, $0.0041$ and $0.00031$ for the sets A, B and C, respectively, to be compared with that of the latter, $\Pi_{c}^{(1)}$, which is  $\rho_{c} = 0.0025$ (cf. Eq.~\eqref{eq:sum2}). There is a strong correlation between these integrated probabilities and the widths of the resonances ($-2 {\rm{Im}} \mathcal{E}_{r}$). They are also correlated with the contributions of the $(a,b)$ scattering states to the resonance wave function expressed by $|\langle c| r \rangle\!\rangle ^{2} - 1|$ (see Table \ref{Table4.2}).
 
 On the other hand, the interference term, $P^{(2)}$, has the same angular dependence as 
the background term, $P^{(0)}$, which is larger by nearly two orders of magnitude, and thus can be 
observed only as a tiny structure of the spectrum in the direction of $p_{z} = 0$. The right panels of Fig.\ref{Fig.4.1} show 
this term multiplied by the phase space factor, $p/2\pi$, for the sets A, B and C, respectively, together with the background term divided by 50 for the illustrative purpose. The interference term is seen to have an oscillatory behavior near the resonance giving rise to a tiny but peculiar structure in the observed spectrum. 
Since the background term is assumed to be known, its subtraction is in principle possible.  According to Eq.~\eqref{eq:sum1}, the net (integrated) contribution of the interference term, $\Pi^{(2)}$, is negative and cancels that of the interaction term, $\Pi^{(1)}_{ab}$.

\begin{figure}
\begin{center}
\includegraphics[bb=0 0 260 168,width=6cm]{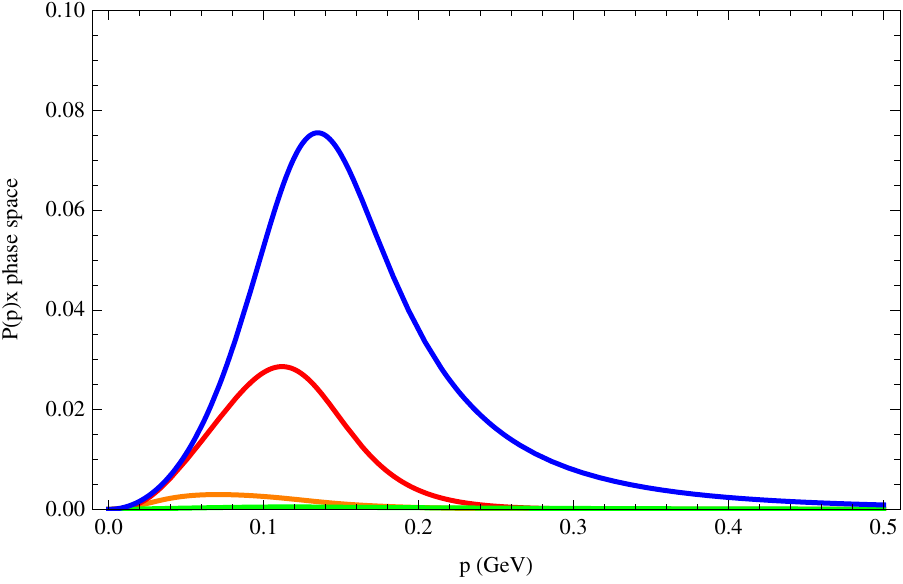}~~%
\includegraphics[bb=0 0 260 163,width=6cm]{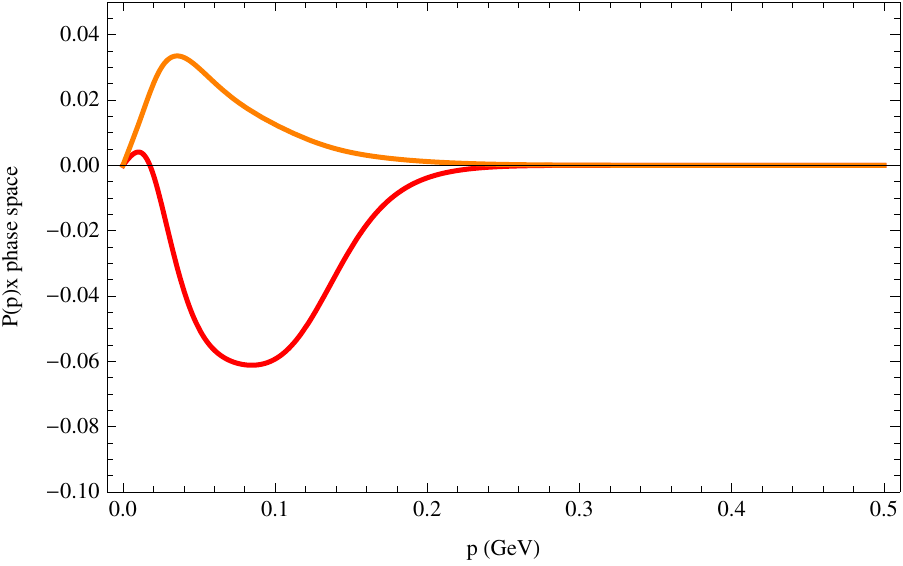}
\end{center}
\caption{
Left: The resonance and non-resonance parts,
$P_{ab}^{(1),r}$ (red), $P_{ab}^{(1),nr}$ (orange), 
$P_{c}^{r}$ (blue) and $P_{c}^{nr}$ (green) for the set B.
Right: The resonance and non-resonance parts, $P_{ab}^{(2),r}$ (red) and $P_{ab}^{(2),nr}$ (orange) for the set B.
}\label{Fig.4.2}
\end{figure}

 As for the resonance and non-resonance parts, the latter for the interaction terms, $P_{ab}^{(1)nr}$ and 
 $P_{c}^{(1)}$, are generally small in this model and are visible in the invariant mass spectra only in the case of the set B.  The structure of the resonance can thus be learnt directly from the observed spectra. On the other hand, the non-resonance part for the interference term, $P^{(2)nr}$, is not small. It is therefore more complicated to analyze the resonance properties through this term, though, as discussed before, its extraction from the observed spectra is difficult anyway.
The left panel of Fig.~\ref{Fig.4.2} shows the interaction terms, $P_{ab} ^{(1),r}$ (red), $P_{ab}^{(1),nr}$ (orange). 
$P_{c} ^{(1)r}$ (blue) and $P_{c} ^{(1)nr}$ (green), while the right panel
of Fig.~\ref{Fig.4.2} shows the interference terms, $P^{(2),r}$ (red) and $P^{(2),nr}$ (orange) for the set B. 

The contributions of the non-resonance parts can be seen more quantitatively from their contributions to the integrated probabilities, $\Pi^{nr}$'s, together with $\Pi^{r}$'s in Table \ref{Table4.4}. They are divided by $\rho_{c}$, which is the total number of $(a,b)$ pairs produced by the interaction in the present model. One sees that for the interference terms, the resonance parts, 
$\Pi^{(2)r}$ is larger in magnitude than $\Pi^{(2)}$ and thus the non-resonance parts have opposite signs. Note that the resonance parts need not satisfy the sum rules Eqs.~\eqref{eq:sum1} and \eqref{eq:sum2}. In the case of set B, however, 
$\Pi^{(2)r}$ is negative and too large in magnitude and makes the summed resonance part, $\Pi^{r} = \Pi_{ab}^{(1)r} + \Pi_{c}^{(1)r} + \Pi^{(2)r}$ negative. Thus the concept of ``integrated (or total) resonance production probability'' becomes unclear for broad resonances.

\begin{table}
\caption{The integrated non-resonance and resonance parts divided by $\rho_{c}$}\label{Table4.4}
\begin{center}
 \begin{tabular} {cccc}
\hline
\hline
 set & A & B & C \\ \hline
$\Pi_{ab}^{(1)nr}/\rho_{c}$ &   $ 0.00 $ & $ 0.18 $ & $ 0.00 $ \\
 $\Pi_{ab}^{(1)r}/\rho_{c}$ & $ 0.86 $ & $ 1.39 $ & $ 0.12 $ \\ \hline
 $\Pi_{c}^{(1)nr}/\rho_{c}$ &  $ 0.00 $ & $ 0.11 $ & $ 0.00 $ \\
 $\Pi_{c}^{(1)r}/\rho_{c}$ & 1.00 & $ 0.89 $ & $ 1.00 $ \\  \hline
 $\Pi^{(2)nr}/\rho_{c}$ & $ 0.11 $ & $ 1.18 $ & $ 0.00 $ \\
 $\Pi^{(2)r}/\rho_{c}$ & $ -0.76 $ & $ -2.78 $ & $ -0.12 $ \\
\hline
\hline
\end{tabular}
\end{center} 
\end{table}
 
Once the resonance parts are obtained from the observed spectra, they are expressed by the resonance wave function, $|r \rangle\!\rangle$, the non-resonance T-matrix, $T^{nr}$, and 
the density matrix, $\hat{\rho}$ as given in Eqs.~\eqref{eq:P1crm}, \eqref{eq:P1abrm} and \eqref{eq:P2rm}. The structure of the resonance in the present model is determined by $|{\bf{k}} \rangle$ and $|c \rangle$ components of its wave function,
$\langle {\bf{k}} | r \rangle\!\rangle $ and $\langle c | r \rangle\!\rangle $, and $\rho_{c}$ which reflects the structure of 
the particle $c$ representing the $\bar{K} N$ bound state with the coupling to the $\pi \Sigma$ channel switched off, and they can be extracted by the analysis described above.

With the resonance wave function, $| r \rangle\!\rangle$, and the density matrix given by Eq.~\eqref{eq:density},
the straightforward extension of Eq.~\eqref{eq:coales} gives the production probability in the coalescence calculation as
\begin{eqnarray}
\tilde{\Pi}^{r} = \tilde{\Pi}^{r} _{c} + \tilde{\Pi}^{r} _{ab}, \hspace{5cm} \\
\tilde{\Pi}^{r} _{c} = \rho_{c} \langle c |r \rangle\!\rangle ^{2}, \hspace{5.1cm} \\
\tilde{\Pi}^{r} _{ab} = \iint \frac{d {\bf{k}} d {\bf{k}}'}{(2\pi)^{6}} \langle -{\bf{k}} |r \rangle\!\rangle \rho ({\bf{k}}, {\bf{k}}') 
\langle {\bf{k}}' | r \rangle\!\rangle  \nonumber  \hspace{1.6cm} \\
= \langle c | r \rangle\!\rangle ^{2} \iint \frac{d {\bf{k}} d {\bf{k}}'}{(2\pi)^{6}} 
\frac{g^{2} v (k) v(k') \rho ({\bf{k}}, {\bf{k}}')}{(\mathcal{E}_{r} - E_{k})(\mathcal{E}_{r} - E_{k'})}, \hspace{0.5cm} 
\end{eqnarray}
where Eq.~\eqref{eq:res1} has been used to get the last expression. Note that these quantities are all complex and thus their physical meaning is not clear unless the real parts are dominant. As discussed in the previous subsections, what can be extracted from the observed spectra to study the structure of the resonance are the resonance parts, $P^{(1)r}$ and $P^{(2)r}$, defined by Eqs.~\eqref{eq:P1rr} and \eqref{eq:P2rr}. In the present model, they can be calculated by Eqs.~\eqref{eq:P1crm}, \eqref{eq:P1abrm} and \eqref{eq:P2rm}, with the decomposition, 
$P^{(1)r} = P_{c}^{(1)r} + P_{ab}^{(1)r}$. 
\begin{table}[!htb]
\caption{Production probabilities with the resonance wave function, $\tilde{\Pi}^{r}$'s, and the integrated resonance parts, $\Pi^{r}$'s, divided by $\rho_{c}$}
\label{Table4.6}
\begin{center}
\begin{tabular} {cccc}
\hline
\hline
set & A & B & C \\
\hline
$\tilde{\Pi}_{c}^{r}/\rho_{c}$ &   $ 1.15 - 0.12i $ & $ 1.76 - 0.39i $ & $ 1.02 - 0.00i $ \\
 $\Pi_{c}^{(1)r}/\rho_{c}$ & $ 1.00 $ & $ 0.89 $ & $ 1.00 $ \\ \hline
 $\tilde{\Pi}_{ab}^{r}/\rho_{c}$ &  $ -0.11 + 0.12i $ & $ -0.38 + 0.05i $ & $ -0.01 + 0.03i $ \\
 $\Pi_{ab}^{r}/\rho_{c}$ & $ -0.01 $ & $ -1.36 $ & $ -0.00 $ \\ \hline
 $\tilde{\Pi}^{r}/\rho_{c}$ & $ 1.04 -0.00i $ & $ 1.36 - 0.34i $ & $ 1.01 + 0.03i $ \\
 $\Pi^{r}/\rho_{c}$ & $ 0.99 $ & $ -0.49 $ & $ 1.00 $ \\
\hline
\hline
\end{tabular}
\end{center} 
\end{table} 

The production probabilities calculated by the above expressions using the resonance wave function are compared with the integrated resonance parts, $\Pi_{c}^{(1)r}$, $\Pi_{ab}^{r} = \Pi_{ab}^{(1)r} + \Pi^{(2)r}$ and $\Pi^{r}$ in Table \ref{Table4.6}.     
It is seen that the correlation between $\tilde{\Pi^{r}}$'s and $\Pi^{r}$'s is very good for the narrow resonance, tolerable for $\Lambda (1405)$ but very poor for the broad resonance. 
 
 \subsection{Summary and discussions}
 
 The coalescence model for resonance particle production in high energy heavy ion collisions is formulated in the way it is observed in experiments. A simple S-wave (Lee type) model where two particles, $a$ and $b$, couple to a particle, $c$, forming a resonance in the $(a,b)$ scattering states, is used to clarify the discussion. The probability, $P({\bf{p}})$,  of finding the two particles with a relative momentum ${\bf{p}}$, which is simply related to the invariant mass spectrum of the two particle system, is calculated in the coalescence model for scattering states. It consists of the background term, $P^{(0)}$, the interaction term, $P^{(1)}$, and the interference term, $P^{(2)}$ (see Eq.~\eqref{eq:P012}). In this model, $P^{(1)}$ can be decomposed into the contribution of the two particle states, $P_{ab}^{(1)}$, and that of the $c$ particle, $P_{c}^{(1)}$, and the former together with the interference term satisfy the sum rule, Eq.~\eqref{eq:sum1}, while the latter satisfies Eq.~\eqref{eq:sum2}. Thus the integrated contribution of the interaction and interference terms gives $\rho_{c}$, the probability of finding the particle $c$ in the source. In the numerical examples, one has in mind 
 $\Lambda (1405)$ as the resonance and $a,b$ is $\pi \Sigma$. $\rho_{c}$ is calculated assuming $c$ to be a 
 $\bar{K} N$ bound state with its coupling to $\pi \Sigma$ states switched off.
This is the procedure used in \ref{subsec:yields} and the previous ExHIC papers to estimate the production yield of $\Lambda (1405)$. In the case of the interaction parameters corresponding to $\Lambda (1405)$ (set A) in this example, the non-resonance parts are small and the integrated sum of $P^{(1)}$ and $P^{(2)}$ can be regarded as the production probability of the resonance  $\Lambda (1405)$. Thus the ExHIC procedure is justified in this example. $P^{(1)}$ and $P^{(2)}$ can be themselves considered as the resonance terms and tell us further details of the resonance such as its $\pi \Sigma$ components. 
 
 In a more general case where the non-resonance parts are not small, another step of separating the resonance parts becomes necessary.  This is the case in the  example of a broad resonance (set B), where the non-resonance part is about the same in magnitude as the resonance part in the interference term (see 
Fig.~\ref{Fig.4.1}
and Table \ref{Table4.4}) and the integrated resonance part, $\Pi^{r}$, becomes negative. Even in such a case, however, the non-resonance part is still small (about $10 \%$) in the interaction term and the information on the structure of the resonance can be extracted through this term. It is then crucial to get the interaction and interference terms separately from the observed invariant mass spectrum. This is in principle possible for anisotropic sources as seen in the discussed examples.  

Although the coalescence calculation for bound states can be straightforwardly extended to resonances, it leads however, to complex values for the production probabilities (see Tables ~\ref{Table4.6}).
In this case,  the calculated probabilities have quantitative meaning for narrow resonances, but this becomes doubtful for broad ones. 
 
It is straightforward to extend the above formulation to more general cases of multichannels with non-separable interactions, though the expressions for the production probabilities become more involved and the numerical calculations based on them would be time-consuming. Such extensions are necessary to use realistic models of resonances and study how their structures are reflected in the production probabilities. For quantitative analyses of the experimental data, the applicability of the coalescence model itself should also be examined in detail.


\section{Hadron-hadron interactions from two particle momentum correlations} 
\label{sec:correlation_result}

Pairwise hadronic interactions as well as quantum statistics produce
a correlation at low relative momenta
in multiparticle production from elementary to heavy ion
collisions~\cite{Koonin:1977fh,Lednicky:1981su,Bauer:1993wq,Lednicky:2005tb,Lisa:2005dd}.
The momentum correlation of identical particles from quantum statistics,
known as the Hanbury-Brown and Twiss (HBT)~\cite{HanburyBrown:1956bqd}
or Goldhaber-Goldhaber-Lee-Pais (GGLP) effect~\cite{Goldhaber:1960sf},
can give information on the size of the emission source
through the (anti-)symmetrization of the two-boson (fermion) wave function.
The quantum statistical effect of stable hadrons,
particularly pions, has been used to estimate the source sizes
created in relativistic nucleus-nucleus collisions~\cite{wiedemann99:_partic,Lisa:2005dd}. 
By comparison,
one expects substantial interaction effects on the correlation function
for particle pairs whose interaction is sufficiently strong
in the range comparable to the effective source size~\cite{Lednicky:1981su,Bauer:1993wq}.
In particular, the correlation function of non-identical pairs
is directly related to the pairwise interaction due to the 
absence of the quantum statistical effect~\cite{Bauer:1993wq}.
Thus, high statistics measurement of the correlation function might
provide information on the pairwise interaction of \emph{any} measurable channel,
including those difficult to perform the scattering experiments. 
In this section, a brief review is given on recent activities on constraining hadron-hadron
interactions through momentum correlations in heavy-ion collisions and their
implications for the interpretation and the possible existence of exotic states.

\subsection{General property of the two-particle momentum correlation function} 
\label{sec:correlation_formalism}

\subsubsection{Formalism}
\label{subsec:correlation_form}

The two-particle momentum correlation function is defined as
the ratio of the two-particle spectrum to
the product of single particle inclusive momentum spectra~\cite{Koonin:1977fh,Lednicky:1981su,Bauer:1993wq,Lisa:2005dd},
\begin{align}
&C(\bm{q},\bm{P})
=\frac{E_1 E_2 dN_{12}/d\bm{p}_1 d\bm{p}_2}
{(E_1 dN_1/d\bm{p}_1)(E_2 dN_2/d\bm{p}_2)}
\ ,\label{Eq:CF}
\\
&P\equiv p_1+p_2\ ,\quad
q^\mu \equiv \frac12\left[
(p_1-p_2)^\mu - \frac{(p_1-p_2)\cdot P}{P^2}\,P^\mu
\right]
=\frac{E'_2p_1^\mu-E'_1p_2^\mu}{M_\mathrm{inv}}
\ ,\label{eq:p_and_q}
\end{align}
where $P$ and $q$ are the center-of-mass and the relative momentum of
the pair, respectively, and $E_i$ ($i=1,2$) is the energy of the hadron
$i$.
In the last equality in Eq.~\eqref{eq:p_and_q}, the relative momentum is
expressed in the center-of-mass frame of the pair (the pair rest frame),
where $E'_i (i=1,2)$ is the energy of the hadron $i$ in this frame and
$M_\mathrm{inv}=E'_1+E'_2$ is the invariant mass.
In the non-relativistic limit,
$E'_1 \to M_1$ and $E'_2 \to M_2$,
the definition of the relative momentum reads
$q=(M_2p_1-M_1p_2)/(M_1+M_2)$.

Assuming independent (chaotic) emission from the source, i.e., particles
are produced with random phases, and the correlation function can be
expressed in terms of the single particle source function
$S(x_i,\bm{p_i})$, 
which describes the emission probability from a space-time point $x_i$ with
momentum $\bm{p_i}$, and the weight factor $|\varphi^{(-)}(\bm{r}, \bm{q})|^2$, 
which depends on the relative coordinate $\bm{r}$ and momentum $\bm{q}$,
\begin{align}
C(\bm{q},\bm{P})
=&\frac{
\int d^4x_1 d^4x_2
S_1(x_1,\bm{p}_1)
S_2(x_2,\bm{p}_2)
\left| \varphi^{(-)}(\bm{r},\bm{q}) \right|^2
}{
\int d^4x_1 
S_1(x_1,\bm{p}_1)
\int d^4x_2
S_2(x_2,\bm{p}_2)
}
\label{Eq:c2formula}
\end{align}
The weight factor $|\varphi^{(-)}(\bm{r},\bm{q})|^2$ can be identified as the
relative wave function of the pairs in the outgoing state, provided
that the difference between the emission times of the two particles is small~\cite{Lednicky:1981su}. 
In general, the emission time difference modifies the relative
coordinate $\bm{r}$ in the relative wave function from the position
difference $\bm{x}_1-\bm{x}_2$.
By using $P$ and $q$, particle momenta are given as
$p_1 = E'_1P/M_\mathrm{inv}+q$ and $p_2 = E'_2P/M_\mathrm{inv}-q$.
Then the free two-particle wave function 
is given as
\begin{align}
&\exp(-ip_1x_1-ip_2x_2)
=\exp\left(-iP\cdot{X}-iq(x_1-x_2)\right)
=\exp\left(-iP\cdot{X}+i\bm{q}\cdot\bm{r}\right)
\ ,\\
&X=\frac{E'_1x_1+E'_2x_2}{M_\mathrm{inv}}
\ ,\quad
\bm{r}=\bm{x}_1-\bm{x}_2-\bm{v}(t_1-t_2)
\ ,\quad
\bm{v}=\bm{P}/\sqrt{M_\mathrm{inv}^2+\bm{P}^2}
\ .\label{eq:relativistic_coordinate}
\end{align}
%
When the interaction between the two particles is switched on,
the relative wave function is modified into a superposition of 
$\exp(i\bm{q}\cdot\bm{r})$, which is a function of $\bm{r}$,
while the center-of-mass wave function $\exp(-iP\cdot X)$ is kept
unchanged. The modification of the relative coordinate in
Eq.~\eqref{eq:relativistic_coordinate} serves as a generalization of the
formula derived in Refs.~\cite{Gong:1991zza,Bauer:1993wq}.

The above formula can be reduced to a convenient form in the pair rest frame
where $P$ and $q$ becomes temporal and spatial four-vectors,
$P=(M_\mathrm{inv},\bm{0})$
and 
$q=(0,\bm{q})$, respectively.
In this case, 
the center-of-mass coordinate $X$ and
relative time $t$ can be integrated out to obtain the relative source function
\begin{align}
S_{12}(\bm{r})
=&\frac{\int dt\, d^4X\,
S_1(X+E'_2x/M_\mathrm{inv},\bm{p}_1)
S_2(X-E'_1x/M_\mathrm{inv},\bm{p}_2)}
{\int dx_1 S_1(x_1,\bm{p}_1) \int dx_2 S_2(x_2,\bm{p}_2)}
\quad(x=x_1-x_2=(t,\bm{r}))
\label{Eq:S12}
\end{align}
and the correlation function is then given by the Koonin-Pratt (KP) formula
\begin{equation}
 C(\bm{q}, \bm{P}) = \int d\bm{r} S_{12}(\bm{r})
  |\varphi^{(-)}(\bm{q},\bm{r})|^2.
 \label{Eq:KP}
\end{equation}
In frames different from the pair rest frame,
one needs to put $x=(t,\bm{r}+\bm{v}t)$ in Eq.~\eqref{Eq:S12}.
In the KP formula, the relative source function $S_{12}(\bm{r})$ can be
interpreted as the relative source distribution integrated over time in
the pair rest frame. 
In particular, for the free (anti-)symmetrized wave
function, this formula reduces to the three-dimensional Fourier
transformation of the source function. It should be noted that the
relative source function $S_{12}(\bm{r})$ depend on $P$ when the
emission point and momentum are correlated. 
In the later discussions, the KP formula, Eq.~\eqref{Eq:KP}, is mainly used.
While there are several conditions under which the KP formula
works~\cite{Anchishkin:1997tb,Lisa:2005dd}, they seem to be satisfied
in high-energy heavy-ion collisions.

\subsubsection{Correlations from strong interactions and quantum statistics}
\label{subsec_strong_quatnum}
Both the pairwise interaction and the quantum statistics modify
the relative wave function from the simple plain wave $e^{i\bm{q\cdot r}}$.
Since the strong interaction is of short range,
the modification of the relative wave function appears mainly in the $s$-wave.
In this case, one can write
the relative wave function in the two-body outgoing state
with an asymptotic relative momentum $\bm{q}$ as
\begin{align}
\varphi^{(-)}(\bm{r},\bm{q})=&\exp(i\bm{q}\cdot\bm{r})-j_0(qr)+\psi(r)
\ ,\label{eq:wavefunc}
\end{align}
where $q=|\bm{q}|$,
$j_0$ is the spherical Bessel function
and $\psi(r)$ is the relative wave function in the $s$-wave,
which is regular at $r\to0$ and has an asymptotic form,
\begin{align}
\psi(r) \to& \psi_\mathrm{asy}(r)
\quad (r\to \infty)\ ,\\
\psi_\mathrm{asy}(r)
=&\frac{e^{-i\delta}}{qr}\sin(qr+\delta)
=\frac{1}{2iqr}\left[e^{iqr}-e^{-2i\delta}e^{-iqr}\right]
\ ,
\label{Eq:asymp}
\end{align}
with $\delta$ being the phase shift.
It should be noted that the above wave function $\psi$ is different
from that appearing in the two-particle scattering by a factor $e^{2i\delta}$.
In the two-body outgoing state,
the coefficient of the outgoing wave is unity
and the incoming spherical wave is modified
in contrast to the scattering of two particles
where the coefficient of the incoming wave is unity and the outgoing
spherical wave is modified.

For illustration, let us consider a spherical and static  Gaussian source,
$S_i(x_i,\bm{p}_i) \propto \delta(t_i-t_0)\,\exp(-\bm{x}_i^2/2R_i^2)$
and simultaneous emission of the pairs. 
Then the correlation function from the non-symmetrized wave function $\varphi^{(-)}$
is obtained as
\begin{align}
C(\bm{q})
=&\int d\bm{r} S_{12}(\bm{r}) 
\left| \varphi^{(-)}(\bm{r},\bm{q}) \right|^2
=1 + \Delta C(\bm{q})
\ ,\\
\Delta C(\bm{q})
=&\int d\bm{r}
S_{12}(\bm{r}) 
\left[
\left| \psi(r) \right|^2
-(j_0(qr))^2
\right]
\ .\label{eq:DeltaC_int}
\end{align}
where 
\begin{equation}
S_{12}(\bm{r})
=\exp(-\bm{r}^2/4R^2)/(4\pi R^2)^{3/2}
\quad (R=\sqrt{(R_1^2+R_2^2)/2})
\ .\label{eq:staticGaussian}
\end{equation}
Since the sum of first two terms in Eq.~\eqref{eq:wavefunc} does not contain
the $s$-wave components,
the cross term involving the third term disappears for a spherical source.
As a result, the effect of interaction on $C(\bm{q})$ appear as the
deviation from unity by the difference of squared wave functions between
the free and $s-$wave, as seen from Eq.~\eqref{eq:DeltaC_int}.

For identical pairs,  the relative wave function is symmetric or antisymmetric
with respect to the exchange of the two-particle spatial coordinates
($\bm{r}\to\,-\bm{r}$). For spin-$1/2$ pairs, the spatial part of the
wave function is symmetric for the spin-singlet ($^1S_0$) state and
antisymmetric for the spin-triplet ($^3S_1$) state, which does not have
$s-$wave interaction. Then the wave function is given as
\begin{align}
\varphi^{(-)}_E(\bm{r},\bm{q})=&\frac{1}{\sqrt{2}}\left(
\varphi(\bm{r})+\varphi(-\bm{r})
\right)
=\sqrt{2}\left(
\cos(\bm{q}\cdot\bm{r})-j_0(qr)+\psi(r)
\right)
\ ,\\
\varphi^{(-)}_O(\bm{r},\bm{q})=&\frac{1}{\sqrt{2}}\left(
\varphi(\bm{r})-\varphi(-\bm{r})
\right)
=\sqrt{2}i\,\sin(\bm{q}\cdot\bm{r})
\ .
\end{align}
The wave functions $\varphi_E$ and $\varphi_O$
have even and odd parities, respectively.

For (anti)symmetrized wave functions,
the KP equation is reduced to the Fourier transform of the source function,
then the correlation function is given by,
\begin{align}
C_E(\bm{q})
=&\int d\bm{r} S_{12}(\bm{r}) 
\left| \varphi_E^{(-)}(\bm{r},\bm{q}) \right|^2
=1 + \mathrm{Re}[\widetilde{S}_{12}(2\bm{q})]+2\Delta C(\bm{q})
\label{Eq:CorrE1}\\
=&1 + \exp(-4q^2R^2)
+2\int d\bm{r} S_{12}(\bm{r}) \left[ \left| \psi(r) \right|^2 -(j_0(qr))^2 \right]
\ ,\label{Eq:CorrE}\\
C_O(\bm{q})
=&\int d\bm{r} S_{12}(\bm{r}) 
\left| \varphi_O^{(-)}(\bm{r},\bm{q}) \right|^2
=1 - \mathrm{Re}[\widetilde{S}_{12}(2\bm{q})]
\label{Eq:CorrO1}\\
=&1 - \exp(-4q^2R^2)
\ ,\label{Eq:CorrO}
\end{align}
where $\widetilde{S}$ denotes the Fourier transform,
$\widetilde{S}(2\bm{q})=\int d\bm{r} S(\bm{r})
\exp(-2i\bm{q}\cdot\bm{r})$.
The second equality in Eqs~\eqref{Eq:CorrE} and \eqref{Eq:CorrO} is
obtained for the Gaussian source \eqref{eq:staticGaussian}.
The Gaussian term in Eqs.~\eqref{Eq:CorrE} and \eqref{Eq:CorrO}
represents the effects from the quantum statistics. 
For the symmetric (asymmetric) wave function, the correlation function
exhibits enhancement (reduction) from unity and its width in $q$ is inversely proportional to the size of the source. In the case of identical interacting particles, Eq.~\eqref{Eq:CorrE1},
the effect of the interaction appears as deviation not from unity but
from the free correlation function. This fact provides an intuitive
understanding of the correlation function as follows~\cite{Morita:2014kza,Morita:2016auo}; 

\begin{itemize}
 \item For large $q$, the wave function rapidly oscillates to give
       $\Delta C(\bm{q})\simeq 0$. Thus one needs to look at small $q$
       to get information on the interaction. 
 \item Weakly attractive interaction gives  $|\psi(r)| > j_0(qr)$
       in the range of the interaction and thus leads to small enhancement
       of the correlation function.
 \item Strongly attractive interaction having an bound state gives a
       node to $\psi(r)$. Since the contribution from the integrand with
       $r \simeq 0$ is suppressed by $r^2$ in $d\bm{r}$, the correlation
       function is also suppressed due to $|\psi(r)| < |j_0(qr)|$. 
       Repulsive interaction also leads to the similar behavior.
\end{itemize}

In reality, the hadron-hadron correlation function  is expressed as
combinations of the above correlation functions.
Considering the spherical Gaussian source and neglecting the Coulomb potential
and channel coupling effects, one may have the following classification 
for non-identical spinless meson pairs, identical spinless meson pairs,
pairs of a spinless meson and a spin-half baryon,
non-identical spin-half baryon pairs,
and 
identical spin-half baryon pairs;
\begin{align}
C_{MM'}(\bm{q})=&1+\Delta C(\bm{q}) \ ,\\
C_{MM}(\bm{q})=&1+\exp(-4q^2R^2)+\Delta C(\bm{q})\ ,\\
C_{MB}(\bm{q})=&1+\Delta C(\bm{q})\ ,\\
C_{BB'}(\bm{q})=&1+\frac{1}{4}\sum_{s=0,1} (2s+1) \Delta C(\bm{q})\ ,
\label{Eq:CFBBp}\\
C_{BB}(\bm{q})
=&\frac{1}{4}C_E(\bm{q})+\frac{3}{4}C_O(\bm{q})
=1-\frac{1}{2}\exp(-4q^2R^2)
+\frac{1}{2}\Delta C(\bm{q})\ ,
\label{Eq:CFBB}
\end{align}
Interaction generally depends on the spin of the pair,
and so does the interaction dependent part of the correlation function,
$\Delta C(\bm{q})$, as found in the $BB'$ pair, Eq.~\eqref{Eq:CFBBp}.
The correlation function of the spin-half baryon pairs is obtained as
the spin-average over the spin-singlet and the triplet states. As a
result, the correlation at $\bm{q}=0$ is not zero like Eq.~\eqref{Eq:CorrO}
but $1/2$ for the non-interacting case ($\Delta C(\bm{q})=0$). 

\subsubsection{Lednick\'{y} and Lyuboshits Model}
\label{subsec_LL}

In order to examine the interaction dependence of the correlation function,
an analytic model developed by Lednick\'{y} and Lyuboshits (LL)~\cite{Lednicky:1981su}
is useful. In the LL model, the correlation function is obtained by
using the asymptotic wave function together with the shape-independent
approximation in the scattering phase shift.
Then the correlation function  is given in terms of the scattering
amplitude and the effective range.

The asymptotic wave function Eq.~\eqref{Eq:asymp} can be rewritten in the following form,
\begin{align}
\psi_\text{asy}(r)=
\calS^{-1}\left[
\frac{\sin{qr}}{qr}+f(q)\frac{e^{iqr}}{r}
\right]
\ ,\label{Eq:LLwf3}
\end{align}
where $f(q)=(\calS-1)/2iq$ is the scattering amplitude 
and $S=e^{2i\delta}$ is the S-matrix.
With the Gaussian source \eqref{eq:staticGaussian},  the integral in the
KP formula for $\psi_\text{asy}$ is reduced to
\begin{align}
\int_0^\infty dr\,S_{12}(r) |\psi_\text{asy}(r)|^2
=
\frac{1}{|\calS|^2}
\left[
\frac{|f(q)|^2}{2R^2}
+\frac{2\text{Re}f(q)}{\sqrt{\pi}R}\,F_1(x)
-\frac{\text{Im}f(q)}{R}\,F_2(x)
+\frac{F_2(x)}{x}
\right]
\ ,
\label{Eq:LLeq1}
\end{align}
where 
$x=2qR$,
$F_1(x)=\int_0^x dt e^{t^2-x^2}/x$ and $F_2(x)=(1-e^{-x^2})/x$.
The use of the asymptotic wave function is well justified when the
source size is sufficiently large compared to the range of the
interaction~\cite{Gmitro:1986ay}. 
In the single channel case,
the deviation from the asymptotic wave function at small $q$ can be
accounted for by using the effective range formula~\cite{roy1967nuclear},
\begin{align}
\lim_{q \to 0} \frac{1}{|f(q)|^2}\,\int_0^\infty r^2 dr \left[
|\psi|^2 - \frac{\sin^2(qr+\delta)}{q^2r^2} \right]
=-\frac12\,r_\text{eff}
\ .
\label{Eq:LLeq2}
\end{align}
The integral in the left hand side of Eq.~\eqref{Eq:LLeq2}
gives the correction to Eq.~\eqref{Eq:LLeq1},
when the integrand is multiplied by the factor $e^{-r^2/4R^2}$.
By using Eqs.~\eqref{Eq:LLeq1} and \eqref{Eq:LLeq2},
one arrives
at the interaction dependent part of the correlation function
in the LL model~\cite{Lednicky:1981su},
\begin{align}
\Delta C^\text{LL}(q)
=&\frac{1}{|\calS|^2}\left[
 \frac{|f(q)|^2}{2R^2}\,F_3\left(\frac{r_\text{eff}}{R}\right)
+\frac{2\text{Re}f(q)}{\sqrt{\pi}R} F_1(x)
-\frac{\text{Im}f(q)}{R} F_2(x)\right]
+\frac{1-|\calS|^2}{|\calS|^2}\frac{F_2(x)}{x}
\ ,\label{Eq:LL}
\end{align}
where $x=2qR$ and the effective range correction appears in 
$F_3(\reff/R)=1-\reff/2\sqrt{\pi}R$.
In the formula given in Ref.~\cite{Lednicky:1981su},
one assumes $\psi^{(-)}_{\bm{q}}=(\psi^{(+)}_{-\bm{q}})^*$ and $|\calS|=1$,
then 
the last term in Eq.~\eqref{Eq:LL} does not exist.

\begin{figure}[tbh]
\begin{center}
\includegraphics[bb=0 0 360 252,width=6cm]{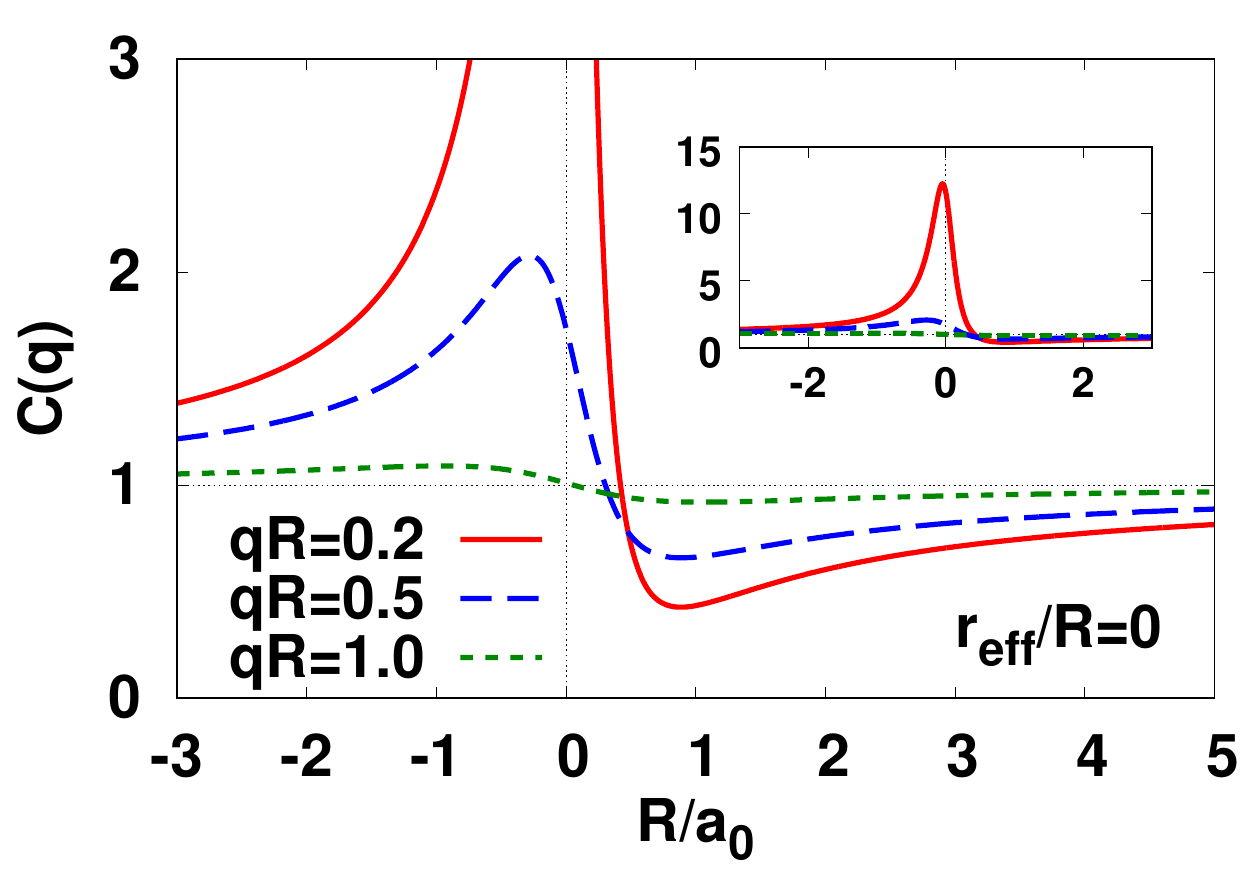}%
\includegraphics[bb=0 0 360 252,width=6cm]{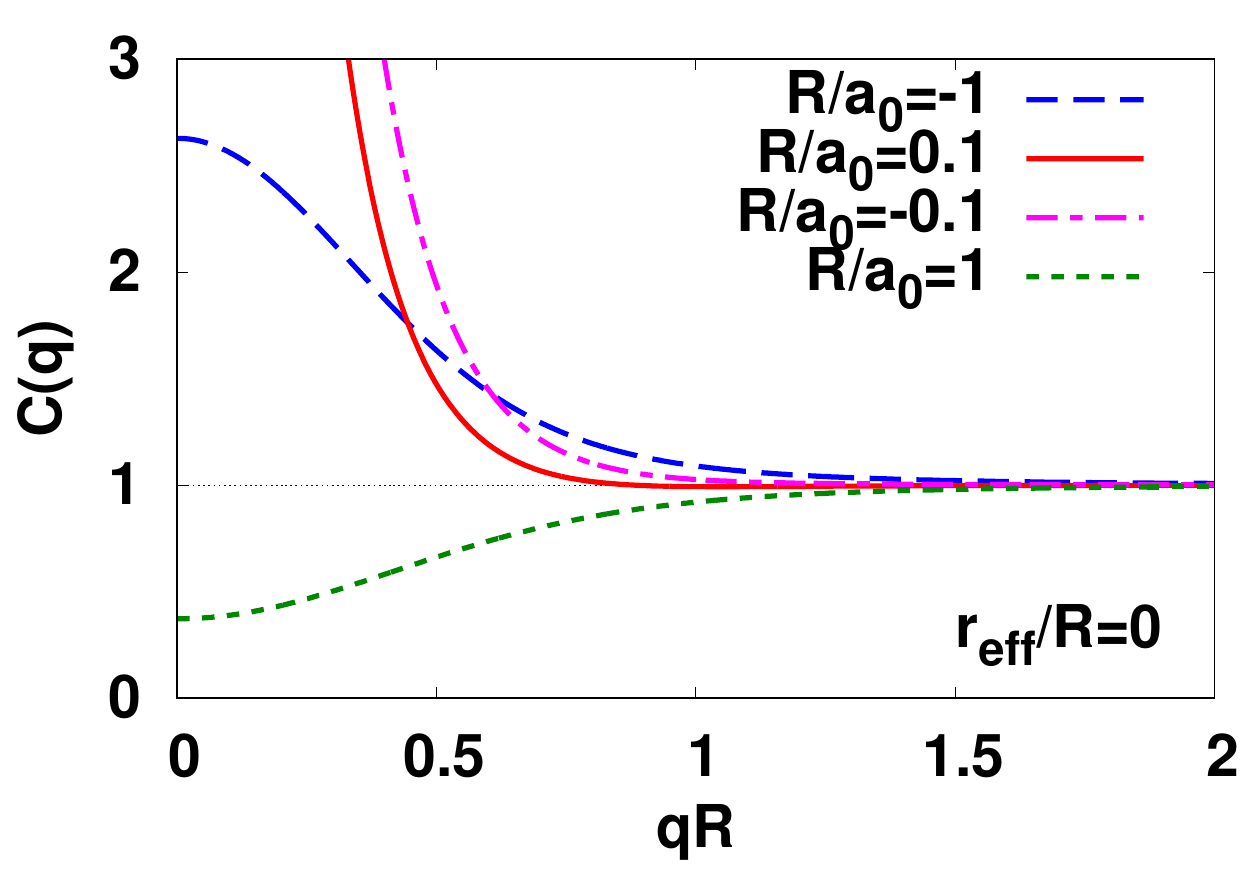}%
\end{center}
\caption{Correlation function in the LL model~\cite{Lednicky:1981su} as
 a function of $qR$ and $R/a_0$ in the case of $r_\mathrm{eff}/R=0$.
}
\label{Fig:CLLex}
\end{figure}

Figure~\ref{Fig:CLLex} displays the interaction dependence of the correlation function in the LL model,
$C(q)=1+\Delta C^\mathrm{LL}(q)$.
The correlation function is given in terms of the scattering amplitude $f(q)$, 
which is known to be well described by the scattering length $a_0$
and the effective range $\reff$ at low energy.
\begin{align}
f=(q\cot\delta-iq)^{-1}
\ ,\quad
q\cot\delta=-1/a_0 + r_\text{eff}q^2/2+\mathcal{O}(q^4)
\ .
\end{align}
Here the effects of Coulomb potential,
channel coupling and the imaginary part of the potential (absorption) have been ignored.
It should be noted that the above discussion is based on the ``nuclear physics'' convention for the scattering length, 
which leads to $\delta \simeq -a_0 q$ at low energy.

The behavior of the correlation function can be understood on the same
footing as discussed above; now the properties of the wave function
is represented by the corresponding scattering length. 
At negative scattering length $a_0<0$,
the correlation function is always enhanced by the attractive interaction.
At positive but small scattering length $R/a_0 \gtrsim 0.453$,
the correlation function is always suppressed.
The positive $a_0$ means
that there is a bound state or the interaction is repulsive,
and the scattering wave function has a node at $r \simeq a_0$ at low energy.
Then the wave function squared is generally suppressed
compared with the free wave function.
At around the unitary limit $1/a_0\simeq 0$,
the correlation function is strongly enhanced at low energy,
while it becomes close to unity at $qR \simeq 1$.
If this qualitative behavior survives other effects,
the correlation function measurement can provide useful information on the scattering length.

\subsubsection{Effect of collectivity}
\label{sec:flow}

In high-energy heavy ion collisions, the hot matter created in the collisions undergoes hydrodynamic expansion,
which eventually affects final particle spectra. 
Assuming a static and spherical Gaussian source, 
Eq.~\eqref{eq:staticGaussian}, is thus a crude approximation, this ignores the
dynamical property of the particle emission sources. 

The effect of the collective expansion can be taken into account by
properly modeling the source function 
$S(x,\bm{p}) = EdN/d\bm{p}d^4x$. 
Denoting the four-velocity of the collective expansion at space-time
point $x$ as $u^\mu = \gamma(1,\bm{v})$, the particle energy at the
local rest frame reads $u\cdot p$. Then, the thermal distribution is
modified into $\exp(-u \cdot p/T)$ with the chemical potential ignored for simplicity.
This factor causes a correlation between the
freeze-out point $x$ and particle momentum $\bm{p}$. As naturally
expected, a fast moving source can produce particle with high momentum
easier than a static source. Also in microscopic approaches for the
collective phenomena such as transport models, the position-momentum
correlation is produced through the multiple scattering of particles~\cite{lin02:_parton_relat_heavy_ion_collid}.

So far the effect of the collective expansion on the correlation
function has been mainly discussed in $\pi^\pm\pi^\pm$ correlation where the final
state interaction is negligible except for the repulsive Coulomb force. 
The apparent dependence of the correlation function on the momentum of
the pion pairs can be understood as a consequence of the expansion and
gives stringent constraints on the property of the hot QCD matter, such
as equation of state and transport coefficients, as well as the dynamics
of the collisions~\cite{Pratt:2008qv}. In short, one may regard the
effect of the expansion as a modification of the source size into
momentum dependent effective source size, often called ``length of
homogeneity''~\cite{Makhlin:1987gm,wiedemann99:_partic}. 
For instance, taking longitudinally expanding boost-invariant source in
which $u^\mu = (\cosh \eta_s, 0, 0, \sinh \eta_s)$ with $\eta_s =
\ln\sqrt{(t+z)/(t-z)}$, one may have an effective source size in the longitudinal
direction~\cite{Makhlin:1987gm}, 
$ R_L \simeq \tau \sqrt{\frac{T}{m_t}}$,
where $m_t = \sqrt{p_t^2+m^2}$ is the transverse mass. Thus, the
source size of heavy or high momentum particles becomes effectively
small. Because one may obtain information on the detailed source shape
by measuring $C(\bm{q})$ as three-dimensional function of $\bm{q}$, the
corresponding source sizes are differently affected by the profile of
the expansion. In the present review, however, we concentrate on the
one-dimensional correlation function of $q = |\bm{q}|$ since the current
statistics of experiments are not sufficient for such analyses in the
specific channels we are interested in. Then, the effect of the
collective expansion might be  regarded as an effective reduction of the
source size $R$. Although rescatterings and resonance decays produce a
non-Gaussian tail in the source function~\cite{Adler:2006as}, this effect
can be safely ignored in the analyses of the final-state interaction
since the dominant part of the source function in the pair rest frame can be
well approximated by a Gaussian~\cite{shapoval15:_proton_lambd_cern_large_hadron_collid} and only small
distance pairs are important.

\subsubsection{Feed-down contribution}
\label{sec:feeddown}

In the discussions above, two particles are assumed to be directly emitted
from the hot matter. This assumption would be valid if one could
remove contributions from the decay of parent particles.
In reality, substantial fraction of observed particles come from decay
of resonances. Strongly decaying short-lived resonances with
lifetimes of $\mathcal{O}(\text{several fm})$ will give the source function an
effective long lifetime and a tail of the spatial distribution, and thus
might influence low $q$ behavior of $C(q)$ through the change of the
source geometry. 

However,  it has been known that long-lived parents give a sharp
correlation near $q\simeq 0$, which cannot be resolved, and 
thus cause an apparent reduction of the
intercept $C(q=0)$~\cite{Grassberger:1976au,Csorgo:1994in,Wiedemann:1996ig}.
With $N_{\text{tot}}^A$ being  the total number of measured particle $A$ of
interest and $N_{\text{res}}= \sum_i N_{i\rightarrow A}$ being the
long-lived parent contribution decaying into $A$, the effective
intercept $\lambda$ is given by
\begin{equation}
 \lambda = \left( 1- \frac{N_\text{res}}{N^A_{\text{tot}}} \right)^2.\label{eq:feeddown_lambda}
\end{equation}
In this case the observed correlation function is expected to take the
following form;
\begin{equation}
 C_{\text{corr}}(q) = 1+\lambda(C_{\text{bare}}(Q)-1),
\label{Eq:corrbare}
\end{equation}
Since $\lambda \leq  1$, the long-lived resonance decay dilutes the
strength of the correlation. 
Practically the same correction should also be applied even without long-lived resonances because the
experiments cannot perfectly identify the particles. Thus the $\lambda$
parameter is often called the ``purity'' parameter, indicating the purity of
particle identification in data samples. When the percentage of the
misidentification is known and estimation of the long-lived resonance
decay contribution is feasible, one may construct a purity-corrected
correlation function by inverting \eqref{Eq:corrbare} as
\begin{equation}
 C_\text{purity-corrected}(q) = 1+
  \frac{C_\text{measured}(q)-1}{\lambda}.\label{eq:c2_puritycorrected}
\end{equation}
This correction serves as a crucial input when one tries to extract the
pairwise interaction, since the overall magnitude is
sensitive to the scattering length as demonstrated in Fig.~\ref{Fig:CLLex} via the inverse of its ratio to the source size.

Decay parents may induce residual correlations to the observed ones. For
example, the observed $pp$ correlation may have been
affected by $p\Lambda$ correlation before the $\Lambda$ decays into the
proton. Introducing a pair fraction (or pair purity) $x_{ij}$, which is defined as a
fraction of $(i,j)$ pairs to total number of the pairs of interest,  one
may include such residual correlations as~\cite{kiesel14:_extrac}
\begin{equation}
 C_{\text{corr}}(q) = 1+ \sum_{i,j} x_{i,j} (C_{i,j}(Q_{i,j})-1).\label{eq:residualc2}
\end{equation}
The fraction parameters $x_{i,j}$ as well as the effective intercept
$\lambda$ can be estimated from experimental data and production models,
and the relative momentum of the parent $Q_{i,j}$ can be obtained from decay kinematics~\cite{kiesel14:_extrac}.
It should be noted that, however, this is also affected by the purity of
the particle identification. 

\subsection{Non-exotic channels}

To begin with, it is instructive to discuss some examples for 
the correlation functions of non-exotic channels whose interaction is
not expected to produce exotic states. 

\subsubsection{$pp$ and $\bar{p}\bar{p}$ correlation}

The correlation method to extract the pairwise interaction, albeit limited to $s$-wave, 
is applicable to any measurable particle species. In particular,
high-energy heavy-ion collisions at the top RHIC energy and the LHC
energies produce as many antimatters as matters. The STAR experiments
reported measurements of $\bar{p}\bar{p}$ correlation as well as $pp$
correlation in Ref.~\cite{Adamczyk:2015hza}.
The measured
$\bar{p}\bar{p}$ correlation is consistent with $pp$ correlation within
errors,
so are the extracted scattering parameters.

The data were analyzed within the LL model (Sec.~\ref{subsec_LL}), but
extended to include the appropriate quantum statistics effect
(Eq.~\eqref{Eq:CFBB}), residual correlation from $p\Lambda$ and $\Lambda\Lambda$
($\bar{p}\bar{\Lambda}$ and $\bar{\Lambda}\bar{\Lambda}$ for
$\bar{p}\bar{p}$ correlation), and the Coulomb repulsion. 
The Coulomb interaction can be taken into account by replacing the
plane-waves in Eq.~\eqref{eq:wavefunc} with the corresponding Coulomb wave
functions and by applying the effective range formula with the Coulomb
interaction. The pair fractions $x_{pp}$, $x_{p\Lambda}$, and
$x_{\Lambda\Lambda}$, are adopted from the THERMINATORS2 model~\cite{Chojnacki:2011hb} which is an extended version of one of the implementations of
statistical models. The extracted low energy scattering parameters of
$\bar{p}\bar{p}$ interactions are 
$a_0 = -7.41 \pm 0.19 (\text{stat.}) \pm 0.36 (\text{sys,})$ fm
and 
$r_\text{eff} = 2.14 \pm 0.27 (\text{stat.}) \pm 1.34 (\text{sys,})$ fm, 
which are consistent with the known values for protons, 
$a_0^{pp} = -7.82$ fm and $r_{\text{eff}}^{pp} = 2.78$ fm. 
The Gaussian radius was also obtained as $R_{\bar{p}\bar{p}}=2.75$ fm
and $R_{pp}=2.8$ fm. These small radii indicate the influence of the
collective expansion. Indeed, similar measurements at LHC~\cite{adam15:_one_pb_pb_nn} show that the Gaussian radii scale with
$m_t$ as indicated in Sec.~\ref{sec:flow}.

\subsubsection{$p\Lambda$ and $p\bar{\Lambda}$ correlations}
\label{subsubsec:plambdac2}

$p\Lambda$ correlation has been measured in several
experiments~\cite{NA49_protonlambda,STAR_p-lambda2006,HADES_p-lambda2016}. 
Since the $p\Lambda$ interaction is rather known from scattering
experiments and hypernuclear data, the $p\Lambda$ correlation measurements in $pA$
and $AA$ have been used to constrain dynamics of the collisions~\cite{wang99:_lambd}.
Nevertheless, high-energy collisions allows for measuring
$p\bar{\Lambda}$ and $\bar{p}\Lambda$ correlations~\cite{STAR_p-lambda2006} which are not known
and serve as inputs for transport model calculations,
and precise analysis to extract the scattering parameters with modern facilities~\cite{STAR_p-lambda2006,HADES_p-lambda2016} provides 
the cross-check with the scattering experiments. 

In Ref.~\cite{STAR_p-lambda2006}, the purity-corrected $p\Lambda$ and
$p\bar{\Lambda}$ correlations and their anti-particle pairs were
reported. The LL model \eqref{Eq:LL} was used with the known scattering
lengths and effective ranges of the $p\Lambda$ interaction in the
spin-triplet $(t)$ and spin-singlet $(s)$ channels
($a_0^t = -1.66$fm, $a_0^s = -2.88$fm, $r_{\text{eff}}^t = 3.78$ fm, and
$r_{\text{eff}}^s = 2.92$ fm~\cite{wang99:_lambd}) to extract the source
size. The measured $p\Lambda$ correlation function is found to fairly
reflect the weakly attracting nature, and the extracted source size is
found to follow the same trend with that in the $pp$ correlation. 
On the other hand, $p\bar{\Lambda}$ and $\bar{p}\Lambda$ correlations exhibit  small
suppression below unity in intermediate range of $q$, $0 < q < 0.15$ GeV.
The LL model was again used with complex scattering length which
accounts for the annihilation effects, under the assumption that the effective
range is zero and spin dependence is neglected. 
While the extracted $\text{Im}a_0$ is found to be comparable with that
of $p\bar{p}$ channel, the source size was found to be significantly
smaller than the result from $p\Lambda$ correlation.
Later refined analyses in
Refs.~\cite{Shapoval:2014yha,shapoval15:_proton_lambd_cern_large_hadron_collid},
which also apply a sophisticated dynamical model incorporating
hydrodynamic expansion and hadronic cascades, 
pointed out that including residual correlations of decay parents (see
\eqref{eq:residualc2}) would cure this problem, although current statistics in
the data do not allow for a precise determination of the scattering
lengths.

\subsection{Exotic channels}

\subsubsection{$\Lambda\Lambda$ correlation}
\label{Sec:LLint}

The measurements of the $\Lambda\Lambda$ correlation function in heavy
ion collisions provide another constraint on the $\Lambda\Lambda$
interaction. The correlation function is also complementary to
direct search of the exotic $H-$dibaryon
from the invariant mass of the decay products 
discussed in Sec.~\ref{sec:dibaryons}.
Indeed STAR collaboration reported the first measurement of the
$\Lambda\Lambda$ correlation in Au+Au collisions at the top RHIC energy~\cite{Adamczyk:2014vca}. The data were corrected only for pair purity
via Eq.~\eqref{eq:c2_puritycorrected}
for the identification (92\%) after rejecting most of $\Lambda$s
from weak decay of higher mass hyperons
by using the distance of closest approach to the primary vertex.
The long-lived resonance contribution from $\Sigma^0$ and a part of
$\Xi$ is still supposed to reduce the correlation strength via Eq.~\eqref{Eq:corrbare}. 

In Ref.~\cite{Adamczyk:2014vca}, the data were analyzed within the LL
model Eq.~\eqref{Eq:LL} with an intercept parameter $\lambda$. Furthermore,
a Gaussian term with two parameters taking account of the residual correlation at large $q$
is included, although its origin has not been understood. Therefore, a
six-parameter fit to the data is made with
\begin{equation}
 C(q) = \mathcal{N}\left[ 1+ \lambda \left( -\frac{1}{2}e^{-4q^2 R^2} +
				      \Delta C^{\text{LL}}(q) \right)+a_{\text{res}}
 e^{-4r_{\text{res}}^2 q^2} \right]\label{eq:c2fit_LL}
\end{equation}
where $\Delta C^{\text{LL}}(q)$ is given by Eq.~\eqref{Eq:LL}.
Optimized parameters given in Ref.~\cite{Adamczyk:2014vca}
are summarized in Table~\ref{Tab:corrpars}.

Although the quality of the fit is quite well
($\chi^2/N_{\text{dof}}\simeq 0.56$), 
the obtained  scattering length\footnote{The opposite sign convention
of the scattering length is adopted in Ref.~\cite{Adamczyk:2014vca}},
$a_0 = 1.10 \pm 0.37^{+0.68}_{-0.08}$ fm,
seems to conflict with the results from the observed double hypernucleus.
Indeed, the $\LL$ bond energy
in $^{~~6}_{\Lambda\Lambda}\mathrm{He}$ is found to be
$\DBLL=
B_{\Lambda\Lambda}(^{~~6}_{\Lambda\Lambda}\mathrm{He})
-2 B_{\Lambda}(^{5}_{\Lambda}\mathrm{He}) \simeq
1.01~\mathrm{MeV}$~\cite{Takahashi:2001nm}.
From $\DBLL(^{~~6}_{\Lambda\Lambda}\mathrm{He})$, the scattering length
and the effective range in the $\LL$ $^1\mathrm{S}_0$ channel 
are suggested as 
$(a_0, \reff)=(-0.77~\fm, 6.59~\fm)$~\cite{Filikhin:2002wm}
or 
$(a_0, \reff)=(-0.575~\fm, 6.45~\fm)$~\cite{Hiyama:2002yj}.
Recent update of the bond energy
due to the update of the $\Xi^-$ mass~\cite{Amsler:2008zzb}
gives $\DBLL(^{~~6}_{\Lambda\Lambda}\mathrm{He})=0.67 \pm 0.17~\MeV$~\cite{Ahn:2013poa},
which suggests $(a_0, \reff) = (-0.44~\fm, 10.1~\fm)$~\cite{Hiyama:2010zzd}.

A detailed investigation of the $\Lambda\Lambda$ correlation function by
making use of the KP formula Eq.~\eqref{Eq:KP} with various $\Lambda\Lambda$
interaction potentials and source functions including collective
expansion in both longitudinal and transverse directions has been carried
out in Ref.~\cite{Morita:2014kza}, 
It was found that after taking into account the correction of
electromagnetic decays from $\Sigma^0$, the scattering length is found
to be consistent with the double hypernuclei. The detailed comparison of
the methods is discussed in Ref.~\cite{Ohnishi:2016elb}, which concludes that
it is crucial to determine the value of $\lambda$. Here we briefly
outline the above points.

\begin{figure}[!t]
 \centering
 \includegraphics[bb=0 0 504 378,width=0.45\textwidth]{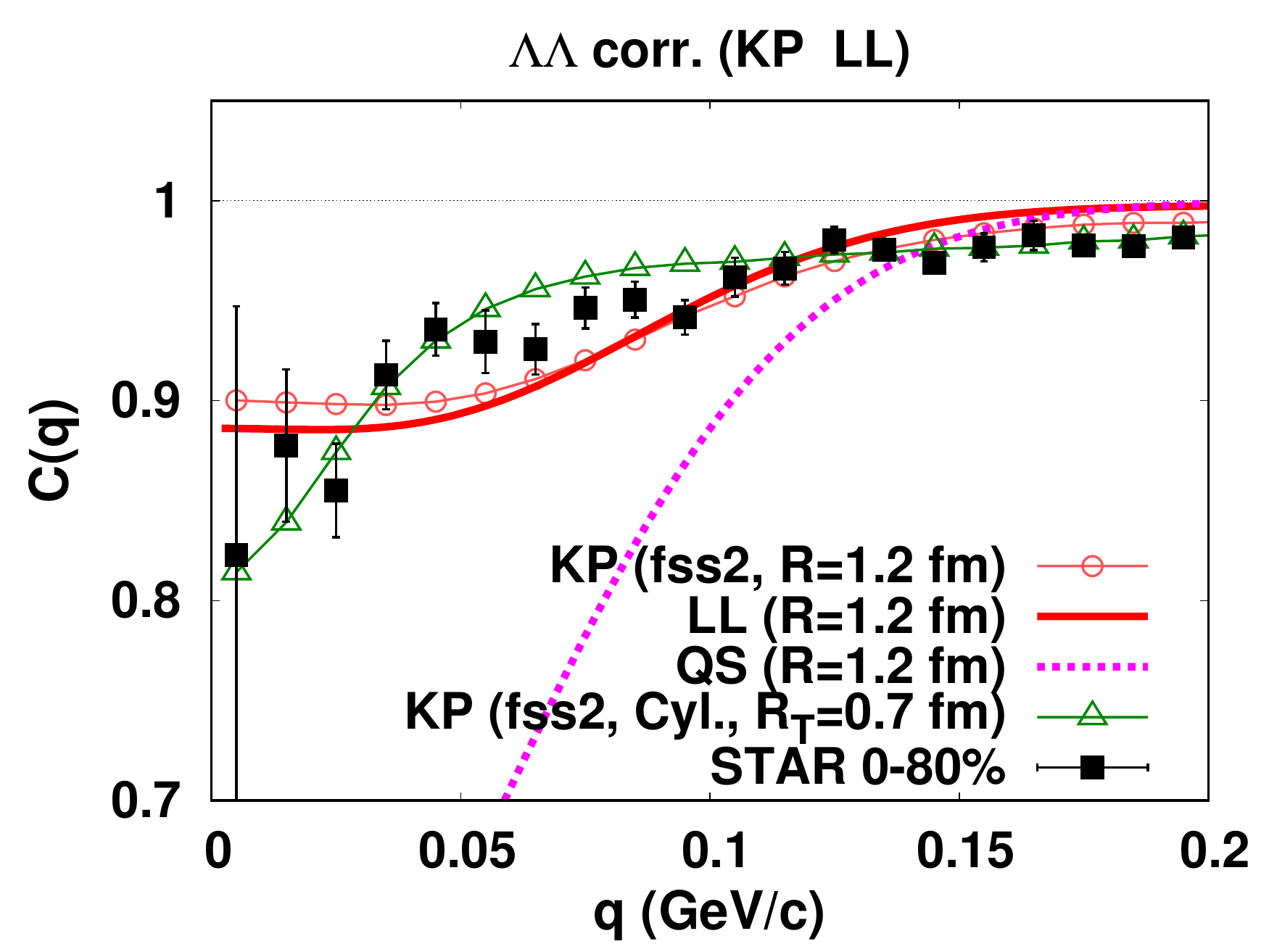}
 \includegraphics[bb=0 0 504 378,width=0.45\textwidth]{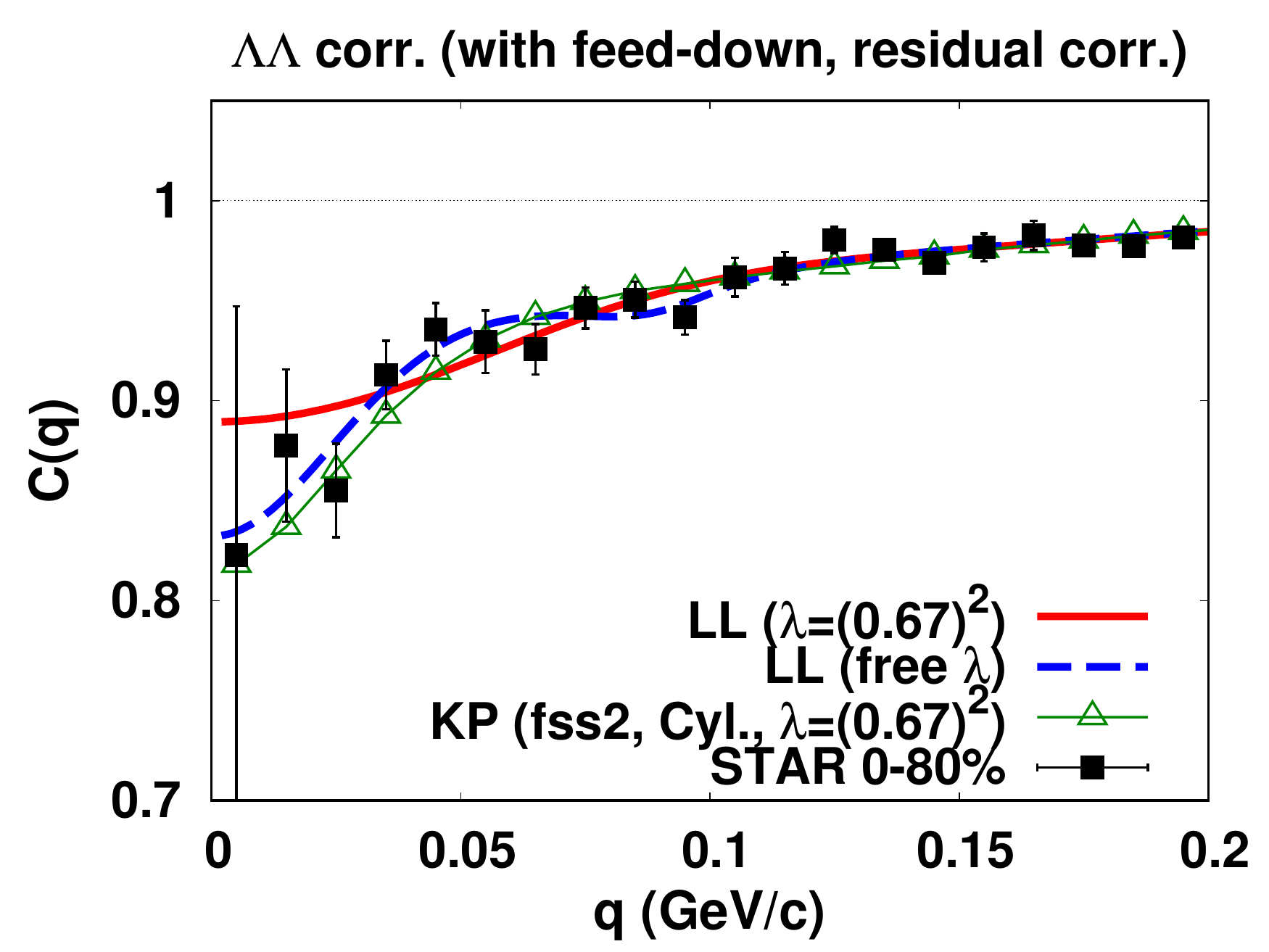}
 \caption{$\Lambda\Lambda$ correlation function with the fss2 $\LL$
 interaction~\cite{Fujiwara:2001xt,Fujiwara:2006yh}, 
 obtained by using the KP and LL formulae in comparison with data~\cite{Adamczyk:2014vca}.
 Left panel shows the results without the feed-down correction and the residual
 correlations.
 Right panel shows the results with the feed-down and residual source
 effects.
 The results in the fixed $\lambda$ case ($\lambda=(0.67)^2$)
 and the free $\lambda$ case are compared.
 Also shown in both panels are
 the results from the cylindrical source including flow effects
 in the KP formula~\cite{Morita:2014kza}.
 }\label{Fig:CLL}
\end{figure}

First, we clarify the difference between the $C(q)$ obtained from the LL formula
Eq.~\eqref{Eq:LL} and the KP formula Eq.~\eqref{Eq:KP}. In the left panel of
Fig.~\ref{Fig:CLL}, $C(q)$ with the fss2 $\Lambda\Lambda$ interaction is
displayed. The corresponding values $a_0 = -0.81$ fm and $r_{\text{eff}}= 3.99$ fm  are used as
inputs for the LL formula. The difference between the two is small, thus
confirming previous studies~\cite{Gmitro:1986ay} that indicate
insensitivity of the correlation to the detailed shape of the wave
function within the interaction range. The difference of $C(q)$ between the
static spherical source (thin red, circles) and the expanding source
(thin  green, triangles) indicates the effect of the collective
expansion. The existence of the fast boost-invariant longitudinal
expansion deforms the source function such that the correlation function
takes a different shape in the best fit to the data~\cite{Morita:2014kza}. Note that such a difference does not take place
in the case of non-identical pairs; as seen in Eqs.~\eqref{Eq:CorrE1}
and \eqref{Eq:CorrO1}, the quantum statistics effect makes $C(q)$ more
sensitive to the source shape through the Fourier transformation. 
 
Second, we estimate the the contribution to $N_{\text{tot}}^{\Lambda}$  with the
help of the statistical model and experimental data, to correct the data
for the long-lived resonance decay via Eqs.~\eqref{eq:feeddown_lambda}
and \eqref{Eq:corrbare}. 
Here $\Sigma^0$ and $\Xi$  are treated as long-lived resonances, since
other decay parents have much shorter lifetime thus only change the
effective source size or have a negligible contribution. 
Adopting data from $p+$Be collisions at $p_{\text{lab}}=28.5$GeV~\cite{Sullivan:1987us},
we take $N_{\Sigma^0}/N_{\Lambda}=0.278$, which is also consistent with
thermal model calculations. 
Taking into account the fact that the $\Xi$ yield in Au+Au collisions at $\sqrt{s_{NN}}=200$
GeV has been shown to be 15\% of total $\Lambda$~\cite{Agakishiev:2011ar}
and the STAR candidate selection with the distance of closest approach less
than 0.4~cm may exclude a part of $\Xi$ decay contributions to $\Lambda$,
we estimate $\lambda=(0.67)^2$. If we take account of the $\Xi$
contribution into the total yields, $\lambda=(0.572)^2$. 
It has been confirmed that the lower value $\lambda=(0.572)^2$ only leads
to small quantitative changes in the following analyses.

\begin{table}[!t]
\caption{Optimized parameters for the $\LL$ correlation
in the fixed and free $\lambda$ cases in the LL model.
Numbers in the parentheses for $\chi^2/\mathrm{DOF}$ and DOF
show those for a given $(1/a_0,\reff)$.
In the fixed $\lambda$ case,
$1/a_0$ and $\reff$ are strongly correlated with $a_\mathrm{res}$.
Errors in the brackets in the fixed $\lambda$ case are those
in the fixed $a_\mathrm{res}$ case.
}
\label{Tab:corrpars}
\centerline{
\begin{tabular}{lccc}
\hline
\hline
		& STAR~\cite{Adamczyk:2014vca}
		& \multicolumn{2}{c}{Ref.~\cite{Ohnishi:2016elb}}
\\
		& (Free $\lambda$)
		& Free $\lambda$ case
		& Fixed $\lambda$ case
		\\
\hline
$\lambda$	& $0.18 \pm 0.05^{+0.12}_{-0.06}$
 		& $0.18 \pm 0.05$
		& $(0.67)^2=0.4489$\\
$1/a_0\ (\fm^{-1})$ &
		& $0.91 \pm 0.20$
		& $-1.26\pm 0.74\ [\pm 0.17]$ \\
$a_0\ (\fm)$	& $1.10 \pm 0.37^{+0.68}_{-0.08}$ & & \\
$\reff\ (\fm)$	& $8.52 \pm 2.56^{+2.09}_{-0.74}$
		& $8.51 \pm 2.14$
		& $1.76\pm 11.62\ [\pm 0.86]$ \\
$R\ (\fm)$	& $2.96\pm0.38^{+0.96}_{-0.02}$
		& $2.88 \pm 0.38$
		& $1.39\pm 0.71\  [\pm 0.17]$ \\
$r_\mathrm{res}\ (\fm)$ & $0.43\pm 0.04^{+0.43}_{-0.03}$
		& $0.43 \pm 0.03$
		& $0.48\pm 0.10\  [\pm 0.02]$ \\
$a_\mathrm{res}\ (\fm)$ & $-0.044 \pm 0.004^{+0.048}_{-0.009}$
		& $-0.045 \pm 0.004$
		& $-0.058\pm 0.069$ [fixed] \\
$\mathcal{N}$	& $1.006\pm 0.001$
		& $1.006 \pm 0.001$
		& $1.006\pm 0.001\ [\pm 0.001]$ \\
\hline
$\chi^2/\mathrm{DOF}$ & $0.56$
		& $0.55 (0.53)$
		& $0.64\ (0.61)\ [0.63]$ \\
DOF		& $43$
		& $43 (45)$
		& $44\ (46)\ [45]$ \\
\hline
\hline
\end{tabular}}
\end{table}

The right panel of Fig.~\ref{Fig:CLL} compares the results
in the fixed $\lambda=(0.67)^2$ (solid line) and free $\lambda$ (dashed line) cases
in the LL model formula with the residual correlation term
Eq.~\eqref{eq:c2fit_LL}. Table \ref{Tab:corrpars} summarizes the results of the
fit to the $\Lambda\Lambda$ correlation data.
The free $\lambda$ case confirms the result obtained by the STAR in Ref.~\cite{Adamczyk:2014vca},
while the fixed $\lambda$ case shows the opposite sign of the scattering
length.
In the free $\lambda$ case where the optimal value is found to be $\lambda\simeq 0.18$,
quantum statistics and the pair purity give $C(q\to0)=1-\lambda/2 \sim 0.91$
while the data show $C(q\to0) \simeq 0.82$.
Thus $C(q)$ needs to be reduced at small $q$ by the $\LL$ interaction
and positive $a_0$ is favored.
By contrast, for a fixed $\lambda=(0.67)^2$,
the corresponding quantum statistical correlation
$C_\sLL(q\to0)=1-\lambda/2\simeq 0.78$
is slightly smaller than the observed correlation.
With the residual source contribution, $a_\text{res}\sim -0.06~\fm$,
the difference from the data becomes more evident.
The $\LL$ interaction needs 
to enhance the correlation,
and the optimal $a_0$ value is found in the negative region,
as concluded in Ref.~\cite{Morita:2014kza}.

One should note that the best fit result of the LL formula in the fixed
$\lambda$ case differs from the KP formula result
from the cylindrical source including flow effects.
This result may indicate
the importance of fixing not only the purity $\lambda$
but also the source geometry including the flow effects.

\begin{figure}[!t]
\begin{center}
\includegraphics[bb=0 0 432 252,width=12cm]{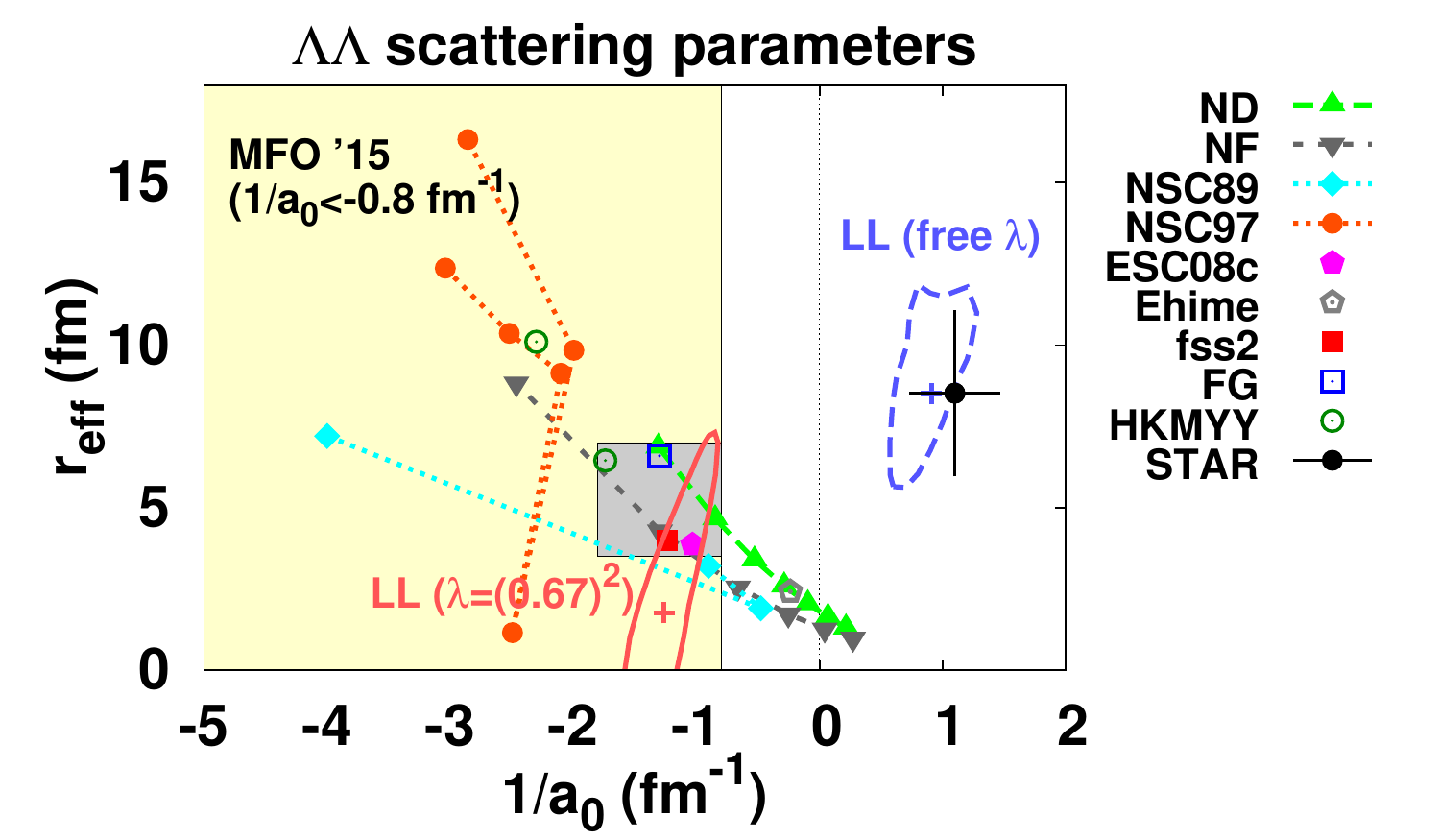}\\[2ex]
\end{center}
\caption{
 Low-energy scattering parameters $(a_0,r_\text{eff})$ of $\LL$.
 Contours show $\chi^2/\text{DOF}=0.65$ ($\lambda=(0.67)^2$, solid contour)
 and $\chi^2/\text{DOF}=0.56$ (free $\lambda$, dashed contour)
 in the LL model analysis of the $\LL$ correlation data.
 Symbols show $(1/a_0,r_\text{eff})$ from $\LL$
 potentials~\cite{Fujiwara:2001xt,Fujiwara:2006yh,Nagels:1976xq,Nagels:1978sc,
 Maessen:1989sx,Rijken:1998yy,Rijken:2006ep,Ueda:1998bz,Filikhin:2002wm,Hiyama:2002yj,Hiyama:2010zzd},
 and shaded areas show the region favored by the $\LL$ correlation data
 in Ref.~\cite{Morita:2014kza}(MFO '15).
 Filled black circle with $xy$ error bar shows
 the analysis result by the STAR collaboration,
 where $\lambda$ is regarded as a free parameter~\cite{Adamczyk:2014vca}.
 Figures are taken from Ref.~\cite{Ohnishi:2016elb} with some modifications.
}
\label{Fig:ar}
\end{figure}

Figure~\ref{Fig:ar} summarizes the constraints from the $\Lambda\Lambda$
correlation data at the present stage and its dependence on the
assumptions made. 
Also shown is the boundary of the favored region, given by $\chi^2/\text{DOF}=0.65$ ($0.56$), in the fixed (free) $\lambda$ case in the LL model.  
The region in the free $\lambda$ case is consistent with that
by the STAR collaboration~\cite{Adamczyk:2014vca}.
As shown in the previous subsection, negative and positive scattering
lengths are favored in the fixed and free $\lambda$ cases, respectively.
It is found that negative scattering lengths are more favored
in the pair purity probability range of $\lambda > 0.35$.
Namely, the $\chi^2/\mathrm{DOF}$ at the negative $a_0$ local minima
is smaller than that at the positive $a_0$ local minima
when $\lambda$ is fixed at a value $\lambda > 0.35$.
The low energy scattering parameters $(1/a_0,r_\text{eff})$ of
several $\LL$ interactions are also shown;
Boson exchange potentials 
(ND, NF, NSC89, NSC97, ESC08, Ehime)~\cite{Nagels:1976xq,Nagels:1978sc,Maessen:1989sx,Rijken:1998yy,Rijken:2006ep,Rijken:2010zzb,Ueda:1998bz}
and Nijmegen-based potentials
fitted to the Nagara data (FG,HKMYY)~\cite{Filikhin:2002wm,Hiyama:2002yj,Hiyama:2010zzd},
in addition to the quark model potential (fss2)~\cite{Fujiwara:2001xt,Fujiwara:2006yh}.
It should be noted that the fixed $\lambda$ region covers recently proposed $\LL$
potentials, fss2 and ESC08~\cite{Fujiwara:2006yh,Rijken:2010zzb}.

The shaded areas in Fig.~\ref{Fig:ar}
show the favored region in the analysis using the KP formula~\cite{Morita:2014kza}.
The dark (grey) shaded area shows the region with $\chi^2/\mathrm{DOF}<5$
from the cylindrical source including flow effects
but without feed-down and residual correlation effects.
The light (yellow) shaded area shows the region with $\chi^2/\mathrm{DOF} \lesssim 1$
under the condition $R > r_\mathrm{res}$
with flow, feed-down and residual correlation effects.
The light shaded area includes
the favored region in the fixed $\lambda$ case in the LL model analysis.

On the basis of the scattering length and the effective range of
the $\Lambda\Lambda$ interaction obtained in the present analyses, the
existence of $H$ particle as a bound state of $\LL$ is not preferred.
This can be understood from the enhanced $\LL$ correlation function
observed in the data compared with the free case.
If there was a bound state in $\LL$ with $H$ being the dominant component,
the correlation function would be suppressed from the free case, 
as seen in the scattering length dependence of the correlation function (Fig.~\ref{Fig:CLLex}).

\begin{figure}[!t]
 \centering
 \includegraphics[bb=0 0 360 252,width=3.375in]{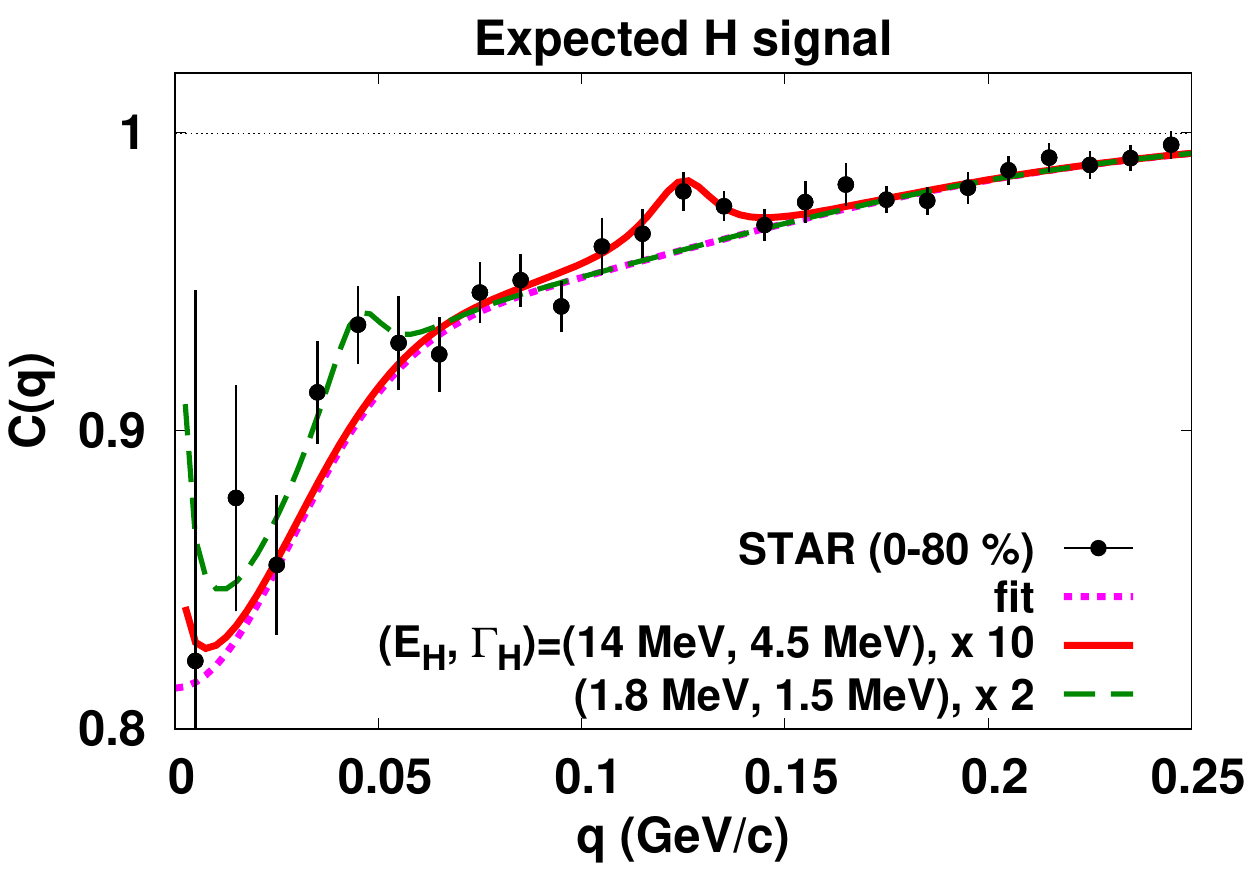}
 \caption{Possible resonance $H$ signal in the $\LL$ correlation function.
 Signal for $(E_H, \Gamma_H)=(14~\MeV, 4.5~\MeV)$
 and $(E_H, \Gamma_H)=(1.8~\MeV, 1.5~\MeV)$ are multiplied by 10 and 2,
 respectively. Figure is taken from Ref.~\cite{Morita:2014kza}
 with some modifications.
}
 \label{fig:H}
\end{figure}

The existence of $H$ as a resonance pole above the $\LL$ threshold
is another interesting possibility,
as discussed in Sec.~\ref{sec:dibaryons}.
The strength of the resonance $H$ signal in the correlation function
is shown in Fig.~\ref{fig:H}~\cite{Morita:2014kza}.
Here the $H$ and $\Lambda$ yields per event per unit rapidity
are evaluated by using the statistical model,
$N_H \simeq 1.3\times 10^{-2}$ and $N_{\Lambda} \simeq 30$~\cite{Cho:2011ew},
and the mass distributions for the $H$ and $\Lambda\Lambda$ pair are
assumed to be the Breit-Wigner and thermal distributions.
Then the resonance $H$ contribution to the $\LL$ relative momentum spectrum
is evaluated to be,
\begin{align}
\Delta C_H=\frac{dN_H/dydq}{dN_{\Lambda\Lambda}/dydq}
\ ,\quad
\frac{dN_H}{dydq}= N_H f_\mathrm{BW}(E_q)
 \frac{q}{\mu} \ ,
\quad
\frac{dN_{\Lambda\Lambda}}{dydq}
=N_{\Lambda\Lambda}\, 
\frac{4\pi q^2\exp(-q^2/2\mu T)}{(2\pi\mu T)^{3/2}} 
\ ,
\end{align}
where $f_\mathrm{BW}(E)=\Gamma_H/[(E-E_H)^2+\Gamma_H^2/4]/2\pi$ is
the Breit-Wigner function,
$N_{\Lambda\Lambda}=N_\Lambda^2$ and $\mu=M_\Lambda/2$.
The bump structures in the data are roughly explained by adding
$\Delta C_H$ multiplied by 10 and 2,
to a simple smooth function fitting the STAR data
for 
$(E_H, \Gamma_H)=(14~\MeV, 4.5~\MeV)$
and
$(E_H, \Gamma_H)=(1.8~\MeV, 1.5~\MeV)$, respectively.
These bumps may come from the statistical fluctuations~\footnote{We thank N.~Shah for this information.},
but it would be possible to confirm or rule out the existence of resonance $H$
with higher statistics.

\subsubsection{$p\Omega$ correlation}
\label{Sec:pOme}

As discussed in Sec.~\ref{sec:dibaryons}, the spin-2 nucleon-Omega
($N\Omega$) state with $S=-3$~\cite{Goldman:1987ma} is the most
promising candidates for bound or resonant dibaryons besides the $H$. 
The measurement of the $p\Omega$ correlation in order to determine the
$N\Omega$ interaction has been recently proposed
by Morita \textit{et al.}~\cite{Morita:2016auo}. 

In Ref.~\cite{Morita:2016auo}, the $p\Omega$ correlation function is
calculated through the KP formula. Since the $p\Omega$ state has either
spin-1 or 2, the wave function can be expressed by the statistical
average, 
\begin{equation}
|\varphi_{p\Omega}(\bm{r}, \bm{q})|^2 = \frac{3}{8}|\varphi_{J=1}(\bm{r},
 \bm{q})|^2  + \frac{5}{8}|\varphi_{J=2}(\bm{r}, \bm{q})|^2
\end{equation}
The interaction in the $^3S_1$ channel is assumed to be a complete
absorption at short distance $r < r_0$, because  there would be a strong coupling to the octet-octet
channels when the spatial distance between $N$ and $\Omega$ becomes
small. This can be modeled by an imaginary potential $V(r;\ ^3S_1)= -i V_0 \theta (r_0-r) $ with $V_0 \rightarrow + \infty$ for the 
strong interaction part. We choose $r_0=2$ fm, because the Coulomb
potential dominates over the strong interaction
for $r> 2$ fm.

The interaction in the $^5S_2$ channel was described by the following
potential with an attractive Gaussian core + an attractive (Yukawa)$^2$ tail with a form factor;
$ V_{N\Omega}(r) = b_1 e^{-b_2 r^2} + b_3 (1-e^{-b_4 r^2}) (e^{-b_5r}/r)^2$,
which well fits the lattice QCD data with heavy quarks ($m_{\pi}$=875 MeV and
$m_K$=916 MeV)~\cite{Etminan:2014tya} with $b_{1,3}<0$ and $b_{2,4,5}>0$.
Assuming that the qualitative form of this attractive potential does not
change even for physical quark masses, a series of potentials 
can be generated by varying the range-parameter at long distance, $b_5$.
Shown in the following are the results for three typical examples:
$V_{\rm I}$ with weaker attraction,
$V_{\rm II}$ with a shallow bound state,
and
$V_{\rm III}$ with stronger attraction.
The binding energies, scattering lengths and effective ranges in
the $^5{\rm S}_2$ $p\Omega$ channel with and without the Coulomb
potential are summarized in Table \ref{tbl:pot}. 

\begin{table}[!b]
 \caption{The binding energy ($E_{\rm B}$), 
 scattering length ($a_0$) and effective range ($r_{\rm eff}$) with and without 
 the Coulomb attraction in the spin-2 $p\Omega$ state. Physical masses of the proton and 
 $\Omega$ are used.}
 \centering
\begin{tabular}[t]{lc|ccc}\hline
 \multicolumn{2}{c|}{Spin-2 $p\Omega$ potentials}  & $V_{\text{I}}$ &$V_{\text{II}}$&$V_{\text{III}}$\\\hline
 & $E_{\rm B}$~[MeV] &$-$& 0.05 &24.8 \\
without Coulomb & $a_0$~[fm]& $-1.0$ & 23.1&1.60 \\
                       & $r_{\text{eff}}$~[fm]&1.15&0.95&0.65 \\ \hline
   & $E_{\rm B}$~[MeV]&$-$&6.3 & 26.9  \\
with Coulomb  &$a_0$~[fm]&$-1.12$ &5.79&1.29 \\
                       &$r_{\text{eff}}$~[fm]&1.16&0.96 &0.65 \\ \hline
\end{tabular}
 \label{tbl:pot}
\end{table}

\begin{figure}[!t]
 \centering
 \includegraphics[bb=0 0 576 360,width=0.45\textwidth]{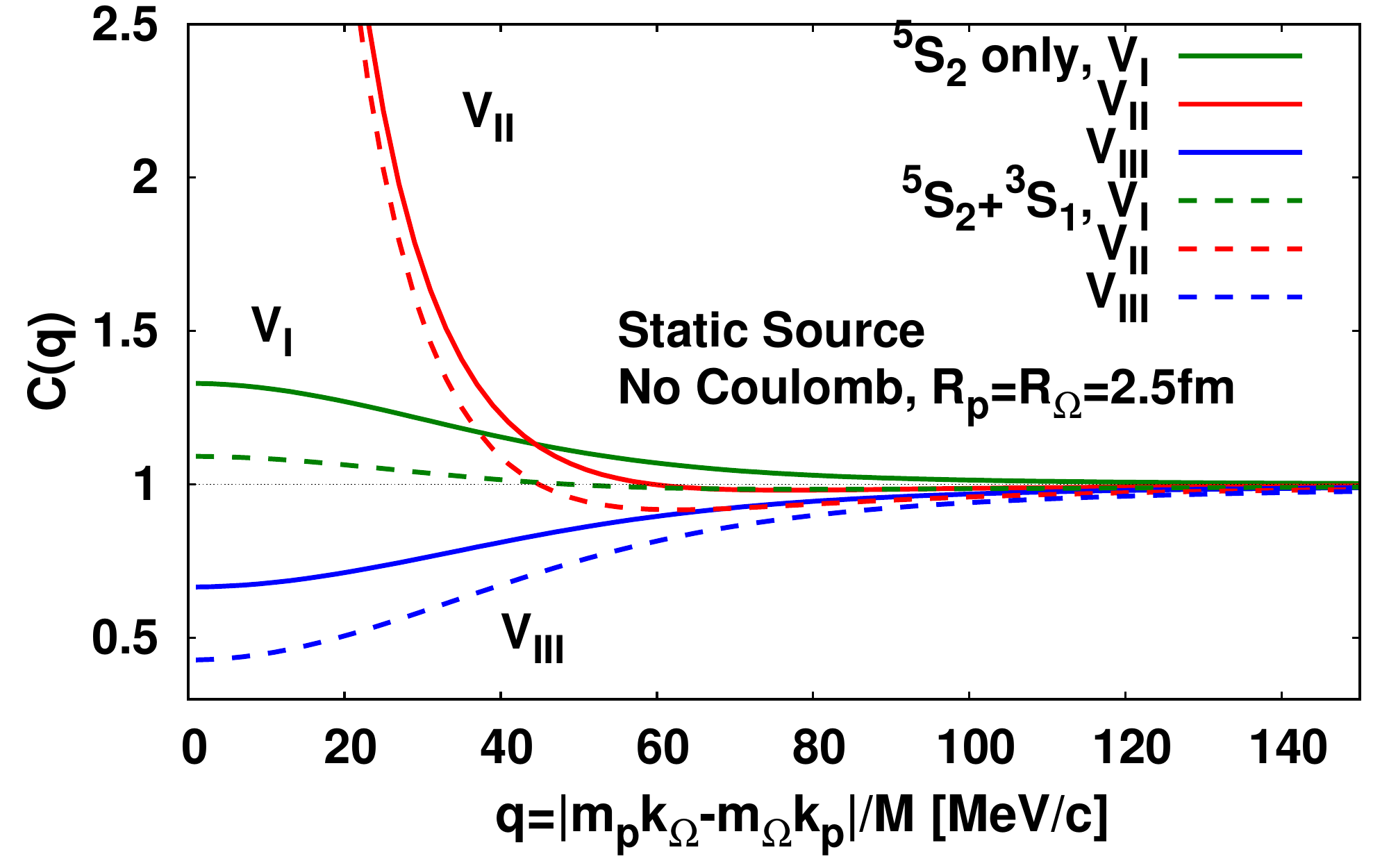}
 \includegraphics[bb=0 0 576 360,width=0.45\textwidth]{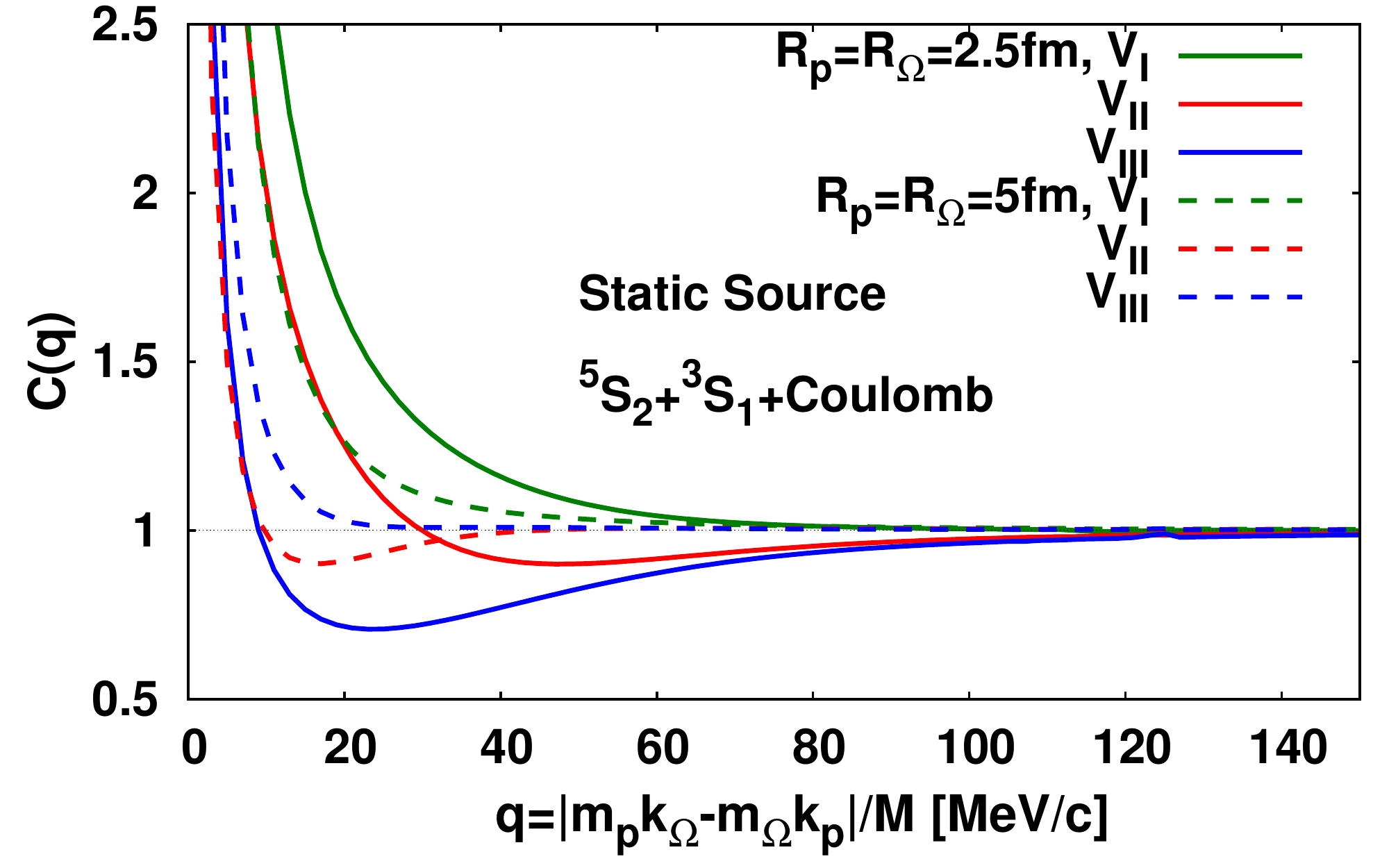}
 \caption{$p\Omega$ correlation function for a static source with 
 $R_p=R_\Omega=2.5$~fm.
 In the left panel,  source sizes are fixed to $R_p=R_\Omega=2.5$~fm and
 the Coulomb interaction is switched off.  Solid (dashed) lines
 denote the correlations with only the $^5{\rm S}_2$ scattering 
  (with both the $^5{\rm S}_2$ scattering and the $^3{\rm S}_1$ absorption).
 In the right panel, the Coulomb interaction is switched on and larger
 $R_p=R_\Omega=5$ fm case is also displayed.}
 \label{fig:c2-pomega_static}
\end{figure}

Figure \ref{fig:c2-pomega_static} shows the $p\Omega$ correlation
function for the various cases.
When the Coulomb interaction is turned off (left panel), the behavior of the $C(q)$
only with $^5S_2$ interaction follows the description given in
Sec.~\ref{subsec_strong_quatnum} and \ref{subsec_LL} (see
Fig.~\ref{Fig:CLLex}).
The effect of absorption in the $^3{\rm S}_1$ channel tends to suppress
the particle correlation as expected from the vanishing wave function
inside the interaction range and resultant scattering length
$a_0 = r_0$. The absorption effect is not negligibly small,
but is not significantly large enough to change the qualitative behavior of
$C(q)$ obtained by the $^5{\rm S}_2$ scattering alone.
One notes that when the imaginary part of the  potential is
finite, the suppression of the correlation becomes the strongest at
non-zero $q$, as seen in the analyses of $p\bar{\Lambda}$ correlation
(see Sec.~\ref{subsubsec:plambdac2}).

When the Coulomb interaction is switched on, the long-range attraction 
gives a strong enhancement of $C(q)$  at small $q$. This is in contrast
to the $pp$ and $\bar{p}\bar{p}$ collisions where the attractive strong
interaction is separated from the Coulomb repulsion, 
as represented by the solid lines
in the right panel of Fig.~\ref{fig:c2-pomega_static}.
The different ordering of the three curves ($R_p = R_\Omega=5$ fm) is
due to the large  reduction of the scattering length for $V_{\rm II}$ by
the Coulomb effect (Table \ref{tbl:pot}).
For larger source size,  $R_{p,\Omega}=5$~fm, 
the correlation function is more sensitive to the long-range part of the interaction as  found for
the proton-proton correlation~\cite{Lednicky:1981su,Bauer:1993wq}. 
As a result, the ordering of the correlation function is further changed
such that $C(q)$ for $V_{\text{II}}$ becomes
the lowest. 

In principle, one may try a full Coulomb correction with source-size
dependence to isolate the effect of strong interaction.
Instead, it has been proposed in Ref.~\cite{Morita:2016auo} to take an ``SL (small-to-large) ratio'' of the correlation functions
for systems with different source sizes,  
\begin{eqnarray}
C_{\rm SL}(q)\equiv \frac{C_{R_{p,\Omega}=2.5{\rm fm}}(q)}{C_{R_{p,\Omega}=5{\rm fm}}(q)},
\end{eqnarray}
as an alternative and model-independent way to handle the Coulomb effect.

\begin{figure}[!t]
 \centering
 \includegraphics[bb=0 0 576 360,width=3.2in]{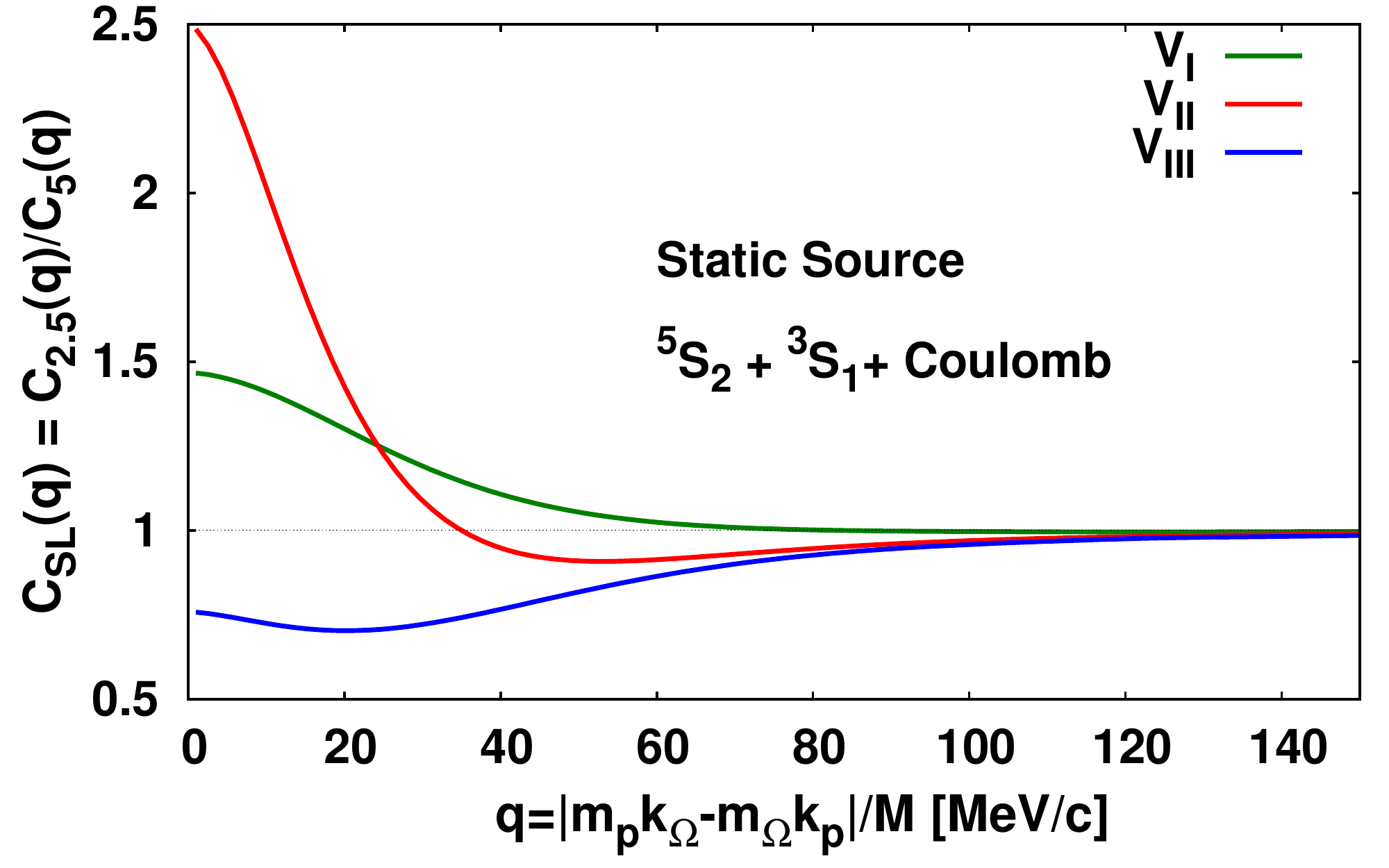}
 \caption{$C_{\rm SL}(q)$ for a static source
 of different source sizes, $R_{p,\Omega}=2.5$ and 5 fm.}
 \label{fig:csl}
\end{figure}

As shown in Fig.~\ref{fig:csl},
an advantage of this ratio is that the effect of the Coulomb interaction 
for small $q$ is largely canceled, so that it has a good 
sensitivity to the strong interaction without much contamination from the 
Coulomb interaction. 
Moreover, taking the ratio of $C(q)$ reduces the apparent reduction of its 
sensitivity to  the strong interaction due to the purity factor.
There are in principle two ways to extract $C_{\rm SL}(q)$ experimentally
in ultrarelativistic heavy ion collisions at RHIC and LHC:  (i)
Comparison of the peripheral and central collisions for the same nuclear system,
 and (ii) comparison of the central collisions with
different system sizes, e.g. central Cu+Cu collisions and central Au+Au
collisions at RHIC.

By using the SL ratio data, it would be possible to guess
the existence or nonexistence of a $N\Omega$ bound state.
Suppressed (slightly enhanced) $C_\mathrm{SL}$ from unity
suggests an attractive $N\Omega$ interaction with (without) a bound state,
and strongly enhanced $C_\mathrm{SL}$ implies a large scattering length
$|a_0|$ of the $N\Omega$ interaction.

\subsubsection{$K^-p$ Correlation}
\label{Sec:Kmp}

For the elucidation of the structure of $\Lambda(1405)$ and the properties of nuclei consisting of $\bar{K}$, the study of the $\bar{K}N$ interaction is essential. Historically, it has been analyzed from the $\bar{K}N$ scattering amplitude constrained by the $\Kmp$ scattering experiments.
While the recent kaonic hydrogen measurement~\cite{Bazzi:2011zj,Bazzi:2012eq} reduces the uncertainty of the subthreshold extrapolation of the $\bar{K}N(I=0)$ amplitude~\cite{Ikeda:2011pi,Ikeda:2012au}, the accuracy in the $I=1$ component is still to be improved. Here we discuss the $\Kmp$ correlation in heavy ion collisions using the potential developed in Ref.~\cite{Miyahara:2015bya}, and demonstrate that the $\Kmp$ correlation can be used as a complementary 
observable to the existing $\bar{K}N$ data.

The correlation function is calculated with the KP formula, as discussed in the previous sections. It should be noted that the $K^{-}p$ channel couples with the $\bar{K}^{0}n$ channel, in contrast to the single-channel problems studied so far. The $\Kmp$ wave function is therefore obtained by solving the coupled-channel Schr\"odinger equation,
\begin{align}
\left(\begin{array}{cc}
-\frac{\nabla^2}{2\mu}+V_{K^-p,K^-p}^{\rm strong}+V^{\rm Coulomb}  &  V_{K^-p,\bar{K}^0n}^{\rm strong}  \\
V_{\bar{K}^0n,K^-p}^{\rm strong}  &  -\frac{\nabla^2}{2\mu}+V_{\bar{K}^0n,\bar{K}^0n}^{\rm strong}
\end{array}\right)
\left(\begin{array}{c}
\psi_{K^-p}({\bm r}) \\
\psi_{\bar{K}^0n}({\bm r})
\end{array}\right)
=E\left(\begin{array}{c}
\psi_{K^-p}({\bm r}) \\
\psi_{\bar{K}^0n}({\bm r})
\end{array}\right).
\end{align}
where $\mu$, $V^{\rm strong}$, and $V^{\rm Coulomb}$ respectively represent the reduced mass, the strong interaction, and the Coulomb interaction. The mass difference between $\Kmp$ and $\Kbn$ is neglected. It should be noted that the Coulomb interaction acts only in the charged $K^{-}p$ channel, and the strong interaction induces the off-diagonal channel coupling between $K^{-}p$ and $\bar{K}^{0}n$.

To extract the physical meaning of the $\Kmp$ correlation function, it is instructive to start from the case without the Coulomb interaction. In this case, the isospin basis, $\bar{K}N(I=0)$ and $\bar{K}N(I=1)$, is adequate as well as the physical basis, $\Kmp$ and $\Kbn$. Considering the relation between these two basis%
\footnote{The phase convention is chosen to be $|K^{-}\rangle=-|I=1/2,I_{3}=-1/2\rangle$.},
\begin{align}
\left(
\begin{array}{c}
|K^-p\rangle \\
|\bar{K}^0n\rangle
\end{array}
\right)
=\frac{1}{\sqrt{2}}
\left(
\begin{array}{cc}
1 & -1 \\
1 & 1
\end{array}
\right)
\left(
\begin{array}{c}
|\bar{K}N^{I=0}\rangle \\
|\bar{K}N^{I=1}\rangle
\end{array}
\right),
\end{align}
the $\bar{K}N$ interactions in the physical basis can be represented by those in the isospin basis $V^{I=0,1}$ as
\begin{align}
\left(\begin{array}{cc}
V_{K^-p,K^-p}^{\rm strong}  &  V_{K^-p,\bar{K}^0n}^{\rm strong}  \\
V_{\bar{K}^0n,K^-p}^{\rm strong}  &  V_{\bar{K}^0n,\bar{K}^0n}^{\rm strong}
\end{array}\right)
 &= \frac{1}{2}
\left(
\begin{array}{cc}
1 & -1 \\
1 & 1
\end{array}
\right)
\left(
\begin{array}{cc}
V^{I=0} & 0 \\
0 & V^{I=1}
\end{array}
\right)
\left(
\begin{array}{cc}
1 & 1 \\
-1 & 1
\end{array}
\right)  \\ \notag
&= \frac{1}{2}
\left(
\begin{array}{cc}
V^{I=0}+V^{I=1} & V^{I=0}-V^{I=1} \\
V^{I=0}-V^{I=1} & V^{I=0}+V^{I=1}
\end{array}
\right).
\end{align}
Using $V^{I=0,1}$ from Ref.~\cite{Miyahara:2015bya}, one can construct the $\bar{K}N$ interaction with the physical basis. 

A general form of the $\bar{K}N$ wave function $\Psi^{(-)}_{{\scriptscriptstyle \bar{K}N},\ell=0}$ can be written as the superposition of the isospin wave function $\psi_{I}(r)$, which has the asymptotic form $e^{-i\delta_I}\sin(qr+\delta_I)/(qr)$, 
\begin{align}
\Psi^{(-)}_{{\scriptscriptstyle \bar{K}N},\ell=0}
&=C_{0}\frac{\chi(\Kmp)+\chi(\Kbn)}{\sqrt{2}}\psi_0(r)
+C_{1}\frac{-\chi(\Kmp)+\chi(\Kbn)}{\sqrt{2}}\psi_1(r)
\ , \\
&=\chi(\Kmp)\psi_\sKmp(r)
+\chi(\Kbn)\psi_\sKbn(r)
,
\end{align}
where $\chi(\Kmp)$ and $\chi(\Kbn)$ represent the isospin wave function of the physical state.
For the wave function used in the correlation function, the $\Kmp$ channel should satisfy the outgoing boundary condition as in Eq.~\eqref{Eq:asymp}. On the other hand, the outgoing wave in the $\Kbn$ channel should disappear. 
From these conditions, the coefficients $C_{0}$ and $C_{1}$ are determined as
$C_{0} = -C_{1} = 1/\sqrt{2}$.
Thus, the asymptotic $\Kmp$ wave function is found to be
\begin{align}
\psi_{\sKmp}(r)
\to&\frac{1}{2iqr}\left[e^{iqr}-\tilde{\calS}^{-1}_\sKmp e^{-iqr}\right]
\ ,\quad
\tilde{\calS}_\sKmp=2\left(\calS_{0}^{-1}+\calS_{1}^{-1}\right)^{-1} ,\quad
\calS_{I}=e^{2i\delta_{I}}.
\label{Eq:wfKmp}
\end{align}
Because of the characteristic boundary condition for the coupled-channel correlation function, the obtained $\tilde{\calS}_\sKmp$ is different from the $S$-matrix in the $\Kmp$ channel $\calS_\sKmp=(\calS_{0}+\calS_{1})/2$ for  usual scattering experiments.

\begin{figure}[bthp]
\begin{center}
 \includegraphics[bb=0 0 540 378,width=0.45\textwidth]{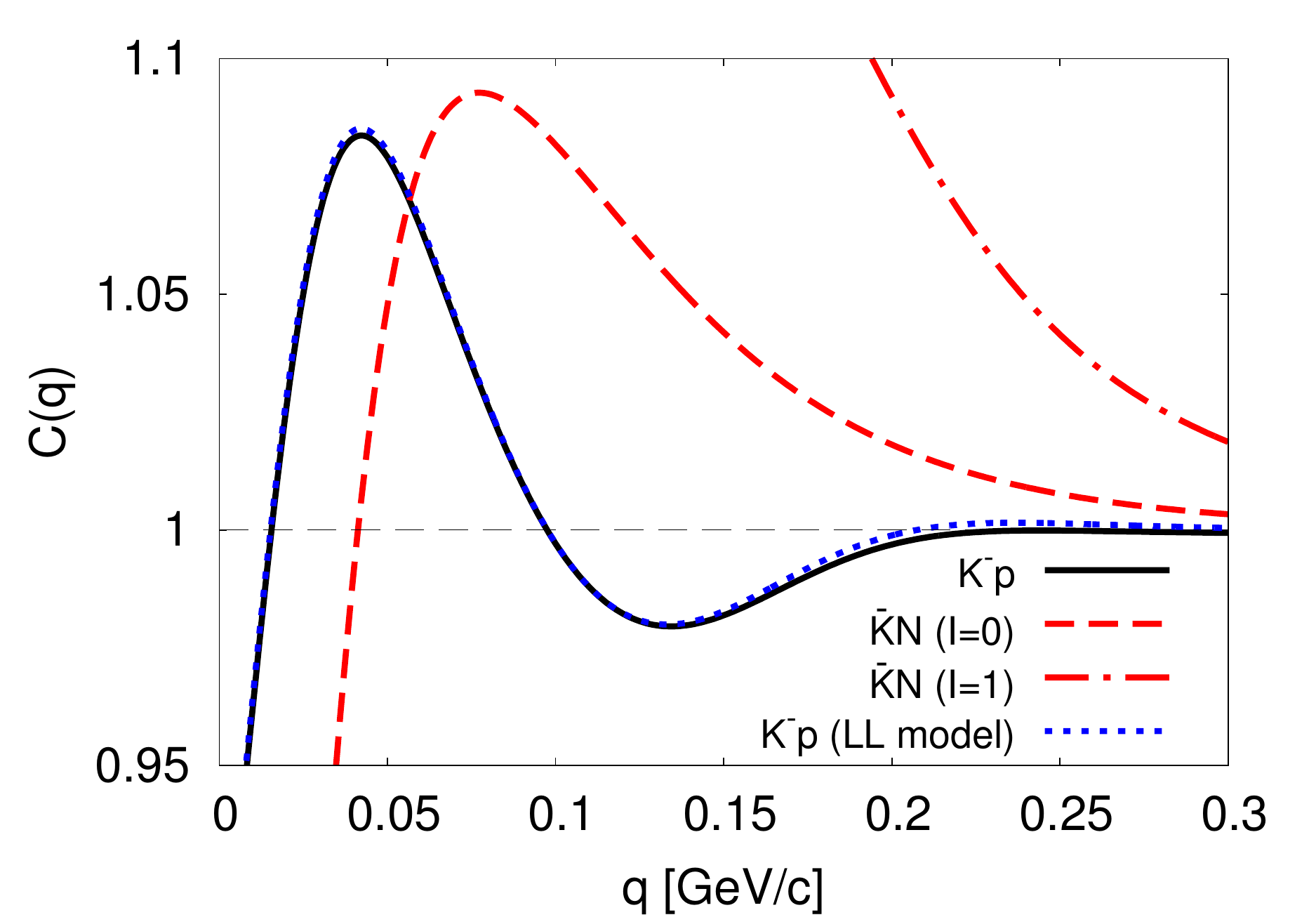}
 \includegraphics[bb=0 0 540 378,width=0.45\textwidth]{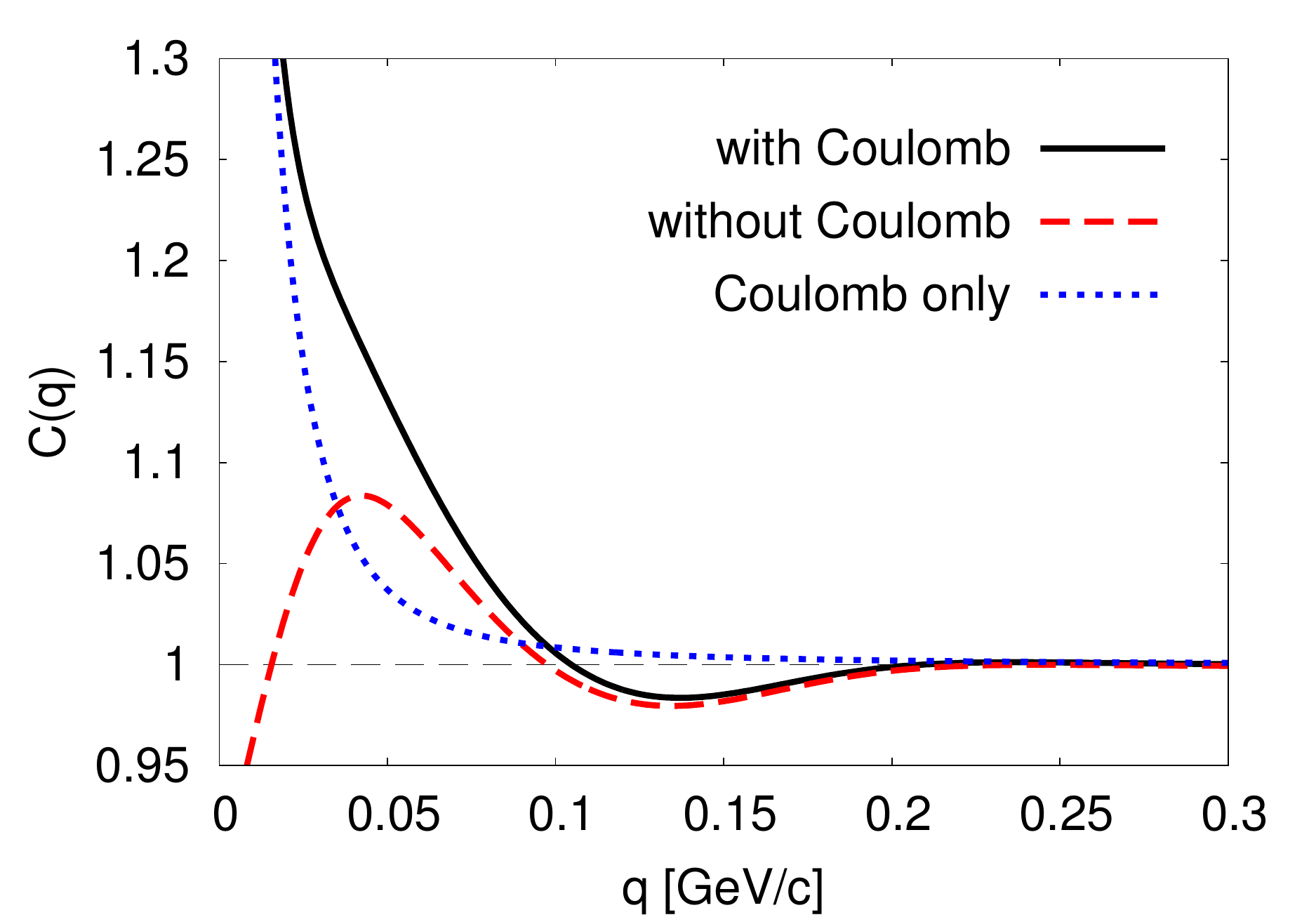}
\end{center}
\caption{$K^{-}p$ correlation function with a static source with $R=3$ fm. 
The left panel shows
the $K^-p$ correlation without the Coulomb function obtained by the potential in Ref.~\cite{Miyahara:2015bya} (solid line) and by the LL model formula (see section~\ref{subsec_LL}) with the same amplitude (dotted line). The correlations of $I=0$ (dashed line) and $I=1$ (dash-dotted line) are also described.
The right panel shows
the $K^-p$ correlation with the Coulomb interaction (solid line), together with the results only with the strong interaction (dashed line) and with the Coulomb interaction (dotted line). 
}
\label{Fig:Kmp}
\end{figure}

The left panel of~\ref{Fig:Kmp}~(left) shows the $K^{-}p$ correlation function without the Coulomb interaction.
The source size of nonidentical particle pairs 
can be estimated as $R=\sqrt{(R_K^2+R_p^2)/2}$.
Considering that the kaon source size in Au+Au collisions at
$\sqrt{s_{{\scriptscriptstyle NN}}}=200~\mathrm{GeV}$ is estimated as $R_K=2$-$5~\fm$~\cite{Csanad:2011bx,Adamczyk:2013wqm}
and the proton source size is expected to be similar,
$R=3.0$ fm is used in this study.
Because of the small interaction range of the $\bar{K}N$ potential (0.4 fm~\cite{Miyahara:2015bya}) owing to the absence of the $\pi$ exchange, the short range details of the $\bar{K}N$ interaction does not affect the correlation function for the source size $R=3.0$ fm. 
Actually, the correlation function is well reproduced by the LL model explained in Sec.~\ref{subsec_LL}, as shown by the dashed line in Fig.~\ref{Fig:Kmp}~(left), which assumes a zero range interaction and uses the asymptotic behavior for the wave function. 

There is another interesting feature, i.e., the existence of the bump and dip structures, around $q\sim0.05$-$0.15$ GeV/$c$, which does not appear in the $K^-p\to K^-p$ scattering amplitude. Its origin seems to be the characteristic isospin combination of $\tilde{\calS}_{K^-p}$ in Eq.~\eqref{Eq:wfKmp}. Especially, the dip structure around $q\sim0.15$ GeV/$c$ is a good example, because the $K^-p$ correlation function is smaller than unity, though both of the $\bar{K}N(I=0)$ and $\bar{K}N(I=1)$ correlation functions are larger than unity in the corresponding energy region [see dashed ($I=0$) and dash-dotted ($I=1$) lines in Fig.~\ref{Fig:Kmp}~(left)], reflecting the attractive $\bar{K}N(I=0,1)$ interaction. 
Thus, the coupled-channel correlation function gives us information 
complementary to that from the $K^-p$ scattering.

For the direct comparison with future experiments, the $K^-p$ correlation with the Coulomb interaction is shown by the solid line in Fig.~\ref{Fig:Kmp}~(right). Similar to the $p\Omega$ correlation in Sec.~\ref{Sec:pOme}, the $K^-p$ correlation is largely enhanced by the Coulomb interaction in the small $q$ region ($q\lesssim0.1$ GeV/$c$). On the other hand, in the relatively higher energy region,
the correlation function is
determined by the strong interaction. As a result, the interesting dip structure in Fig.~\ref{Fig:Kmp}~(left) is kept in the case with the Coulomb interaction in Fig.~\ref{Fig:Kmp}~(right).

It should be noted that the $\Lambda(1520)$ effect, which appears in the $d$-wave $\bar{K}N(I=0)$ scattering, is not included in the above results. Because the $\Lambda(1520)$ energy region corresponds to $q\sim 0.24$ GeV/$c$ and the width of $\Lambda(1520)$ is not very large ($\sim 15$ MeV), the inclusion of the $\Lambda(1520)$ would not affect very much the dip structure around $q\sim0.15$ GeV/$c$. Thus, the interesting feature of the isospin interference is expected to be seen in actual measurements. 

\section{Summary} 
\label{sec:summary}
High-energy heavy-ion collisions provide a unique opportunity to study the properties of the high energy-density QCD matter formed at the early stage of the collisions.  At the same time, these experiments can also be used to investigate the hadronic interactions at low energies because the final stages of heavy ion collision is an excellent environment where heavy hadrons and composite particles including the light (anti)nuclei can be produced.

In this article, we have summarized the present status of recently observed exotic hadrons that can potentially also be measured in a heavy ion collision.  We have also reviewed the current understanding on the production mechanisms of hadronic resonances and bound states as well as hadrons of multiquark configurations in heavy ion collision.  We have further reported  the yields of these particles in relativistic heavy ion collisions and how they can be used to discriminate between different configurations for their structures that otherwise would not be possible by simply considering their quantum numbers.  We have finally discussed the current status of two-particle correlation measurements in relativistic heavy ion collisions and how these studies can shed light on the interactions between the particles involved in the measurements.  Further theoretical and experimental studies along these directions will open up a new window for understanding the properties of QCD at low energy from high energy heavy ion collisions.

\section*{Acknowledgments}

This work was supported in part by the Grants-in-Aid for Scientific
Research on Innovative Areas from MEXT (Grants No. 24105008 and
No. 24105001), by JSPS KAKENHI (the Grant-in-Aid for Scientific
Research from Japan Society for the Promotion of Science (JSPS))
with Grants No.~24740152 and No.~16K17694 (Tetsuo Hyodo),
No.~16K05349 (Kenji Morita), No.~25247036 and No.~15K17641
(Shigehiro Yasui), No.~15K05079, No.~15H03663, No.~16K05350 (Akira Ohnishi),
by the Yukawa International Program for
Quark-Hadron Sciences (YIPQS), by the Korea National Research
Foundation under the grant number KRF-2011-0030621 and the Korean
ministry of education under the grant number 2016R1D1A1B03930089
(Su Houng Lee), by the National Research Foundation of Korea
(NRF) grant funded by the Korea government (MSIP) (No.
2016R1C1B1016270) and 2015 Research Grant from Kangwon National
University (Sungtae Cho), and by FAPESP and CNPq-Brazil, by  US Department of Energy under Contract No. $\mathrm{DE}$-$\mathrm{SC0015266}$ and the Welch Foundation under Grant No. A-1358.





\bibliographystyle{elsarticle-num} 


%
%
%


\begin{thebibliography}{100}
\expandafter\ifx\csname url\endcsname\relax
  \def\url#1{\texttt{#1}}\fi
\expandafter\ifx\csname urlprefix\endcsname\relax\def\urlprefix{URL }\fi
\expandafter\ifx\csname href\endcsname\relax
  \def\href#1#2{#2} \def\path#1{#1}\fi

\bibitem{Braun-Munzinger:2014pya}
P.~Braun-Munzinger, B.~Friman, J.~Stachel, {Proceedings, 24th International
  Conference on Ultra-Relativistic Nucleus-Nucleus Collisions (Quark Matter
  2014)}, Nucl. Phys. A931 (2014) pp.1--1266.

\bibitem{Abelev:2010rv}
B.~I. Abelev, et~al., {Observation of an Antimatter Hypernucleus}, Science 328
  (2010) 58--62.
\newblock \href {http://arxiv.org/abs/1003.2030} {\path{arXiv:1003.2030}},
  \href {http://dx.doi.org/10.1126/science.1183980}
  {\path{doi:10.1126/science.1183980}}.

\bibitem{Agakishiev:2011ib}
H.~A. {\it et al.}, {Observation of the antimatter helium-4 nucleus}, Nature
  473 (2011) 353.
\newblock \href {http://arxiv.org/abs/1103.3312} {\path{arXiv:1103.3312}}.

\bibitem{Adam:2015vda}
J.~Adam, et~al., {Production of light nuclei and anti-nuclei in pp and Pb-Pb
  collisions at energies available at the CERN Large Hadron Collider}, Phys.
  Rev. C93~(2) (2016) 024917.
\newblock \href {http://arxiv.org/abs/1506.08951} {\path{arXiv:1506.08951}},
  \href {http://dx.doi.org/10.1103/PhysRevC.93.024917}
  {\path{doi:10.1103/PhysRevC.93.024917}}.

\bibitem{Adam:2016bpr}
J.~Adam, et~al., {Production of K$^{*}$ (892)$^{0}$ and $\phi $ (1020) in p-Pb
  collisions at $\sqrt{s_{{\text {NN}}}}$ = 5.02 TeV}, Eur. Phys. J. C76~(5)
  (2016) 245.
\newblock \href {http://arxiv.org/abs/1601.07868} {\path{arXiv:1601.07868}},
  \href {http://dx.doi.org/10.1140/epjc/s10052-016-4088-7}
  {\path{doi:10.1140/epjc/s10052-016-4088-7}}.

\bibitem{KanadaEn'yo:2006zk}
{Y. Kanada-En'yo and B. M\"{u}ller}, {Suppression of p-wave baryons in quark
  recombination}, Phys.Rev. C74 (2006) 061901.
\newblock \href {http://arxiv.org/abs/nucl-th/0608015}
  {\path{arXiv:nucl-th/0608015}}.

\bibitem{Cho:2014xha}
S.~Cho, {Enhanced production of $\psi(2S)$ mesons in heavy ion collisions},
  Phys. Rev. C91~(5) (2015) 054914.
\newblock \href {http://arxiv.org/abs/1408.4756} {\path{arXiv:1408.4756}},
  \href {http://dx.doi.org/10.1103/PhysRevC.91.054914}
  {\path{doi:10.1103/PhysRevC.91.054914}}.

\bibitem{Aubert:2003fg}
B.~Aubert, et~al., {Observation of a narrow meson decaying to $D_s^+ \pi^0$ at
  a mass of 2.32-GeV/c$^2$}, Phys. Rev. Lett. 90 (2003) 242001.
\newblock \href {http://arxiv.org/abs/hep-ex/0304021}
  {\path{arXiv:hep-ex/0304021}}, \href
  {http://dx.doi.org/10.1103/PhysRevLett.90.242001}
  {\path{doi:10.1103/PhysRevLett.90.242001}}.

\bibitem{Choi:2003ue}
S.~K. Choi, et~al., {Observation of a narrow charmonium - like state in
  exclusive B+- ---> K+- pi+ pi- J / psi decays}, Phys. Rev. Lett. 91 (2003)
  262001.
\newblock \href {http://arxiv.org/abs/hep-ex/0309032}
  {\path{arXiv:hep-ex/0309032}}, \href
  {http://dx.doi.org/10.1103/PhysRevLett.91.262001}
  {\path{doi:10.1103/PhysRevLett.91.262001}}.

\bibitem{Aaij:2015tga}
R.~Aaij, et~al., {Observation of $J/\psi p$ Resonances Consistent with
  Pentaquark States in $\Lambda_b^0 \to J/\psi K^- p$ Decays}, Phys. Rev. Lett.
  115 (2015) 072001.
\newblock \href {http://arxiv.org/abs/1507.03414} {\path{arXiv:1507.03414}},
  \href {http://dx.doi.org/10.1103/PhysRevLett.115.072001}
  {\path{doi:10.1103/PhysRevLett.115.072001}}.

\bibitem{Jaffe:1976ig}
R.~L. Jaffe, {Multi-Quark Hadrons. 1. The Phenomenology of (2 Quark 2
  anti-Quark) Mesons}, Phys.Rev. D15 (1977) 267.
\newblock \href {http://dx.doi.org/10.1103/PhysRevD.15.267}
  {\path{doi:10.1103/PhysRevD.15.267}}.

\bibitem{Jaffe:1976ih}
R.~L. Jaffe, {Multi-Quark Hadrons. 2. Methods}, Phys.Rev. D15 (1977) 281.
\newblock \href {http://dx.doi.org/10.1103/PhysRevD.15.281}
  {\path{doi:10.1103/PhysRevD.15.281}}.

\bibitem{Pisarski:2016ukx}
R.~D. Pisarski, V.~V. Skokov, {How tetraquarks can generate a second chiral
  phase transition}, Phys. Rev. D94~(5) (2016) 054008.
\newblock \href {http://arxiv.org/abs/1606.04111} {\path{arXiv:1606.04111}},
  \href {http://dx.doi.org/10.1103/PhysRevD.94.054008}
  {\path{doi:10.1103/PhysRevD.94.054008}}.

\bibitem{Cho:2010db}
S.~Cho, et~al., {Multi-quark hadrons from Heavy Ion Collisions}, Phys.Rev.Lett.
  106 (2011) 212001.
\newblock \href {http://arxiv.org/abs/1011.0852} {\path{arXiv:1011.0852}},
  \href {http://dx.doi.org/10.1103/PhysRevLett.106.212001}
  {\path{doi:10.1103/PhysRevLett.106.212001}}.

\bibitem{Cho:2011ew}
S.~Cho, et~al., {Exotic Hadrons in Heavy Ion Collisions}, Phys.Rev. C84 (2011)
  064910.
\newblock \href {http://arxiv.org/abs/1107.1302} {\path{arXiv:1107.1302}},
  \href {http://dx.doi.org/10.1103/PhysRevC.84.064910}
  {\path{doi:10.1103/PhysRevC.84.064910}}.

\bibitem{Adamczyk:2015hza}
L.~Adamczyk, et~al., {Measurement of Interaction between Antiprotons}, Nature
  527 (2015) 345--348.
\newblock \href {http://arxiv.org/abs/1507.07158} {\path{arXiv:1507.07158}},
  \href {http://dx.doi.org/10.1038/nature15724}
  {\path{doi:10.1038/nature15724}}.

\bibitem{NA49_protonlambda}
T.~Anticic, et~al., {Proton - Lambda Correlations in Central Pb+Pb Collisions
  at $\sqrt{s_{NN}} = 17.3$ GeV}, Phys. Rev. C83 (2011) 054906.
\newblock \href {http://arxiv.org/abs/1103.3395} {\path{arXiv:1103.3395}},
  \href {http://dx.doi.org/10.1103/PhysRevC.83.054906}
  {\path{doi:10.1103/PhysRevC.83.054906}}.

\bibitem{STAR_p-lambda2006}
J.~Adams, et~al., Proton-$\lambda$ correlations in central au+au collisions at
  $\sqrt{s_{NN}}=200$ gev, Phys. Rev. C 74 (2006) 064906.

\bibitem{HADES_p-lambda2016}
J.~Adamczewski-Musch, et~al., $\lambda p$ interaction studied via femtoscopy in
  $p+nb$ reactions at $\sqrt{s_{NN}}=3.18$ gev, Phys. Rev. C 94 (2016) 025201.

\bibitem{Ahn:1998fj}
J.~K. Ahn, et~al., {Enhanced Lambda Lambda production near threshold in the
  C-12(K-,K+) reaction}, Phys. Lett. B 444 (1998) 267--272.
\newblock \href {http://dx.doi.org/10.1016/S0370-2693(98)01416-6}
  {\path{doi:10.1016/S0370-2693(98)01416-6}}.

\bibitem{Yoon:2007aq}
C.~J. Yoon, et~al., {Search for the H-dibaryon resonance in C-12 (K-, K+ Lambda
  Lambda X)}, Phys. Rev. C 75 (2007) 022201.
\newblock \href {http://dx.doi.org/10.1103/PhysRevC.75.022201}
  {\path{doi:10.1103/PhysRevC.75.022201}}.

\bibitem{Adamczyk:2014vca}
L.~Adamczyk, et~al., {$\Lambda\Lambda$ Correlation Function in Au+Au collisions
  at $\sqrt{s_{NN}}=$ 200 GeV}, Phys. Rev. Lett. 114~(2) (2015) 022301.
\newblock \href {http://arxiv.org/abs/1408.4360} {\path{arXiv:1408.4360}},
  \href {http://dx.doi.org/10.1103/PhysRevLett.114.022301}
  {\path{doi:10.1103/PhysRevLett.114.022301}}.

\bibitem{Adam:2015nca}
J.~Adam, et~al., {Search for weakly decaying $\bar{\Lambda\mathrm{n}}$ and
  $\Lambda\Lambda $ exotic bound states in central Pb-Pb collisions at
  $\sqrt{s_{\rm NN}}$ = 2.76 TeV}, Phys. Lett. B752 (2016) 267--277.
\newblock \href {http://arxiv.org/abs/1506.07499} {\path{arXiv:1506.07499}},
  \href {http://dx.doi.org/10.1016/j.physletb.2015.11.048}
  {\path{doi:10.1016/j.physletb.2015.11.048}}.

\bibitem{Kim:2013vym}
B.~H. Kim, et~al., {Search for an $H$-dibaryon with mass near $2m_\Lambda$ in
  $\Upsilon(1S)$ and $\Upsilon(2S)$ decays}, Phys. Rev. Lett. 110~(22) (2013)
  222002.
\newblock \href {http://arxiv.org/abs/1302.4028} {\path{arXiv:1302.4028}},
  \href {http://dx.doi.org/10.1103/PhysRevLett.110.222002}
  {\path{doi:10.1103/PhysRevLett.110.222002}}.

\bibitem{Koonin:1977fh}
S.~E. Koonin, {Proton Pictures of High-Energy Nuclear Collisions}, Phys. Lett.
  B 70 (1977) 43--47.
\newblock \href {http://dx.doi.org/10.1016/0370-2693(77)90340-9}
  {\path{doi:10.1016/0370-2693(77)90340-9}}.

\bibitem{Lednicky:1981su}
R.~Lednicky, V.~L. Lyuboshits, {Final State Interaction Effect on Pairing
  Correlations Between Particles with Small Relative Momenta}, Sov. J. Nucl.
  Phys. 35 (1982) 770, [Yad. Fiz.35,1316(1981)].

\bibitem{Bauer:1993wq}
W.~Bauer, C.~K. Gelbke, S.~Pratt, {Hadronic interferometry in heavy ion
  collisions}, Ann. Rev. Nucl. Part. Sci. 42 (1992) 77--100.
\newblock \href {http://dx.doi.org/10.1146/annurev.ns.42.120192.000453}
  {\path{doi:10.1146/annurev.ns.42.120192.000453}}.

\bibitem{Lednicky:2005tb}
R.~Lednicky, {Finite-size effects on two-particle production in continuous and
  discrete spectrum}, Phys. Part. Nucl. 40 (2009) 307--352.
\newblock \href {http://arxiv.org/abs/nucl-th/0501065}
  {\path{arXiv:nucl-th/0501065}}, \href
  {http://dx.doi.org/10.1134/S1063779609030034}
  {\path{doi:10.1134/S1063779609030034}}.

\bibitem{Lisa:2005dd}
M.~A. Lisa, S.~Pratt, R.~Soltz, U.~Wiedemann, {Femtoscopy in relativistic heavy
  ion collisions}, Ann. Rev. Nucl. Part. Sci. 55 (2005) 357--402.
\newblock \href {http://arxiv.org/abs/nucl-ex/0505014}
  {\path{arXiv:nucl-ex/0505014}}, \href
  {http://dx.doi.org/10.1146/annurev.nucl.55.090704.151533}
  {\path{doi:10.1146/annurev.nucl.55.090704.151533}}.

\bibitem{Pratt:2008qv}
S.~Pratt, {Resolving the HBT Puzzle in Relativistic Heavy Ion Collision}, Phys.
  Rev. Lett. 102 (2009) 232301.
\newblock \href {http://arxiv.org/abs/0811.3363} {\path{arXiv:0811.3363}},
  \href {http://dx.doi.org/10.1103/PhysRevLett.102.232301}
  {\path{doi:10.1103/PhysRevLett.102.232301}}.

\bibitem{Greiner:1989ig}
C.~Greiner, B.~Muller, {Pair Correlations of Neutral Strange Particles Emitted
  in Relativistic Heavy Ion Collisions}, Phys. Lett. B 219 (1989) 199--204.
\newblock \href {http://dx.doi.org/10.1016/0370-2693(89)90377-8}
  {\path{doi:10.1016/0370-2693(89)90377-8}}.

\bibitem{Ohnishi:1998at}
A.~Ohnishi, Y.~Hirata, Y.~Nara, S.~Shinmura, Y.~Akaishi, {Can we extract
  Lambda-Lambda interaction from two particle momentum correlation?}, Nucl.
  Phys. A 670 (2000) 297--300.
\newblock \href {http://arxiv.org/abs/nucl-th/9903021}
  {\path{arXiv:nucl-th/9903021}}, \href
  {http://dx.doi.org/10.1016/S0375-9474(00)00117-2}
  {\path{doi:10.1016/S0375-9474(00)00117-2}}.

\bibitem{Kisiel:2014mma}
A.~Kisiel, H.~Zbroszczyk, M.~Szyma\'{n}ski, {Extracting baryon-antibaryon
  strong interaction potentials from p$\bar{\Lambda}$ femtoscopic correlation
  functions}, Phys. Rev. C 89~(5) (2014) 054916.
\newblock \href {http://arxiv.org/abs/1403.0433} {\path{arXiv:1403.0433}},
  \href {http://dx.doi.org/10.1103/PhysRevC.89.054916}
  {\path{doi:10.1103/PhysRevC.89.054916}}.

\bibitem{Shapoval:2014yha}
V.~M. Shapoval, B.~Erazmus, R.~Lednicky, {\relax Yu}.~M. Sinyukov, {Extracting
  $p\Lambda$ scattering lengths from heavy ion collisions}, Phys. Rev. C 92~(3)
  (2015) 034910.
\newblock \href {http://arxiv.org/abs/1405.3594} {\path{arXiv:1405.3594}},
  \href {http://dx.doi.org/10.1103/PhysRevC.92.034910}
  {\path{doi:10.1103/PhysRevC.92.034910}}.

\bibitem{Ohnishi:2016elb}
A.~Ohnishi, K.~Morita, K.~Miyahara, T.~Hyodo, {Hadron-hadron correlation and
  interaction from heavy-ion collisions}, Nucl. Phys. A 954 (2016) 294--307.
\newblock \href {http://arxiv.org/abs/1603.05761} {\path{arXiv:1603.05761}},
  \href {http://dx.doi.org/10.1016/j.nuclphysa.2016.05.010}
  {\path{doi:10.1016/j.nuclphysa.2016.05.010}}.

\bibitem{Morita:2014kza}
K.~Morita, T.~Furumoto, A.~Ohnishi, {$\Lambda\Lambda$ interaction from
  relativistic heavy-ion collisions}, Phys. Rev. C 91~(2) (2015) 024916.
\newblock \href {http://arxiv.org/abs/1408.6682} {\path{arXiv:1408.6682}},
  \href {http://dx.doi.org/10.1103/PhysRevC.91.024916}
  {\path{doi:10.1103/PhysRevC.91.024916}}.

\bibitem{Morita:2016auo}
K.~Morita, A.~Ohnishi, F.~Etminan, T.~Hatsuda, {Probing Multi-Strange Dibaryon
  with Proton-Omega Correlation in High-energy Heavy Ion Collisions}, Phys.
  Rev. C 94~(3) (2016) 031901.
\newblock \href {http://arxiv.org/abs/1605.06765} {\path{arXiv:1605.06765}},
  \href {http://dx.doi.org/10.1103/PhysRevC.94.031901}
  {\path{doi:10.1103/PhysRevC.94.031901}}.

\bibitem{Takahashi:2001nm}
H.~Takahashi, et~al., {Observation of a (Lambda Lambda)He-6 double
  hypernucleus}, Phys. Rev. Lett. 87 (2001) 212502.
\newblock \href {http://dx.doi.org/10.1103/PhysRevLett.87.212502}
  {\path{doi:10.1103/PhysRevLett.87.212502}}.

\bibitem{Weinstein:1982gc}
J.~D. Weinstein, N.~Isgur, {Do Multi-Quark Hadrons Exist?}, Phys.Rev.Lett. 48
  (1982) 659.
\newblock \href {http://dx.doi.org/10.1103/PhysRevLett.48.659}
  {\path{doi:10.1103/PhysRevLett.48.659}}.

\bibitem{Weinstein:1983gd}
J.~D. Weinstein, N.~Isgur, {The q q anti-q anti-q System in a Potential Model},
  Phys.Rev. D27 (1983) 588.
\newblock \href {http://dx.doi.org/10.1103/PhysRevD.27.588}
  {\path{doi:10.1103/PhysRevD.27.588}}.

\bibitem{Oller:1997ng}
J.~Oller, E.~Oset, J.~Pelaez, {Nonperturbative approach to effective chiral
  Lagrangians and meson interactions}, Phys.Rev.Lett. 80 (1998) 3452--3455.
\newblock \href {http://arxiv.org/abs/hep-ph/9803242}
  {\path{arXiv:hep-ph/9803242}}, \href
  {http://dx.doi.org/10.1103/PhysRevLett.80.3452}
  {\path{doi:10.1103/PhysRevLett.80.3452}}.

\bibitem{Oller:1998hw}
J.~Oller, E.~Oset, J.~Pelaez, {Meson meson interaction in a nonperturbative
  chiral approach}, Phys.Rev. D59 (1999) 074001.
\newblock \href {http://arxiv.org/abs/hep-ph/9804209}
  {\path{arXiv:hep-ph/9804209}}, \href
  {http://dx.doi.org/10.1103/PhysRevD.59.074001, 10.1103/PhysRevD.60.099906 ,
  10.1103/PhysRevD.60.099906, 10.1103/PhysRevD.75.099903}
  {\path{doi:10.1103/PhysRevD.59.074001, 10.1103/PhysRevD.60.099906 ,
  10.1103/PhysRevD.60.099906, 10.1103/PhysRevD.75.099903}}.

\bibitem{Oller:1998zr}
J.~Oller, E.~Oset, {N/D description of two meson amplitudes and chiral
  symmetry}, Phys.Rev. D60 (1999) 074023.
\newblock \href {http://arxiv.org/abs/hep-ph/9809337}
  {\path{arXiv:hep-ph/9809337}}, \href
  {http://dx.doi.org/10.1103/PhysRevD.60.074023}
  {\path{doi:10.1103/PhysRevD.60.074023}}.

\bibitem{Hyodo:2013nka}
T.~Hyodo, {Structure and compositeness of hadron resonances}, Int. J. Mod.
  Phys. A28 (2013) 1330045.
\newblock \href {http://arxiv.org/abs/1310.1176} {\path{arXiv:1310.1176}},
  \href {http://dx.doi.org/10.1142/S0217751X13300457}
  {\path{doi:10.1142/S0217751X13300457}}.

\bibitem{Sekihara:2014kya}
T.~Sekihara, T.~Hyodo, D.~Jido, {Comprehensive analysis of the wave function of
  a hadronic resonance and its compositeness}, PTEP 2015~(6) (2014) 063D04.
\newblock \href {http://arxiv.org/abs/1411.2308} {\path{arXiv:1411.2308}},
  \href {http://dx.doi.org/10.1093/ptep/ptv081}
  {\path{doi:10.1093/ptep/ptv081}}.

\bibitem{Sekihara:2014qxa}
T.~Sekihara, S.~Kumano, {Constraint on $K\bar K$ compositeness of the
  $a_0(980)$ and $f_0(980)$ resonances from their mixing intensity}, Phys. Rev.
  D92~(3) (2015) 034010.
\newblock \href {http://arxiv.org/abs/1409.2213} {\path{arXiv:1409.2213}},
  \href {http://dx.doi.org/10.1103/PhysRevD.92.034010}
  {\path{doi:10.1103/PhysRevD.92.034010}}.

\bibitem{Sekihara:2012xp}
T.~Sekihara, T.~Hyodo, {Size measurement of dynamically generated hadronic
  resonances with finite boxes}, Phys.Rev. C87~(4) (2013) 045202.
\newblock \href {http://arxiv.org/abs/1209.0577} {\path{arXiv:1209.0577}},
  \href {http://dx.doi.org/10.1103/PhysRevC.87.045202}
  {\path{doi:10.1103/PhysRevC.87.045202}}.

\bibitem{Agashe:2014kda}
C.~Patrignani, et~al., {Review of Particle Physics}, Chin. Phys. C40~(10)
  (2016) 100001.
\newblock \href {http://dx.doi.org/10.1088/1674-1137/40/10/100001}
  {\path{doi:10.1088/1674-1137/40/10/100001}}.

\bibitem{Isgur:1978xj}
N.~Isgur, G.~Karl, {P Wave Baryons in the Quark Model}, Phys. Rev. D18 (1978)
  4187.
\newblock \href {http://dx.doi.org/10.1103/PhysRevD.18.4187}
  {\path{doi:10.1103/PhysRevD.18.4187}}.

\bibitem{Dalitz:1959dn}
R.~Dalitz, S.~Tuan, {A possible resonant state in pion-hyperon scattering},
  Phys. Rev. Lett. 2 (1959) 425--428.
\newblock \href {http://dx.doi.org/10.1103/PhysRevLett.2.425}
  {\path{doi:10.1103/PhysRevLett.2.425}}.

\bibitem{Dalitz:1960du}
R.~Dalitz, S.~Tuan, {The phenomenological description of -K -nucleon reaction
  processes}, Annals Phys. 10 (1960) 307--351.
\newblock \href {http://dx.doi.org/10.1016/0003-4916(60)90001-4}
  {\path{doi:10.1016/0003-4916(60)90001-4}}.

\bibitem{Kaiser:1995eg}
N.~Kaiser, P.~Siegel, W.~Weise, {Chiral dynamics and the low-energy kaon -
  nucleon interaction}, Nucl. Phys. A594 (1995) 325--345.
\newblock \href {http://arxiv.org/abs/nucl-th/9505043}
  {\path{arXiv:nucl-th/9505043}}, \href
  {http://dx.doi.org/10.1016/0375-9474(95)00362-5}
  {\path{doi:10.1016/0375-9474(95)00362-5}}.

\bibitem{Oset:1998it}
E.~Oset, A.~Ramos, {Nonperturbative chiral approach to s wave anti-K N
  interactions}, Nucl. Phys. A635 (1998) 99--120.
\newblock \href {http://arxiv.org/abs/nucl-th/9711022}
  {\path{arXiv:nucl-th/9711022}}, \href
  {http://dx.doi.org/10.1016/S0375-9474(98)00170-5}
  {\path{doi:10.1016/S0375-9474(98)00170-5}}.

\bibitem{Oller:2000fj}
J.~Oller, U.~G. Meissner, {Chiral dynamics in the presence of bound states:
  Kaon nucleon interactions revisited}, Phys. Lett. B500 (2001) 263--272.
\newblock \href {http://arxiv.org/abs/hep-ph/0011146}
  {\path{arXiv:hep-ph/0011146}}, \href
  {http://dx.doi.org/10.1016/S0370-2693(01)00078-8}
  {\path{doi:10.1016/S0370-2693(01)00078-8}}.

\bibitem{Lutz:2001yb}
M.~Lutz, E.~Kolomeitsev, {Relativistic chiral SU(3) symmetry, large N(c) sum
  rules and meson baryon scattering}, Nucl. Phys. A700 (2002) 193--308.
\newblock \href {http://arxiv.org/abs/nucl-th/0105042}
  {\path{arXiv:nucl-th/0105042}}, \href
  {http://dx.doi.org/10.1016/S0375-9474(01)01312-4}
  {\path{doi:10.1016/S0375-9474(01)01312-4}}.

\bibitem{Hyodo:2011ur}
T.~Hyodo, D.~Jido, {The nature of the Lambda(1405) resonance in chiral
  dynamics}, Prog. Part. Nucl. Phys. 67 (2012) 55--98.
\newblock \href {http://arxiv.org/abs/1104.4474} {\path{arXiv:1104.4474}},
  \href {http://dx.doi.org/10.1016/j.ppnp.2011.07.002}
  {\path{doi:10.1016/j.ppnp.2011.07.002}}.

\bibitem{Kamiya:2016jqc}
Y.~Kamiya, K.~Miyahara, S.~Ohnishi, Y.~Ikeda, T.~Hyodo, E.~Oset, W.~Weise,
  {Antikaon-nucleon interaction and Lambda(1405) in chiral SU(3) dynamics}
  (2016).
\newblock \href {http://arxiv.org/abs/1602.08852} {\path{arXiv:1602.08852}}.

\bibitem{Hall:2014uca}
J.~M.~M. Hall, W.~Kamleh, D.~B. Leinweber, B.~J. Menadue, B.~J. Owen, A.~W.
  Thomas, R.~D. Young, {Lattice QCD Evidence that the Lambda(1405) Resonance is
  an Antikaon-Nucleon Molecule}, Phys. Rev. Lett. 114~(13) (2015) 132002.
\newblock \href {http://arxiv.org/abs/1411.3402} {\path{arXiv:1411.3402}},
  \href {http://dx.doi.org/10.1103/PhysRevLett.114.132002}
  {\path{doi:10.1103/PhysRevLett.114.132002}}.

\bibitem{Kamiya:2015aea}
Y.~Kamiya, T.~Hyodo, {Structure of near-threshold quasibound states}, Phys.
  Rev. C93~(3) (2016) 035203.
\newblock \href {http://arxiv.org/abs/1509.00146} {\path{arXiv:1509.00146}},
  \href {http://dx.doi.org/10.1103/PhysRevC.93.035203}
  {\path{doi:10.1103/PhysRevC.93.035203}}.

\bibitem{Minireview}
U.-G. Meissner, T.~Hyodo, \textit{Pole Structure of the $\Lambda(1405)$ Region}
  in Review of Particle Physics, Chin. Phys. C40~(10) (2016) 100001.

\bibitem{Jido:2003cb}
D.~Jido, J.~Oller, E.~Oset, A.~Ramos, U.~Meissner, {Chiral dynamics of the two
  Lambda(1405) states}, Nucl. Phys. A725 (2003) 181--200.
\newblock \href {http://arxiv.org/abs/nucl-th/0303062}
  {\path{arXiv:nucl-th/0303062}}, \href
  {http://dx.doi.org/10.1016/S0375-9474(03)01598-7}
  {\path{doi:10.1016/S0375-9474(03)01598-7}}.

\bibitem{Hyodo:2007jq}
T.~Hyodo, W.~Weise, {Effective anti-K N interaction based on chiral SU(3)
  dynamics}, Phys. Rev. C 77 (2008) 035204.
\newblock \href {http://arxiv.org/abs/0712.1613} {\path{arXiv:0712.1613}},
  \href {http://dx.doi.org/10.1103/PhysRevC.77.035204}
  {\path{doi:10.1103/PhysRevC.77.035204}}.

\bibitem{Bazzi:2011zj}
M.~Bazzi, G.~Beer, L.~Bombelli, A.~Bragadireanu, M.~Cargnelli, et~al., {A New
  Measurement of Kaonic Hydrogen X rays}, Phys. Lett. B704 (2011) 113--117.
\newblock \href {http://arxiv.org/abs/1105.3090} {\path{arXiv:1105.3090}},
  \href {http://dx.doi.org/10.1016/j.physletb.2011.09.011}
  {\path{doi:10.1016/j.physletb.2011.09.011}}.

\bibitem{Bazzi:2012eq}
M.~Bazzi, G.~Beer, L.~Bombelli, A.~Bragadireanu, M.~Cargnelli, et~al., {Kaonic
  hydrogen X-ray measurement in SIDDHARTA}, Nucl. Phys. A881 (2012) 88--97.
\newblock \href {http://arxiv.org/abs/1201.4635} {\path{arXiv:1201.4635}},
  \href {http://dx.doi.org/10.1016/j.nuclphysa.2011.12.008}
  {\path{doi:10.1016/j.nuclphysa.2011.12.008}}.

\bibitem{Ikeda:2011pi}
Y.~Ikeda, T.~Hyodo, W.~Weise, {Improved constraints on chiral SU(3) dynamics
  from kaonic hydrogen}, Phys. Lett. B706 (2011) 63--67.
\newblock \href {http://arxiv.org/abs/1109.3005} {\path{arXiv:1109.3005}},
  \href {http://dx.doi.org/10.1016/j.physletb.2011.10.068}
  {\path{doi:10.1016/j.physletb.2011.10.068}}.

\bibitem{Ikeda:2012au}
Y.~Ikeda, T.~Hyodo, W.~Weise, {Chiral SU(3) theory of antikaon-nucleon
  interactions with improved threshold constraints}, Nucl. Phys. A881 (2012)
  98--114.
\newblock \href {http://arxiv.org/abs/1201.6549} {\path{arXiv:1201.6549}},
  \href {http://dx.doi.org/10.1016/j.nuclphysa.2012.01.029}
  {\path{doi:10.1016/j.nuclphysa.2012.01.029}}.

\bibitem{Sekihara:2008qk}
T.~Sekihara, T.~Hyodo, D.~Jido, {Electromagnetic mean squared radii of
  Lambda(1405) in chiral dynamics}, Phys. Lett. B669 (2008) 133--138.
\newblock \href {http://arxiv.org/abs/0803.4068} {\path{arXiv:0803.4068}},
  \href {http://dx.doi.org/10.1016/j.physletb.2008.09.023}
  {\path{doi:10.1016/j.physletb.2008.09.023}}.

\bibitem{Sekihara:2010uz}
T.~Sekihara, T.~Hyodo, D.~Jido, {Internal structure of resonant Lambda(1405)
  state in chiral dynamics}, Phys. Rev. C 83 (2011) 055202.
\newblock \href {http://arxiv.org/abs/1012.3232} {\path{arXiv:1012.3232}},
  \href {http://dx.doi.org/10.1103/PhysRevC.83.055202}
  {\path{doi:10.1103/PhysRevC.83.055202}}.

\bibitem{Miyahara:2015bya}
K.~Miyahara, T.~Hyodo, {Structure of $\Lambda$(1405) and construction of
  $\overline{K}N$ local potential based on chiral SU(3) dynamics}, Phys. Rev.
  C93~(1) (2016) 015201.
\newblock \href {http://arxiv.org/abs/1506.05724} {\path{arXiv:1506.05724}},
  \href {http://dx.doi.org/10.1103/PhysRevC.93.015201}
  {\path{doi:10.1103/PhysRevC.93.015201}}.

\bibitem{Thomas:1973uh}
D.~W. Thomas, A.~Engler, H.~E. Fisk, R.~W. Kraemer, {Strange particle
  production from pi- p interactions at 1.69 gev/c}, Nucl. Phys. B56 (1973)
  15--45.
\newblock \href {http://dx.doi.org/10.1016/0550-3213(73)90217-4}
  {\path{doi:10.1016/0550-3213(73)90217-4}}.

\bibitem{Hemingway:1984pz}
R.~J. Hemingway, {Production of $\Lambda(1405)$ in $K^- p$ Reactions at
  4.2-{GeV}/$c$}, Nucl. Phys. B253 (1985) 742--752.
\newblock \href {http://dx.doi.org/10.1016/0550-3213(85)90556-5}
  {\path{doi:10.1016/0550-3213(85)90556-5}}.

\bibitem{Niiyama:2008rt}
M.~Niiyama, et~al., {Photoproduction of Lambda(1405) and Sigma0(1385) on the
  proton at E(gamma) = 1.5-2.4-GeV}, Phys. Rev. C78 (2008) 035202.
\newblock \href {http://arxiv.org/abs/0805.4051} {\path{arXiv:0805.4051}},
  \href {http://dx.doi.org/10.1103/PhysRevC.78.035202}
  {\path{doi:10.1103/PhysRevC.78.035202}}.

\bibitem{Moriya:2013eb}
K.~Moriya, et~al., {Measurement of the Sigma pi photoproduction line shapes
  near the Lambda(1405)}, Phys. Rev. C87~(3) (2013) 035206.
\newblock \href {http://arxiv.org/abs/1301.5000} {\path{arXiv:1301.5000}},
  \href {http://dx.doi.org/10.1103/PhysRevC.87.035206}
  {\path{doi:10.1103/PhysRevC.87.035206}}.

\bibitem{Moriya:2013hwg}
K.~Moriya, et~al., {Differential Photoproduction Cross Sections of the
  $\Sigma^0(1385)$, $\Lambda(1405)$, and $\Lambda(1520)$}, Phys. Rev. C88
  (2013) 045201, [Addendum: Phys. Rev.C88,no.4,049902(2013)].
\newblock \href {http://arxiv.org/abs/1305.6776} {\path{arXiv:1305.6776}},
  \href {http://dx.doi.org/10.1103/PhysRevC.88.049902,
  10.1103/PhysRevC.88.045201} {\path{doi:10.1103/PhysRevC.88.049902,
  10.1103/PhysRevC.88.045201}}.

\bibitem{Agakishiev:2012xk}
G.~Agakishiev, et~al., {Baryonic resonances close to the $\bar{K}N$ threshold:
  the case of $\Lambda$(1405) in $pp$ collisions}, Phys. Rev. C87 (2013)
  025201.
\newblock \href {http://arxiv.org/abs/1208.0205} {\path{arXiv:1208.0205}},
  \href {http://dx.doi.org/10.1103/PhysRevC.87.025201}
  {\path{doi:10.1103/PhysRevC.87.025201}}.

\bibitem{Moriya:2014kpv}
K.~Moriya, et~al., {Spin and parity measurement of the Lambda(1405) baryon},
  Phys. Rev. Lett. 112~(8) (2014) 082004.
\newblock \href {http://arxiv.org/abs/1402.2296} {\path{arXiv:1402.2296}},
  \href {http://dx.doi.org/10.1103/PhysRevLett.112.082004}
  {\path{doi:10.1103/PhysRevLett.112.082004}}.

\bibitem{Jaffe:1976yi}
R.~L. Jaffe, {Perhaps a Stable Dihyperon}, Phys. Rev. Lett. 38 (1977) 195--198,
  [Erratum: Phys. Rev. Lett.38,617(1977)].
\newblock \href {http://dx.doi.org/10.1103/PhysRevLett.38.195}
  {\path{doi:10.1103/PhysRevLett.38.195}}.

\bibitem{Yamamoto:2000wf}
K.~Yamamoto, et~al., {Search for double-Lambda hypernuclei and the H dibaryon
  in the (K-,K+) reaction on C-12}, Phys. Lett. B478 (2000) 401--407.
\newblock \href {http://dx.doi.org/10.1016/S0370-2693(00)00299-9}
  {\path{doi:10.1016/S0370-2693(00)00299-9}}.

\bibitem{Ahn:2013poa}
J.~K. Ahn, et~al., {Double-$\Lambda$ hypernuclei observed in a hybrid emulsion
  experiment}, Phys. Rev. C88~(1) (2013) 014003.
\newblock \href {http://dx.doi.org/10.1103/PhysRevC.88.014003}
  {\path{doi:10.1103/PhysRevC.88.014003}}.

\bibitem{Beane:2010hg}
S.~R. Beane, et~al., {Evidence for a Bound H-dibaryon from Lattice QCD}, Phys.
  Rev. Lett. 106 (2011) 162001.
\newblock \href {http://arxiv.org/abs/1012.3812} {\path{arXiv:1012.3812}},
  \href {http://dx.doi.org/10.1103/PhysRevLett.106.162001}
  {\path{doi:10.1103/PhysRevLett.106.162001}}.

\bibitem{Inoue:2010es}
T.~Inoue, N.~Ishii, S.~Aoki, T.~Doi, T.~Hatsuda, Y.~Ikeda, K.~Murano,
  H.~Nemura, K.~Sasaki, {Bound H-dibaryon in Flavor SU(3) Limit of Lattice
  QCD}, Phys. Rev. Lett. 106 (2011) 162002.
\newblock \href {http://arxiv.org/abs/1012.5928} {\path{arXiv:1012.5928}},
  \href {http://dx.doi.org/10.1103/PhysRevLett.106.162002}
  {\path{doi:10.1103/PhysRevLett.106.162002}}.

\bibitem{Inoue:2011ai}
T.~Inoue, S.~Aoki, T.~Doi, T.~Hatsuda, Y.~Ikeda, N.~Ishii, K.~Murano,
  H.~Nemura, K.~Sasaki, {Two-Baryon Potentials and H-Dibaryon from 3-flavor
  Lattice QCD Simulations}, Nucl. Phys. A881 (2012) 28--43.
\newblock \href {http://arxiv.org/abs/1112.5926} {\path{arXiv:1112.5926}},
  \href {http://dx.doi.org/10.1016/j.nuclphysa.2012.02.008}
  {\path{doi:10.1016/j.nuclphysa.2012.02.008}}.

\bibitem{Takeuchi:1990qj}
S.~Takeuchi, M.~Oka, {Can the H particle survive instantons?}, Phys. Rev. Lett.
  66 (1991) 1271--1274.
\newblock \href {http://dx.doi.org/10.1103/PhysRevLett.66.1271}
  {\path{doi:10.1103/PhysRevLett.66.1271}}.

\bibitem{Kobayashi:1970ji}
M.~Kobayashi, T.~Maskawa, {Chiral symmetry and eta-x mixing}, Prog. Theor.
  Phys. 44 (1970) 1422--1424.
\newblock \href {http://dx.doi.org/10.1143/PTP.44.1422}
  {\path{doi:10.1143/PTP.44.1422}}.

\bibitem{'tHooft:1976fv}
G.~'t~Hooft, {Computation of the Quantum Effects Due to a Four-Dimensional
  Pseudoparticle}, Phys. Rev. D14 (1976) 3432--3450, [Erratum: Phys.
  Rev.D18,2199(1978)].
\newblock \href {http://dx.doi.org/10.1103/PhysRevD.18.2199.3,
  10.1103/PhysRevD.14.3432} {\path{doi:10.1103/PhysRevD.18.2199.3,
  10.1103/PhysRevD.14.3432}}.

\bibitem{Shanahan:2011su}
P.~E. Shanahan, A.~W. Thomas, R.~D. Young, {Mass of the H-dibaryon}, Phys. Rev.
  Lett. 107 (2011) 092004.
\newblock \href {http://arxiv.org/abs/1106.2851} {\path{arXiv:1106.2851}},
  \href {http://dx.doi.org/10.1103/PhysRevLett.107.092004}
  {\path{doi:10.1103/PhysRevLett.107.092004}}.

\bibitem{Haidenbauer:2011za}
J.~Haidenbauer, U.~G. Meissner, {Exotic bound states of two baryons in light of
  chiral effective field theory}, Nucl. Phys. A881 (2012) 44--61.
\newblock \href {http://arxiv.org/abs/1111.4069} {\path{arXiv:1111.4069}},
  \href {http://dx.doi.org/10.1016/j.nuclphysa.2012.01.021}
  {\path{doi:10.1016/j.nuclphysa.2012.01.021}}.

\bibitem{Yamaguchi:2016kxa}
Y.~Yamaguchi, T.~Hyodo, {Quark mass dependence of H-dibaryon in
  $\Lambda\Lambda$ scattering} (2016).
\newblock \href {http://arxiv.org/abs/1607.04053} {\path{arXiv:1607.04053}}.

\bibitem{Park:2016cmg}
W.~Park, A.~Park, S.~H. Lee, {Dibaryons with two strange quarks and total spin
  zero in a constituent quark model}, Phys. Rev. D93~(7) (2016) 074007.
\newblock \href {http://arxiv.org/abs/1602.05017} {\path{arXiv:1602.05017}},
  \href {http://dx.doi.org/10.1103/PhysRevD.93.074007}
  {\path{doi:10.1103/PhysRevD.93.074007}}.

\bibitem{Oka:1988yq}
M.~Oka, {Flavor Octet Dibaryons in the Quark Model}, Phys. Rev. D38 (1988) 298.
\newblock \href {http://dx.doi.org/10.1103/PhysRevD.38.298}
  {\path{doi:10.1103/PhysRevD.38.298}}.

\bibitem{Gal:2015rev}
A.~Gal, {Meson assisted dibaryons}, Acta Phys. Polon. B47 (2016) 471.
\newblock \href {http://arxiv.org/abs/1511.06605} {\path{arXiv:1511.06605}},
  \href {http://dx.doi.org/10.5506/APhysPolB.47.471}
  {\path{doi:10.5506/APhysPolB.47.471}}.

\bibitem{Oka:1981rj}
M.~Oka, K.~Yazaki, {Short Range Part of Baryon Baryon Interaction in a Quark
  Model. 2. Numerical Results for S-Wave}, Prog. Theor. Phys. 66 (1981)
  572--587.
\newblock \href {http://dx.doi.org/10.1143/PTP.66.572}
  {\path{doi:10.1143/PTP.66.572}}.

\bibitem{Goldman:1987ma}
J.~T. Goldman, K.~Maltman, G.~J. Stephenson, Jr., K.~E. Schmidt, F.~Wang,
  {STRANGENESS -3 DIBARYONS}, Phys. Rev. Lett. 59 (1987) 627.
\newblock \href {http://dx.doi.org/10.1103/PhysRevLett.59.627}
  {\path{doi:10.1103/PhysRevLett.59.627}}.

\bibitem{Etminan:2014tya}
F.~Etminan, H.~Nemura, S.~Aoki, T.~Doi, T.~Hatsuda, Y.~Ikeda, T.~Inoue,
  N.~Ishii, K.~Murano, K.~Sasaki, {Spin-2 $N\Omega$ dibaryon from Lattice QCD},
  Nucl. Phys. A928 (2014) 89--98.
\newblock \href {http://arxiv.org/abs/1403.7284} {\path{arXiv:1403.7284}},
  \href {http://dx.doi.org/10.1016/j.nuclphysa.2014.05.014}
  {\path{doi:10.1016/j.nuclphysa.2014.05.014}}.

\bibitem{Abashian:1960zz}
A.~Abashian, N.~E. Booth, K.~M. Crowe, {Possible Anomaly in Meson Production in
  p+d Collisions}, Phys.Rev.Lett. 5 (1960) 258--260.
\newblock \href {http://dx.doi.org/10.1103/PhysRevLett.5.258}
  {\path{doi:10.1103/PhysRevLett.5.258}}.

\bibitem{Booth:1961zz}
N.~E. Booth, A.~Abashian, K.~M. Crowe, {Anomaly in Meson Production in p+d
  Collisions}, Phys.Rev.Lett. 7 (1961) 35--39.
\newblock \href {http://dx.doi.org/10.1103/PhysRevLett.7.35}
  {\path{doi:10.1103/PhysRevLett.7.35}}.

\bibitem{Bashkanov:2008ih}
M.~Bashkanov, C.~Bargholtz, M.~Berlowski, D.~Bogoslawsky, H.~Calen, et~al.,
  {Double-Pionic Fusion of Nuclear Systems and the ABC Effect: Approaching a
  Puzzle by Exclusive and Kinematically Complete Measurements}, Phys.Rev.Lett.
  102 (2009) 052301.
\newblock \href {http://arxiv.org/abs/0806.4942} {\path{arXiv:0806.4942}},
  \href {http://dx.doi.org/10.1103/PhysRevLett.102.052301}
  {\path{doi:10.1103/PhysRevLett.102.052301}}.

\bibitem{Adlarson:2011bh}
P.~Adlarson, et~al., {ABC Effect in Basic Double-Pionic Fusion --- Observation
  of a new resonance?}, Phys. Rev. Lett. 106 (2011) 242302.
\newblock \href {http://arxiv.org/abs/1104.0123} {\path{arXiv:1104.0123}},
  \href {http://dx.doi.org/10.1103/PhysRevLett.106.242302}
  {\path{doi:10.1103/PhysRevLett.106.242302}}.

\bibitem{Adlarson:2012fe}
P.~Adlarson, et~al., {Isospin Decomposition of the Basic Double-Pionic Fusion
  in the Region of the ABC Effect}, Phys. Lett. B721 (2013) 229--236.
\newblock \href {http://arxiv.org/abs/1212.2881} {\path{arXiv:1212.2881}},
  \href {http://dx.doi.org/10.1016/j.physletb.2013.03.019}
  {\path{doi:10.1016/j.physletb.2013.03.019}}.

\bibitem{Adlarson:2014pxj}
P.~Adlarson, et~al., {Evidence for a New Resonance from Polarized
  Neutron-Proton Scattering}, Phys. Rev. Lett. 112~(20) (2014) 202301.
\newblock \href {http://arxiv.org/abs/1402.6844} {\path{arXiv:1402.6844}},
  \href {http://dx.doi.org/10.1103/PhysRevLett.112.202301}
  {\path{doi:10.1103/PhysRevLett.112.202301}}.

\bibitem{Adlarson:2014ozl}
P.~Adlarson, et~al., {Neutron-proton scattering in the context of the d* (2380)
  resonance}, Phys. Rev. C90~(3) (2014) 035204.
\newblock \href {http://arxiv.org/abs/1408.4928} {\path{arXiv:1408.4928}},
  \href {http://dx.doi.org/10.1103/PhysRevC.90.035204}
  {\path{doi:10.1103/PhysRevC.90.035204}}.

\bibitem{Kukulin:2008sx}
V.~I. Kukulin, P.~Grabmayr, A.~Faessler, K.~U. Abraamyan, M.~Bashkanov,
  H.~Clement, T.~Skorodko, V.~N. Pomerantsev, {Experimental and theoretical
  evidences for an intermediate sigma-dressed dibaryon in the NN interaction},
  Annals Phys. 325 (2010) 1173--1189.
\newblock \href {http://arxiv.org/abs/0807.0192} {\path{arXiv:0807.0192}},
  \href {http://dx.doi.org/10.1016/j.aop.2010.03.011}
  {\path{doi:10.1016/j.aop.2010.03.011}}.

\bibitem{Adlarson:2014tcn}
P.~Adlarson, et~al., {Measurement of the $np \to np\pi^0\pi^0$ Reaction in
  Search for the Recently Observed $d^*(2380)$ Resonance}, Phys. Lett. B743
  (2015) 325--332.
\newblock \href {http://arxiv.org/abs/1409.2659} {\path{arXiv:1409.2659}},
  \href {http://dx.doi.org/10.1016/j.physletb.2015.02.067}
  {\path{doi:10.1016/j.physletb.2015.02.067}}.

\bibitem{Chen:2014vha}
H.-X. Chen, E.-L. Cui, W.~Chen, T.~G. Steele, S.-L. Zhu, {QCD sum rule study of
  the d*(2380)}, Phys. Rev. C91~(2) (2015) 025204.
\newblock \href {http://arxiv.org/abs/1410.0394} {\path{arXiv:1410.0394}},
  \href {http://dx.doi.org/10.1103/PhysRevC.91.025204}
  {\path{doi:10.1103/PhysRevC.91.025204}}.

\bibitem{Park:2015nha}
W.~Park, A.~Park, S.~H. Lee, {Dibaryons in a constituent quark model}, Phys.
  Rev. D92~(1) (2015) 014037.
\newblock \href {http://arxiv.org/abs/1506.01123} {\path{arXiv:1506.01123}},
  \href {http://dx.doi.org/10.1103/PhysRevD.92.014037}
  {\path{doi:10.1103/PhysRevD.92.014037}}.

\bibitem{Bevan:2014iga}
A.~J. Bevan, et~al., {The Physics of the B Factories}, Eur. Phys. J. C74 (2014)
  3026.
\newblock \href {http://arxiv.org/abs/1406.6311} {\path{arXiv:1406.6311}},
  \href {http://dx.doi.org/10.1140/epjc/s10052-014-3026-9}
  {\path{doi:10.1140/epjc/s10052-014-3026-9}}.

\bibitem{Besson:2003cp}
D.~Besson, et~al., {Observation of a narrow resonance of mass 2.46-GeV/c**2
  decaying to D*+(s) pi0 and confirmation of the D*(sJ)(2317) state}, Phys.
  Rev. D68 (2003) 032002, [Erratum: Phys. Rev.D75,119908(2007)].
\newblock \href {http://arxiv.org/abs/hep-ex/0305100}
  {\path{arXiv:hep-ex/0305100}}, \href
  {http://dx.doi.org/10.1103/PhysRevD.68.032002, 10.1103/PhysRevD.75.119908}
  {\path{doi:10.1103/PhysRevD.68.032002, 10.1103/PhysRevD.75.119908}}.

\bibitem{Krokovny:2003zq}
P.~Krokovny, et~al., {Observation of the D(sJ)(2317) and D(sJ)(2457) in B
  decays}, Phys. Rev. Lett. 91 (2003) 262002.
\newblock \href {http://arxiv.org/abs/hep-ex/0308019}
  {\path{arXiv:hep-ex/0308019}}, \href
  {http://dx.doi.org/10.1103/PhysRevLett.91.262002}
  {\path{doi:10.1103/PhysRevLett.91.262002}}.

\bibitem{Vaandering:2004ix}
E.~W. Vaandering, {Charmed hadron spectroscopy from FOCUS}, in: {QCD and high
  energy hadronic interactions. Proceedings, 39th Rencontres de Moriond, La
  Thuile, Italy, March 28-April 2, 2004}, 2004, pp. 127--132.
\newblock \href {http://arxiv.org/abs/hep-ex/0406044}
  {\path{arXiv:hep-ex/0406044}}.

\bibitem{Godfrey:1985xj}
S.~Godfrey, N.~Isgur, {Mesons in a Relativized Quark Model with
  Chromodynamics}, Phys. Rev. D32 (1985) 189--231.
\newblock \href {http://dx.doi.org/10.1103/PhysRevD.32.189}
  {\path{doi:10.1103/PhysRevD.32.189}}.

\bibitem{Godfrey:1986wj}
S.~Godfrey, R.~Kokoski, {The Properties of p Wave Mesons with One Heavy Quark},
  Phys. Rev. D43 (1991) 1679--1687.
\newblock \href {http://dx.doi.org/10.1103/PhysRevD.43.1679}
  {\path{doi:10.1103/PhysRevD.43.1679}}.

\bibitem{Dai:2003yg}
Y.-B. Dai, C.-S. Huang, C.~Liu, S.-L. Zhu, {Understanding the D+(sJ)(2317) and
  D+(sJ)(2460) with sum rules in HQET}, Phys. Rev. D68 (2003) 114011.
\newblock \href {http://arxiv.org/abs/hep-ph/0306274}
  {\path{arXiv:hep-ph/0306274}}, \href
  {http://dx.doi.org/10.1103/PhysRevD.68.114011}
  {\path{doi:10.1103/PhysRevD.68.114011}}.

\bibitem{Bali:2003jv}
G.~S. Bali, {The D+(sJ)(2317): What can the lattice say?}, Phys. Rev. D68
  (2003) 071501.
\newblock \href {http://arxiv.org/abs/hep-ph/0305209}
  {\path{arXiv:hep-ph/0305209}}, \href
  {http://dx.doi.org/10.1103/PhysRevD.68.071501}
  {\path{doi:10.1103/PhysRevD.68.071501}}.

\bibitem{Dougall:2003hv}
A.~Dougall, R.~D. Kenway, C.~M. Maynard, C.~McNeile, {The Spectrum of D(s)
  mesons from lattice QCD}, Phys. Lett. B569 (2003) 41--44.
\newblock \href {http://arxiv.org/abs/hep-lat/0307001}
  {\path{arXiv:hep-lat/0307001}}, \href
  {http://dx.doi.org/10.1016/j.physletb.2003.07.017}
  {\path{doi:10.1016/j.physletb.2003.07.017}}.

\bibitem{Hayashigaki:2004gq}
A.~Hayashigaki, K.~Terasaki, {Charmed-meson spectroscopy in QCD sum rule}
  (2004).
\newblock \href {http://arxiv.org/abs/hep-ph/0411285}
  {\path{arXiv:hep-ph/0411285}}.

\bibitem{Narison:2003td}
S.~Narison, {Open charm and beauty chiral multiplets in QCD}, Phys. Lett. B605
  (2005) 319--325.
\newblock \href {http://arxiv.org/abs/hep-ph/0307248}
  {\path{arXiv:hep-ph/0307248}}, \href
  {http://dx.doi.org/10.1016/j.physletb.2004.11.002}
  {\path{doi:10.1016/j.physletb.2004.11.002}}.

\bibitem{Barnes:2003dj}
T.~Barnes, F.~E. Close, H.~J. Lipkin, {Implications of a DK molecule at
  2.32-GeV}, Phys. Rev. D68 (2003) 054006.
\newblock \href {http://arxiv.org/abs/hep-ph/0305025}
  {\path{arXiv:hep-ph/0305025}}, \href
  {http://dx.doi.org/10.1103/PhysRevD.68.054006}
  {\path{doi:10.1103/PhysRevD.68.054006}}.

\bibitem{Szczepaniak:2003vy}
A.~P. Szczepaniak, {Description of the D*(s)(2320) resonance as the D pi atom},
  Phys. Lett. B567 (2003) 23--26.
\newblock \href {http://arxiv.org/abs/hep-ph/0305060}
  {\path{arXiv:hep-ph/0305060}}, \href
  {http://dx.doi.org/10.1016/S0370-2693(03)00865-7}
  {\path{doi:10.1016/S0370-2693(03)00865-7}}.

\bibitem{Kolomeitsev:2003ac}
E.~E. Kolomeitsev, M.~F.~M. Lutz, {On Heavy light meson resonances and chiral
  symmetry}, Phys. Lett. B582 (2004) 39--48.
\newblock \href {http://arxiv.org/abs/hep-ph/0307133}
  {\path{arXiv:hep-ph/0307133}}, \href
  {http://dx.doi.org/10.1016/j.physletb.2003.10.118}
  {\path{doi:10.1016/j.physletb.2003.10.118}}.

\bibitem{Guo:2006fu}
F.-K. Guo, P.-N. Shen, H.-C. Chiang, R.-G. Ping, B.-S. Zou, {Dynamically
  generated 0+ heavy mesons in a heavy chiral unitary approach}, Phys. Lett.
  B641 (2006) 278--285.
\newblock \href {http://arxiv.org/abs/hep-ph/0603072}
  {\path{arXiv:hep-ph/0603072}}, \href
  {http://dx.doi.org/10.1016/j.physletb.2006.08.064}
  {\path{doi:10.1016/j.physletb.2006.08.064}}.

\bibitem{Faessler:2007gv}
A.~Faessler, T.~Gutsche, V.~E. Lyubovitskij, Y.-L. Ma, {Strong and radiative
  decays of the D(s0)*(2317) meson in the DK-molecule picture}, Phys. Rev. D76
  (2007) 014005.
\newblock \href {http://arxiv.org/abs/0705.0254} {\path{arXiv:0705.0254}},
  \href {http://dx.doi.org/10.1103/PhysRevD.76.014005}
  {\path{doi:10.1103/PhysRevD.76.014005}}.

\bibitem{Gamermann:2006nm}
D.~Gamermann, E.~Oset, D.~Strottman, M.~J. Vicente~Vacas, {Dynamically
  generated open and hidden charm meson systems}, Phys. Rev. D76 (2007) 074016.
\newblock \href {http://arxiv.org/abs/hep-ph/0612179}
  {\path{arXiv:hep-ph/0612179}}, \href
  {http://dx.doi.org/10.1103/PhysRevD.76.074016}
  {\path{doi:10.1103/PhysRevD.76.074016}}.

\bibitem{vanBeveren:2003kd}
E.~van Beveren, G.~Rupp, {Observed D(s)(2317) and tentative D(2030) as the
  charmed cousins of the light scalar nonet}, Phys. Rev. Lett. 91 (2003)
  012003.
\newblock \href {http://arxiv.org/abs/hep-ph/0305035}
  {\path{arXiv:hep-ph/0305035}}, \href
  {http://dx.doi.org/10.1103/PhysRevLett.91.012003}
  {\path{doi:10.1103/PhysRevLett.91.012003}}.

\bibitem{Cheng:2003kg}
H.-Y. Cheng, W.-S. Hou, {B decays as spectroscope for charmed four quark
  states}, Phys. Lett. B566 (2003) 193--200.
\newblock \href {http://arxiv.org/abs/hep-ph/0305038}
  {\path{arXiv:hep-ph/0305038}}, \href
  {http://dx.doi.org/10.1016/S0370-2693(03)00834-7}
  {\path{doi:10.1016/S0370-2693(03)00834-7}}.

\bibitem{Terasaki:2003qa}
K.~Terasaki, {BABAR resonance as a new window of hadron physics}, Phys. Rev.
  D68 (2003) 011501.
\newblock \href {http://arxiv.org/abs/hep-ph/0305213}
  {\path{arXiv:hep-ph/0305213}}, \href
  {http://dx.doi.org/10.1103/PhysRevD.68.011501}
  {\path{doi:10.1103/PhysRevD.68.011501}}.

\bibitem{Maiani:2004vq}
L.~Maiani, F.~Piccinini, A.~D. Polosa, V.~Riquer, {Diquark-antidiquarks with
  hidden or open charm and the nature of X(3872)}, Phys. Rev. D71 (2005)
  014028.
\newblock \href {http://arxiv.org/abs/hep-ph/0412098}
  {\path{arXiv:hep-ph/0412098}}, \href
  {http://dx.doi.org/10.1103/PhysRevD.71.014028}
  {\path{doi:10.1103/PhysRevD.71.014028}}.

\bibitem{Bracco:2005kt}
M.~E. Bracco, A.~Lozea, R.~D. Matheus, F.~S. Navarra, M.~Nielsen,
  {Disentangling two- and four-quark state pictures of the charmed scalar
  mesons}, Phys. Lett. B624 (2005) 217--222.
\newblock \href {http://arxiv.org/abs/hep-ph/0503137}
  {\path{arXiv:hep-ph/0503137}}, \href
  {http://dx.doi.org/10.1016/j.physletb.2005.08.037}
  {\path{doi:10.1016/j.physletb.2005.08.037}}.

\bibitem{Browder:2003fk}
T.~E. Browder, S.~Pakvasa, A.~A. Petrov, {Comment on the new D(s)(*)+ pi0
  resonances}, Phys. Lett. B578 (2004) 365--368.
\newblock \href {http://arxiv.org/abs/hep-ph/0307054}
  {\path{arXiv:hep-ph/0307054}}, \href
  {http://dx.doi.org/10.1016/j.physletb.2003.10.067}
  {\path{doi:10.1016/j.physletb.2003.10.067}}.

\bibitem{Albaladejo:2016hae}
M.~Albaladejo, D.~Jido, J.~Nieves, E.~Oset, {$D^*_{s0}(2317)$ and $\textit{DK}$
  scattering in B decays from BaBar and LHCb data}, Eur. Phys. J. C76~(6)
  (2016) 300.
\newblock \href {http://arxiv.org/abs/1604.01193} {\path{arXiv:1604.01193}},
  \href {http://dx.doi.org/10.1140/epjc/s10052-016-4144-3}
  {\path{doi:10.1140/epjc/s10052-016-4144-3}}.

\bibitem{Torres:2014vna}
A.~Mart\'inez~Torres, E.~Oset, S.~Prelovsek, A.~Ramos, {Reanalysis of lattice
  QCD spectra leading to the $D_{s0}^*(2317)$ and $D_{s1}^*(2460)$}, JHEP 05
  (2015) 153.
\newblock \href {http://arxiv.org/abs/1412.1706} {\path{arXiv:1412.1706}},
  \href {http://dx.doi.org/10.1007/JHEP05(2015)153}
  {\path{doi:10.1007/JHEP05(2015)153}}.

\bibitem{Swanson:2006st}
E.~S. Swanson, {The New heavy mesons: A Status report}, Phys. Rept. 429 (2006)
  243--305.
\newblock \href {http://arxiv.org/abs/hep-ph/0601110}
  {\path{arXiv:hep-ph/0601110}}, \href
  {http://dx.doi.org/10.1016/j.physrep.2006.04.003}
  {\path{doi:10.1016/j.physrep.2006.04.003}}.

\bibitem{Nielsen:2009uh}
M.~Nielsen, F.~S. Navarra, S.~H. Lee, {New Charmonium States in QCD Sum Rules:
  A Concise Review}, Phys. Rept. 497 (2010) 41--83.
\newblock \href {http://arxiv.org/abs/0911.1958} {\path{arXiv:0911.1958}},
  \href {http://dx.doi.org/10.1016/j.physrep.2010.07.005}
  {\path{doi:10.1016/j.physrep.2010.07.005}}.

\bibitem{Brambilla:2010cs}
N.~Brambilla, S.~Eidelman, B.~Heltsley, R.~Vogt, G.~Bodwin, et~al., {Heavy
  quarkonium: progress, puzzles, and opportunities}, Eur. Phys. J. C 71 (2011)
  1534.
\newblock \href {http://arxiv.org/abs/1010.5827} {\path{arXiv:1010.5827}},
  \href {http://dx.doi.org/10.1140/epjc/s10052-010-1534-9}
  {\path{doi:10.1140/epjc/s10052-010-1534-9}}.

\bibitem{Hosaka:2016pey}
A.~Hosaka, T.~Iijima, K.~Miyabayashi, Y.~Sakai, S.~Yasui, {Exotic hadrons with
  heavy flavors: X, Y, Z, and related states}, PTEP 2016~(6) (2016) 062C01.
\newblock \href {http://arxiv.org/abs/1603.09229} {\path{arXiv:1603.09229}},
  \href {http://dx.doi.org/10.1093/ptep/ptw045}
  {\path{doi:10.1093/ptep/ptw045}}.

\bibitem{Acosta:2003zx}
D.~Acosta, et~al., {Observation of the narrow state $X(3872) \to J/\psi \pi^+
  \pi^-$ in $\bar{p}p$ collisions at $\sqrt{s} = 1.96$ TeV}, Phys. Rev. Lett.
  93 (2004) 072001.
\newblock \href {http://arxiv.org/abs/hep-ex/0312021}
  {\path{arXiv:hep-ex/0312021}}, \href
  {http://dx.doi.org/10.1103/PhysRevLett.93.072001}
  {\path{doi:10.1103/PhysRevLett.93.072001}}.

\bibitem{Abazov:2004kp}
V.~M. Abazov, et~al., {Observation and properties of the $X(3872)$ decaying to
  $J/\psi \pi^+ \pi^-$ in $p\bar{p}$ collisions at $\sqrt{s} = 1.96$ TeV},
  Phys. Rev. Lett. 93 (2004) 162002.
\newblock \href {http://arxiv.org/abs/hep-ex/0405004}
  {\path{arXiv:hep-ex/0405004}}, \href
  {http://dx.doi.org/10.1103/PhysRevLett.93.162002}
  {\path{doi:10.1103/PhysRevLett.93.162002}}.

\bibitem{Aubert:2004ns}
B.~Aubert, et~al., {Study of the $B \to J/\psi K^- \pi^+ \pi^-$ decay and
  measurement of the $B \to X(3872) K^-$ branching fraction}, Phys. Rev. D71
  (2005) 071103.
\newblock \href {http://arxiv.org/abs/hep-ex/0406022}
  {\path{arXiv:hep-ex/0406022}}, \href
  {http://dx.doi.org/10.1103/PhysRevD.71.071103}
  {\path{doi:10.1103/PhysRevD.71.071103}}.

\bibitem{Aaij:2011sn}
R.~Aaij, et~al., {Observation of $X(3872) $ production in $pp$ collisions at
  $\sqrt{s}=7$ TeV}, Eur. Phys. J. C72 (2012) 1972.
\newblock \href {http://arxiv.org/abs/1112.5310} {\path{arXiv:1112.5310}},
  \href {http://dx.doi.org/10.1140/epjc/s10052-012-1972-7}
  {\path{doi:10.1140/epjc/s10052-012-1972-7}}.

\bibitem{Chatrchyan:2013cld}
S.~Chatrchyan, et~al., {Measurement of the X(3872) production cross section via
  decays to J/psi pi pi in pp collisions at sqrt(s) = 7 TeV}, JHEP 04 (2013)
  154.
\newblock \href {http://arxiv.org/abs/1302.3968} {\path{arXiv:1302.3968}},
  \href {http://dx.doi.org/10.1007/JHEP04(2013)154}
  {\path{doi:10.1007/JHEP04(2013)154}}.

\bibitem{Aubert:2008ae}
B.~Aubert, et~al., {Evidence for $X(3872) \to \psi_{2S} \gamma$ in $B^\pm \to
  X_{3872} K^\pm$ decays, and a study of $B \to c \bar{c} \gamma K$}, Phys.
  Rev. Lett. 102 (2009) 132001.
\newblock \href {http://arxiv.org/abs/0809.0042} {\path{arXiv:0809.0042}},
  \href {http://dx.doi.org/10.1103/PhysRevLett.102.132001}
  {\path{doi:10.1103/PhysRevLett.102.132001}}.

\bibitem{Bhardwaj:2011dj}
V.~Bhardwaj, et~al., {Observation of $X(3872)\to J/\psi \gamma$ and search for
  $X(3872)\to\psi'\gamma$ in B decays}, Phys. Rev. Lett. 107 (2011) 091803.
\newblock \href {http://arxiv.org/abs/1105.0177} {\path{arXiv:1105.0177}},
  \href {http://dx.doi.org/10.1103/PhysRevLett.107.091803}
  {\path{doi:10.1103/PhysRevLett.107.091803}}.

\bibitem{Aaij:2013zoa}
R.~Aaij, et~al., {Determination of the X(3872) meson quantum numbers}, Phys.
  Rev. Lett. 110 (2013) 222001.
\newblock \href {http://arxiv.org/abs/1302.6269} {\path{arXiv:1302.6269}},
  \href {http://dx.doi.org/10.1103/PhysRevLett.110.222001}
  {\path{doi:10.1103/PhysRevLett.110.222001}}.

\bibitem{Dong:2008gb}
Y.-b. Dong, A.~Faessler, T.~Gutsche, V.~E. Lyubovitskij, {Estimate for the
  X(3872)$\rightarrow \gamma J/\psi$ decay width}, Phys. Rev. D77 (2008)
  094013.
\newblock \href {http://arxiv.org/abs/0802.3610} {\path{arXiv:0802.3610}},
  \href {http://dx.doi.org/10.1103/PhysRevD.77.094013}
  {\path{doi:10.1103/PhysRevD.77.094013}}.

\bibitem{Bignamini:2009sk}
C.~Bignamini, B.~Grinstein, F.~Piccinini, A.~D. Polosa, C.~Sabelli, {Is the
  X(3872) Production Cross Section at Tevatron Compatible with a Hadron
  Molecule Interpretation?}, Phys. Rev. Lett. 103 (2009) 162001.
\newblock \href {http://arxiv.org/abs/0906.0882} {\path{arXiv:0906.0882}},
  \href {http://dx.doi.org/10.1103/PhysRevLett.103.162001}
  {\path{doi:10.1103/PhysRevLett.103.162001}}.

\bibitem{Padmanath:2015era}
M.~Padmanath, C.~B. Lang, S.~Prelovsek, {X(3872) and Y(4140) using
  diquark-antidiquark operators with lattice QCD}, Phys. Rev. D92~(3) (2015)
  034501.
\newblock \href {http://arxiv.org/abs/1503.03257} {\path{arXiv:1503.03257}},
  \href {http://dx.doi.org/10.1103/PhysRevD.92.034501}
  {\path{doi:10.1103/PhysRevD.92.034501}}.

\bibitem{Choi:2007wga}
S.~Choi, et~al., {Observation of a resonance-like structure in the pi+-
  psi-prime mass distribution in exclusive B ---> K pi+- psi-prime decays},
  Phys. Rev. Lett. 100 (2008) 142001.
\newblock \href {http://arxiv.org/abs/0708.1790} {\path{arXiv:0708.1790}},
  \href {http://dx.doi.org/10.1103/PhysRevLett.100.142001}
  {\path{doi:10.1103/PhysRevLett.100.142001}}.

\bibitem{Aubert:2008aa}
B.~Aubert, et~al., {Search for the Z(4430)- at BABAR}, Phys. Rev. D 79 (2009)
  112001.
\newblock \href {http://arxiv.org/abs/0811.0564} {\path{arXiv:0811.0564}},
  \href {http://dx.doi.org/10.1103/PhysRevD.79.112001}
  {\path{doi:10.1103/PhysRevD.79.112001}}.

\bibitem{Mizuk:2009da}
R.~Mizuk, et~al., {Dalitz analysis of B ---> K pi+ psi-prime decays and the
  Z(4430)+}, Phys. Rev. D 80 (2009) 031104.
\newblock \href {http://arxiv.org/abs/0905.2869} {\path{arXiv:0905.2869}},
  \href {http://dx.doi.org/10.1103/PhysRevD.80.031104}
  {\path{doi:10.1103/PhysRevD.80.031104}}.

\bibitem{Chilikin:2013tch}
K.~Chilikin, et~al., {Experimental constraints on the spin and parity of the
  $Z$(4430)$^+$}, Phys. Rev. D 88~(7) (2013) 074026.
\newblock \href {http://arxiv.org/abs/1306.4894} {\path{arXiv:1306.4894}},
  \href {http://dx.doi.org/10.1103/PhysRevD.88.074026}
  {\path{doi:10.1103/PhysRevD.88.074026}}.

\bibitem{Aaij:2014jqa}
R.~Aaij, et~al., {Observation of the resonant character of the $Z(4430)^-$
  state}, Phys. Rev. Lett. 112~(22) (2014) 222002.
\newblock \href {http://arxiv.org/abs/1404.1903} {\path{arXiv:1404.1903}},
  \href {http://dx.doi.org/10.1103/PhysRevLett.112.222002}
  {\path{doi:10.1103/PhysRevLett.112.222002}}.

\bibitem{Mizuk:2008me}
R.~Mizuk, et~al., {Observation of two resonance-like structures in the pi+
  chi(c1) mass distribution in exclusive anti-B0 ---> K- pi+ chi(c1) decays},
  Phys. Rev. D 78 (2008) 072004.
\newblock \href {http://arxiv.org/abs/0806.4098} {\path{arXiv:0806.4098}},
  \href {http://dx.doi.org/10.1103/PhysRevD.78.072004}
  {\path{doi:10.1103/PhysRevD.78.072004}}.

\bibitem{Lees:2011ik}
J.~Lees, et~al., {Search for the $Z_1(4050)^+$ and $Z_2(4250)^+$ states in
  $\bar B^0 \to \chi_{c1} K^- \pi^+$ and $B^+ \to \chi_{c1} K^0_S \pi^+$},
  Phys. Rev. D 85 (2012) 052003.
\newblock \href {http://arxiv.org/abs/1111.5919} {\path{arXiv:1111.5919}},
  \href {http://dx.doi.org/10.1103/PhysRevD.85.052003}
  {\path{doi:10.1103/PhysRevD.85.052003}}.

\bibitem{Ablikim:2013mio}
M.~Ablikim, et~al., {Observation of a Charged Charmoniumlike Structure in
  $e^+e^-$ $\rightarrow$ $\pi^+\pi^-$ $J/\psi$ at $\sqrt{s}$=4.26 GeV}, Phys.
  Rev. Lett. 110 (2013) 252001.
\newblock \href {http://arxiv.org/abs/1303.5949} {\path{arXiv:1303.5949}},
  \href {http://dx.doi.org/10.1103/PhysRevLett.110.252001}
  {\path{doi:10.1103/PhysRevLett.110.252001}}.

\bibitem{Liu:2013dau}
Z.~Liu, et~al., {Study of $e^+e^- \rightarrow \pi^+ \pi^- J/\psi$ and
  Observation of a Charged Charmoniumlike State at Belle}, Phys. Rev. Lett. 110
  (2013) 252002.
\newblock \href {http://arxiv.org/abs/1304.0121} {\path{arXiv:1304.0121}},
  \href {http://dx.doi.org/10.1103/PhysRevLett.110.252002}
  {\path{doi:10.1103/PhysRevLett.110.252002}}.

\bibitem{Xiao:2013iha}
T.~Xiao, S.~Dobbs, A.~Tomaradze, K.~K. Seth, {Observation of the Charged Hadron
  $Z_c^{\pm}(3900)$ and Evidence for the Neutral $Z_c^0(3900)$ in $e^+e^-\to
  \pi\pi J/\psi$ at $\sqrt{s}=4170$ MeV}, Phys. Lett. B727 (2013) 366--370.
\newblock \href {http://arxiv.org/abs/1304.3036} {\path{arXiv:1304.3036}},
  \href {http://dx.doi.org/10.1016/j.physletb.2013.10.041}
  {\path{doi:10.1016/j.physletb.2013.10.041}}.

\bibitem{Ablikim:2013wzq}
M.~Ablikim, et~al., {Observation of a Charged Charmoniumlike Structure
  $Z_c$(4020) and Search for the $Z_c$(3900) in $e^+e^- \to \pi^+\pi^-h_c$},
  Phys. Rev. Lett. 111~(24) (2013) 242001.
\newblock \href {http://arxiv.org/abs/1309.1896} {\path{arXiv:1309.1896}},
  \href {http://dx.doi.org/10.1103/PhysRevLett.111.242001}
  {\path{doi:10.1103/PhysRevLett.111.242001}}.

\bibitem{Ablikim:2013xfr}
M.~Ablikim, et~al., {Observation of a charged $(D\bar{D}^{*})^\pm$ mass peak in
  $e^{+}e^{-} \to \pi D\bar{D}^{*}$ at $\sqrt{s} =$ 4.26 GeV}, Phys. Rev. Lett.
  112~(2) (2014) 022001.
\newblock \href {http://arxiv.org/abs/1310.1163} {\path{arXiv:1310.1163}},
  \href {http://dx.doi.org/10.1103/PhysRevLett.112.022001}
  {\path{doi:10.1103/PhysRevLett.112.022001}}.

\bibitem{Prelovsek:2013xba}
S.~Prelovsek, L.~Leskovec, {Search for $Z^{+}_{c}$(3900) in the $1^{+-}$
  Channel on the Lattice}, Phys. Lett. B727 (2013) 172--176.
\newblock \href {http://arxiv.org/abs/1308.2097} {\path{arXiv:1308.2097}},
  \href {http://dx.doi.org/10.1016/j.physletb.2013.10.009}
  {\path{doi:10.1016/j.physletb.2013.10.009}}.

\bibitem{Prelovsek:2014swa}
S.~Prelovsek, C.~B. Lang, L.~Leskovec, D.~Mohler, {Study of the $Z_c^+$ channel
  using lattice QCD}, Phys. Rev. D91~(1) (2015) 014504.
\newblock \href {http://arxiv.org/abs/1405.7623} {\path{arXiv:1405.7623}},
  \href {http://dx.doi.org/10.1103/PhysRevD.91.014504}
  {\path{doi:10.1103/PhysRevD.91.014504}}.

\bibitem{Ikeda:2016zwx}
Y.~Ikeda, S.~Aoki, T.~Doi, S.~Gongyo, T.~Hatsuda, T.~Inoue, T.~Iritani,
  N.~Ishii, K.~Murano, K.~Sasaki, {Fate of the Tetraquark Candidate Zc(3900) in
  Lattice QCD} (2016).
\newblock \href {http://arxiv.org/abs/1602.03465} {\path{arXiv:1602.03465}}.

\bibitem{Ablikim:2013emm}
M.~Ablikim, et~al., {Observation of a charged charmoniumlike structure in
  $e^+e^- \to (D^{*} \bar{D}^{*})^{\pm} \pi^\mp$ at $\sqrt{s}=4.26$GeV}, Phys.
  Rev. Lett. 112~(13) (2014) 132001.
\newblock \href {http://arxiv.org/abs/1308.2760} {\path{arXiv:1308.2760}},
  \href {http://dx.doi.org/10.1103/PhysRevLett.112.132001}
  {\path{doi:10.1103/PhysRevLett.112.132001}}.

\bibitem{Chilikin:2014bkk}
K.~Chilikin, et~al., {Observation of a new charged charmoniumlike state in
  $\bar{B}^0 \rightarrow J/\psi K^-\pi^+$ decays}, Phys. Rev. D 90~(11) (2014)
  112009.
\newblock \href {http://arxiv.org/abs/1408.6457} {\path{arXiv:1408.6457}},
  \href {http://dx.doi.org/10.1103/PhysRevD.90.112009}
  {\path{doi:10.1103/PhysRevD.90.112009}}.

\bibitem{Wang:2014hta}
X.~L. Wang, et~al., {Measurement of $e^+e^- \to \pi^+\pi^-\psi(2S)$ via Initial
  State Radiation at Belle}, Phys. Rev. D91 (2015) 112007.
\newblock \href {http://arxiv.org/abs/1410.7641} {\path{arXiv:1410.7641}},
  \href {http://dx.doi.org/10.1103/PhysRevD.91.112007}
  {\path{doi:10.1103/PhysRevD.91.112007}}.

\bibitem{Brodsky:2014xia}
S.~J. Brodsky, D.~S. Hwang, R.~F. Lebed, {Dynamical Picture for the Formation
  and Decay of the Exotic XYZ Mesons}, Phys. Rev. Lett. 113~(11) (2014) 112001.
\newblock \href {http://arxiv.org/abs/1406.7281} {\path{arXiv:1406.7281}},
  \href {http://dx.doi.org/10.1103/PhysRevLett.113.112001}
  {\path{doi:10.1103/PhysRevLett.113.112001}}.

\bibitem{Belle:2011aa}
A.~Bondar, et~al., {Observation of two charged bottomonium-like resonances in
  Y(5S) decays}, Phys. Rev. Lett. 108 (2012) 122001.
\newblock \href {http://arxiv.org/abs/1110.2251} {\path{arXiv:1110.2251}},
  \href {http://dx.doi.org/10.1103/PhysRevLett.108.122001}
  {\path{doi:10.1103/PhysRevLett.108.122001}}.

\bibitem{Krokovny:2013mgx}
P.~Krokovny, et~al., {First observation of the $Z \frac{0}{b}$(10610) in a
  Dalitz analysis of $\Upsilon$(10860) $\to \Upsilon$(nS)$\pi^0 \pi^0$}, Phys.
  Rev. D88~(5) (2013) 052016.
\newblock \href {http://arxiv.org/abs/1308.2646} {\path{arXiv:1308.2646}},
  \href {http://dx.doi.org/10.1103/PhysRevD.88.052016}
  {\path{doi:10.1103/PhysRevD.88.052016}}.

\bibitem{Bondar:2011ev}
A.~E. Bondar, A.~Garmash, A.~I. Milstein, R.~Mizuk, M.~B. Voloshin, {Heavy
  quark spin structure in $Z_b$ resonances}, Phys. Rev. D84 (2011) 054010.
\newblock \href {http://arxiv.org/abs/1105.4473} {\path{arXiv:1105.4473}},
  \href {http://dx.doi.org/10.1103/PhysRevD.84.054010}
  {\path{doi:10.1103/PhysRevD.84.054010}}.

\bibitem{Voloshin:2011qa}
M.~B. Voloshin, {Radiative transitions from Upsilon(5S) to molecular
  bottomonium}, Phys. Rev. D84 (2011) 031502.
\newblock \href {http://arxiv.org/abs/1105.5829} {\path{arXiv:1105.5829}},
  \href {http://dx.doi.org/10.1103/PhysRevD.84.031502}
  {\path{doi:10.1103/PhysRevD.84.031502}}.

\bibitem{Ohkoda:2011vj}
S.~Ohkoda, Y.~Yamaguchi, S.~Yasui, K.~Sudoh, A.~Hosaka, {Exotic Mesons with
  Hidden Bottom near Thresholds}, Phys. Rev. D86 (2012) 014004.
\newblock \href {http://arxiv.org/abs/1111.2921} {\path{arXiv:1111.2921}},
  \href {http://dx.doi.org/10.1103/PhysRevD.86.014004}
  {\path{doi:10.1103/PhysRevD.86.014004}}.

\bibitem{Ohkoda:2012rj}
S.~Ohkoda, Y.~Yamaguchi, S.~Yasui, A.~Hosaka, {Decays and productions via
  bottomonium for $Z_b$ resonances and other B anti-B molecules}, Phys. Rev.
  D86 (2012) 117502.
\newblock \href {http://arxiv.org/abs/1210.3170} {\path{arXiv:1210.3170}},
  \href {http://dx.doi.org/10.1103/PhysRevD.86.117502}
  {\path{doi:10.1103/PhysRevD.86.117502}}.

\bibitem{Ohkoda:2013cea}
S.~Ohkoda, S.~Yasui, A.~Hosaka, {Decays of $Z_b \to \Upsilon \pi$ via triangle
  diagrams in heavy meson molecules}, Phys. Rev. D89~(7) (2014) 074029.
\newblock \href {http://arxiv.org/abs/1310.3029} {\path{arXiv:1310.3029}},
  \href {http://dx.doi.org/10.1103/PhysRevD.89.074029}
  {\path{doi:10.1103/PhysRevD.89.074029}}.

\bibitem{Dong:2012hc}
Y.~Dong, A.~Faessler, T.~Gutsche, V.~E. Lyubovitskij, {Decays of Zb(+) and
  Zb'(+) as Hadronic Molecules}, J. Phys. G40 (2013) 015002.
\newblock \href {http://arxiv.org/abs/1203.1894} {\path{arXiv:1203.1894}},
  \href {http://dx.doi.org/10.1088/0954-3899/40/1/015002}
  {\path{doi:10.1088/0954-3899/40/1/015002}}.

\bibitem{Swanson:2014tra}
E.~S. Swanson, {$Z_b$ and $Z_c$ Exotic States as Coupled Channel Cusps}, Phys.
  Rev. D91~(3) (2015) 034009.
\newblock \href {http://arxiv.org/abs/1409.3291} {\path{arXiv:1409.3291}},
  \href {http://dx.doi.org/10.1103/PhysRevD.91.034009}
  {\path{doi:10.1103/PhysRevD.91.034009}}.

\bibitem{Maeda:2015hxa}
S.~Maeda, M.~Oka, A.~Yokota, E.~Hiyama, Y.-R. Liu, {A model of charmed
  baryon-nucleon potential and two- and three-body bound states with charmed
  baryon}, PTEP 2016~(2) (2016) 023D02.
\newblock \href {http://arxiv.org/abs/1509.02445} {\path{arXiv:1509.02445}},
  \href {http://dx.doi.org/10.1093/ptep/ptv194}
  {\path{doi:10.1093/ptep/ptv194}}.

\bibitem{Diakonov:1997mm}
D.~Diakonov, V.~Petrov, M.~V. Polyakov, {Exotic anti-decuplet of baryons:
  Prediction from chiral solitons}, Z. Phys. A359 (1997) 305--314.
\newblock \href {http://arxiv.org/abs/hep-ph/9703373}
  {\path{arXiv:hep-ph/9703373}}, \href
  {http://dx.doi.org/10.1007/s002180050406} {\path{doi:10.1007/s002180050406}}.

\bibitem{Jido:2008kp}
D.~Jido, Y.~Kanada-En'yo, {K anti-K N molecule state with I = 1/2 and J**P =
  1/2+ studied with three-body calculation}, Phys. Rev. C78 (2008) 035203.
\newblock \href {http://arxiv.org/abs/0806.3601} {\path{arXiv:0806.3601}},
  \href {http://dx.doi.org/10.1103/PhysRevC.78.035203}
  {\path{doi:10.1103/PhysRevC.78.035203}}.

\bibitem{Akaishi:2002bg}
Y.~Akaishi, T.~Yamazaki, {Nuclear anti-K bound states in light nuclei}, Phys.
  Rev. C65 (2002) 044005.
\newblock \href {http://dx.doi.org/10.1103/PhysRevC.65.044005}
  {\path{doi:10.1103/PhysRevC.65.044005}}.

\bibitem{Zhang:2000sv}
Z.~Y. Zhang, Y.~W. Yu, C.~R. Ching, T.~H. Ho, Z.-D. Lu, {Suggesting a di-omega
  dibaryon search in heavy ion collision experiments}, Phys. Rev. C61 (2000)
  065204.
\newblock \href {http://dx.doi.org/10.1103/PhysRevC.61.065204}
  {\path{doi:10.1103/PhysRevC.61.065204}}.

\bibitem{Zouzou:1986qh}
S.~Zouzou, B.~Silvestre-Brac, C.~Gignoux, J.~M. Richard, {FOUR QUARK BOUND
  STATES}, Z. Phys. C30 (1986) 457.
\newblock \href {http://dx.doi.org/10.1007/BF01557611}
  {\path{doi:10.1007/BF01557611}}.

\bibitem{Yasui:2009bz}
S.~Yasui, K.~Sudoh, {Exotic nuclei with open heavy flavor mesons}, Phys. Rev.
  D80 (2009) 034008.
\newblock \href {http://arxiv.org/abs/0906.1452} {\path{arXiv:0906.1452}},
  \href {http://dx.doi.org/10.1103/PhysRevD.80.034008}
  {\path{doi:10.1103/PhysRevD.80.034008}}.

\bibitem{Yamaguchi:2011xb}
Y.~Yamaguchi, S.~Ohkoda, S.~Yasui, A.~Hosaka, {Exotic baryons from a heavy
  meson and a nucleon -- Negative parity states --}, Phys. Rev. D84 (2011)
  014032.
\newblock \href {http://arxiv.org/abs/1105.0734} {\path{arXiv:1105.0734}},
  \href {http://dx.doi.org/10.1103/PhysRevD.84.014032}
  {\path{doi:10.1103/PhysRevD.84.014032}}.

\bibitem{Lipkin:1987sk}
H.~J. Lipkin, {New Possibilities for Exotic Hadrons: Anticharmed Strange
  Baryons}, Phys. Lett. B195 (1987) 484--488.
\newblock \href {http://dx.doi.org/10.1016/0370-2693(87)90055-4}
  {\path{doi:10.1016/0370-2693(87)90055-4}}.

\bibitem{Gignoux:1987cn}
C.~Gignoux, B.~Silvestre-Brac, J.~M. Richard, {Possibility of Stable Multi -
  Quark Baryons}, Phys. Lett. B193 (1987) 323.
\newblock \href {http://dx.doi.org/10.1016/0370-2693(87)91244-5}
  {\path{doi:10.1016/0370-2693(87)91244-5}}.

\bibitem{Lee:2009rt}
S.~H. Lee, S.~Yasui, {Stable multiquark states with heavy quarks in a diquark
  model}, Eur. Phys. J. C64 (2009) 283--295.
\newblock \href {http://arxiv.org/abs/0901.2977} {\path{arXiv:0901.2977}},
  \href {http://dx.doi.org/10.1140/epjc/s10052-009-1140-x}
  {\path{doi:10.1140/epjc/s10052-009-1140-x}}.

\bibitem{Yamaguchi:2013hsa}
Y.~Yamaguchi, S.~Yasui, A.~Hosaka, {Exotic dibaryons with a heavy antiquark},
  Nucl. Phys. A927 (2014) 110--118.
\newblock \href {http://arxiv.org/abs/1309.4324} {\path{arXiv:1309.4324}},
  \href {http://dx.doi.org/10.1016/j.nuclphysa.2014.04.002}
  {\path{doi:10.1016/j.nuclphysa.2014.04.002}}.

\bibitem{Wu:2010jy}
J.-J. Wu, R.~Molina, E.~Oset, B.~S. Zou, {Prediction of narrow $N^*$ and
  $\Lambda^*$ resonances with hidden charm above 4 GeV}, Phys. Rev. Lett. 105
  (2010) 232001.
\newblock \href {http://arxiv.org/abs/1007.0573} {\path{arXiv:1007.0573}},
  \href {http://dx.doi.org/10.1103/PhysRevLett.105.232001}
  {\path{doi:10.1103/PhysRevLett.105.232001}}.

\bibitem{Yuan:2012wz}
S.~G. Yuan, K.~W. Wei, J.~He, H.~S. Xu, B.~S. Zou, {Study of $qqqc\bar{c}$ five
  quark system with three kinds of quark-quark hyperfine interaction}, Eur.
  Phys. J. A48 (2012) 61.
\newblock \href {http://arxiv.org/abs/1201.0807} {\path{arXiv:1201.0807}},
  \href {http://dx.doi.org/10.1140/epja/i2012-12061-2}
  {\path{doi:10.1140/epja/i2012-12061-2}}.

\bibitem{Yang:2011wz}
Z.-C. Yang, Z.-F. Sun, J.~He, X.~Liu, S.-L. Zhu, {The possible hidden-charm
  molecular baryons composed of anti-charmed meson and charmed baryon}, Chin.
  Phys. C36 (2012) 6--13.
\newblock \href {http://arxiv.org/abs/1105.2901} {\path{arXiv:1105.2901}},
  \href {http://dx.doi.org/10.1088/1674-1137/36/1/002,
  10.1088/1674-1137/36/3/006} {\path{doi:10.1088/1674-1137/36/1/002,
  10.1088/1674-1137/36/3/006}}.

\bibitem{Xiao:2013yca}
C.~W. Xiao, J.~Nieves, E.~Oset, {Combining heavy quark spin and local hidden
  gauge symmetries in the dynamical generation of hidden charm baryons}, Phys.
  Rev. D88 (2013) 056012.
\newblock \href {http://arxiv.org/abs/1304.5368} {\path{arXiv:1304.5368}},
  \href {http://dx.doi.org/10.1103/PhysRevD.88.056012}
  {\path{doi:10.1103/PhysRevD.88.056012}}.

\bibitem{Uchino:2015uha}
T.~Uchino, W.-H. Liang, E.~Oset, {Baryon states with hidden charm in the
  extended local hidden gauge approach}, Eur. Phys. J. A52~(3) (2016) 43.
\newblock \href {http://arxiv.org/abs/1504.05726} {\path{arXiv:1504.05726}},
  \href {http://dx.doi.org/10.1140/epja/i2016-16043-0}
  {\path{doi:10.1140/epja/i2016-16043-0}}.

\bibitem{Chen:2015loa}
R.~Chen, X.~Liu, X.-Q. Li, S.-L. Zhu, {Identifying exotic hidden-charm
  pentaquarks}, Phys. Rev. Lett. 115~(13) (2015) 132002.
\newblock \href {http://arxiv.org/abs/1507.03704} {\path{arXiv:1507.03704}},
  \href {http://dx.doi.org/10.1103/PhysRevLett.115.132002}
  {\path{doi:10.1103/PhysRevLett.115.132002}}.

\bibitem{Roca:2015dva}
L.~Roca, J.~Nieves, E.~Oset, {LHCb pentaquark as a
  $\bar{D}^*\Sigma_c-\bar{D}^*\Sigma_c^*$ molecular state}, Phys. Rev. D92~(9)
  (2015) 094003.
\newblock \href {http://arxiv.org/abs/1507.04249} {\path{arXiv:1507.04249}},
  \href {http://dx.doi.org/10.1103/PhysRevD.92.094003}
  {\path{doi:10.1103/PhysRevD.92.094003}}.

\bibitem{He:2015cea}
J.~He, {$\bar{D}\Sigma^*_c$ and $\bar{D}^*\Sigma_c$ interactions and the LHCb
  hidden-charmed pentaquarks}, Phys. Lett. B753 (2016) 547--551.
\newblock \href {http://arxiv.org/abs/1507.05200} {\path{arXiv:1507.05200}},
  \href {http://dx.doi.org/10.1016/j.physletb.2015.12.071}
  {\path{doi:10.1016/j.physletb.2015.12.071}}.

\bibitem{Huang:2015uda}
H.~Huang, C.~Deng, J.~Ping, F.~Wang, {Possible pentaquarks with heavy quarks}
  (2015).
\newblock \href {http://arxiv.org/abs/1510.04648} {\path{arXiv:1510.04648}}.

\bibitem{Meissner:2015mza}
U.-G. Meissner, J.~A. Oller, {Testing the $\chi_{c1}\, p$ composite nature of
  the $P_c(4450)$}, Phys. Lett. B751 (2015) 59--62.
\newblock \href {http://arxiv.org/abs/1507.07478} {\path{arXiv:1507.07478}},
  \href {http://dx.doi.org/10.1016/j.physletb.2015.10.015}
  {\path{doi:10.1016/j.physletb.2015.10.015}}.

\bibitem{Xiao:2015fia}
C.~W. Xiao, U.~G. Meissner, {J/psi(etac)N and Upsilon(etab)N cross sections},
  Phys. Rev. D92~(11) (2015) 114002.
\newblock \href {http://arxiv.org/abs/1508.00924} {\path{arXiv:1508.00924}},
  \href {http://dx.doi.org/10.1103/PhysRevD.92.114002}
  {\path{doi:10.1103/PhysRevD.92.114002}}.

\bibitem{Chen:2016heh}
R.~Chen, X.~Liu, S.-L. Zhu, {Hidden-charm molecular pentaquarks and their
  charm-strange partners}, Nucl. Phys. A954 (2016) 406--421.
\newblock \href {http://arxiv.org/abs/1601.03233} {\path{arXiv:1601.03233}},
  \href {http://dx.doi.org/10.1016/j.nuclphysa.2016.04.012}
  {\path{doi:10.1016/j.nuclphysa.2016.04.012}}.

\bibitem{Shimizu:2016rrd}
Y.~Shimizu, D.~Suenaga, M.~Harada, {Coupled channel analysis of molecule
  picture of $P_{c}(4380)$}, Phys. Rev. D93~(11) (2016) 114003.
\newblock \href {http://arxiv.org/abs/1603.02376} {\path{arXiv:1603.02376}},
  \href {http://dx.doi.org/10.1103/PhysRevD.93.114003}
  {\path{doi:10.1103/PhysRevD.93.114003}}.

\bibitem{Maiani:2015vwa}
L.~Maiani, A.~D. Polosa, V.~Riquer, {The New Pentaquarks in the Diquark Model},
  Phys. Lett. B749 (2015) 289--291.
\newblock \href {http://arxiv.org/abs/1507.04980} {\path{arXiv:1507.04980}},
  \href {http://dx.doi.org/10.1016/j.physletb.2015.08.008}
  {\path{doi:10.1016/j.physletb.2015.08.008}}.

\bibitem{Anisovich:2015cia}
V.~V. Anisovich, M.~A. Matveev, J.~Nyiri, A.~V. Sarantsev, A.~N. Semenova,
  {Pentaquarks and resonances in the $pJ/\psi$ spectrum} (2015).
\newblock \href {http://arxiv.org/abs/1507.07652} {\path{arXiv:1507.07652}}.

\bibitem{Ghosh:2015ksa}
R.~Ghosh, A.~Bhattacharya, B.~Chakrabarti, {The masses of $P_{c}^{*}(4380)$ and
  $P_{c}^{*}(4450)$ in the quasi particle diquark model} (2015).
\newblock \href {http://arxiv.org/abs/1508.00356} {\path{arXiv:1508.00356}}.

\bibitem{Wang:2015epa}
Z.-G. Wang, {Analysis of $P_c(4380)$ and $P_c(4450)$ as pentaquark states in
  the diquark model with QCD sum rules}, Eur. Phys. J. C76~(2) (2016) 70.
\newblock \href {http://arxiv.org/abs/1508.01468} {\path{arXiv:1508.01468}},
  \href {http://dx.doi.org/10.1140/epjc/s10052-016-3920-4}
  {\path{doi:10.1140/epjc/s10052-016-3920-4}}.

\bibitem{Wang:2015ixb}
Z.-G. Wang, {Analysis of the ${\frac{3}{2}}^{\pm}$ pentaquark states in the
  diquark-diquark-antiquark model with QCD sum rules} (2015).
\newblock \href {http://arxiv.org/abs/1512.04763} {\path{arXiv:1512.04763}}.

\bibitem{Chen:2015moa}
H.-X. Chen, W.~Chen, X.~Liu, T.~G. Steele, S.-L. Zhu, {Towards exotic
  hidden-charm pentaquarks in QCD}, Phys. Rev. Lett. 115~(17) (2015) 172001.
\newblock \href {http://arxiv.org/abs/1507.03717} {\path{arXiv:1507.03717}},
  \href {http://dx.doi.org/10.1103/PhysRevLett.115.172001}
  {\path{doi:10.1103/PhysRevLett.115.172001}}.

\bibitem{Lebed:2015tna}
R.~F. Lebed, {The Pentaquark Candidates in the Dynamical Diquark Picture},
  Phys. Lett. B749 (2015) 454--457.
\newblock \href {http://arxiv.org/abs/1507.05867} {\path{arXiv:1507.05867}},
  \href {http://dx.doi.org/10.1016/j.physletb.2015.08.032}
  {\path{doi:10.1016/j.physletb.2015.08.032}}.

\bibitem{Zhu:2015bba}
R.~Zhu, C.-F. Qiao, {Pentaquark states in a diquark-triquark model}, Phys.
  Lett. B756 (2016) 259--264.
\newblock \href {http://arxiv.org/abs/1510.08693} {\path{arXiv:1510.08693}},
  \href {http://dx.doi.org/10.1016/j.physletb.2016.03.022}
  {\path{doi:10.1016/j.physletb.2016.03.022}}.

\bibitem{Chen:2016otp}
H.-X. Chen, E.-L. Cui, W.~Chen, T.~G. Steele, X.~Liu, S.-L. Zhu, {QCD sum rule
  study of hidden-charm pentaquarks} (2016).
\newblock \href {http://arxiv.org/abs/1602.02433} {\path{arXiv:1602.02433}}.

\bibitem{Scoccola:2015nia}
N.~N. Scoccola, D.~O. Riska, M.~Rho, {Pentaquark candidates P$_c^+$(4380) and
  P$_c^+$(4450) within the soliton picture of baryons}, Phys. Rev. D92~(5)
  (2015) 051501.
\newblock \href {http://arxiv.org/abs/1508.01172} {\path{arXiv:1508.01172}},
  \href {http://dx.doi.org/10.1103/PhysRevD.92.051501}
  {\path{doi:10.1103/PhysRevD.92.051501}}.

\bibitem{Mironov:2015ica}
A.~Mironov, A.~Morozov, {Is the pentaquark doublet a hadronic molecule?}, JETP
  Lett. 102~(5) (2015) 271--273, [Pisma Zh. Eksp. Teor.
  Fiz.102,no.5,302(2015)].
\newblock \href {http://arxiv.org/abs/1507.04694} {\path{arXiv:1507.04694}},
  \href {http://dx.doi.org/10.7868/S0370274X15170038,
  10.1134/S0021364015170099} {\path{doi:10.7868/S0370274X15170038,
  10.1134/S0021364015170099}}.

\bibitem{Guo:2015umn}
F.-K. Guo, U.-G. Meissner, W.~Wang, Z.~Yang, {How to reveal the exotic nature
  of the P$_c$(4450)}, Phys. Rev. D92~(7) (2015) 071502.
\newblock \href {http://arxiv.org/abs/1507.04950} {\path{arXiv:1507.04950}},
  \href {http://dx.doi.org/10.1103/PhysRevD.92.071502}
  {\path{doi:10.1103/PhysRevD.92.071502}}.

\bibitem{Liu:2015fea}
X.-H. Liu, Q.~Wang, Q.~Zhao, {Understanding the newly observed heavy pentaquark
  candidates}, Phys. Lett. B757 (2016) 231--236.
\newblock \href {http://arxiv.org/abs/1507.05359} {\path{arXiv:1507.05359}},
  \href {http://dx.doi.org/10.1016/j.physletb.2016.03.089}
  {\path{doi:10.1016/j.physletb.2016.03.089}}.

\bibitem{Mikhasenko:2015vca}
M.~Mikhasenko, {A triangle singularity and the LHCb pentaquarks} (2015).
\newblock \href {http://arxiv.org/abs/1507.06552} {\path{arXiv:1507.06552}}.

\bibitem{Chen:2016qju}
H.-X. Chen, W.~Chen, X.~Liu, S.-L. Zhu, {The hidden-charm pentaquark and
  tetraquark states}, Phys. Rept. 639 (2016) 1--121.
\newblock \href {http://arxiv.org/abs/1601.02092} {\path{arXiv:1601.02092}},
  \href {http://dx.doi.org/10.1016/j.physrep.2016.05.004}
  {\path{doi:10.1016/j.physrep.2016.05.004}}.

\bibitem{Burns:2015dwa}
T.~J. Burns, {Phenomenology of P$_{c}$(4380)$^{+}$, P$_{c}$(4450)$^{+}$ and
  related states}, Eur. Phys. J. A51~(11) (2015) 152.
\newblock \href {http://arxiv.org/abs/1509.02460} {\path{arXiv:1509.02460}},
  \href {http://dx.doi.org/10.1140/epja/i2015-15152-6}
  {\path{doi:10.1140/epja/i2015-15152-6}}.

\bibitem{Oset:2016lyh}
E.~Oset, et~al., {Weak decays of heavy hadrons into dynamically generated
  resonances}, Int. J. Mod. Phys. E25 (2016) 1630001.
\newblock \href {http://arxiv.org/abs/1601.03972} {\path{arXiv:1601.03972}},
  \href {http://dx.doi.org/10.1142/S0218301316300010}
  {\path{doi:10.1142/S0218301316300010}}.

\bibitem{D0:2016mwd}
V.~M. Abazov, et~al., {Evidence for a $B_s^0 \pi^\pm$ state}, Phys. Rev. Lett.
  117~(2) (2016) 022003.
\newblock \href {http://arxiv.org/abs/1602.07588} {\path{arXiv:1602.07588}},
  \href {http://dx.doi.org/10.1103/PhysRevLett.117.022003}
  {\path{doi:10.1103/PhysRevLett.117.022003}}.

\bibitem{Aaij:2016iev}
R.~Aaij, et~al., {Search for structure in the $B_s^0\pi^\pm$ invariant mass
  spectrum} (2016).
\newblock \href {http://arxiv.org/abs/1608.00435} {\path{arXiv:1608.00435}}.

\bibitem{Agaev:2016mjb}
S.~S. Agaev, K.~Azizi, H.~Sundu, {Mass and decay constant of the newly observed
  exotic $X(5568)$ state}, Phys. Rev. D93~(7) (2016) 074024.
\newblock \href {http://arxiv.org/abs/1602.08642} {\path{arXiv:1602.08642}},
  \href {http://dx.doi.org/10.1103/PhysRevD.93.074024}
  {\path{doi:10.1103/PhysRevD.93.074024}}.

\bibitem{Zanetti:2016wjn}
C.~M. Zanetti, M.~Nielsen, K.~P. Khemchandani, {QCD sum rule study of a charged
  bottom-strange scalar meson}, Phys. Rev. D93~(9) (2016) 096011.
\newblock \href {http://arxiv.org/abs/1602.09041} {\path{arXiv:1602.09041}},
  \href {http://dx.doi.org/10.1103/PhysRevD.93.096011}
  {\path{doi:10.1103/PhysRevD.93.096011}}.

\bibitem{Wang:2016mee}
Z.-G. Wang, {Analysis of the $X(5568)$ as scalar tetraquark state in the
  diquark-antidiquark model with QCD sum rules}, Commun. Theor. Phys. 66~(3)
  (2016) 335--339.
\newblock \href {http://arxiv.org/abs/1602.08711} {\path{arXiv:1602.08711}},
  \href {http://dx.doi.org/10.1088/0253-6102/66/3/335}
  {\path{doi:10.1088/0253-6102/66/3/335}}.

\bibitem{Chen:2016mqt}
W.~Chen, H.-X. Chen, X.~Liu, T.~G. Steele, S.-L. Zhu, {Decoding the $X(5568)$
  as a fully open-flavor $su\bar b\bar d$ tetraquark state}, Phys. Rev. Lett.
  117~(2) (2016) 022002.
\newblock \href {http://arxiv.org/abs/1602.08916} {\path{arXiv:1602.08916}},
  \href {http://dx.doi.org/10.1103/PhysRevLett.117.022002}
  {\path{doi:10.1103/PhysRevLett.117.022002}}.

\bibitem{Agaev:2016ijz}
S.~S. Agaev, K.~Azizi, H.~Sundu, {Width of the exotic $X_b(5568)$ state through
  its strong decay to $B_s^{0} \pi^{+}$}, Phys. Rev. D93~(11) (2016) 114007.
\newblock \href {http://arxiv.org/abs/1603.00290} {\path{arXiv:1603.00290}},
  \href {http://dx.doi.org/10.1103/PhysRevD.93.114007}
  {\path{doi:10.1103/PhysRevD.93.114007}}.

\bibitem{Dias:2016dme}
J.~M. Dias, K.~P. Khemchandani, A.~Martinez~Torres, M.~Nielsen, C.~M. Zanetti,
  {A QCD sum rule calculation of the $X^\pm(5568) \to B_{s}^0\pi^\pm$ decay
  width}, Phys. Lett. B758 (2016) 235--238.
\newblock \href {http://arxiv.org/abs/1603.02249} {\path{arXiv:1603.02249}},
  \href {http://dx.doi.org/10.1016/j.physletb.2016.05.015}
  {\path{doi:10.1016/j.physletb.2016.05.015}}.

\bibitem{Wang:2016wkj}
Z.-G. Wang, {Analysis of the strong decay $X(5568) \rightarrow B_s^0\pi ^+$
  with QCD sum rules}, Eur. Phys. J. C76~(5) (2016) 279.
\newblock \href {http://arxiv.org/abs/1603.02498} {\path{arXiv:1603.02498}},
  \href {http://dx.doi.org/10.1140/epjc/s10052-016-4133-6}
  {\path{doi:10.1140/epjc/s10052-016-4133-6}}.

\bibitem{Tang:2016pcf}
L.~Tang, C.-F. Qiao, {Tetraquark States with Open Flavors} (2016).
\newblock \href {http://arxiv.org/abs/1603.04761} {\path{arXiv:1603.04761}}.

\bibitem{Agaev:2016urs}
S.~S. Agaev, K.~Azizi, H.~Sundu, {Exploring $X(5568)$ as a meson molecule}
  (2016).
\newblock \href {http://arxiv.org/abs/1603.02708} {\path{arXiv:1603.02708}}.

\bibitem{Albuquerque:2016nlw}
R.~Albuquerque, S.~Narison, A.~Rabemananjara, D.~Rabetiarivony, {Nature of the
  X(5568) : a critical Laplace sum rule analysis at N2LO}, Int. J. Mod. Phys.
  A31~(17) (2016) 1650093.
\newblock \href {http://arxiv.org/abs/1604.05566} {\path{arXiv:1604.05566}},
  \href {http://dx.doi.org/10.1142/S0217751X16500937}
  {\path{doi:10.1142/S0217751X16500937}}.

\bibitem{Wang:2016tsi}
W.~Wang, R.~Zhu, {Can $X(5568)$ be a tetraquark state?}, Chin. Phys. C40~(9)
  (2016) 093101.
\newblock \href {http://arxiv.org/abs/1602.08806} {\path{arXiv:1602.08806}},
  \href {http://dx.doi.org/10.1088/1674-1137/40/9/093101}
  {\path{doi:10.1088/1674-1137/40/9/093101}}.

\bibitem{Liu:2016ogz}
Y.-R. Liu, X.~Liu, S.-L. Zhu, {$X(5568)$ and and its partner states}, Phys.
  Rev. D93~(7) (2016) 074023.
\newblock \href {http://arxiv.org/abs/1603.01131} {\path{arXiv:1603.01131}},
  \href {http://dx.doi.org/10.1103/PhysRevD.93.074023}
  {\path{doi:10.1103/PhysRevD.93.074023}}.

\bibitem{Xiao:2016mho}
C.-J. Xiao, D.-Y. Chen, {Possible $B^{(\ast)} \bar{K}$ hadronic molecule state}
  (2016).
\newblock \href {http://arxiv.org/abs/1603.00228} {\path{arXiv:1603.00228}}.

\bibitem{Stancu:2016sfd}
F.~Stancu, {X(5568) as a ${su}\bar{d}\bar{b}$ tetraquark in a simple quark
  model}, J. Phys. G43~(10) (2016) 105001.
\newblock \href {http://arxiv.org/abs/1603.03322} {\path{arXiv:1603.03322}},
  \href {http://dx.doi.org/10.1088/0954-3899/43/10/105001}
  {\path{doi:10.1088/0954-3899/43/10/105001}}.

\bibitem{Lu:2016zhe}
Q.-F. Lu, Y.-B. Dong, {Masses of open charm and bottom tetraquark states in
  relativized quark model} (2016).
\newblock \href {http://arxiv.org/abs/1603.06417} {\path{arXiv:1603.06417}}.

\bibitem{Chen:2016npt}
X.~Chen, J.~Ping, {Is the exotic $X(5568)$ a bound state?}, Eur. Phys. J.
  C76~(6) (2016) 351.
\newblock \href {http://arxiv.org/abs/1604.05651} {\path{arXiv:1604.05651}},
  \href {http://dx.doi.org/10.1140/epjc/s10052-016-4210-x}
  {\path{doi:10.1140/epjc/s10052-016-4210-x}}.

\bibitem{Ali:2016gdg}
A.~Ali, L.~Maiani, A.~D. Polosa, V.~Riquer, {$B_c^\pm$ decays into
  tetraquarks}, Phys. Rev. D94~(3) (2016) 034036.
\newblock \href {http://arxiv.org/abs/1604.01731} {\path{arXiv:1604.01731}},
  \href {http://dx.doi.org/10.1103/PhysRevD.94.034036}
  {\path{doi:10.1103/PhysRevD.94.034036}}.

\bibitem{He:2016yhd}
X.-G. He, P.~Ko, {Flavor $SU(3)$ properties of beauty tetraquark states with
  three different light quarks}, Phys. Lett. B761 (2016) 92--97.
\newblock \href {http://arxiv.org/abs/1603.02915} {\path{arXiv:1603.02915}},
  \href {http://dx.doi.org/10.1016/j.physletb.2016.08.005}
  {\path{doi:10.1016/j.physletb.2016.08.005}}.

\bibitem{Liu:2016xly}
X.-H. Liu, G.~Li, {Could the observation of $X(5568)$ be resulted by the near
  threshold rescattering effects?} (2016).
\newblock \href {http://arxiv.org/abs/1603.00708} {\path{arXiv:1603.00708}}.

\bibitem{Albaladejo:2016eps}
M.~Albaladejo, J.~Nieves, E.~Oset, Z.-F. Sun, X.~Liu, {Can $X(5568)$ be
  described as a $B_s\pi$, $B\bar{K}$ resonant state?}, Phys. Lett. B757 (2016)
  515--519.
\newblock \href {http://arxiv.org/abs/1603.09230} {\path{arXiv:1603.09230}},
  \href {http://dx.doi.org/10.1016/j.physletb.2016.04.033}
  {\path{doi:10.1016/j.physletb.2016.04.033}}.

\bibitem{Burns:2016gvy}
T.~J. Burns, E.~S. Swanson, {Interpreting the X (5568)}, Phys. Lett. B760
  (2016) 627--633.
\newblock \href {http://arxiv.org/abs/1603.04366} {\path{arXiv:1603.04366}},
  \href {http://dx.doi.org/10.1016/j.physletb.2016.07.049}
  {\path{doi:10.1016/j.physletb.2016.07.049}}.

\bibitem{Guo:2016nhb}
F.-K. Guo, U.-G. Meissner, B.-S. Zou, {How the X(5568) challenges our
  understanding of QCD}, Commun. Theor. Phys. 65~(5) (2016) 593--595.
\newblock \href {http://arxiv.org/abs/1603.06316} {\path{arXiv:1603.06316}},
  \href {http://dx.doi.org/10.1088/0253-6102/65/5/593}
  {\path{doi:10.1088/0253-6102/65/5/593}}.

\bibitem{Cassing:2009vt}
W.~Cassing, E.~L. Bratkovskaya, {Parton-Hadron-String Dynamics: an off-shell
  transport approach for relativistic energies}, Nucl. Phys. A831 (2009)
  215--242.
\newblock \href {http://arxiv.org/abs/0907.5331} {\path{arXiv:0907.5331}},
  \href {http://dx.doi.org/10.1016/j.nuclphysa.2009.09.007}
  {\path{doi:10.1016/j.nuclphysa.2009.09.007}}.

\bibitem{Lin:2004en}
Z.-W. Lin, C.~M. Ko, B.-A. Li, B.~Zhang, S.~Pal, {A Multi-phase transport model
  for relativistic heavy ion collisions}, Phys. Rev. C72 (2005) 064901.
\newblock \href {http://arxiv.org/abs/nucl-th/0411110}
  {\path{arXiv:nucl-th/0411110}}, \href
  {http://dx.doi.org/10.1103/PhysRevC.72.064901}
  {\path{doi:10.1103/PhysRevC.72.064901}}.

\bibitem{Hirano:2005wx}
T.~Hirano, M.~Gyulassy, {Perfect fluidity of the quark gluon plasma core as
  seen through its dissipative hadronic corona}, Nucl. Phys. A769 (2006)
  71--94.
\newblock \href {http://arxiv.org/abs/nucl-th/0506049}
  {\path{arXiv:nucl-th/0506049}}, \href
  {http://dx.doi.org/10.1016/j.nuclphysa.2006.02.005}
  {\path{doi:10.1016/j.nuclphysa.2006.02.005}}.

\bibitem{Song:2007ux}
H.~Song, U.~W. Heinz, {Causal viscous hydrodynamics in 2+1 dimensions for
  relativistic heavy-ion collisions}, Phys. Rev. C77 (2008) 064901.
\newblock \href {http://arxiv.org/abs/0712.3715} {\path{arXiv:0712.3715}},
  \href {http://dx.doi.org/10.1103/PhysRevC.77.064901}
  {\path{doi:10.1103/PhysRevC.77.064901}}.

\bibitem{Andronic:2005yp}
A.~Andronic, P.~Braun-Munzinger, J.~Stachel, {Hadron production in central
  nucleus-nucleus collisions at chemical freeze-out}, Nucl. Phys. A772 (2006)
  167--199.
\newblock \href {http://arxiv.org/abs/nucl-th/0511071}
  {\path{arXiv:nucl-th/0511071}}, \href
  {http://dx.doi.org/10.1016/j.nuclphysa.2006.03.012}
  {\path{doi:10.1016/j.nuclphysa.2006.03.012}}.

\bibitem{Cho:2015exb}
S.~Cho, T.~Song, S.~H. Lee, {Freeze-out conditions for production of
  resonances, hadronic molecules, and light nuclei} (2015).
\newblock \href {http://arxiv.org/abs/1511.08019} {\path{arXiv:1511.08019}}.

\bibitem{Andronic:2012dm}
A.~Andronic, P.~Braun-Munzinger, K.~Redlich, J.~Stachel, {The statistical model
  in Pb-Pb collisions at the LHC}, Nucl. Phys. A904-905 (2013) 535c--538c.
\newblock \href {http://arxiv.org/abs/1210.7724} {\path{arXiv:1210.7724}},
  \href {http://dx.doi.org/10.1016/j.nuclphysa.2013.02.070}
  {\path{doi:10.1016/j.nuclphysa.2013.02.070}}.

\bibitem{Stachel:2013zma}
J.~Stachel, A.~Andronic, P.~Braun-Munzinger, K.~Redlich, {Confronting LHC data
  with the statistical hadronization model}, J. Phys. Conf. Ser. 509 (2014)
  012019.
\newblock \href {http://arxiv.org/abs/1311.4662} {\path{arXiv:1311.4662}},
  \href {http://dx.doi.org/10.1088/1742-6596/509/1/012019}
  {\path{doi:10.1088/1742-6596/509/1/012019}}.

\bibitem{Borsanyi:2010cj}
S.~Borsanyi, G.~Endrodi, Z.~Fodor, A.~Jakovac, S.~D. Katz, S.~Krieg, C.~Ratti,
  K.~K. Szabo, {The QCD equation of state with dynamical quarks}, JHEP 11
  (2010) 077.
\newblock \href {http://arxiv.org/abs/1007.2580} {\path{arXiv:1007.2580}},
  \href {http://dx.doi.org/10.1007/JHEP11(2010)077}
  {\path{doi:10.1007/JHEP11(2010)077}}.

\bibitem{Greco:2003mm}
V.~Greco, C.~M. Ko, P.~Levai, {Parton coalescence at RHIC}, Phys. Rev. C68
  (2003) 034904.
\newblock \href {http://arxiv.org/abs/nucl-th/0305024}
  {\path{arXiv:nucl-th/0305024}}, \href
  {http://dx.doi.org/10.1103/PhysRevC.68.034904}
  {\path{doi:10.1103/PhysRevC.68.034904}}.

\bibitem{Molnar:2003ff}
D.~Molnar, S.~A. Voloshin, {Elliptic flow at large transverse momenta from
  quark coalescence}, Phys. Rev. Lett. 91 (2003) 092301.
\newblock \href {http://arxiv.org/abs/nucl-th/0302014}
  {\path{arXiv:nucl-th/0302014}}, \href
  {http://dx.doi.org/10.1103/PhysRevLett.91.092301}
  {\path{doi:10.1103/PhysRevLett.91.092301}}.

\bibitem{Hwa:2003bn}
R.~C. Hwa, C.~B. Yang, {Scaling distributions of quarks, mesons and proton for
  all p(T), energy and centrality}, Phys. Rev. C67 (2003) 064902.
\newblock \href {http://arxiv.org/abs/nucl-th/0302006}
  {\path{arXiv:nucl-th/0302006}}, \href
  {http://dx.doi.org/10.1103/PhysRevC.67.064902}
  {\path{doi:10.1103/PhysRevC.67.064902}}.

\bibitem{Greco:2003xt}
C.~M.~K. V.~Greco, P.~Levai, {Parton coalescence and antiproton/pion anomaly at
  RHIC}, Phys.Rev.Lett. 90 (2003) 202302.
\newblock \href {http://arxiv.org/abs/0301093} {\path{arXiv:0301093}}, \href
  {http://dx.doi.org/10.1103/PhysRevLett.90.202302}
  {\path{doi:10.1103/PhysRevLett.90.202302}}.

\bibitem{Fries:2003vb}
R.~J. Fries, B.~Muller, C.~Nonaka, S.~A. Bass, {Hadronization in heavy ion
  collisions: Recombination and fragmentation of partons}, Phys.Rev.Lett. 90
  (2003) 202303.
\newblock \href {http://arxiv.org/abs/nucl-th/0301087}
  {\path{arXiv:nucl-th/0301087}}, \href
  {http://dx.doi.org/10.1103/PhysRevLett.90.202303}
  {\path{doi:10.1103/PhysRevLett.90.202303}}.

\bibitem{Fries:2003kq}
R.~J. Fries, B.~Muller, C.~Nonaka, S.~A. Bass, {Hadron production in heavy ion
  collisions: Fragmentation and recombination from a dense parton phase}, Phys.
  Rev. C68 (2003) 044902.
\newblock \href {http://arxiv.org/abs/nucl-th/0306027}
  {\path{arXiv:nucl-th/0306027}}, \href
  {http://dx.doi.org/10.1103/PhysRevC.68.044902}
  {\path{doi:10.1103/PhysRevC.68.044902}}.

\bibitem{Chen:2003tn}
L.~W. Chen, V.~Greco, C.~M. Ko, S.~H. Lee, W.~Liu, {Pentaquark baryon
  production at the Relativistic Heavy Ion Collider}, Phys. Lett. B601 (2004)
  34--40.
\newblock \href {http://arxiv.org/abs/nucl-th/0308006}
  {\path{arXiv:nucl-th/0308006}}, \href
  {http://dx.doi.org/10.1016/j.physletb.2004.09.027}
  {\path{doi:10.1016/j.physletb.2004.09.027}}.

\bibitem{Chen:2007zp}
L.~Chen, C.~Ko, W.~Liu, M.~Nielsen, {D(sJ)(2317) meson production at RHIC},
  Phys.Rev. C76 (2007) 014906.
\newblock \href {http://arxiv.org/abs/0705.1697} {\path{arXiv:0705.1697}},
  \href {http://dx.doi.org/10.1103/PhysRevC.76.014906}
  {\path{doi:10.1103/PhysRevC.76.014906}}.

\bibitem{Oh:2009zj}
Y.~Oh, C.~M. Ko, S.~H. Lee, S.~Yasui, {Heavy baryon/meson ratios in
  relativistic heavy ion collisions}, Phys. Rev. C79 (2009) 044905.
\newblock \href {http://arxiv.org/abs/0901.1382} {\path{arXiv:0901.1382}},
  \href {http://dx.doi.org/10.1103/PhysRevC.79.044905}
  {\path{doi:10.1103/PhysRevC.79.044905}}.

\bibitem{Lee:2007wr}
S.~H. Lee, K.~Ohnishi, S.~Yasui, I.-K. Yoo, C.-M. Ko, {Lambda(c) enhancement
  from strongly coupled quark-gluon plasma}, Phys. Rev. Lett. 100 (2008)
  222301.
\newblock \href {http://arxiv.org/abs/0709.3637} {\path{arXiv:0709.3637}},
  \href {http://dx.doi.org/10.1103/PhysRevLett.100.222301}
  {\path{doi:10.1103/PhysRevLett.100.222301}}.

\bibitem{Song:2015sfa}
T.~Song, H.~Berrehrah, D.~Cabrera, J.~M. Torres-Rincon, L.~Tolos, W.~Cassing,
  E.~Bratkovskaya, {Tomography of the Quark-Gluon-Plasma by Charm Quarks},
  Phys. Rev. C92~(1) (2015) 014910.
\newblock \href {http://arxiv.org/abs/1503.03039} {\path{arXiv:1503.03039}},
  \href {http://dx.doi.org/10.1103/PhysRevC.92.014910}
  {\path{doi:10.1103/PhysRevC.92.014910}}.

\bibitem{Song:2015ykw}
T.~Song, H.~Berrehrah, D.~Cabrera, W.~Cassing, E.~Bratkovskaya, {Charm
  production in Pb + Pb collisions at energies available at the CERN Large
  Hadron Collider}, Phys. Rev. C93~(3) (2016) 034906.
\newblock \href {http://arxiv.org/abs/1512.00891} {\path{arXiv:1512.00891}},
  \href {http://dx.doi.org/10.1103/PhysRevC.93.034906}
  {\path{doi:10.1103/PhysRevC.93.034906}}.

\bibitem{Sjostrand:2006za}
T.~Sjostrand, S.~Mrenna, P.~Z. Skands, {PYTHIA 6.4 Physics and Manual}, JHEP 05
  (2006) 026.
\newblock \href {http://arxiv.org/abs/hep-ph/0603175}
  {\path{arXiv:hep-ph/0603175}}, \href
  {http://dx.doi.org/10.1088/1126-6708/2006/05/026}
  {\path{doi:10.1088/1126-6708/2006/05/026}}.

\bibitem{Cacciari:1998it}
M.~Cacciari, M.~Greco, P.~Nason, {The P(T) spectrum in heavy flavor
  hadroproduction}, JHEP 05 (1998) 007.
\newblock \href {http://arxiv.org/abs/hep-ph/9803400}
  {\path{arXiv:hep-ph/9803400}}, \href
  {http://dx.doi.org/10.1088/1126-6708/1998/05/007}
  {\path{doi:10.1088/1126-6708/1998/05/007}}.

\bibitem{Cacciari:2001td}
M.~Cacciari, S.~Frixione, P.~Nason, {The p(T) spectrum in heavy flavor
  photoproduction}, JHEP 03 (2001) 006.
\newblock \href {http://arxiv.org/abs/hep-ph/0102134}
  {\path{arXiv:hep-ph/0102134}}, \href
  {http://dx.doi.org/10.1088/1126-6708/2001/03/006}
  {\path{doi:10.1088/1126-6708/2001/03/006}}.

\bibitem{Eskola:2009uj}
K.~J. Eskola, H.~Paukkunen, C.~A. Salgado, {EPS09: A New Generation of NLO and
  LO Nuclear Parton Distribution Functions}, JHEP 04 (2009) 065.
\newblock \href {http://arxiv.org/abs/0902.4154} {\path{arXiv:0902.4154}},
  \href {http://dx.doi.org/10.1088/1126-6708/2009/04/065}
  {\path{doi:10.1088/1126-6708/2009/04/065}}.

\bibitem{Cao:2015hia}
S.~Cao, G.-Y. Qin, S.~A. Bass, {Energy loss, hadronization and hadronic
  interactions of heavy flavors in relativistic heavy-ion collisions}, Phys.
  Rev. C92~(2) (2015) 024907.
\newblock \href {http://arxiv.org/abs/1505.01413} {\path{arXiv:1505.01413}},
  \href {http://dx.doi.org/10.1103/PhysRevC.92.024907}
  {\path{doi:10.1103/PhysRevC.92.024907}}.

\bibitem{Bondorf:1978kz}
J.~P. Bondorf, S.~I.~A. Garpman, J.~Zimanyi, {A Simple Analytic Hydrodynamic
  Model for Expanding Fireballs}, Nucl. Phys. A296 (1978) 320--332.
\newblock \href {http://dx.doi.org/10.1016/0375-9474(78)90076-3}
  {\path{doi:10.1016/0375-9474(78)90076-3}}.

\bibitem{Becattini:2014hla}
F.~Becattini, E.~Grossi, M.~Bleicher, J.~Steinheimer, R.~Stock, {Centrality
  dependence of hadronization and chemical freeze-out conditions in heavy ion
  collisions at $\sqrt s_{NN}$ = 2.76 TeV}, Phys. Rev. C90~(5) (2014) 054907.
\newblock \href {http://arxiv.org/abs/1405.0710} {\path{arXiv:1405.0710}},
  \href {http://dx.doi.org/10.1103/PhysRevC.90.054907}
  {\path{doi:10.1103/PhysRevC.90.054907}}.

\bibitem{Cho:2015qca}
S.~Cho, S.~H. Lee, {Reduction of the $K^*$ meson abundance in heavy ion
  collisions} (2015).
\newblock \href {http://arxiv.org/abs/1509.04092} {\path{arXiv:1509.04092}}.

\bibitem{Sato:1981ez}
H.~Sato, K.~Yazaki, {On the coalescence model for high-energy nuclear
  reactions}, Phys. Lett. B98 (1981) 153--157.
\newblock \href {http://dx.doi.org/10.1016/0370-2693(81)90976-X}
  {\path{doi:10.1016/0370-2693(81)90976-X}}.

\bibitem{Lee:1954iq}
T.~D. Lee, {Some Special Examples in Renormalizable Field Theory}, Phys. Rev.
  95 (1954) 1329--1334, [,11(1954)].
\newblock \href {http://dx.doi.org/10.1103/PhysRev.95.1329}
  {\path{doi:10.1103/PhysRev.95.1329}}.

\bibitem{HanburyBrown:1956bqd}
R.~Hanbury~Brown, R.~Q. Twiss, {A Test of a new type of stellar interferometer
  on Sirius}, Nature 178 (1956) 1046--1048.
\newblock \href {http://dx.doi.org/10.1038/1781046a0}
  {\path{doi:10.1038/1781046a0}}.

\bibitem{Goldhaber:1960sf}
G.~Goldhaber, S.~Goldhaber, W.-Y. Lee, A.~Pais, {Influence of Bose-Einstein
  statistics on the anti-proton proton annihilation process}, Phys. Rev. 120
  (1960) 300--312.
\newblock \href {http://dx.doi.org/10.1103/PhysRev.120.300}
  {\path{doi:10.1103/PhysRev.120.300}}.

\bibitem{wiedemann99:_partic}
U.~A. Wiedemann, U.~Heinz, Particle interferometry for relativistic heavy-ion
  collisions, Phys. Rept. 319 (1999) 145.

\bibitem{Gong:1991zza}
W.~G. Gong, W.~Bauer, C.~K. Gelbke, S.~Pratt, {Space-time evolution of nuclear
  reactions probed by two-proton intensity interferometry}, Phys. Rev. C43
  (1991) 781--800.
\newblock \href {http://dx.doi.org/10.1103/PhysRevC.43.781}
  {\path{doi:10.1103/PhysRevC.43.781}}.

\bibitem{Anchishkin:1997tb}
D.~Anchishkin, U.~W. Heinz, P.~Renk, {Final state interactions in two particle
  interferometry}, Phys. Rev. C57 (1998) 1428--1439.
\newblock \href {http://arxiv.org/abs/nucl-th/9710051}
  {\path{arXiv:nucl-th/9710051}}, \href
  {http://dx.doi.org/10.1103/PhysRevC.57.1428}
  {\path{doi:10.1103/PhysRevC.57.1428}}.

\bibitem{Gmitro:1986ay}
M.~Gmitro, J.~Kvasil, R.~Lednick\'{y}, V.~L. Lyuboshits, {On the Sensitivity of
  Nucleon-nucleon Correlations to the Form of Short Range Potential}, Czech. J.
  Phys. B36 (1986) 1281.
\newblock \href {http://dx.doi.org/10.1007/BF01598029}
  {\path{doi:10.1007/BF01598029}}.

\bibitem{roy1967nuclear}
R.~Roy, B.~Nigam,
  \href{https://books.google.co.jp/books?id=kAFRAAAAMAAJ}{Nuclear physics:
  theory and experiment}, Wiley, 1967.
\newline\urlprefix\url{https://books.google.co.jp/books?id=kAFRAAAAMAAJ}

\bibitem{lin02:_parton_relat_heavy_ion_collid}
Z.~W. Lin, C.~M. Ko, S.~Pal, Partonic effects on pion interferometry at the
  relativistic heavy-ion collider, Phys. Rev. Lett. 89 (2002) 152301.

\bibitem{Makhlin:1987gm}
A.~N. Makhlin, {\relax Yu}.~M. Sinyukov, {Hydrodynamics of Hadron Matter Under
  Pion Interferometric Microscope}, Z. Phys. C39 (1988) 69.
\newblock \href {http://dx.doi.org/10.1007/BF01560393}
  {\path{doi:10.1007/BF01560393}}.

\bibitem{Adler:2006as}
S.~S. Adler, et~al., {Evidence for a long-range component in the pion emission
  source in Au + Au collisions at s(NN)**(1/2) = 200-GeV}, Phys. Rev. Lett. 98
  (2007) 132301.
\newblock \href {http://arxiv.org/abs/nucl-ex/0605032}
  {\path{arXiv:nucl-ex/0605032}}, \href
  {http://dx.doi.org/10.1103/PhysRevLett.98.132301}
  {\path{doi:10.1103/PhysRevLett.98.132301}}.

\bibitem{shapoval15:_proton_lambd_cern_large_hadron_collid}
V.~M. Shapoval, Y.~M. Sinyukov, V.~Y. Naboka, Proton-$\lambda$ correlation
  functions at energies available at the {CERN} large hadron collider taking
  into account residual correlations, Phys. Rev. C 92 (2015) 044910.

\bibitem{Grassberger:1976au}
P.~Grassberger, {Interference Effects from Inclusive Resonance Production},
  Nucl. Phys. B120 (1977) 231.
\newblock \href {http://dx.doi.org/10.1016/0550-3213(77)90042-6}
  {\path{doi:10.1016/0550-3213(77)90042-6}}.

\bibitem{Csorgo:1994in}
T.~Csorgo, B.~Lorstad, J.~Zimanyi, {Bose-Einstein correlations for systems with
  large halo}, Z. Phys. C71 (1996) 491--497.
\newblock \href {http://arxiv.org/abs/hep-ph/9411307}
  {\path{arXiv:hep-ph/9411307}}, \href {http://dx.doi.org/10.1007/BF02907008,
  10.1007/s002880050195} {\path{doi:10.1007/BF02907008,
  10.1007/s002880050195}}.

\bibitem{Wiedemann:1996ig}
U.~A. Wiedemann, U.~W. Heinz, {Resonance contributions to HBT correlation
  radii}, Phys. Rev. C56 (1997) 3265--3286.
\newblock \href {http://arxiv.org/abs/nucl-th/9611031}
  {\path{arXiv:nucl-th/9611031}}, \href
  {http://dx.doi.org/10.1103/PhysRevC.56.3265}
  {\path{doi:10.1103/PhysRevC.56.3265}}.

\bibitem{kiesel14:_extrac}
A.~Kiesel, H.~Zbroszczyk, M.~Szyma\'{n}ski, Extracting baryon-antibaryon
  strong-interaction potentials from $p\bar{Lambda}$ femtoscopic correlation
  functions, Phys. Rev. C 89 (2014) 054916.

\bibitem{Chojnacki:2011hb}
M.~Chojnacki, A.~Kisiel, W.~Florkowski, W.~Broniowski, {THERMINATOR 2: THERMal
  heavy IoN generATOR 2}, Comput. Phys. Commun. 183 (2012) 746--773.
\newblock \href {http://arxiv.org/abs/1102.0273} {\path{arXiv:1102.0273}},
  \href {http://dx.doi.org/10.1016/j.cpc.2011.11.018}
  {\path{doi:10.1016/j.cpc.2011.11.018}}.

\bibitem{adam15:_one_pb_pb_nn}
J.~Adam, et~al., One-dimensional pion, kaon, and proton femtoscopy in pb-pb
  collisions at $\sqrt{s_{NN}}$=2.76 TeV, Phys. Rev. C 92 (2015) 054908.

\bibitem{wang99:_lambd}
F.~Wang, S.~Pratt, Lambda-proton correlations in relativistic heavy ion
  collisions, Phys. Rev. Lett. 83 (1999) 3138.

\bibitem{Filikhin:2002wm}
I.~N. Filikhin, A.~Gal, {Faddeev-Yakubovsky calculations for light lambda
  lambda hypernuclei}, Nucl. Phys. A 707 (2002) 491--509.
\newblock \href {http://arxiv.org/abs/nucl-th/0203036}
  {\path{arXiv:nucl-th/0203036}}, \href
  {http://dx.doi.org/10.1016/S0375-9474(02)01008-4}
  {\path{doi:10.1016/S0375-9474(02)01008-4}}.

\bibitem{Hiyama:2002yj}
E.~Hiyama, M.~Kamimura, T.~Motoba, T.~Yamada, Y.~Yamamoto, {Four-body cluster
  structure of A = 7 -10 double Lambda hypernuclei}, Phys. Rev. C 66 (2002)
  024007.
\newblock \href {http://arxiv.org/abs/nucl-th/0204059}
  {\path{arXiv:nucl-th/0204059}}, \href
  {http://dx.doi.org/10.1103/PhysRevC.66.024007}
  {\path{doi:10.1103/PhysRevC.66.024007}}.

\bibitem{Amsler:2008zzb}
C.~Amsler, et~al., {Review of Particle Physics}, Phys. Lett. B667 (2008)
  1--1340.
\newblock \href {http://dx.doi.org/10.1016/j.physletb.2008.07.018}
  {\path{doi:10.1016/j.physletb.2008.07.018}}.

\bibitem{Hiyama:2010zzd}
E.~Hiyama, M.~Kamimura, Y.~Yamamoto, T.~Motoba, {Five-body cluster structure of
  double-$\Lambda$ hypernucleus $^{11}_{\Lambda \Lambda}$Be}, Phys. Rev. Lett.
  104 (2010) 212502.
\newblock \href {http://arxiv.org/abs/1006.2626} {\path{arXiv:1006.2626}},
  \href {http://dx.doi.org/10.1103/PhysRevLett.104.212502}
  {\path{doi:10.1103/PhysRevLett.104.212502}}.

\bibitem{Fujiwara:2001xt}
Y.~Fujiwara, M.~Kohno, C.~Nakamoto, Y.~Suzuki, {Interactions between octet
  baryons in the SU(6) quark model}, Phys. Rev. C64 (2001) 054001.
\newblock \href {http://arxiv.org/abs/nucl-th/0106052}
  {\path{arXiv:nucl-th/0106052}}, \href
  {http://dx.doi.org/10.1103/PhysRevC.64.054001}
  {\path{doi:10.1103/PhysRevC.64.054001}}.

\bibitem{Fujiwara:2006yh}
Y.~Fujiwara, Y.~Suzuki, C.~Nakamoto, {Baryon-baryon interactions in the SU(6)
  quark model and their applications to light nuclear systems}, Prog. Part.
  Nucl. Phys. 58 (2007) 439--520.
\newblock \href {http://arxiv.org/abs/nucl-th/0607013}
  {\path{arXiv:nucl-th/0607013}}, \href
  {http://dx.doi.org/10.1016/j.ppnp.2006.08.001}
  {\path{doi:10.1016/j.ppnp.2006.08.001}}.

\bibitem{Sullivan:1987us}
M.~W. Sullivan, et~al., {Measurement of the Ratio of $\Sigma^0$ to $\Lambda^0$
  Inclusive Production From 28.5-{GeV}/$c$ Protons on Beryllium}, Phys. Rev.
  D36 (1987) 674.
\newblock \href {http://dx.doi.org/10.1103/PhysRevD.36.674}
  {\path{doi:10.1103/PhysRevD.36.674}}.

\bibitem{Agakishiev:2011ar}
G.~Agakishiev, et~al., {Strangeness Enhancement in Cu+Cu and Au+Au Collisions
  at $\sqrt{s_{NN}} = 200$ GeV}, Phys. Rev. Lett. 108 (2012) 072301.
\newblock \href {http://arxiv.org/abs/1107.2955} {\path{arXiv:1107.2955}},
  \href {http://dx.doi.org/10.1103/PhysRevLett.108.072301}
  {\path{doi:10.1103/PhysRevLett.108.072301}}.

\bibitem{Nagels:1976xq}
M.~M. Nagels, T.~A. Rijken, J.~J. de~Swart, {Baryon Baryon Scattering in a One
  Boson Exchange Potential Approach. 2. Hyperon-Nucleon Scattering}, Phys. Rev.
  D15 (1977) 2547.
\newblock \href {http://dx.doi.org/10.1103/PhysRevD.15.2547}
  {\path{doi:10.1103/PhysRevD.15.2547}}.

\bibitem{Nagels:1978sc}
M.~M. Nagels, T.~A. Rijken, J.~J. de~Swart, {Baryon Baryon Scattering in a One
  Boson Exchange Potential Approach. 3. A Nucleon-Nucleon and Hyperon - Nucleon
  Analysis Including Contributions of a Nonet of Scalar Mesons}, Phys. Rev. D20
  (1979) 1633.
\newblock \href {http://dx.doi.org/10.1103/PhysRevD.20.1633}
  {\path{doi:10.1103/PhysRevD.20.1633}}.

\bibitem{Maessen:1989sx}
P.~M.~M. Maessen, T.~A. Rijken, J.~J. de~Swart, {Soft Core Baryon Baryon One
  Boson Exchange Models. 2. Hyperon - Nucleon Potential}, Phys. Rev. C40 (1989)
  2226--2245.
\newblock \href {http://dx.doi.org/10.1103/PhysRevC.40.2226}
  {\path{doi:10.1103/PhysRevC.40.2226}}.

\bibitem{Rijken:1998yy}
T.~A. Rijken, V.~G.~J. Stoks, Y.~Yamamoto, {Soft core hyperon - nucleon
  potentials}, Phys. Rev. C59 (1999) 21--40.
\newblock \href {http://arxiv.org/abs/nucl-th/9807082}
  {\path{arXiv:nucl-th/9807082}}, \href
  {http://dx.doi.org/10.1103/PhysRevC.59.21}
  {\path{doi:10.1103/PhysRevC.59.21}}.

\bibitem{Rijken:2006ep}
T.~A. Rijken, Y.~Yamamoto, {Extended-soft-core baryon-baryon model. II.
  Hyperon-nucleon interaction}, Phys. Rev. C73 (2006) 044008.
\newblock \href {http://arxiv.org/abs/nucl-th/0603042}
  {\path{arXiv:nucl-th/0603042}}, \href
  {http://dx.doi.org/10.1103/PhysRevC.73.044008}
  {\path{doi:10.1103/PhysRevC.73.044008}}.

\bibitem{Ueda:1998bz}
T.~Ueda, K.~Tominaga, M.~Yamaguchi, N.~Kijima, D.~Okamoto, K.~Miyagawa,
  T.~Yamada, {Lambda N and Lambda Lambda interactions in an OBE model and
  hypernuclei}, Prog. Theor. Phys. 99 (1998) 891--896.
\newblock \href {http://dx.doi.org/10.1143/PTP.99.891}
  {\path{doi:10.1143/PTP.99.891}}.

\bibitem{Rijken:2010zzb}
T.~A. Rijken, M.~M. Nagels, Y.~Yamamoto, {Baryon-baryon interactions: Nijmegen
  extended-soft-core models}, Prog. Theor. Phys. Suppl. 185 (2010) 14--71.
\newblock \href {http://dx.doi.org/10.1143/PTPS.185.14}
  {\path{doi:10.1143/PTPS.185.14}}.

\bibitem{Csanad:2011bx}
M.~Csanad, {Femtoscopic results in Au+Au and p+p from PHENIX at RHIC}, Phys.
  Part. Nucl. Lett. 8 (2011) 934--937.
\newblock \href {http://arxiv.org/abs/1101.2086} {\path{arXiv:1101.2086}},
  \href {http://dx.doi.org/10.1134/S1547477111090123}
  {\path{doi:10.1134/S1547477111090123}}.

\bibitem{Adamczyk:2013wqm}
L.~Adamczyk, et~al., {Freeze-out dynamics via charged kaon femtoscopy in
  $\sqrt{{s}_{NN}}$ = 200 GeV central Au + Au collisions}, Phys. Rev. C88~(3)
  (2013) 034906.
\newblock \href {http://arxiv.org/abs/1302.3168} {\path{arXiv:1302.3168}},
  \href {http://dx.doi.org/10.1103/PhysRevC.88.034906}
  {\path{doi:10.1103/PhysRevC.88.034906}}.

\end{thebibliography}

\end{document}